\documentclass{aastex63}
\received{\today}
\revised{November XX, 2020}
\accepted{November YY, 2020}
\submitjournal{ApJ}

\shorttitle{Neutrino transfer in a deformed remnant from neutron star merger}
\shortauthors{Sumiyoshi et al.}
\graphicspath{{./}{figures/}}

\begin{document}

\title{Properties of neutrino transfer in a deformed remnant of neutron star merger}

\correspondingauthor{Kohsuke Sumiyoshi}
\email{sumi@numazu-ct.ac.jp}

\author[0000-0002-9224-9449]{Kohsuke Sumiyoshi}
\affil{National Institute of Technology, Numazu College, 
Ooka 3600, Numazu, Shizuoka, 410-8501, Japan}

\author[0000-0001-6467-4969]{Sho Fujibayashi}
\affiliation{Max Planck Institute for Gravitational Physics 
(Albert Einstein Institute), \\
Am M\"uhlenberg 1, Potsdam-Golm, D-14476, Germany}
\author[0000-0002-2648-3835]{Yuichiro Sekiguchi}
\affiliation{Toho University, 
Funabashi, Chiba, 274-8510, Japan}
\affiliation{Center for Gravitational Physics, Yukawa Institute for Theoretical Physics, \\
Kyoto University, Kyoto, 606-8502, Japan}


\author[0000-0002-4979-5671]{Masaru Shibata}
\affiliation{Max Planck Institute for Gravitational Physics 
(Albert Einstein Institute), \\
Am M\"uhlenberg 1, Potsdam-Golm, D-14476, Germany}
\affiliation{Center for Gravitational Physics, Yukawa Institute for Theoretical Physics, \\
Kyoto University, Kyoto, 606-8502, Japan}



\begin{abstract}

We study properties of neutrino transfer 
in a remnant of neutron star merger, consisting of a massive neutron star and a surrounding torus. 
We perform numerical simulations of the neutrino transfer by solving the Boltzmann equation with momentum-space angles and energies of neutrinos for snapshots of the merger remnant having elongated shapes.  The evaluation of the neutrino distributions in the multi-dimensions enable us to provide the detailed information of angle and energy spectra and neutrino reaction rates.  We demonstrate features of asymmetric neutrino fluxes from the deformed remnant and investigate the neutrino emission region by determining the neutrinosphere for each energy.  We examine the emission and absorption of neutrinos to identify important ingredients of heating rates through neutrino irradiation.  We show that the contributions of $\mu$- and $\tau$-types neutrinos are important for the heating in the region above the massive neutron star.  We also examine the angle moments and the Eddington tensor calculated directly by the neutrino distribution functions and compare them with those obtained by a moment closure approach, which is often used in the study of neutrino-radiation hydrodynamics.  We show that the components of the Eddington tensor have non-monotonic behaviors and the approximation of the closure relation may become inaccurate for high energy neutrinos, whose fluxes are highly aspherical due to the extended merger remnant.  

\end{abstract}



\section{Introduction} \label{sec:intro}
Neutron star mergers attract a surge of interests as a target of multi-messenger observations to reveal the fate of compact objects and its diverse roles in the Universe \citep{Abbott2017b,Villar2017a} (See also references in \citet{Radice2018a,Shibata2019}) since the first detection of gravitational waves \citep{Abbott2017a}.  The dynamics of the merger and the evolution of the merger remnant is essential to pin down the elusive problems; the origin of heavy elements through $r$-process nucleosynthesis \citep{Lattimer1974a,Symbalisty1982a,Eichler1989a,Meyer1989a,Freiburghaus1999a,Wanajo2014a}, kilonova/macronova \citep{Li1998a,Kulkarni2005a,Metzger2010a}, and the possible connection to the short-hard gamma ray bursts \citep{Eichler1989a}, to name a few.  Mass ejection from the merger remnant is one of the keys to determine the composition of elements, the jet formation in the heated region, and the associated eletromagnetic radiation.  

Neutrinos can play significant roles for the dynamics through the energy transport and the conversion of composition in the ejected material.  In core-collapse supernovae, it is well known that they are essential to drive the core collapse, bounce and explosion, which provide the burst of neutrino emission and the influence on the explosive nucleosynthesis (for a review, see \citet{jan12a}).  In neutron star mergers, neutrinos are influential in providing the heating around the merger remnant and modifying the compositional balance between neutrons and protons, which are crucial for the $r$-process nucleosynthesis.  The prediction of neutrino emissions from the neutron star mergers is also interesting as an observational signature \citep{ruf97,ros03,sek11,Albert2017a,abe18,kyu18}.  

Obtaining knowledge of transport and reactions of neutrinos 
is, therefore, mandatory to investigate quantitatively the dynamics and properties of neutron star mergers and core-collapse supernovae albeit its role has different impacts.  The complication arises in highly deformed structures of the merger remnant composed of a massive neutron star and a torus, which can be opaque to neutrinos.  Resulting neutrino emission can be largely asymmetric and propagation of neutrinos may become non-trivial.  It is problematic to describe the neutrino emission and absorption in and around such a highly deformed remnant.  Neutrino emission can be aspherical and provides vastly different angle and energy spectra.  Pair annihilation is known to be sensitive to anisotropic neutrino fluxes \citep{Birkl2007a,Liu2010a,Zalamea2011a} and different treatments may provide different magnitude of heating.  Neutrino irradiation with different energy spectra can drive a shift of electron fraction, which is the key quantity in the $r$-process nucleosynthesis.  

While the importance of neutrinos is obvious, numerical simulations of neutron star merger have been performed with approximate or even simple methods of neutrino transport.  This is because numerical-relativity simulations in full space dimensions is highly demanding with microphysics even with a simplified neutrino treatment.  
Moreover, long-term simulations are necessary to study gravitational waves from the merger, the mass ejection from the merger remnant, and so on.  
Therefore,  
it is meaningful to assess the influence of neutrino reactions and transport in order to make further quantitative studies.  

There have been continuous efforts in the numerical studies of neutron star merger to implement neutrino transfer with various levels of approximations.  Studies at the early frontiers \citep{ruf96,ros03} adopt a leakage scheme for neutrino emissions, which is originally introduced in supernova studies \citep{vrip81}, to consider the time scale of emission depending on the environment.  Many of modern studies utilize advanced methods of the leakage scheme to capture detail variations 
\citep{sek10,ocon10,gal13,fou16,rad16,perego16,fuj17,fuj18,pul19}.  Sophisticated treatments with the moment method with formulae of the closure relation \citep{shi11,sek12,jus15} and the Monte Carlo method \citep{sher15,fou18,mil19,fou20} are used to explore detailed dynamics in recent numerical simulations.  Evaluation of the advantage and disadvantage of these methods has been conducted recently by comparisons of approximations in recent studies \citep{end20}.  

The Monte Carlo method is one of the promising methods to provide the detailed information of the neutrino transfer \citep{jan89}.  It describes the solution of the Boltzmann equation by a sampling approach and provides the neutrino distribution in full dimension.  General relativistic calculations are recently realized \citep{fou18,mil19} (See also \citet{aka20}) and applied to numerical simulations of neutron star merger through coupling with hydrodynamics \citep{fou20} (See also \citet{abd12}).  
The distribution in space, angle, and energy obtained by tracking the particle transport with reactions can be used to provide the elaborate information of observational signals of supernova neutrinos \citep{kei03,kat20} and to validate the closure relations for high angle moments \citep{fou18,fou18b}.  It is advantageous to obtain accurate angle distributions at large distances in contrast to the limited ability of the discrete ordinate methods \citep{yam99}.  It is also suitable to describe the crossing of two beams, which is important for merger remnants  (See, for example, \citet{fou18}).  On the other hand, it requires a large number of sampling to reduce the random noise, careful prescriptions to describe frequent reactions in the diffusion limit, and has only restricted capability in the rapidly varying background matter.  In these aspects, the Monte Carlo method is a complementary approach to the discrete ordinate method for the solution of the Boltzmann equation.  

The direct evaluation of the neutrino transfer by the Boltzmann equation requires high demand of computation in multi-dimensional space and has been applied to the neutron star merger in the limited study \citep{des09} based on \citet{ott08}.  While the applications of general relativistic neutrino radiation hydrodynamics with the Boltzmann equation under the spherical symmetry \citep{yam97,yam99,lie04} have been made in the studies of core-collapse supernovae \citep{lie01,sum05,sum07,fis10,fis11}, the neutrino radiation hydrodynamics by the discrete ordinate (S$_n$) method for the Boltzmann equation for axially symmetric case has been performed in a study by \citet{ott08}.  Numerical simulations of core-collapse supernovae in multi-dimensions have been extensively performed using sophisticated approximations of neutrino transfer \citep{jan12a,bur13,jan16,mul16}.  In order to follow a long term evolution over several hundred milliseconds to find out the outcome of explosion with limited computational resources, there have been steady progress of numerical methods of neutrino transfer from the leakage scheme \citep{vrip81} to the diffusion approximations \citep{bur06a,lie09}, variable Eddington factor \citep{ram02} and two moment schemes \citep{kur12,jus15}. In addition, the ray-by-ray methods \citep{bura06} to efficiently handle  directional variations have been often adopted to describe the multi-dimensional neutrino-transfer.  

Recently, the solution of neutrino transfer by the Boltzmann equation in 6 dimensions (6D) has become possible \citep{sum12,sum15}.  The solver of the Boltzmann equation in 6D is applied to neutrino-radiation hydrodynamics in core-collapse supernovae under axial symmetry \citep{nag14,nag16,nag17,nag18,har19} and in three spatial dimensions (3D) \citep{iwa20}.  In the analysis by the 6D Boltzmann equation \citep{sum15}, the feature of neutrino transfer in the profiles of core-collapse supernovae in three dimensions has been revealed having the detailed information of neutrino distributions in neutrino angles and energy in 3D space.  

In this study, we apply the numerical code to solve the 6D Boltzmann equation to investigate the neutrino transfer in the matter profile taken from the numerical simulations of the neutron star merger.  We performed numerical simulations for the fixed background to obtain the stationary state of neutrino distribution functions by solving the Boltzmann equation.  We obtain all information of neutrino distribution functions in 5D (2D for space and 3D for neutrinos) to reveal the properties of neutrino transport and reactions.  The simulation covers the whole regions from diffusion, intermediate and to transparent regimes so that it is possible to analyze the behavior in a seamless manner.  

We examine the basic feature of neutrino transfer; neutrino number density, flux distributions, angle moments, neutrino emission region such as neutrinospheres and angular variations of luminosities.  We explore the neutrino emission inside a oblate-shaped remnant neutron star with a geometrically thick torus and the neutrino heating rates around them.  
We reveal that the neutrino fluxes are highly aspherical and focused in the region above the neutron star due to the deformed neutrinosphere elongated along the equator.  
We analyse the contributions of each neutrino reaction to find the important reaction for cooling and heating.  
We show that pair annihilation of $\mu$ and $\tau$ neutrinos plays an important role in the neutrino heating in the region above the merger remnant.  

We examine the properties of the Eddington tensor through comparisons  between evaluation by the neutrino distribution functions and a closure relation which is used in the moment formalism.  
We find that the closure relation provides a good approximation in most of the region, while it can be erroneous for high energy neutrinos.  
Deviations of the Eddington tensor by the closure relation are seen in the regions where neutrino emission from both the neutron star and the torus is important.  

The information of neutrino transfer obtained by the Boltzmann equation will be helpful to examine the approximations used in other studies of the neutron star merger.  The present analysis will be also used to validate the approximate methods and to develop a new closure relation in future.  We plan to compare the neutrino quantities such as neutrino luminosity in the Boltzmann evaluation and dynamical simulations more intensively in a separate paper.  

We organize this paper as follows.  We explain the profiles of the merger remnant used for the simulation of neutrino transfer in \S \ref{sec:profiles}.  We briefly describe the numerical treatment of neutrino transfer by solving the Boltzmann equation in \S \ref{sec:boltzmann}.  We report the basic features of neutrino transfer in the deformed remnant in \S \ref{sec:nu-transfer}.  In \S \ref{sec:moment_evaluation}, we analyze the angle moments (\S \ref{sec:nu-transfer_moments}) and the Eddington tensor together with comparisons with those by the closure relation to examine the validity of the approximation (\S \ref{sec:nu-transfer_tensor}).  We report the general trend of neutrino transfer in the profiles for selected time slices taken from the numerical-relativity simulation in \S \ref{sec:evolution}.  
We summarize the paper in \S \ref{sec:summary} with some discussions.  
We provide the details of the Eddington tensor in Appendix \ref{sec:append_eddington} and the angular resolution in Appendix \ref{sec:append_resolution}.  

\section{Profiles of the merger remnant}
\label{sec:profiles}

We utilize the matter profiles from the numerical relativity simulations of binary neutron star mergers which are composed of a massive neutron star with a surrounding torus.  
We take four snapshots from the time evolution in a 2D numerical simulation of radiation hydrodynamics under axial symmetry \citep{fuj17}.  The initial profile (0 ms) in the 2D simulation \citep{fuj17} is set up  based on a result of 3D simulations \citep{sek15} at $\sim50$ ms after the onset of the merger for the equal mass model with the total mass of 2.7$M_{\odot}$.  The axially symmetric profile is constructed through an average over azimuthal angle from the 3D profile which is quasi-stationary and nearly axisymmetric
\citep{fuj17}.  We adopt the initial profile (0 ms) and the following profiles at 30, 65 and 135 ms from the 2D simulation. The dynamical simulations were performed using the DD2 equation of state (EOS) by \citet{ban14}.  We adopt the same DD2 EOS for simulations of the neutrino transfer.  


In the 2D simulation, Einstein's equation is solved with a version of the puncture-Baumgarte-Shapiro-Shibata-Nakamura formalism \citep{shibata1995a,baumgarte1999,marronetti2008}.
The radiation transfer equation for neutrinos is solved by employing a leakage-based scheme.
In this scheme, neutrinos are decomposed into the ``trapped" and ``streaming" neutrinos.
The trapped neutrinos are assumed to couple the fluid tightly and advected as a part of the fluid.
To solve the streaming neutrinos, we employ Thorne’s truncated moment formalism \citep{Thorne1981a,Shibata2011b} in an energy-integrated manner with a closure relation \citep{lev84,gonzalez2007a}.
The detailed description of these schemes is found in \cite{sek10} and \cite{fuj17}.
For the weak interaction rates, we adopt the rates in \cite{Fuller1985a} for the electron and positron capture processes, those in \cite{Cooperstein1986a} for pair-production processes, those in \cite{ruf96} for plasmon decay, and those in \cite{bur06} for nucleon-nucleon bremsstrahlung.


We use the central part of the profile up to 250 km for the simulation of neutrino transfer by the Boltzmann equation.  We remap the rest-mass density, temperature and electron fraction at the cell center of 128 and 48 grids for the radial and polar coordinates, respectively.  (See details on numerical mesh in \S \ref{sec:boltzmann}.)  Note that we cover 90 degrees of polar angles and the reflection symmetry is imposed with respect to $Z=0$ as in the original simulations\footnote{We show contour plots in the cylindrical coordinates ($R$, $Z$) hereafter.  Note that the simulation of the 6D Boltzmann equation is done in the spherical polar coordinates.  }.  

\begin{figure}[ht!]
\plotone{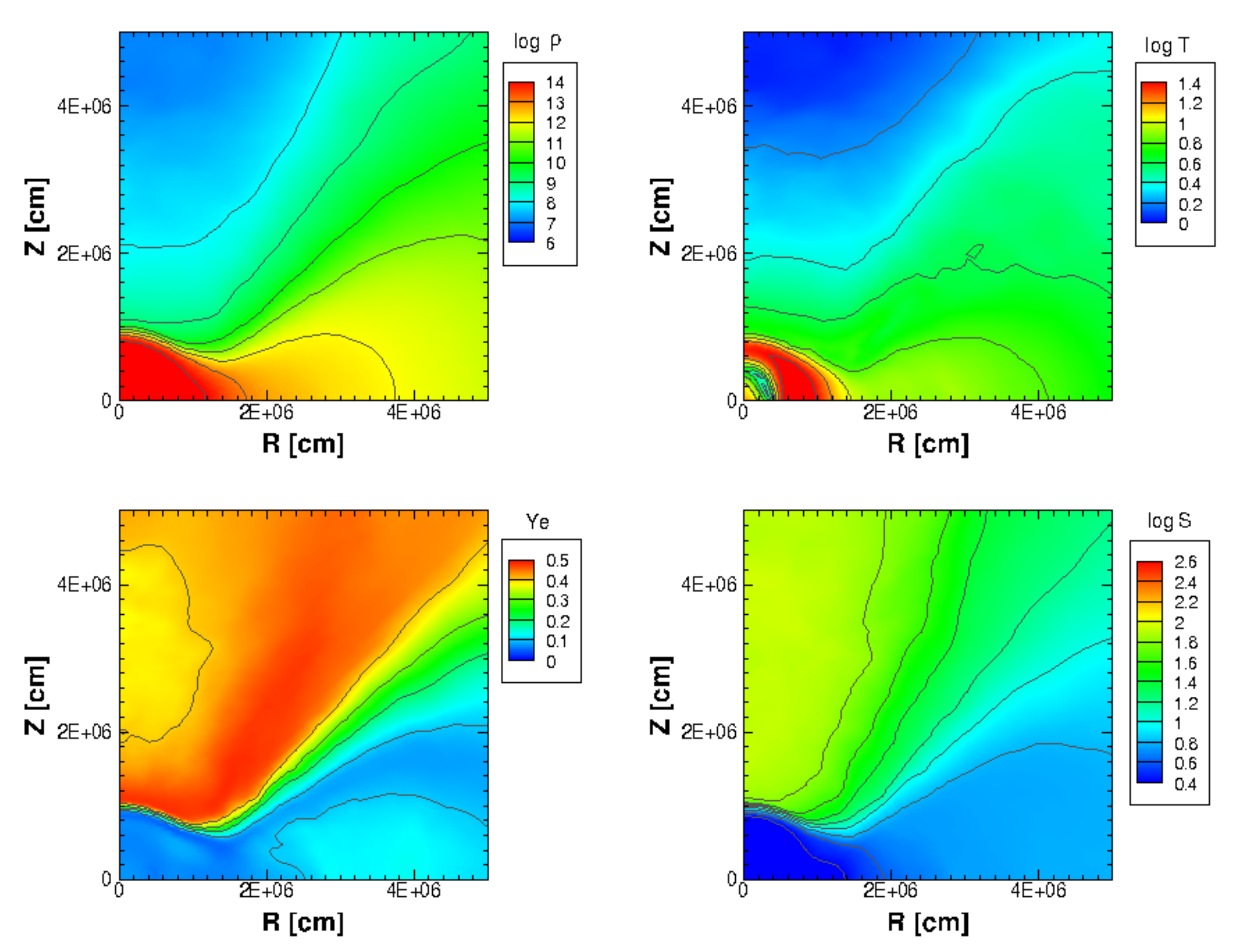}
\caption{Profiles of hydrodynamics quantities in the remnant of binary neutron star merger at 0 ms.  The rest-mass density (upper left) in units of g\,cm$^{-3}$, temperature (upper right) in MeV, electron fraction (lower left) and entropy per baryon (lower right) in units of $k_B$ are shown in the plane of $R$ and $Z$ axes.  Note that the rest-mass density, temperature, and entropy per baryon are plotted in log scale.  
\label{fig:hydro_set_t000}}
\end{figure}

We show the profile of merger remnant at 0 ms ($\sim 50$\, ms after the merger) in Fig. \ref{fig:hydro_set_t000}.  The remnant neutron star at the center has an oblate shape with an extended torus.  The high matter density region ($\gtrsim 10^{11}$ g\,cm$^{-3}$) extends over 60 km along the equator.  This extended region leads to trapping of neutrinos with energy of $\sim10$~MeV and contributes asymmetric neutrino emissions as we will see below.  The high temperature region appears at off-center locations originated in the dynamics of neutron star merger.  The electron fraction, which is equal to the total proton fraction under the charge neutrality, is low below 0.1 in the remnant neutron star and the torus.  The electron fraction is high in the region above the merger remnant.  The entropy per baryon is low in the merger remnant and high in the region above it.  These thermodynamical and chemical conditions are inherited from the neutron star merger and the subsequent evolution.  

\begin{figure}[ht!]
\epsscale{0.37}
\plotone{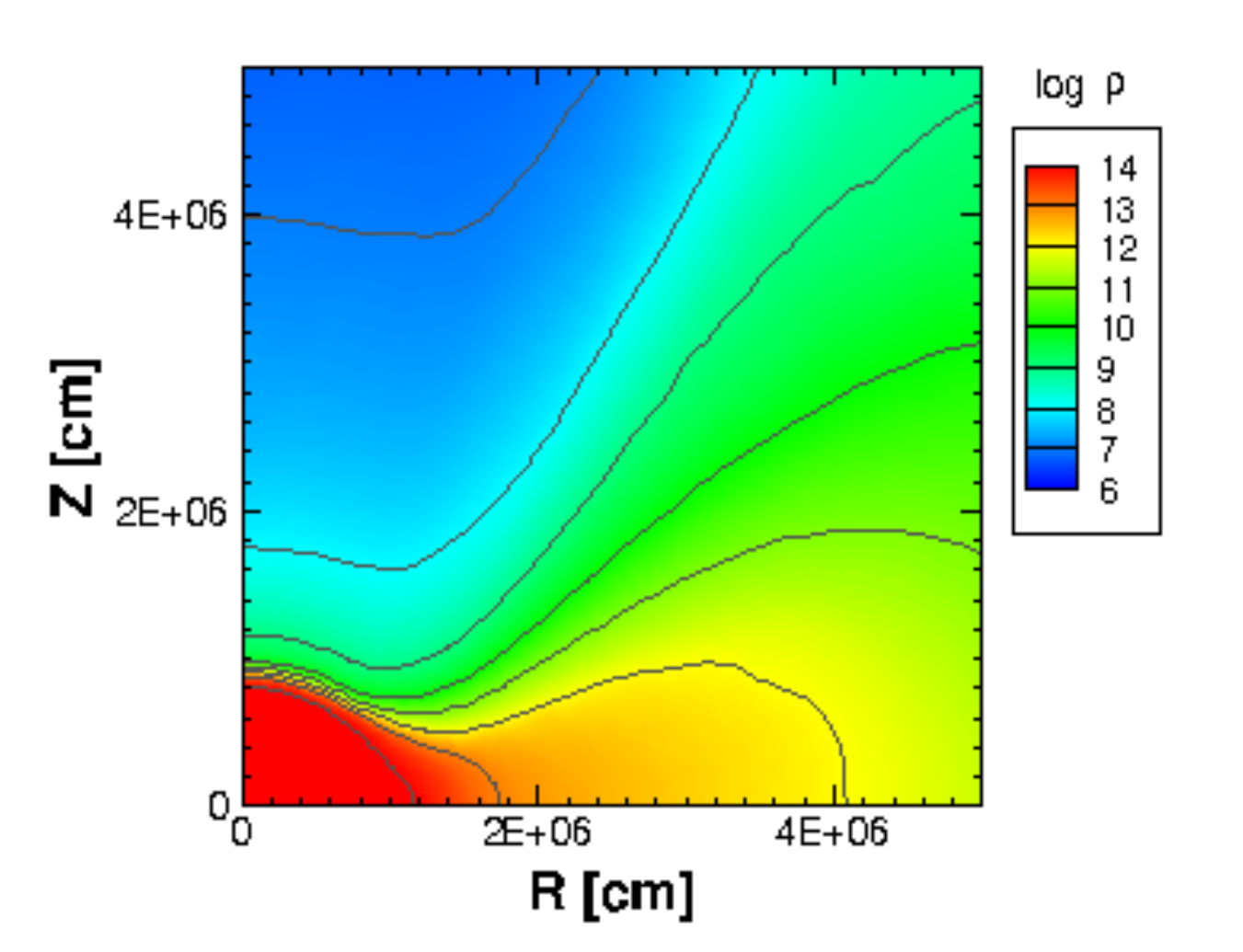}
\plotone{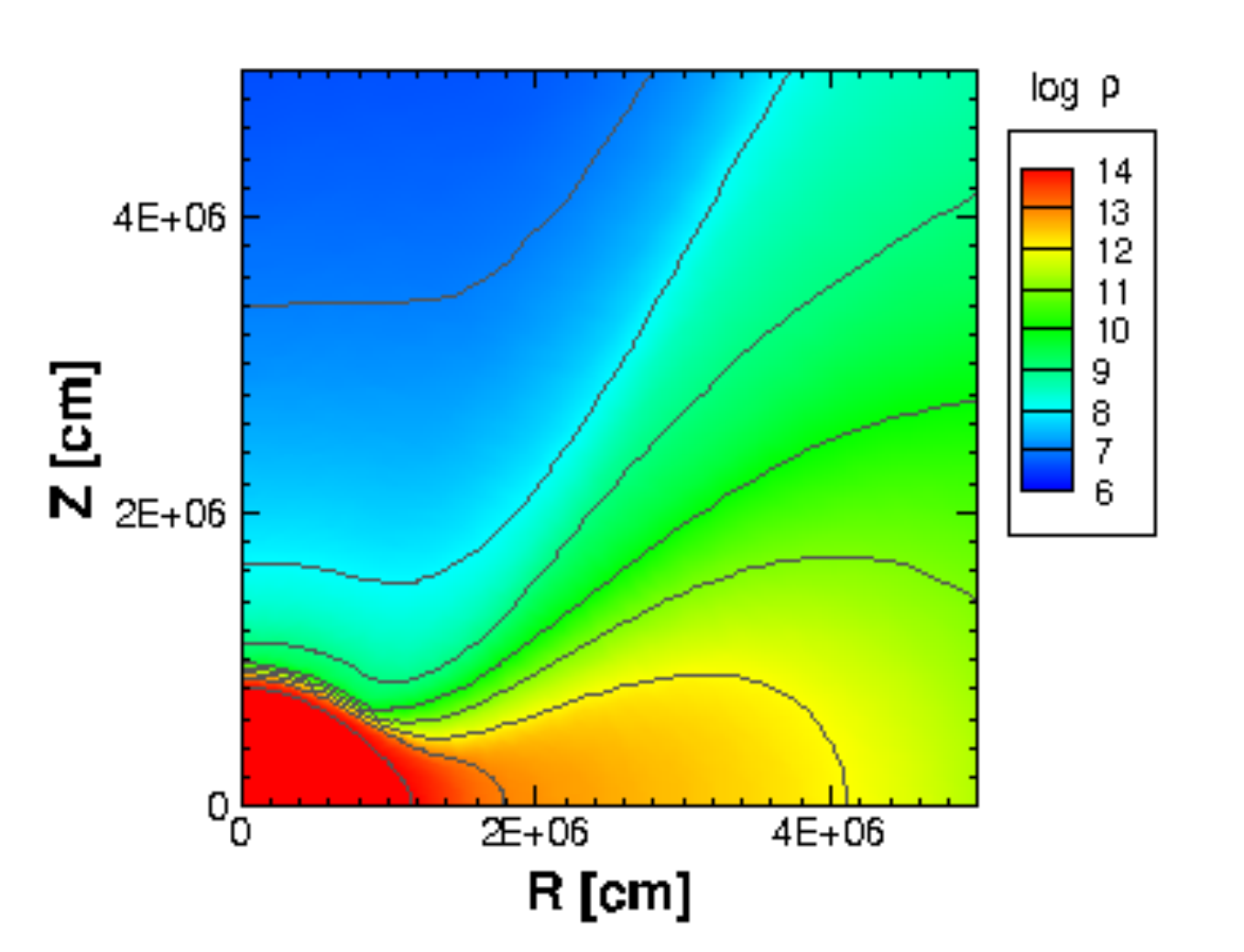}
\plotone{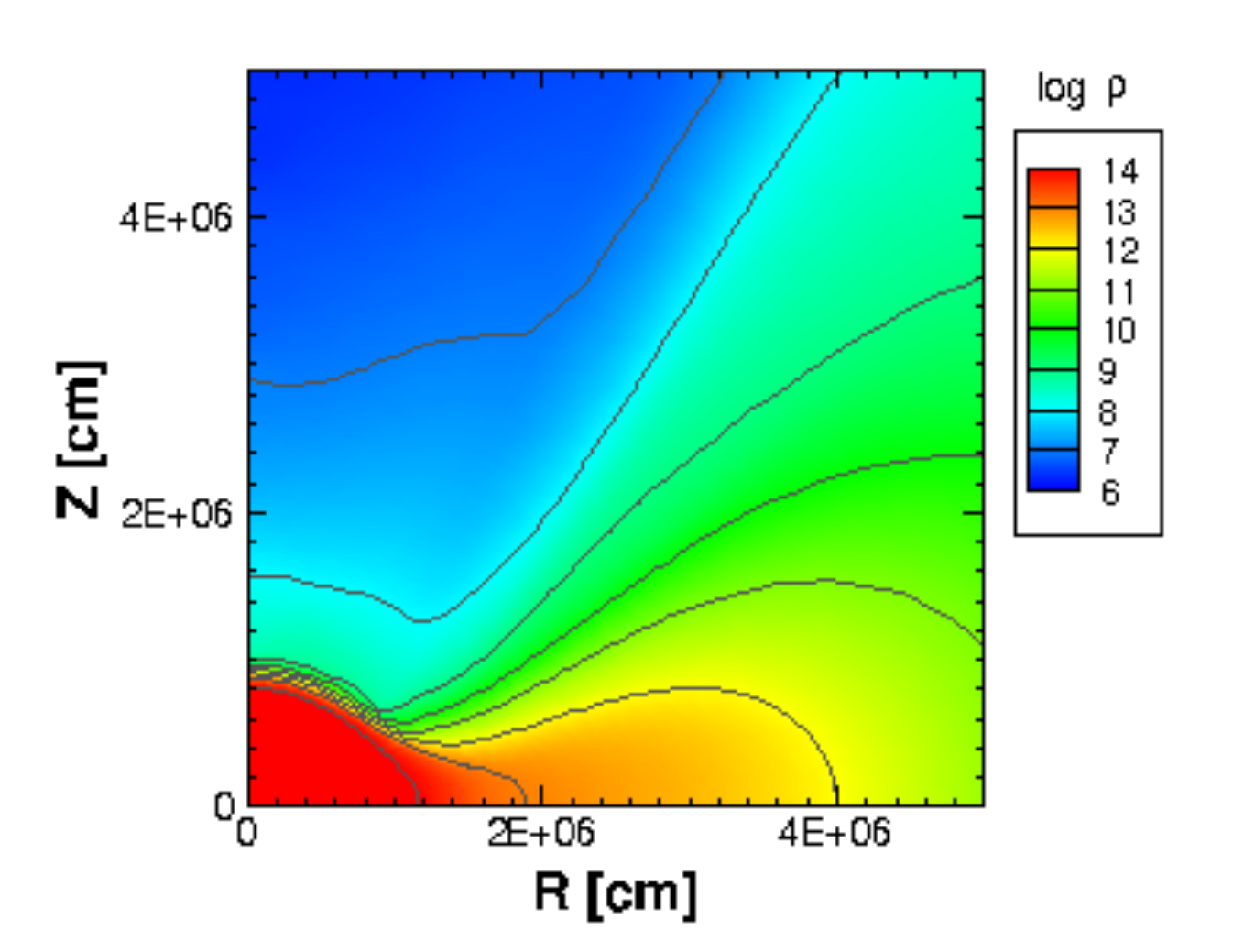}
\plotone{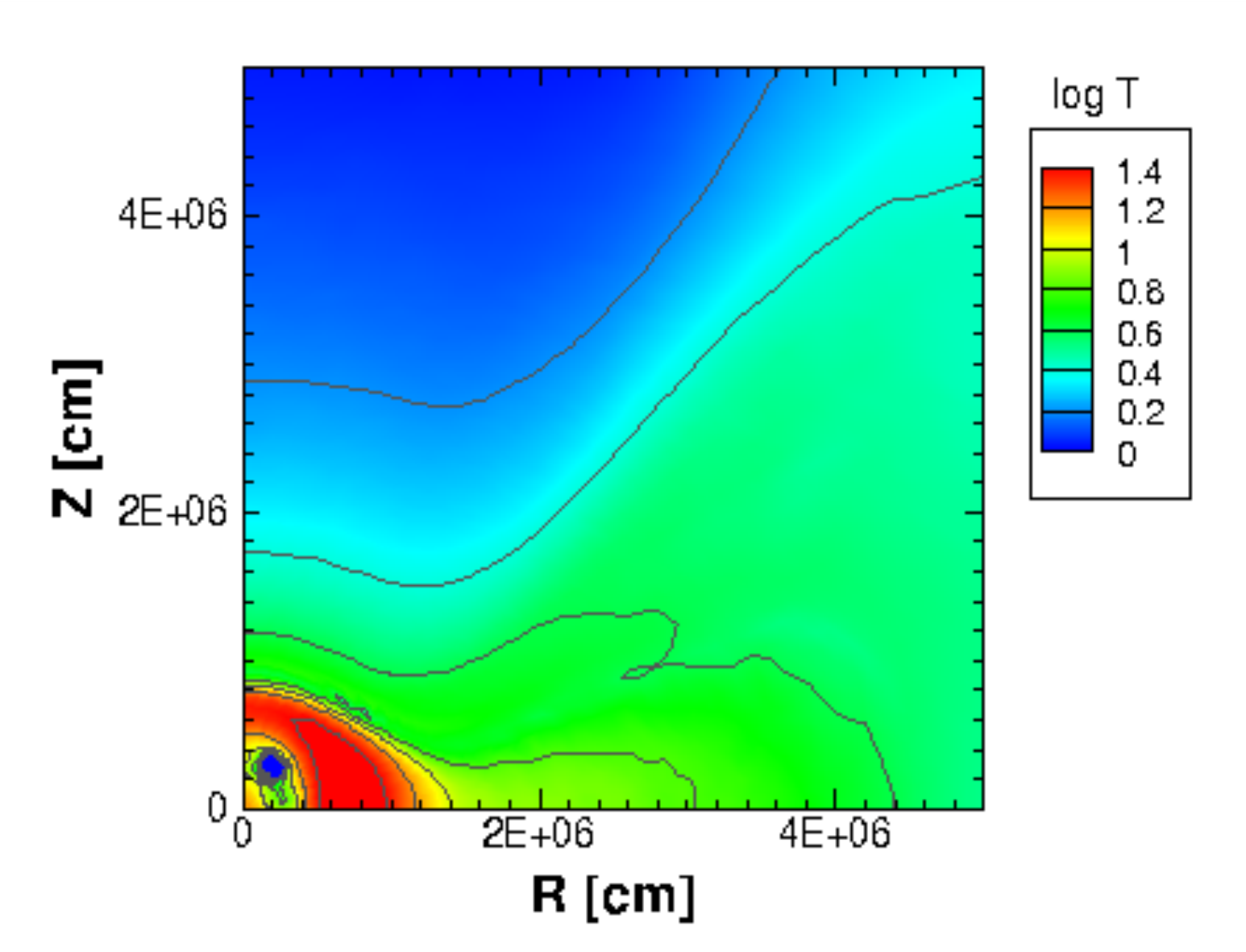}
\plotone{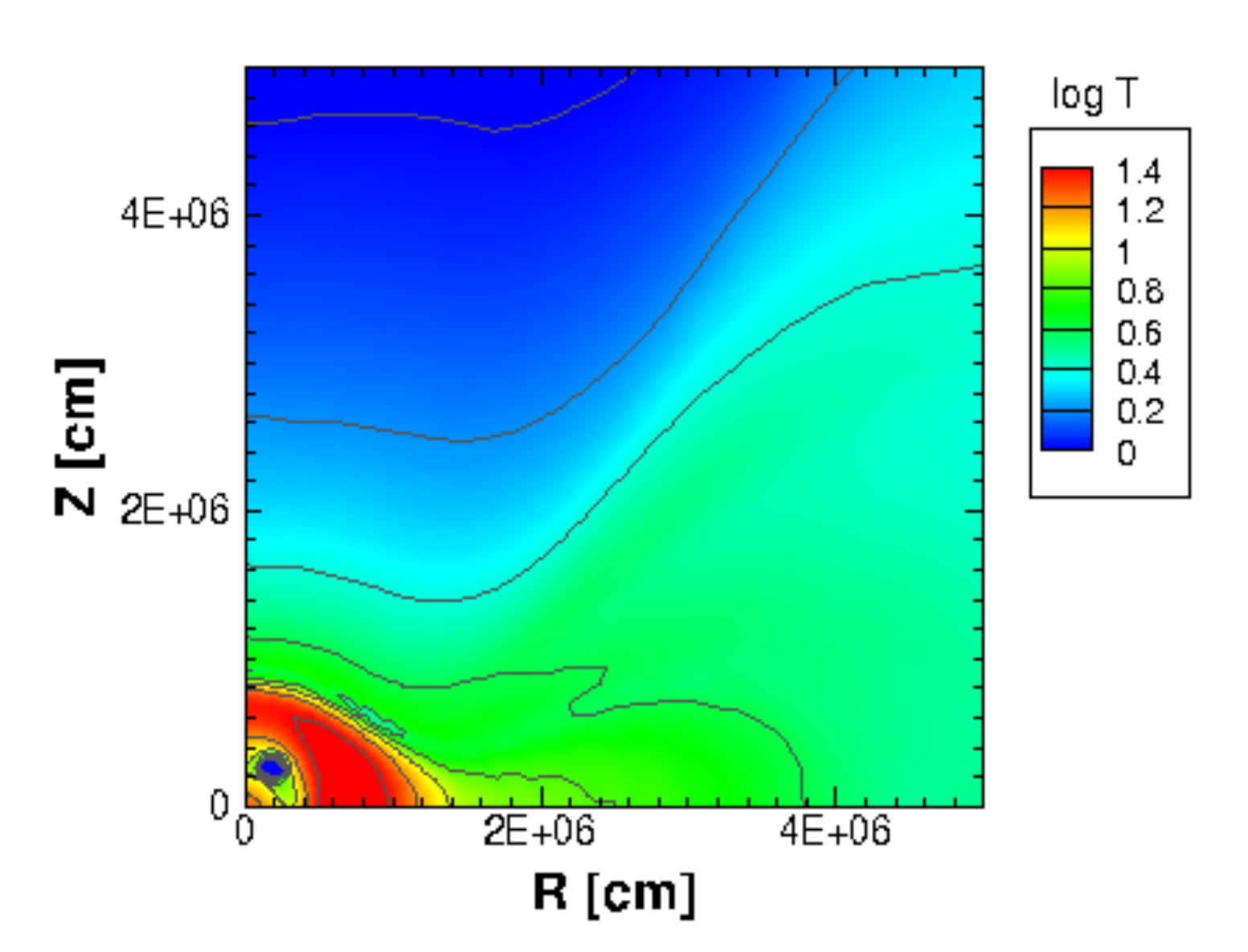}
\plotone{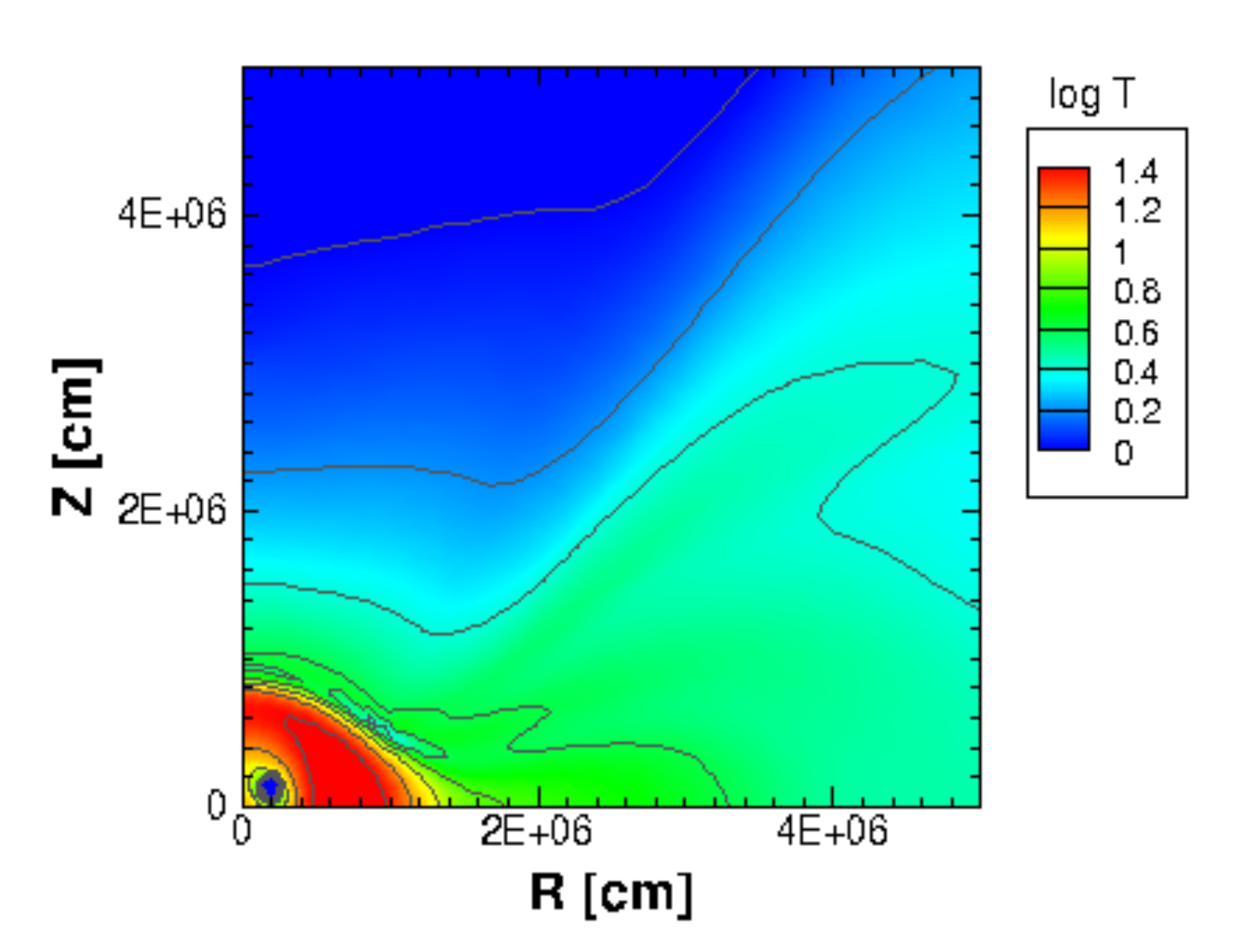}
\caption{Profiles of rest-mass density (top) and temperature (bottom) in the merger remnant at 30, 65 and 135 ms are shown in the plane of $R$ and $Z$ axes.  The same units as in Fig. \ref{fig:hydro_set_t000} are used.  
\label{fig:hydro_set_txxx}}
\end{figure}

The deformed structure of the merger remnant  persists in a quasi-static manner over 100 ms.  Figure \ref{fig:hydro_set_txxx} shows the profiles of rest-mass density and temperature at 30, 65 and 135 ms.  The remnant neutron star remains approximately stationary 
while the torus gradually shrinks and becomes more compact because of the cooling by neutrino emission.  The peak of temperature is located at the off-center region.  
The rest-mass density above the neutron star gradually decreases due to the launch of mass ejection along the $z$-axis.  The change of hydrodynamical quantities of the neutron star and the torus is relatively slow, therefore, the stationary simulations of neutrino transfer by fixing the matter profiles as the background in each snapshot can be made as a good approximation.  

\section{Neutrino transfer by the Boltzmann equation}\label{sec:boltzmann}

We adopt the numerical code 
to solve the 6D Boltzmann equation \citep{sum12}.  
The code solves the time evolution of neutrino distribution functions 
in 6D (3D in spherical coordinate system and 
3D for two angles and energy of neutrinos).  
The 6D Boltzmann equation is solved directly 
by the S$_n$ method for multi-energy bins.  
We obtain stationary state of neutrino distributions 
for fixed matter profiles 
by following their evolution for sufficient time periods over $\sim$10 ms.  
The same code is applied to study the neutrino transfer 
in 3D supernova cores and the details of procedures 
can be found in \citet{sum15}.  
The solver of the 6D Boltzmann equation 
is extended to the special relativistic version \citep{nag14}, 
however, 
the velocity dependent terms are dropped in the current study 
for fixed backgrounds.  
The simulation of the Boltzmann equation is done in the flat space though we take into account the geometric factor for evaluation by the volume integration (see for extensions toward the general relativistic simulation \citep{nag17}).  
The neutrino-radiation hydrodynamics 
using the solver of the relativistic 6D Boltzmann equation 
is applied to 
dynamical simulations of core-collapse supernovae in 2D \citep{nag18,nag19a,har19,nag19b,har20} and in 3D \citep{iwa20}.  

We treat three species of neutrinos; $\nu_e$, $\bar{\nu}_e$ and $\nu_\mu$.  
The $\mu$-type neutrino, $\nu_\mu$, 
is a representative of the group of four species 
of (anti-)neutrinos of heavy flavors.  
The basic set of neutrino reactions for supernovae 
is implemented in the collision term 
with angle- and energy-dependent expressions.  
The standard form of the Bruenn's reaction rates \citep{bru85} 
is employed together with 
extended rates for pair processes \citep{sum05,sum12}.  
We note that the pair process is calculated 
using the distribution of the counterpart neutrino.  
In order to make the Boltzmann equation linear 
in neutrino distribution function and reduce the computational cost, 
the counterpart neutrino distribution function at the previous step 
is inserted with the dependence of polar angle ($\theta_{\nu}$) of neutrino direction and 
the average over azimuthal angle ($\phi_{\nu}$) with respect to the radial coordinate.  
The use of distribution functions for the counterpart neutrinos 
at the previous step is validated 
in the study of 3D supernova cores because of the environment at high matter density and temperature.  
The tables of DD2 EOS 
by \citet{ban14} 
is used to match with the original simulations.  

The numbers of grid points for space 
are 128, 48 for radial and polar angle coordinate.  
The 14 energy grid points are placed 
to logarithmically cover neutrino energies 
up to 300 MeV.  
High angular resolution is necessary to capture the neutrino transfer in the deformed neutron star with torus.  We set the numbers of grid points for neutrino angle as 56 and 12 for polar ($\theta_{\nu}$) and aziumthal ($\phi_{\nu}$) angles with respect to the radial coordinate.  The angular resolution is higher than that in the case of 2D/3D core-collapse supernovae \citep{sum15}.  This is important for the convergence of angle distributions in non-spherical situations and integrated quantities such as heating rates along the $z$-axis.  We describe the detailed examinations in Appendix \ref{sec:append_resolution}.  



\section{Feature of neutrino transfer} \label{sec:nu-transfer}

\subsection{Neutrino density, flux and average energy} \label{sec:nu-transfer_distrib}

\begin{figure}[ht!]
\epsscale{0.37}
\plotone{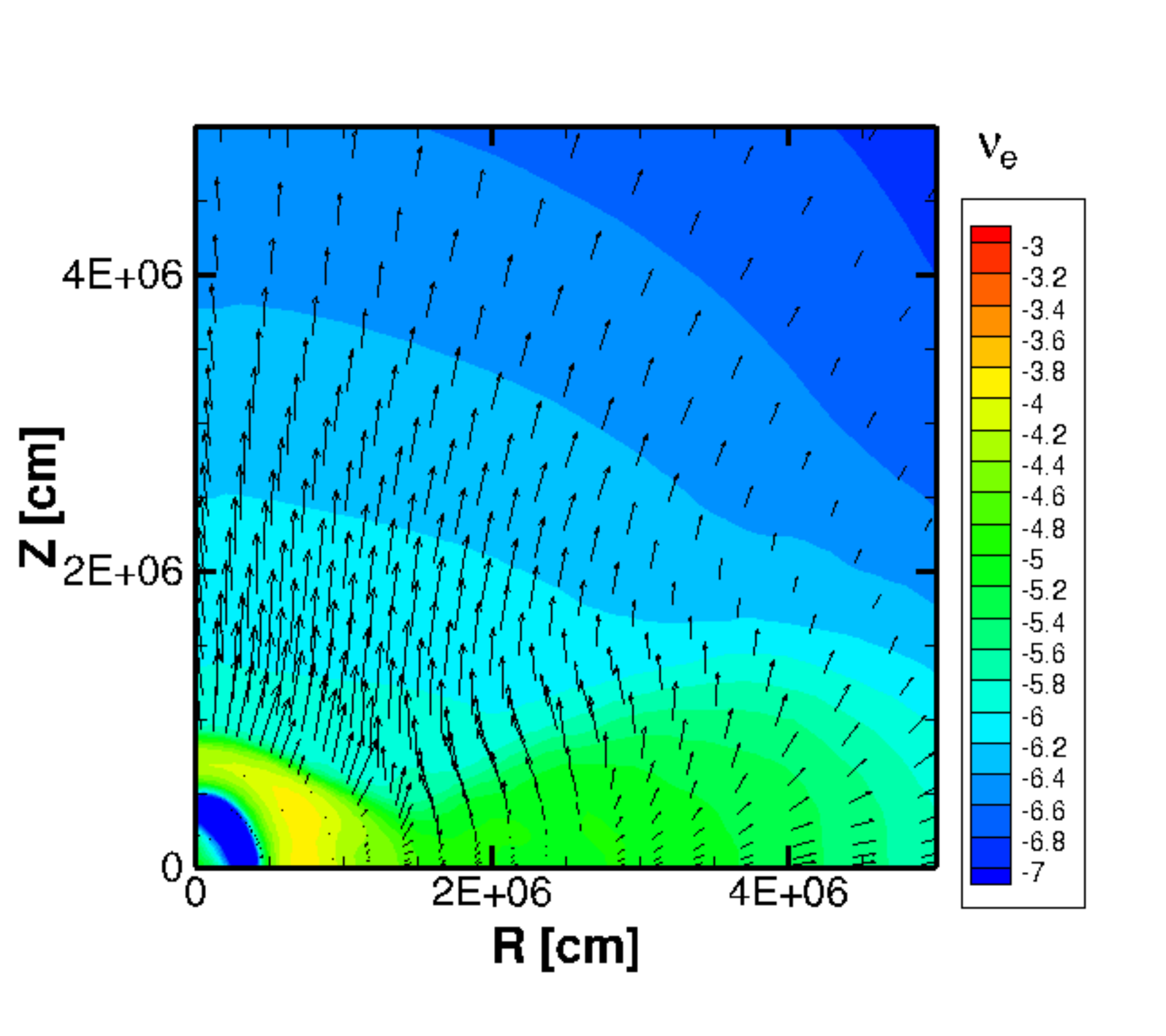}
\plotone{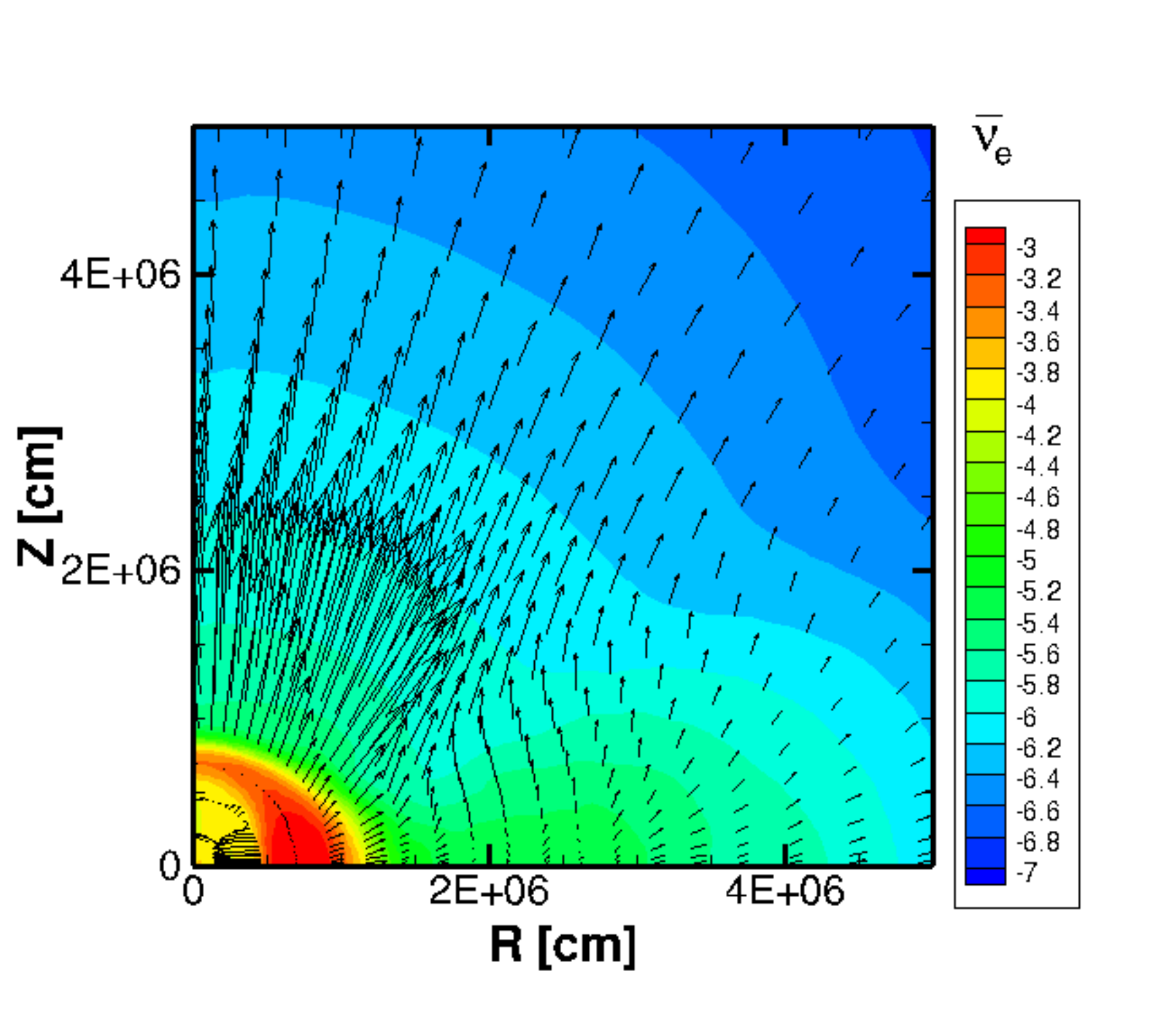}
\plotone{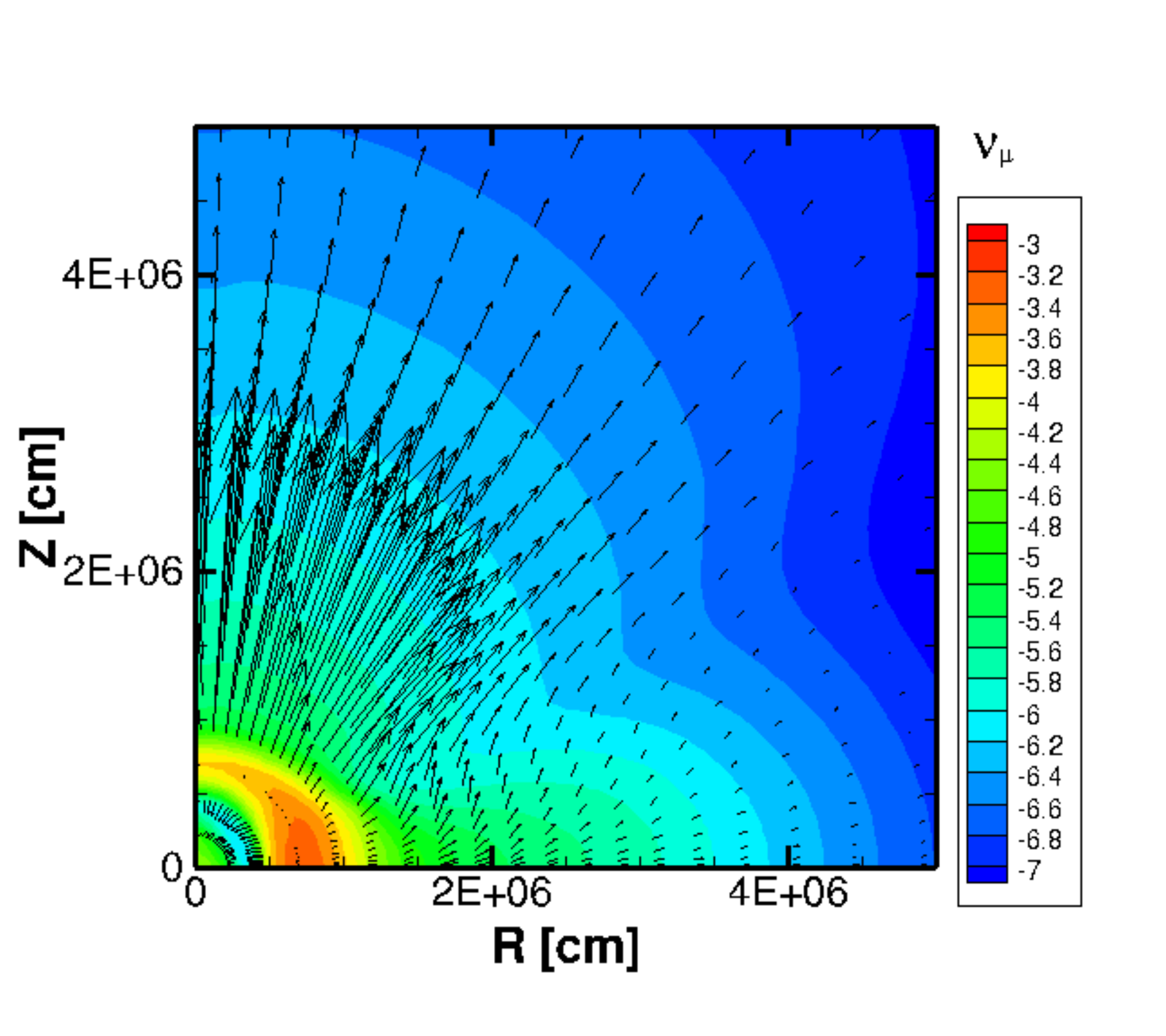}
\caption{Neutrino number density and flux are shown for electron-type neutrino, $\nu_e$ (left), electron-type anti-neutrino, $\bar{\nu}_e$ (middle), and $\mu$-type neutrino, $\nu_\mu$ (right).  The neutrino number densities in the log scale of fm$^{-3}$ are plotted by color maps.  
The vectors of neutrino flux are plotted by arrows whose length are proportional to the magnitude of flux.  
\label{fig:nu_densflux_t000iny}}
\end{figure}

The neutrino distributions have novel features due to highly deformed shape of the merger remnant.  Combination of the elongated neutron star and the extended torus contributes to the unique characteristic features of neutrino transfer.  
We show in Fig. \ref{fig:nu_densflux_t000iny} the number density and flux of neutrinos for three species.  The remnant neutron star abundantly contains neutrinos not at center but at off-center region where the temperature is high.  Among the species, the electron-type anti-neutrinos, $\bar{\nu}_e$, is the most abundant and the $\mu$-type neutrinos, $\nu_\mu$, is the next.  These species are produced by the pair-process under such high temperature environment.  The electron-type neutrinos, $\nu_e$, on the other hand, are produced through electron captures under the high matter density environment.  The distribution of $\nu_e$ is extended in the geometrically thick torus along the equator.  The neutrino emission from the neutron star contributes to large neutrino fluxes in the region above it.  
The neutrino fluxes toward the equatorial region are suppressed since the geometrically thick torus is opaque to neutrinos.  
The neutrino emission from the torus contributes to the non-radial fluxes and leads to the concentration of neutrino fluxes above the merger remnant.  
%
Such a configuration results in the enhancement of heating 
in the region above the merger remnant 
(see \S~\ref{sec:nu-transfer_heat}).  

\begin{figure}[ht!]
\epsscale{0.35}
\plotone{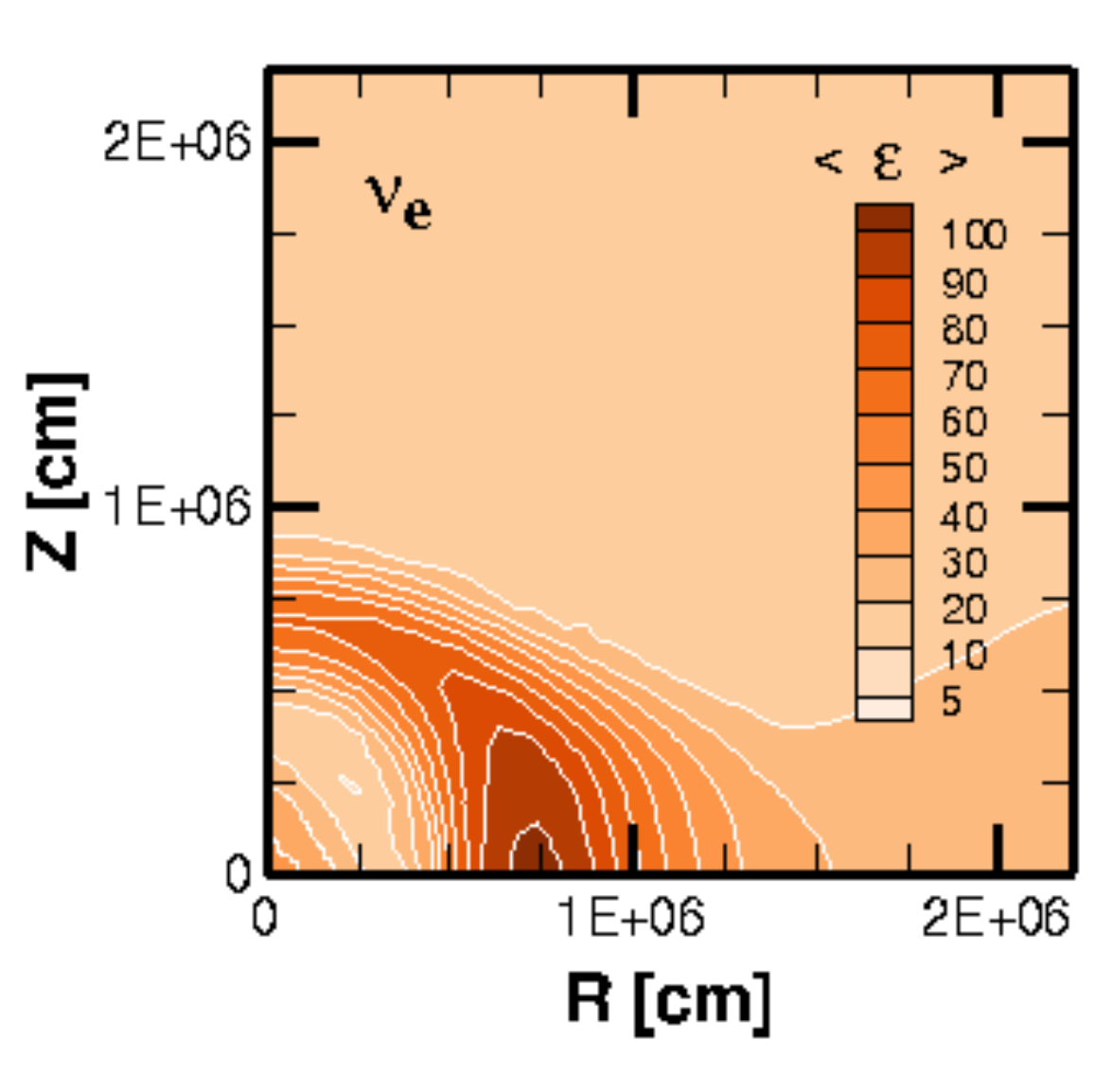}
\plotone{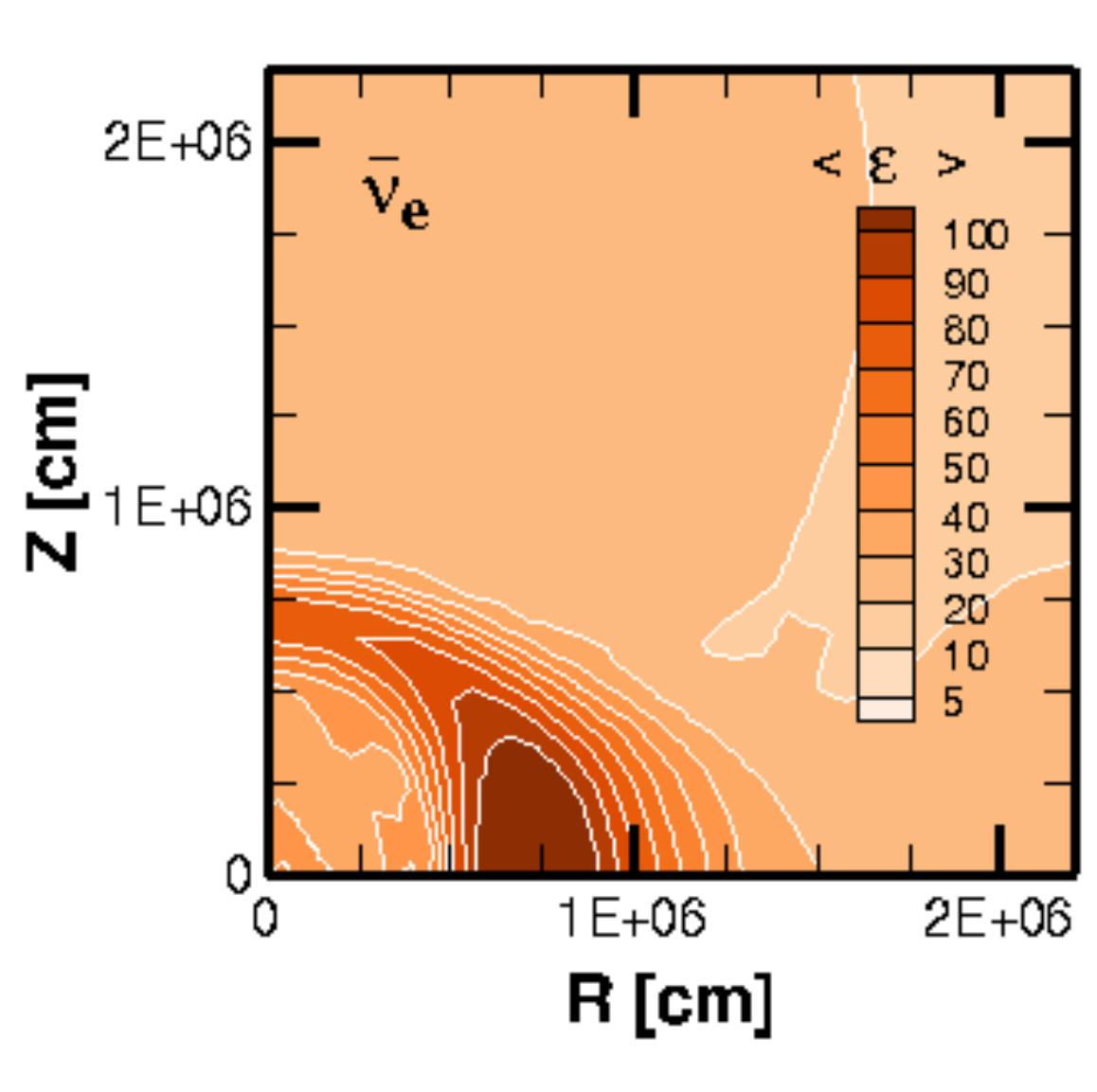}
\plotone{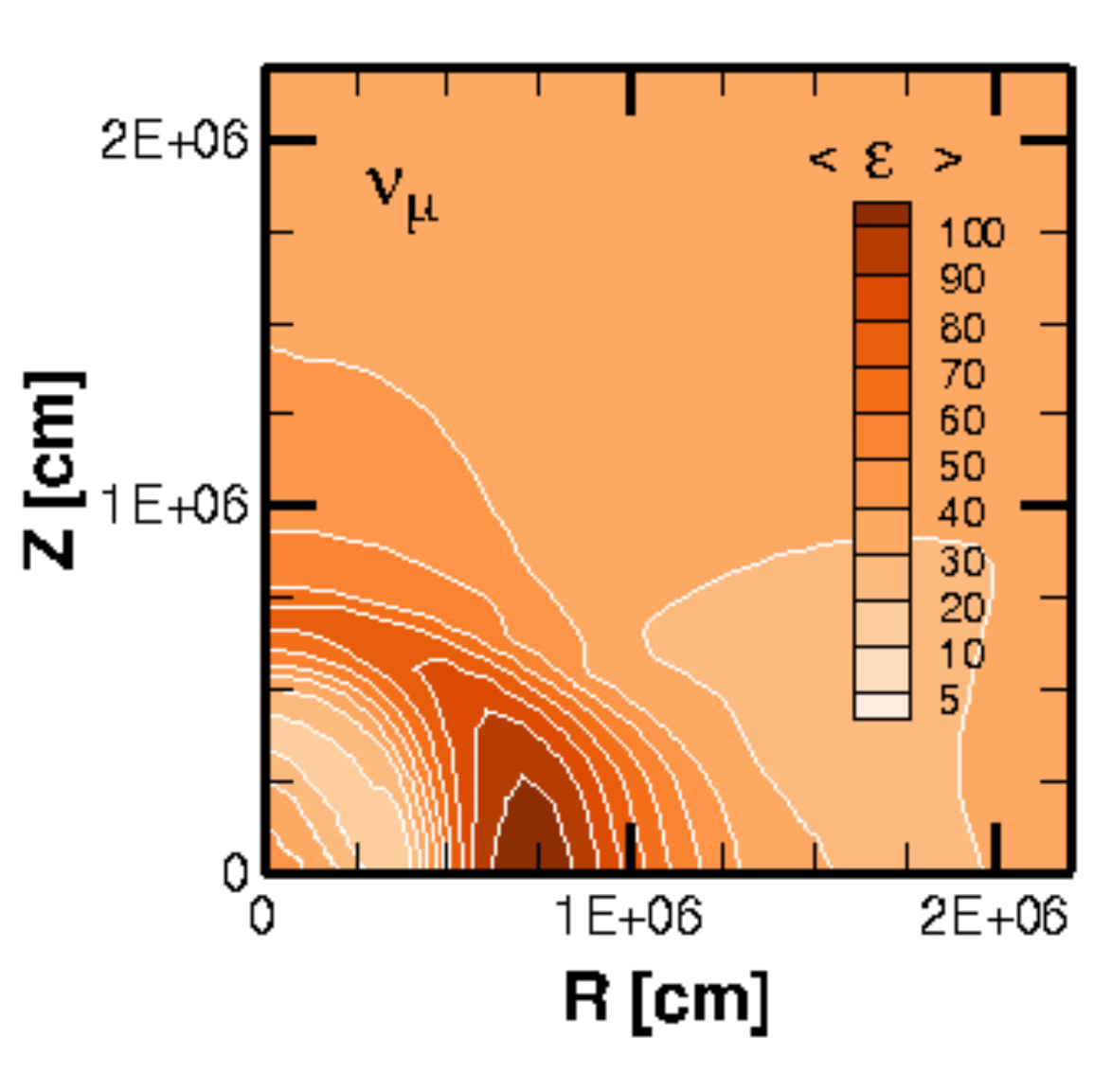}
\caption{Average energies of neutrinos [MeV] are plotted for $\nu_e$ (left), $\bar{\nu}_e$ (middle) and $\nu_\mu$ (right).  
\label{fig:nu_eave_t000iny}}
\end{figure}
\begin{figure}[ht!]
\epsscale{0.5}
\plotone{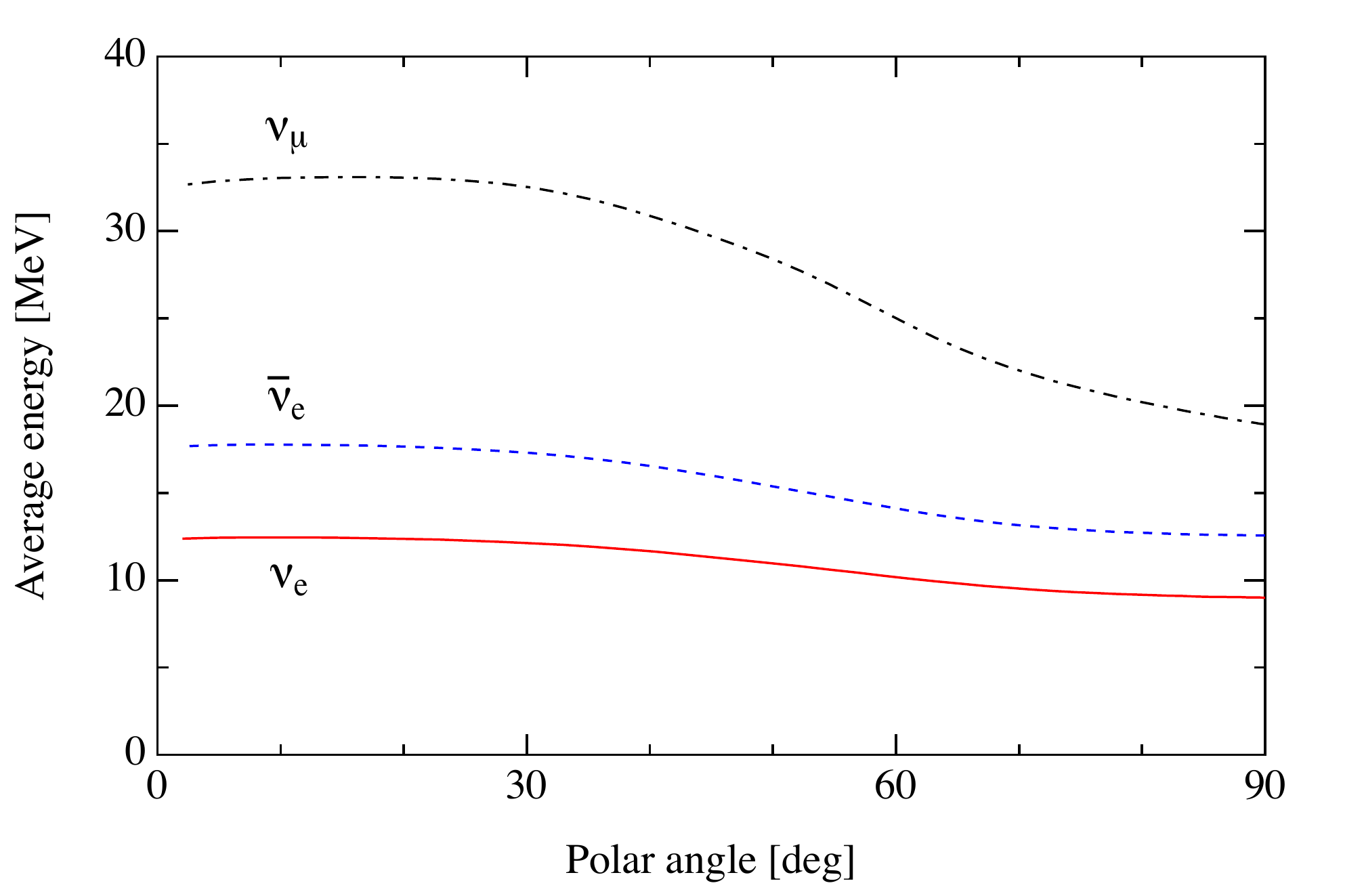}
\plotone{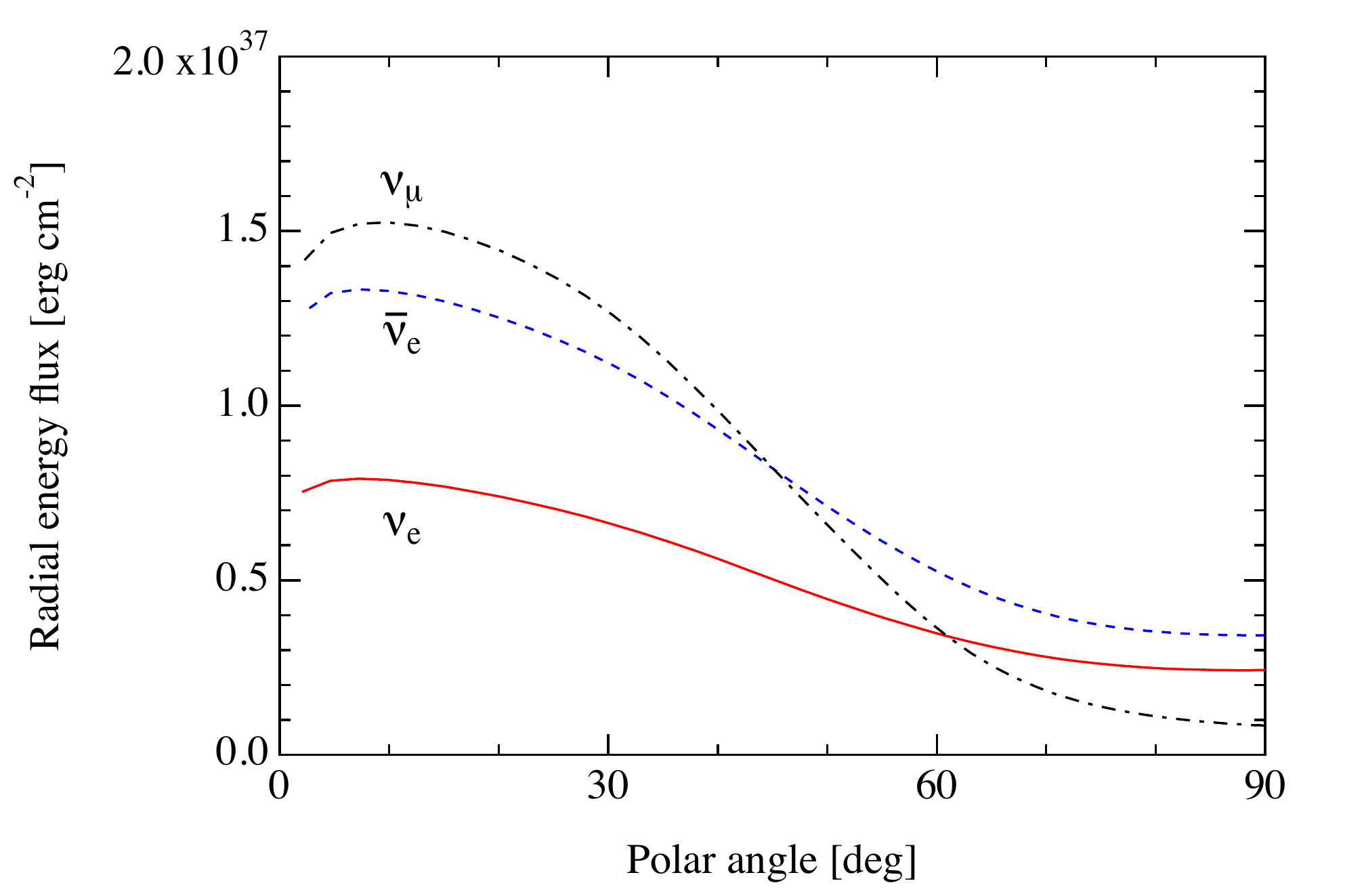}
\caption{Average energies (left) and radial energy flux (right) of neutrinos are plotted as functions of the polar angle for $\nu_e$ (solid line), $\bar{\nu}_e$ (dashed line) and $\nu_\mu$ (dash-dotted line).  
\label{fig:nu_eave_t000iny_theta}}
\end{figure}

We display in Fig. \ref{fig:nu_eave_t000iny} the average energies of neutrinos in the central region.  We show the first moment of energy in these plots.  The average energies of three species are high above 100 MeV in the high temperature region.  
Moreover, the enhancement of the average energy for $\nu_{\mu}$ in the region above the neutron star is noticeable.  
In the left panel of Fig. \ref{fig:nu_eave_t000iny_theta}, the average energies of neutrinos evaluated at the radius of 250 km, which is the outer boundary, are plotted as a function of the polar angle.  
The average energies of neutrinos emitted from the merger remnant have a definite hierarchy among species.  The average energy of $\nu_\mu$ is distinctively higher than those of $\nu_e$ and $\bar{\nu}_e$ and shows strong dependence on the polar angle.  

\begin{figure}[ht!]
\epsscale{0.35}
\plotone{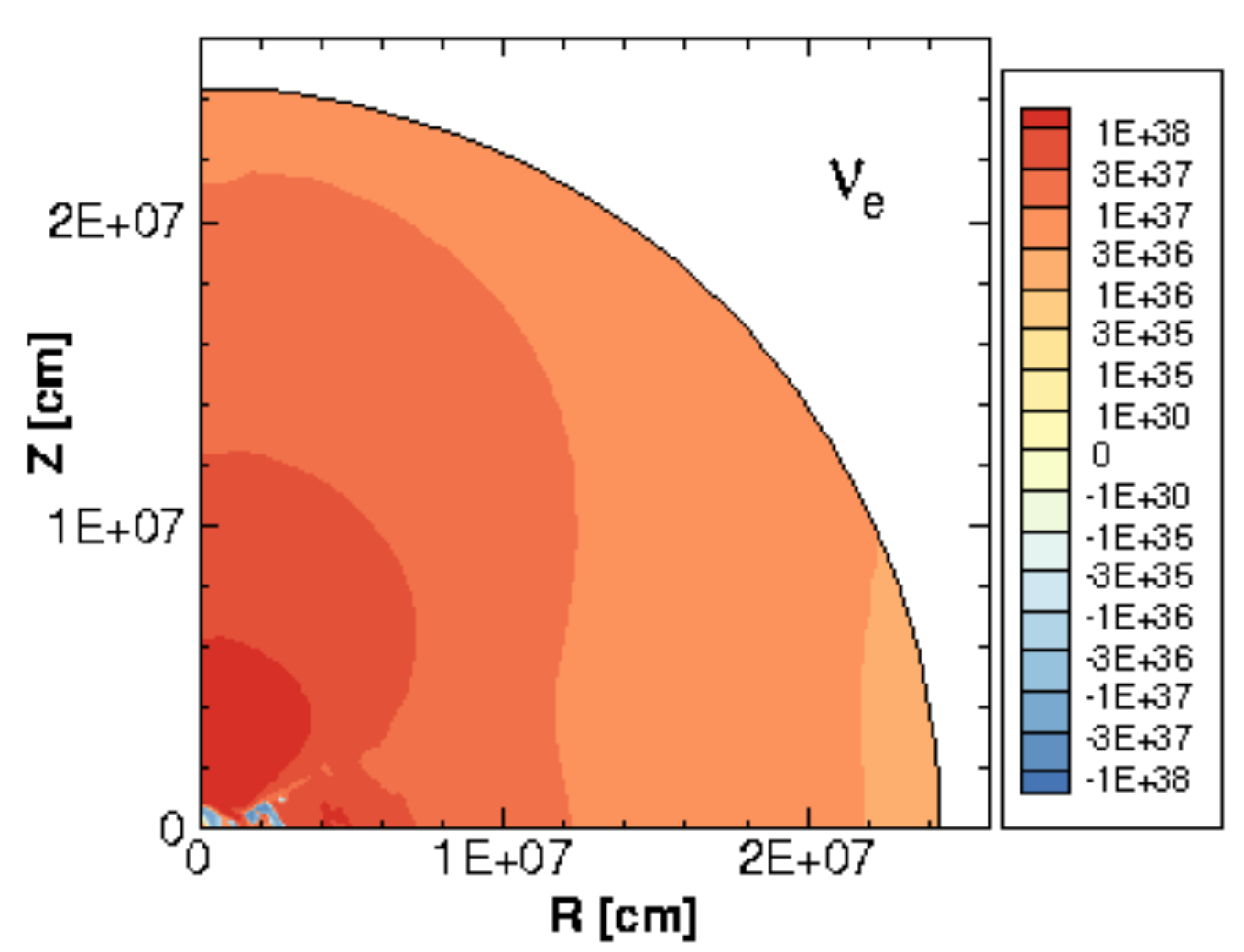}
\plotone{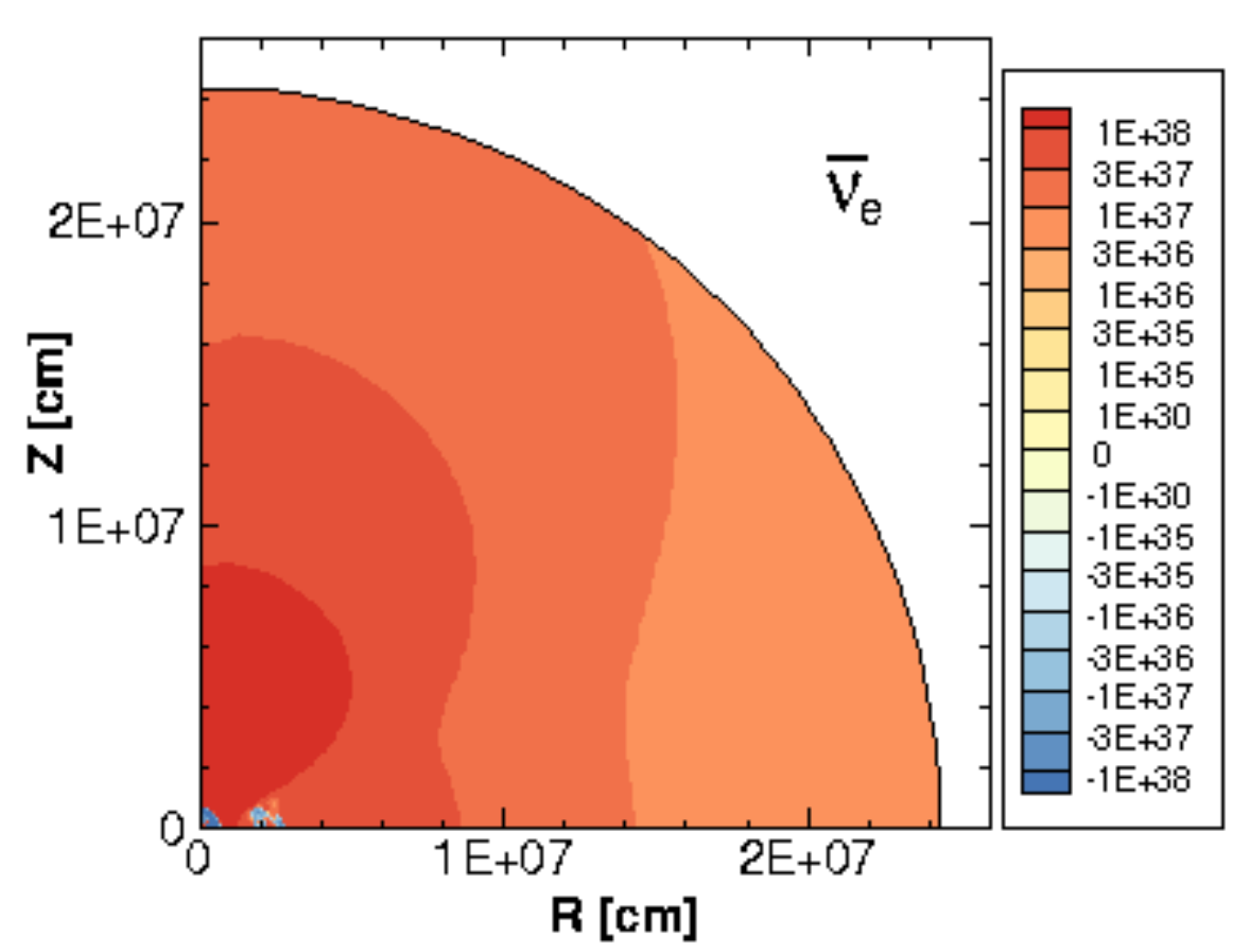}
\plotone{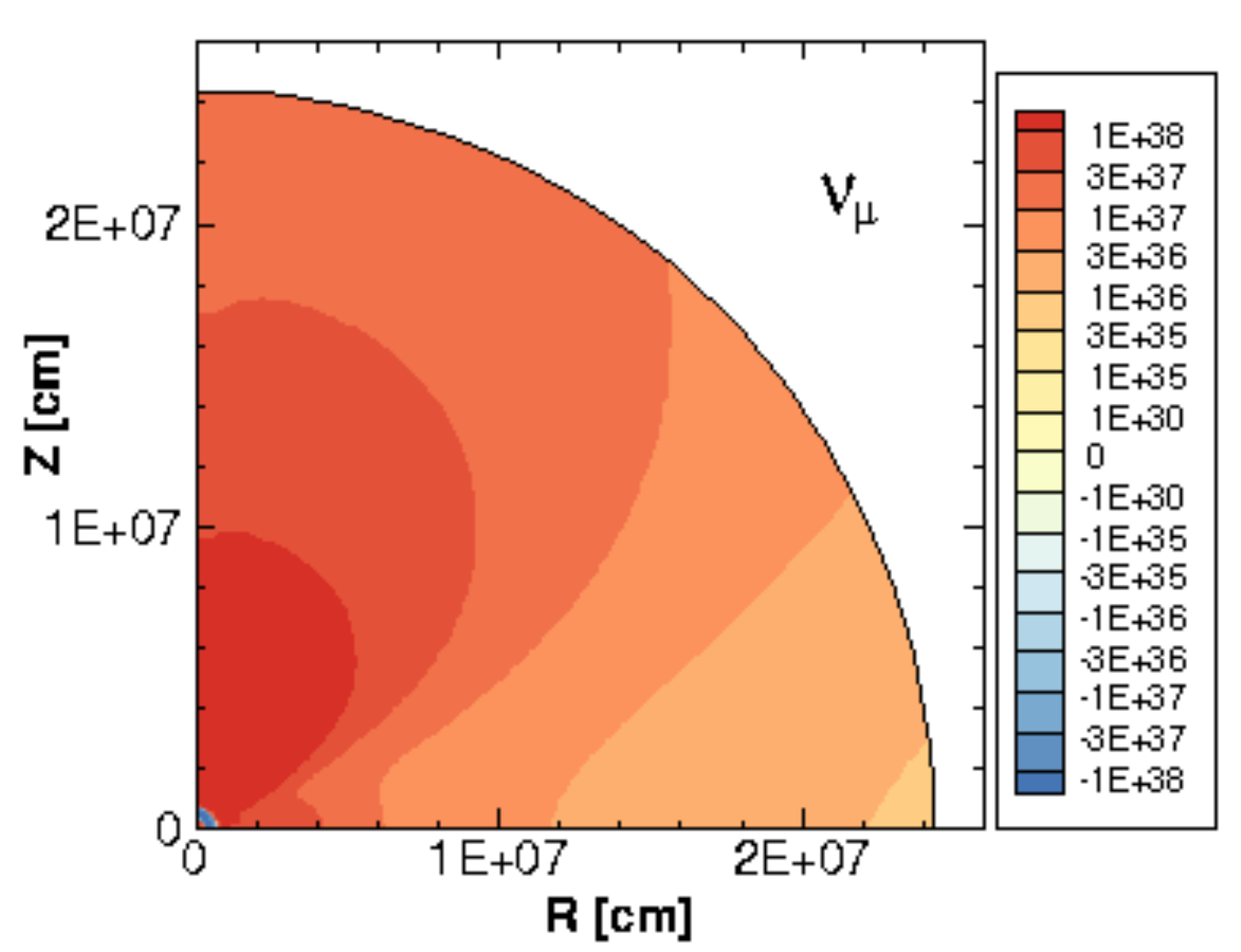}
\plotone{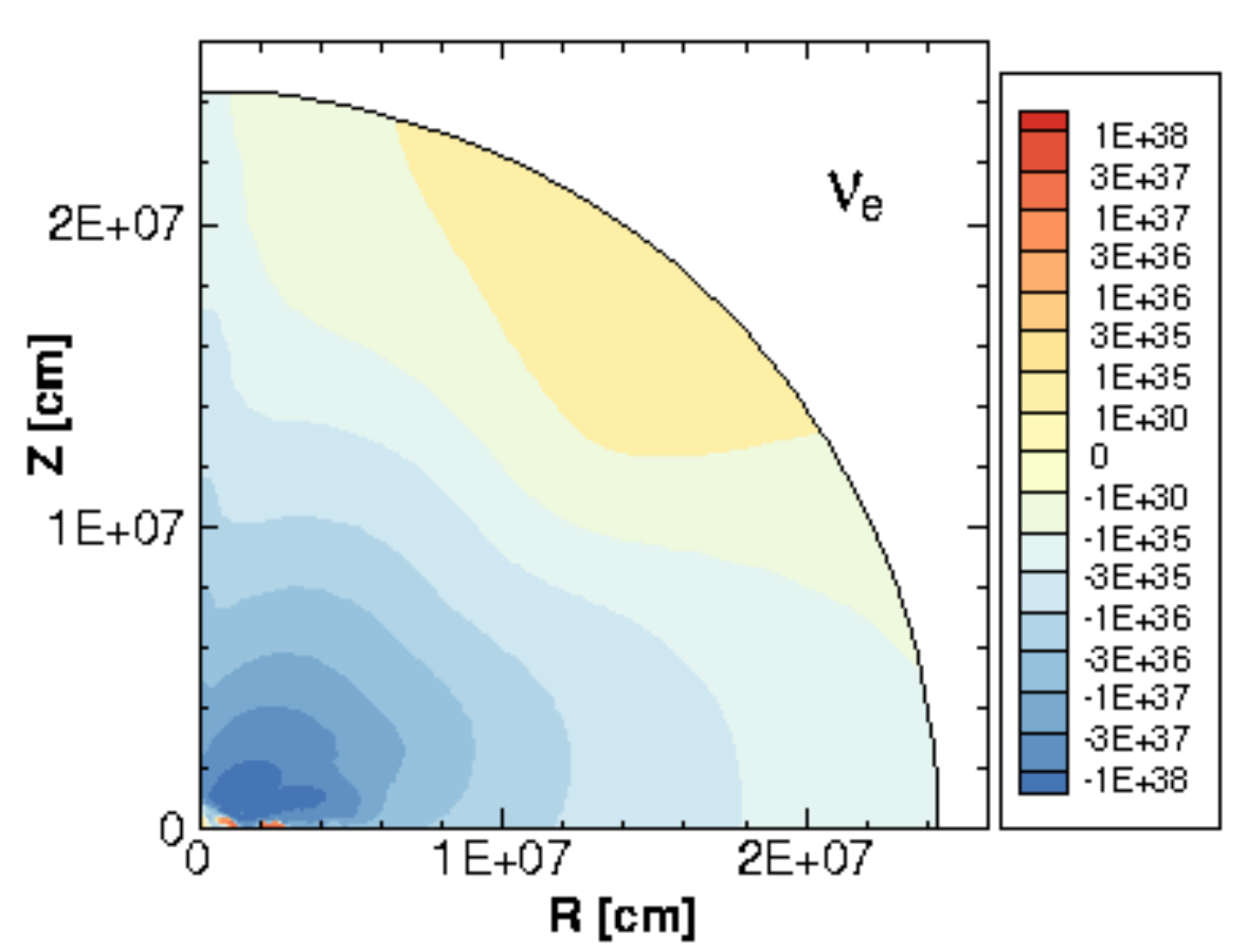}
\plotone{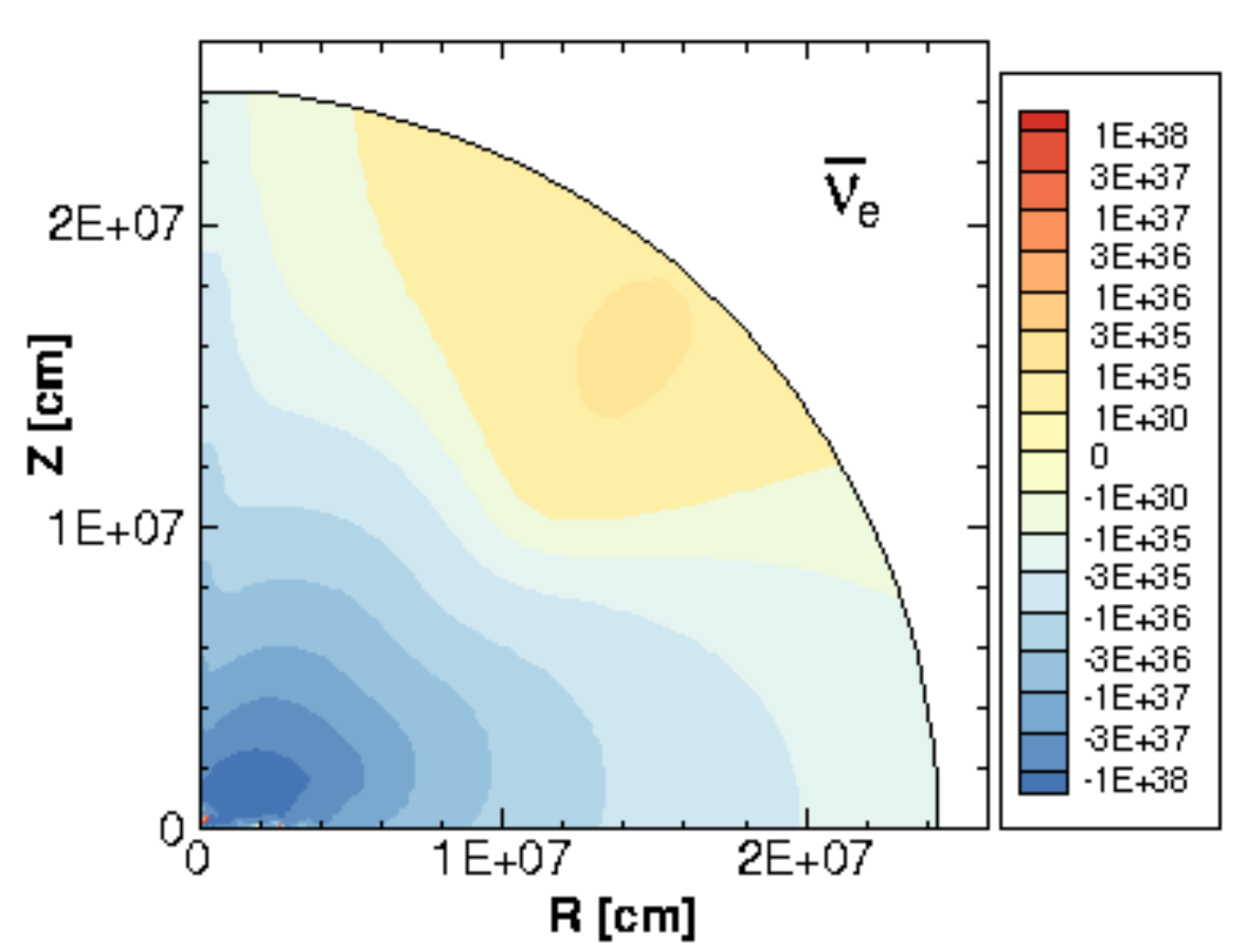}
\plotone{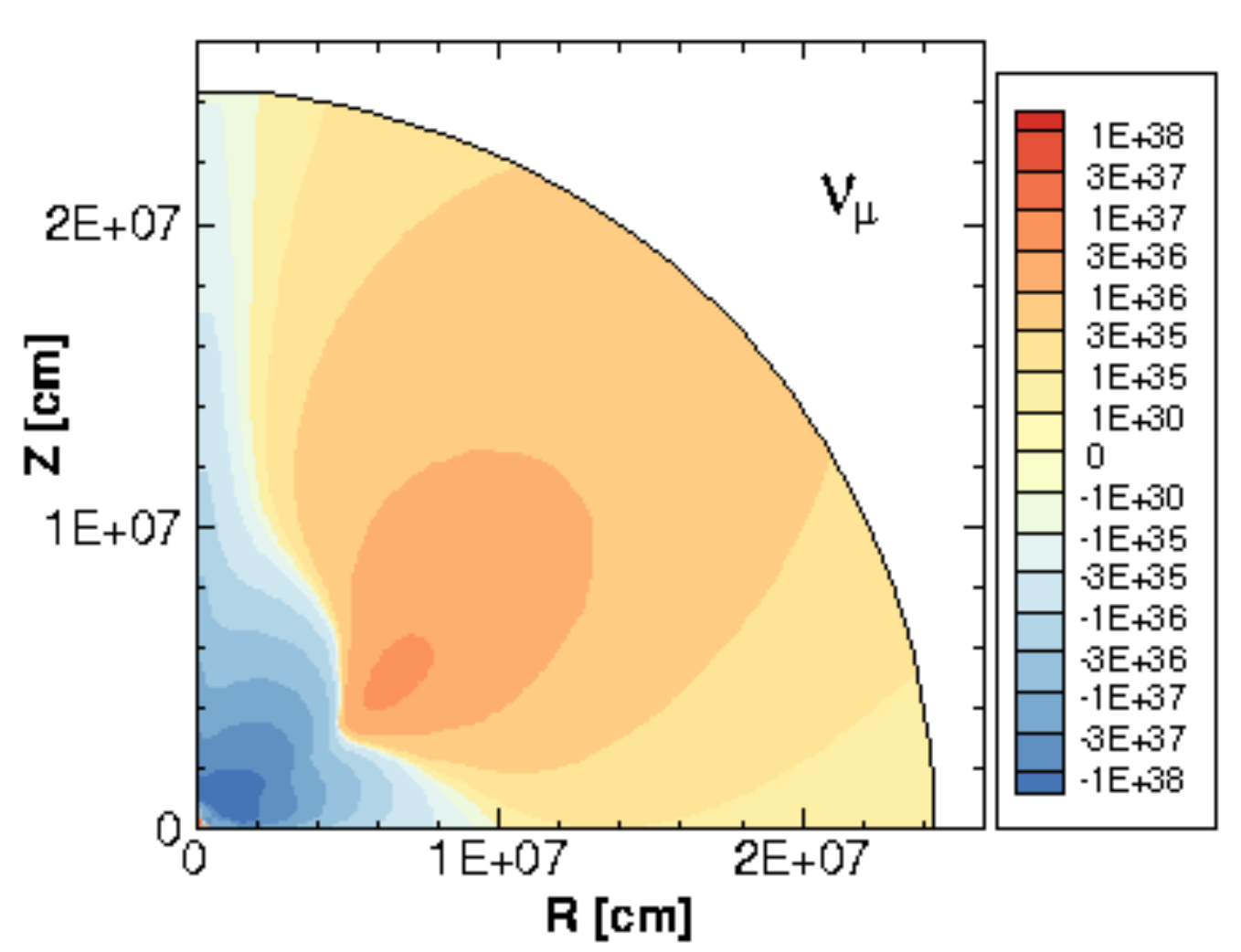}
\caption{The radial (upper panels) and polar (lower panels) components of the neutrino energy fluxes [erg cm$^{-2}$ s$^{-1}$] are plotted for $\nu_e$ (left), $\bar{\nu}_e$ (middle) and $\nu_\mu$ (right).  
\label{fig:nu_eflux_t000iny}}
\end{figure}

Figure \ref{fig:nu_eflux_t000iny} shows the radial and polar components of the energy fluxes for the three species.  The fluxes are highly asymmetric with non-zero polar components.  The radial component is dominant in the region above the neutron star for all species.  
The magnitude of polar component is comparable or even larger with respect to that of the radial one.  
The energy fluxes of $\nu_e$ and $\bar{\nu}_e$ are widely extended due to the emission by charged current reactions from both the neutron star and the torus.   The energy flux of $\nu_\mu$, on the other hand, is more focused due to the emission by pair productions from the high temperature region of the neutron star with minor contributions from the torus.  
The contribution of the torus as a peripheral source is most remarkable for $\nu_e$, moderate for $\bar{\nu}_e$, and minor for $\nu_\mu$ as we will see the emission rates in Fig. \ref{fig:nu_emission_t000_xxxx} (\S~\ref{sec:nu-transfer_emis}).  In addition, the torus plays a role of a shield toward the equator for $\nu_\mu$ having only the central source.  

In the right panel of Fig. \ref{fig:nu_eave_t000iny_theta}, the radial energy fluxes of neutrinos evaluated at the radius of 250 km are plotted as a function of the polar angle.  The dependence of energy fluxes on the polar angle is strong especially for $\nu_\mu$.  The integration of the radial energy fluxes over solid angle provides the total luminosities for the whole direction of $3.2\times10^{52}$, $4.9\times10^{52}$ and $4.2\times10^{52}$ ~erg s$^{-1}$ for $\nu_e$, $\bar{\nu}_e$ and $\nu_\mu$, respectively.  
The luminosities of $\bar{\nu}_e$ and $\nu_\mu$ are high due to the thermal processes, especially through the Bremsstrahlung.  
The luminosity in \cite{fuj17} for each species is $\sim 2\times10^{52}$, $\sim 3\times10^{52}$, $\sim 6\times10^{51}$\,erg\,s$^{-1}$, respectively.
The luminosity is larger by several tens of percent for $\nu_e$ and $\bar{\nu}_e$, and remarkably, larger by a factor of $\sim 7$ for $\nu_\mu$.
The large difference in the luminosity for $\nu_\mu$ is also found in \cite{fou20}, in which the neutrino transfer with the Monte-Carlo method is compared to the moment formalism in binary neutron star merger.
Thus, this difference would be because of the different level of the sophistication of neutrino transfer.
We will investigate the difference closely in our future work.
%

\subsection{Neutrinosphere} \label{sec:nu-transfer_nu-sphere}

\begin{figure}[ht!]
\epsscale{0.37}
\plotone{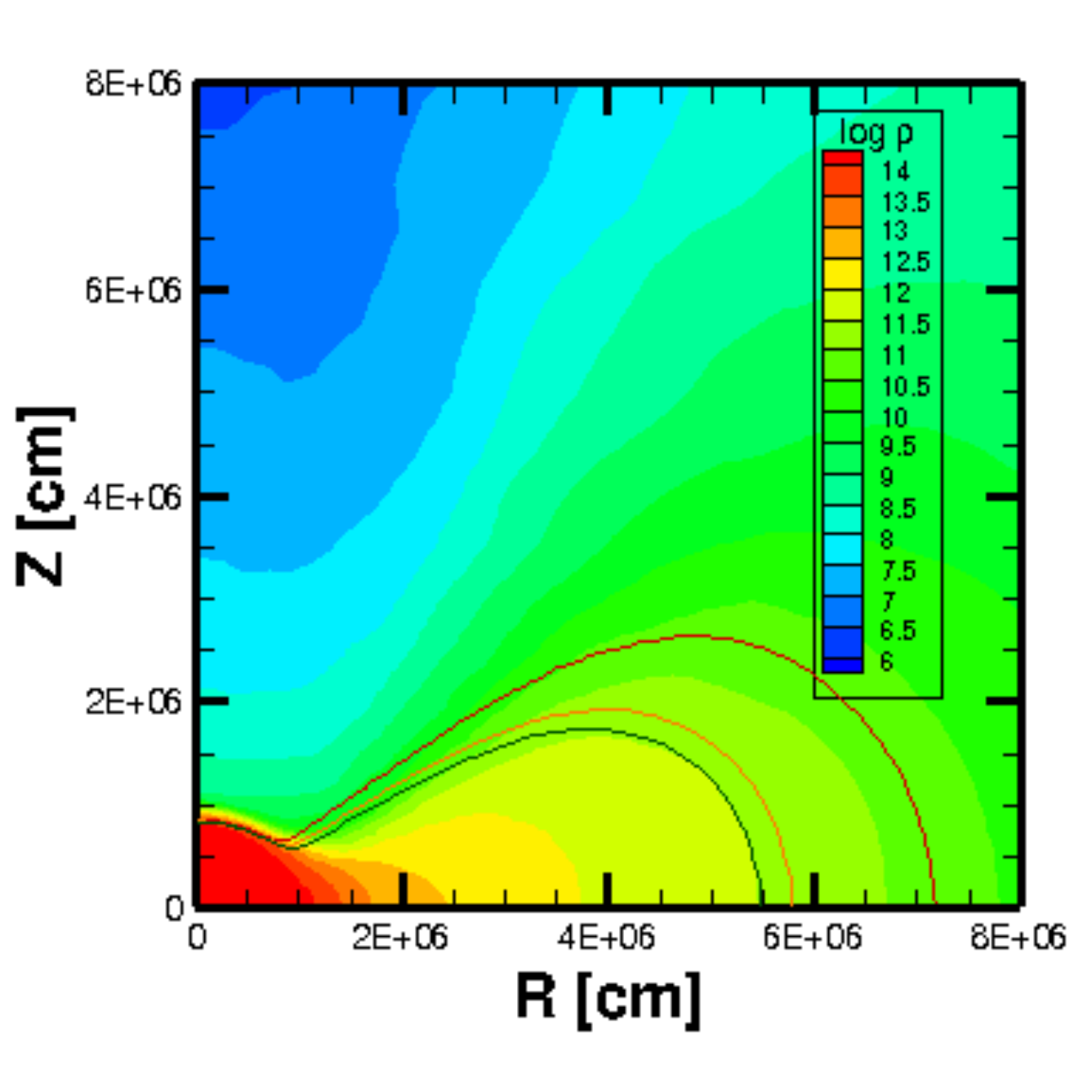}
\plotone{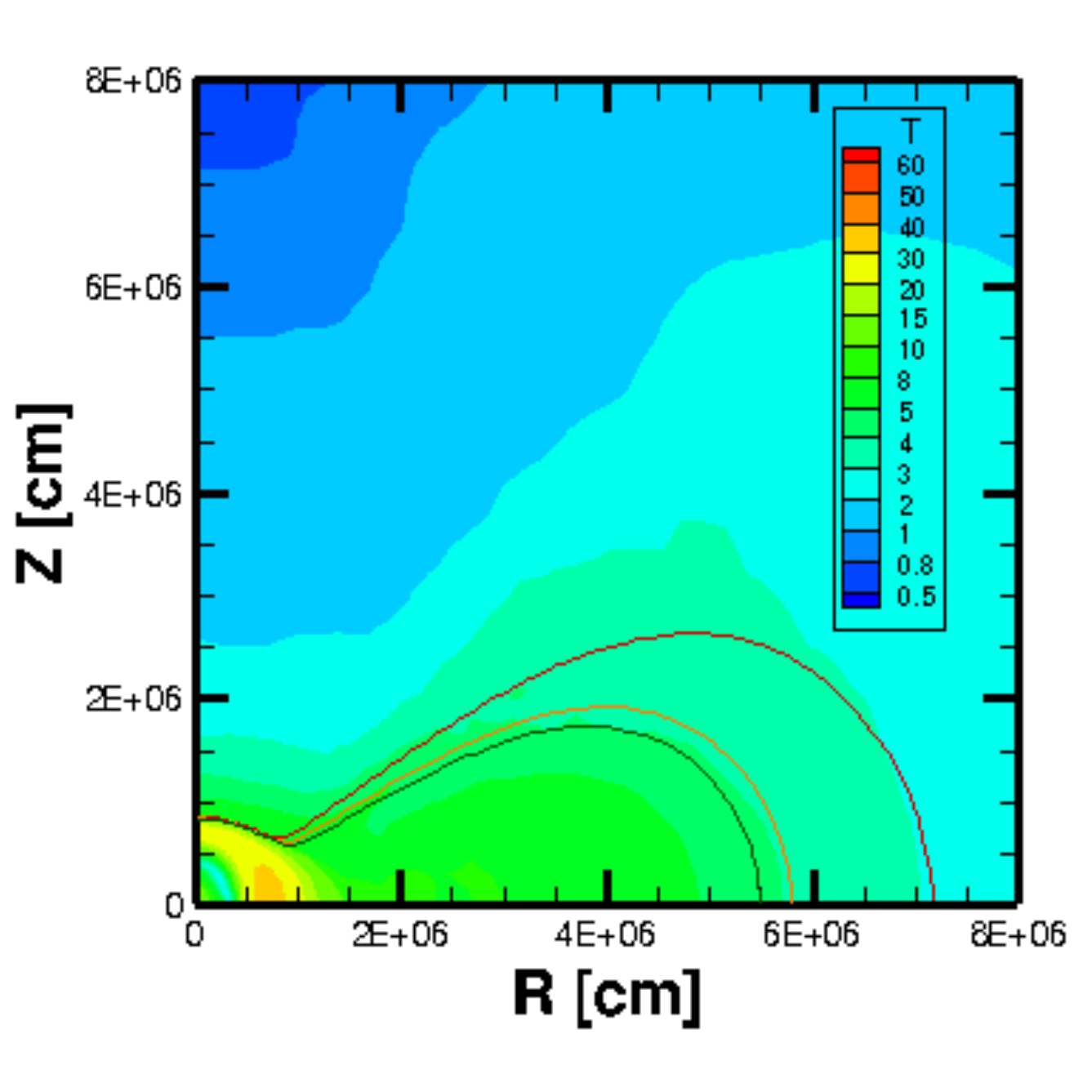}
\plotone{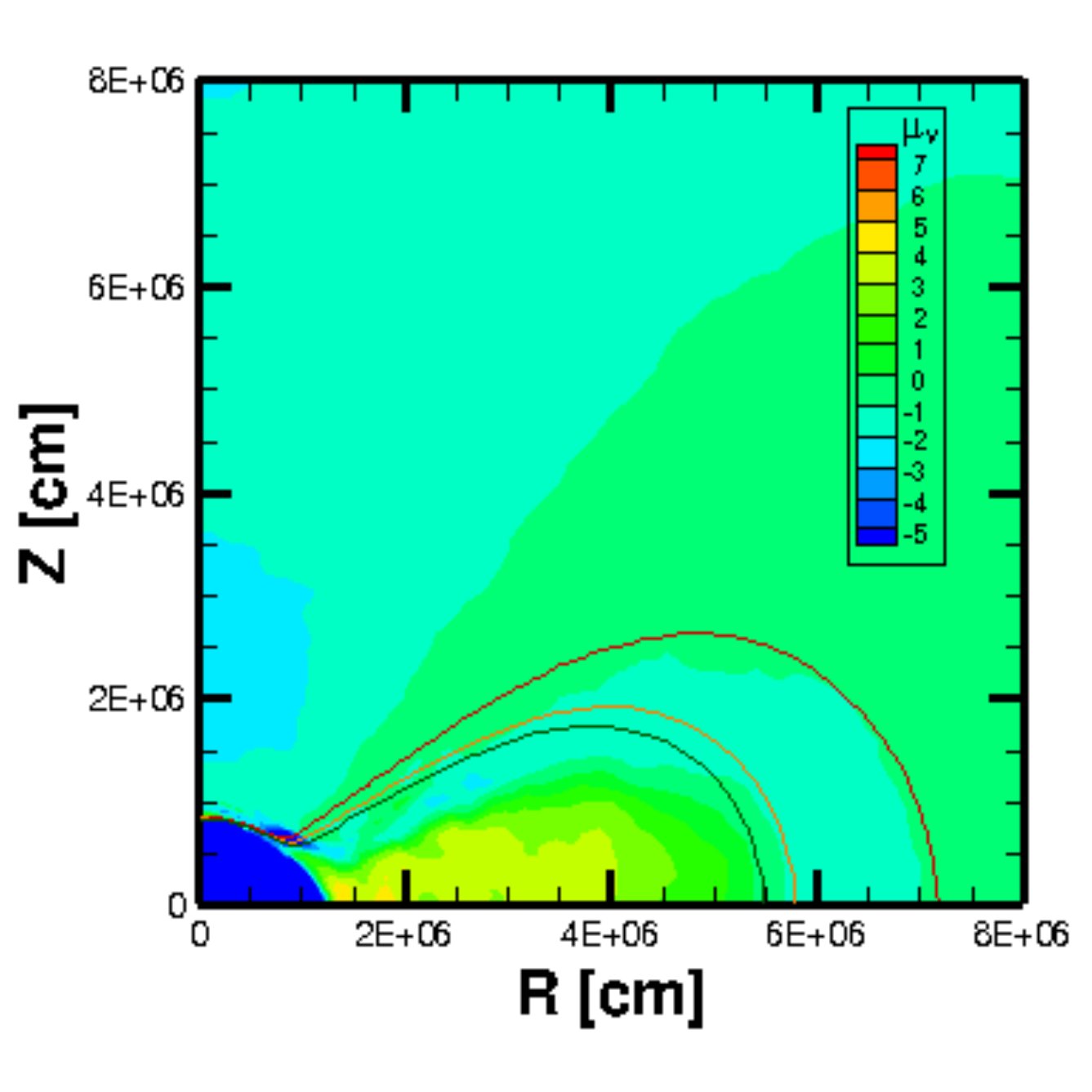}
\caption{Locations of the neutrinosphere are drawn on the contour plots of rest-mass density [g\,cm$^{-3}$] (left), temperature [MeV] (middle) and neutrino chemical potential in [MeV] (right).  The locations of neutrinosphere are evaluated for neutrino energy of 13 MeV.  Three solid lines correspond to the neutrinosphere for $\nu_\mu$, $\bar{\nu}_e$ and $\nu_e$ in the order from inside to outside.  
\label{fig:nu_sphere_xxxx_t000ie06}}
\end{figure}

The region of neutrino emission has a deformed shape due to the deformed nature of the merger remnant.  The locations of neutrinosphere are shown in Fig. \ref{fig:nu_sphere_xxxx_t000ie06} on top of the rest-mass density profiles.  The neutrinosphere is defined by the location for which the optical depth is 2/3.  Here, the optical depth is evaluated by the integral of the inverse mean free path along the radial coordinate from outside using the mean free path for forward angle\footnote{We adopted the radial direction in the current analysis, though it would be interesting to examine other cases along the non-radial trajectories.  }.  

The shape of neutrinosphere for three species are highly deformed reflecting the matter distribution.  The neutrinosphere for neutrino energy of 13 MeV roughly follows the rest-mass density contour at around 10$^{11}$--10$^{12}$ g\,cm$^{-3}$.  
This explains why the neutrino fluxes are focused in the region above the neutron star.  It is remarkable that the neutrinosphere extends to the regions of a low temperature and a positive neutrino chemical potential.  

\begin{figure}[ht!]
\epsscale{0.37}
\plotone{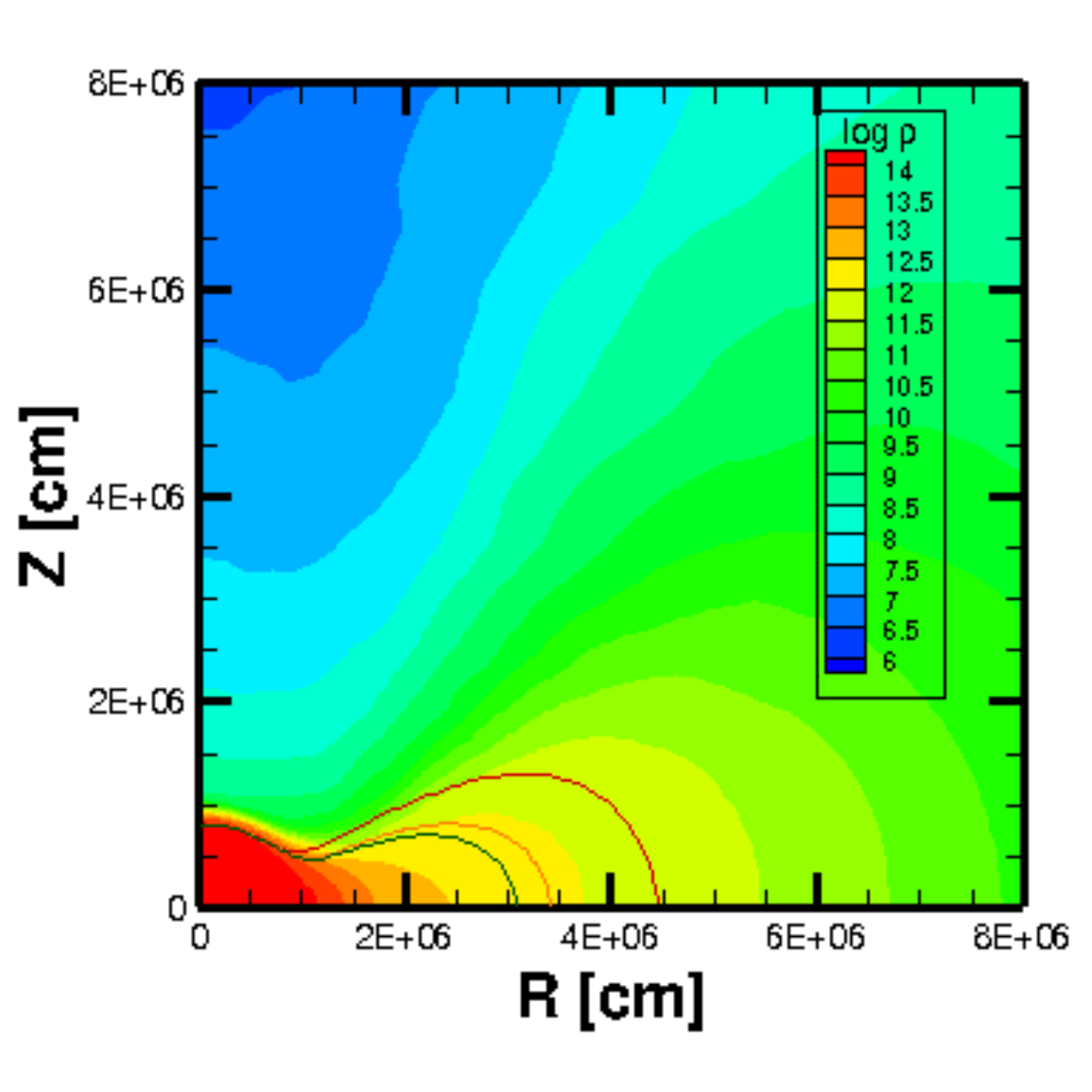}
\plotone{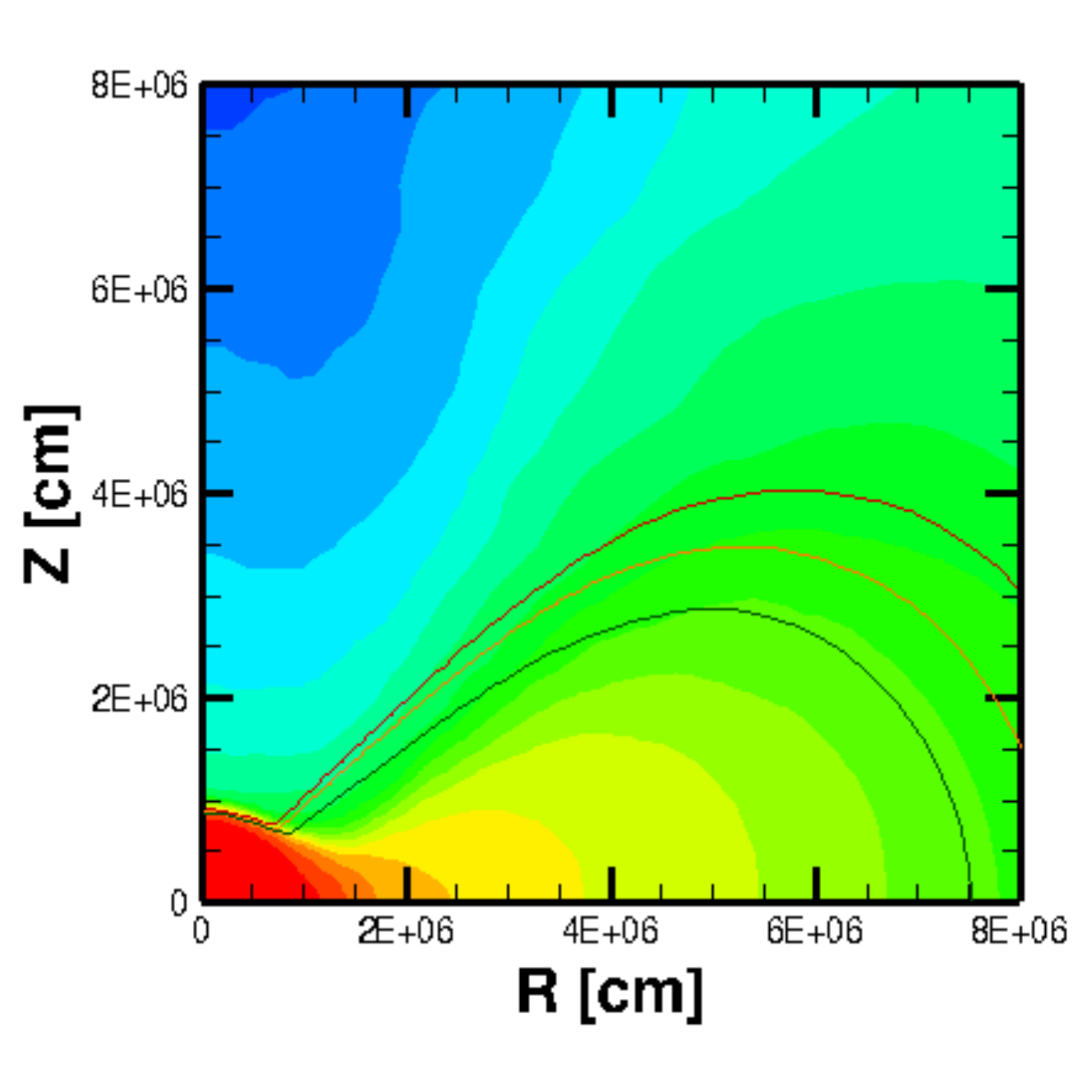}
\plotone{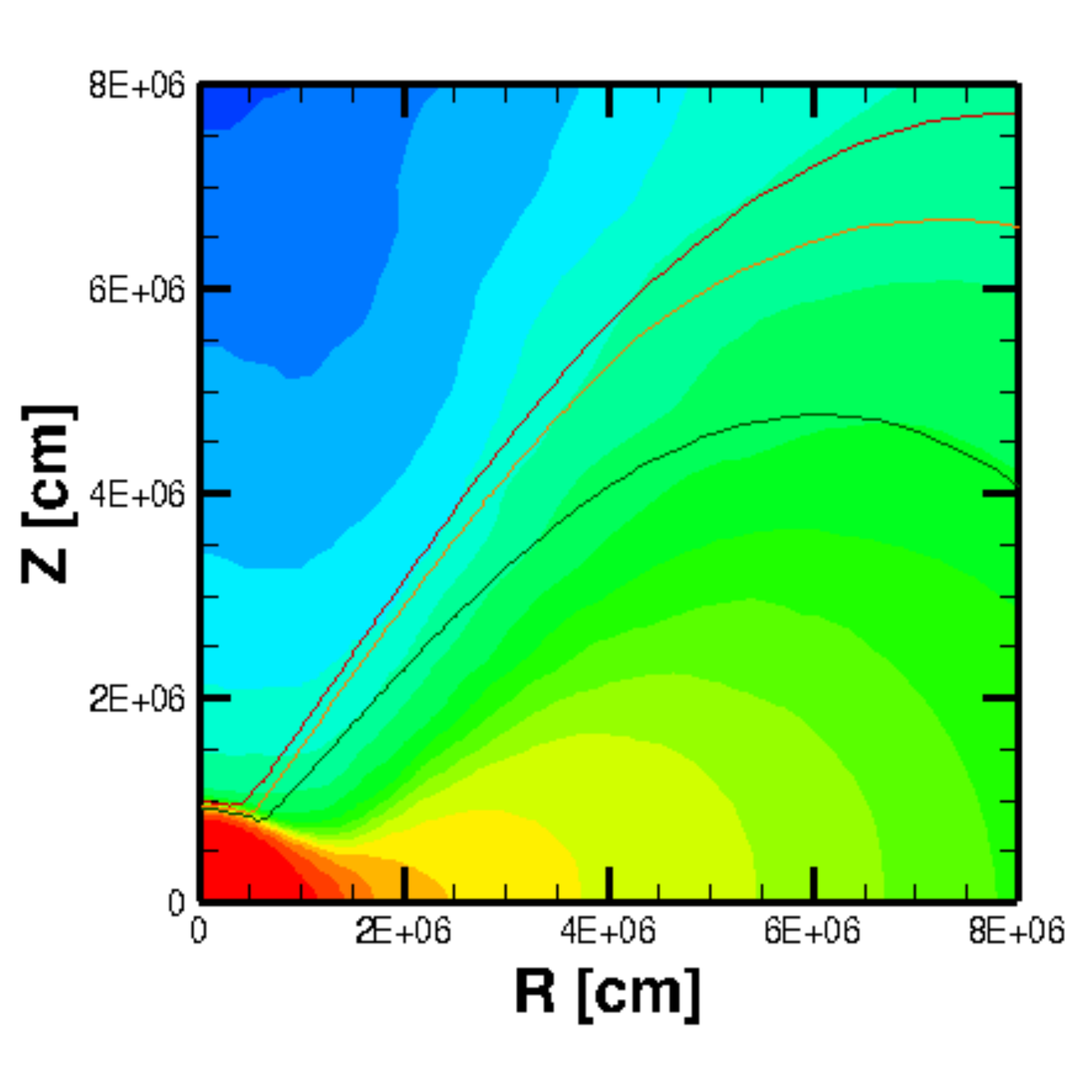}
\caption{Locations of the neutrinosphere for neutrino energy of 4.9 (left), 34 (middle), and 89 (right) MeV are drawn on the contour plots of rest-mass density.  
The lines denote the locations in the same way as Fig. \ref{fig:nu_sphere_xxxx_t000ie06}.  
\label{fig:nu_sphere_dens_t000iexx}}
\end{figure}

The degree of deformation and equatorial extension of neutrinosphere strongly depends on the neutrino energy.  Figure \ref{fig:nu_sphere_dens_t000iexx} displays the locations of neutrinosphere for different energies.  The shapes of the neutrinospheres are more compact for 4.9 MeV and more extended for 34 and 89 MeV as compared with those for 13 MeV in Fig. \ref{fig:nu_sphere_xxxx_t000ie06}.  The positions of neutrinosphere above the remnant neutron star are rather close for different energies because of steep matter density gradient.  Fluxes of high energy neutrinos tend to be focused along the $z$-axis as a consequence.  The strong dependence of neutrino flux on the energy may have influence on the composition of ejecta and resulting product of nucleosynthesis.  

\subsection{Heating rates} \label{sec:nu-transfer_heat}

\begin{figure}[ht!]
\epsscale{0.5}
\plotone{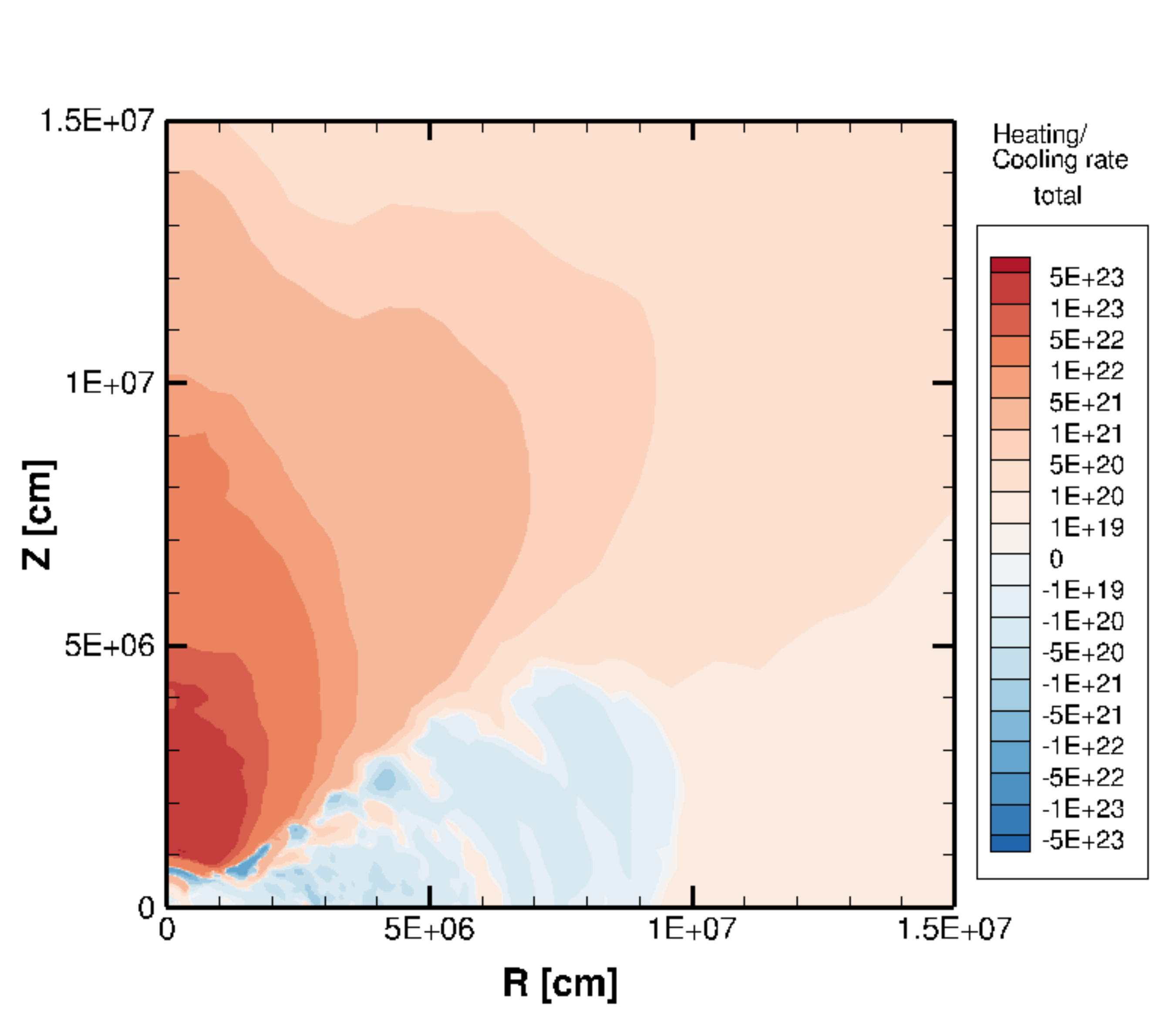}
\caption{Specific heating and cooling rates [erg g$^{-1}$ s$^{-1}$] are shown by contour plots in reddish (heating) and bluish (cooling) colors.  
\label{fig:nu_heating_t000}}
\end{figure}

Energy transfer to matter through neutrino reactions 
plays an important role in the mass ejection from the merger remnant and the resulting nucleosynthesis.  We show the contour map of the specific heating rate in Fig. \ref{fig:nu_heating_t000}.  The heating is most efficient in the region above the remnant neutron star.  The specific heating rate is larger than 10$^{23}$ erg g$^{-1}$ s$^{-1}$ at the radius around 10--40 km.  The region of strong heating extends over 45 degrees.  The cooling proceeds in the limited region of the neutron star and torus.  The total heating rate 
amounts to 1.2$\times 10^{52}$~erg s$^{-1}$ by volume integral of local heating rates in the heating region.  
The corresponding value in \cite{fuj17} is $\sim3\times10^{51}$\,erg\,s$^{-1}$, and thus, the heating rate in the simulation of the Boltzmann equation is $\sim4$ times larger than that in the simulation with energy-integrated leakage-based scheme.
This difference is due to the stronger pair annihilation heating because of the larger luminosity in the simulation of the Boltzmann equation (see \S \ref{sec:nu-transfer_distrib} and \S \ref{sec:nu-transfer_emis}).



\begin{figure}[ht!]
\epsscale{0.37}
\plotone{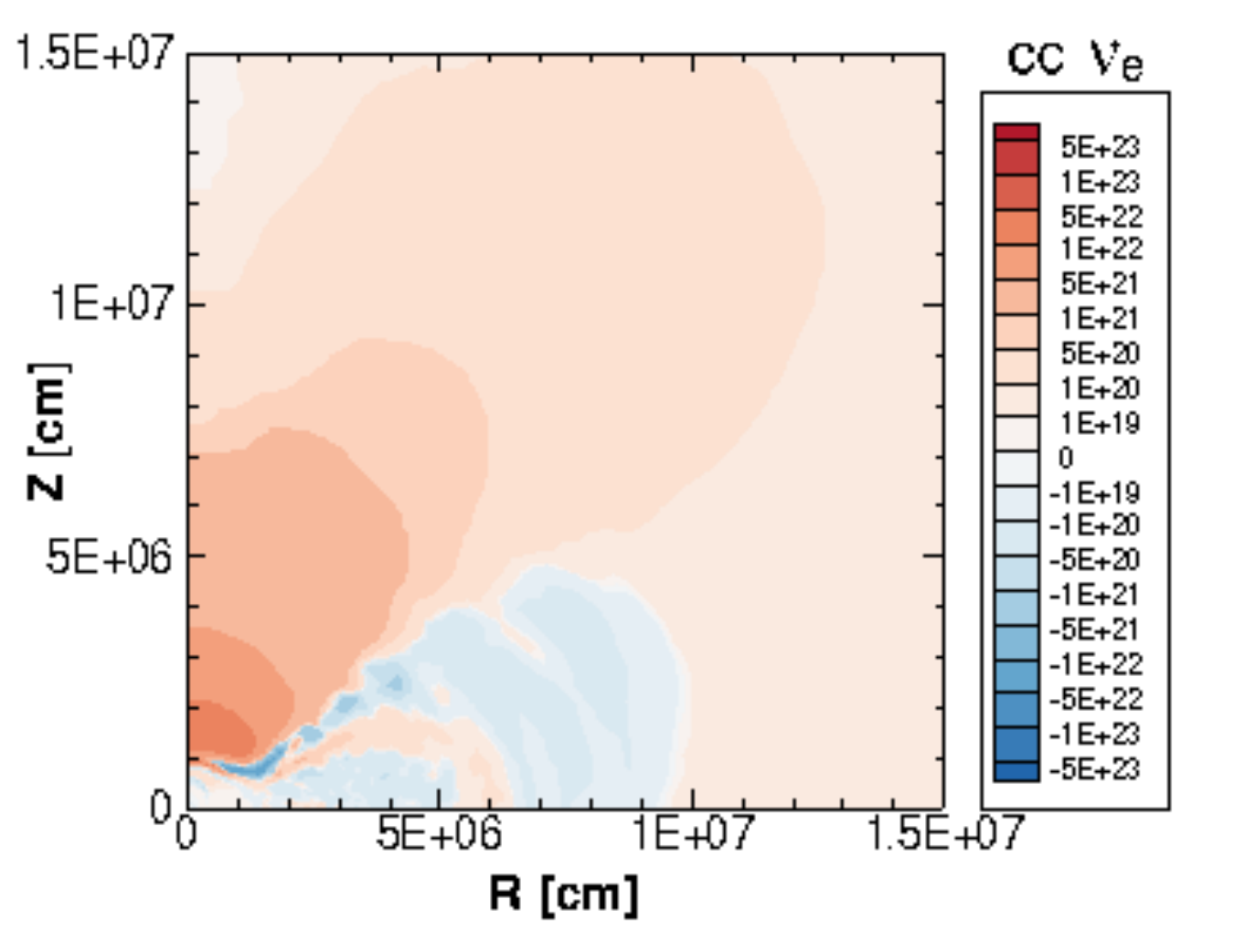}
\plotone{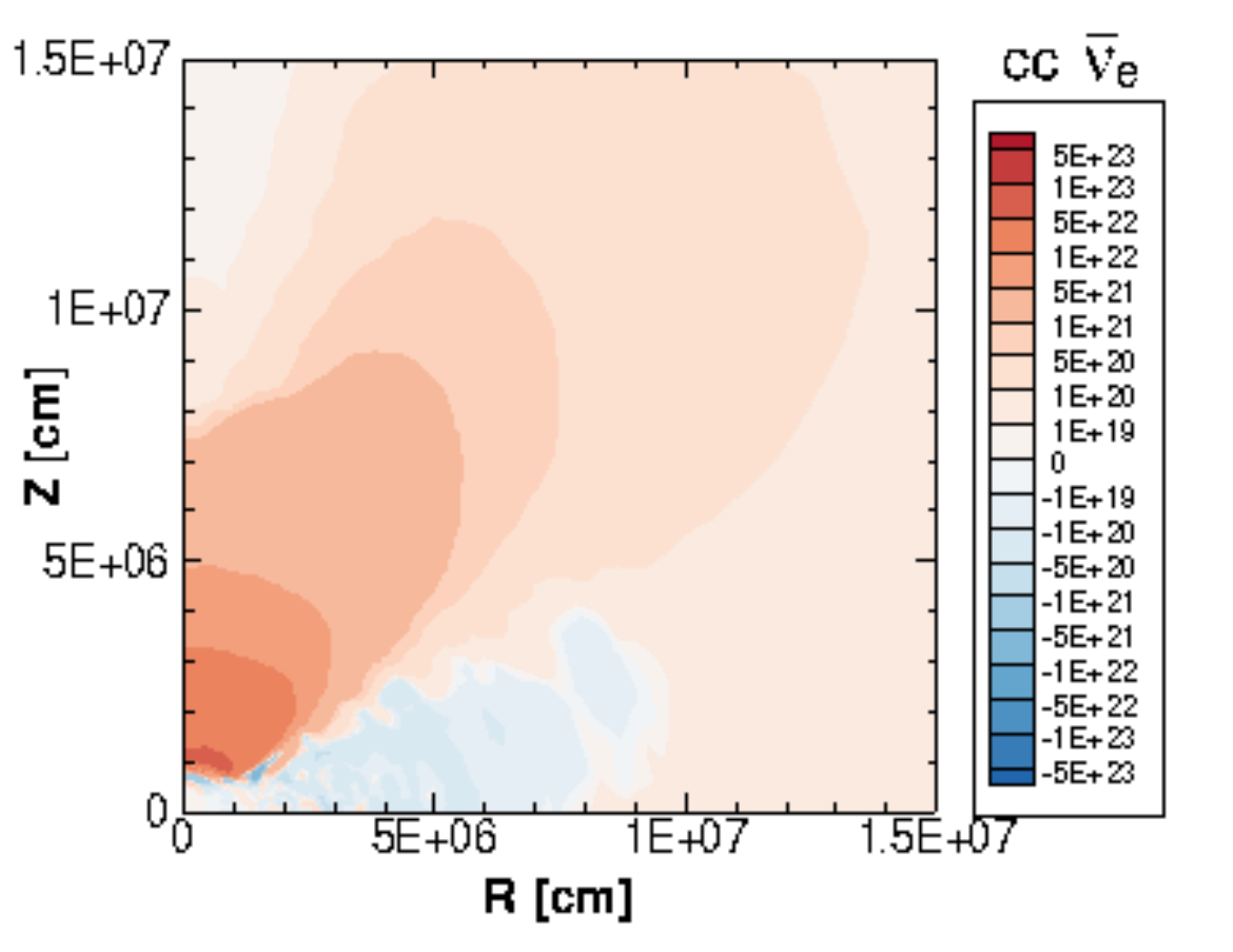}\\
\plotone{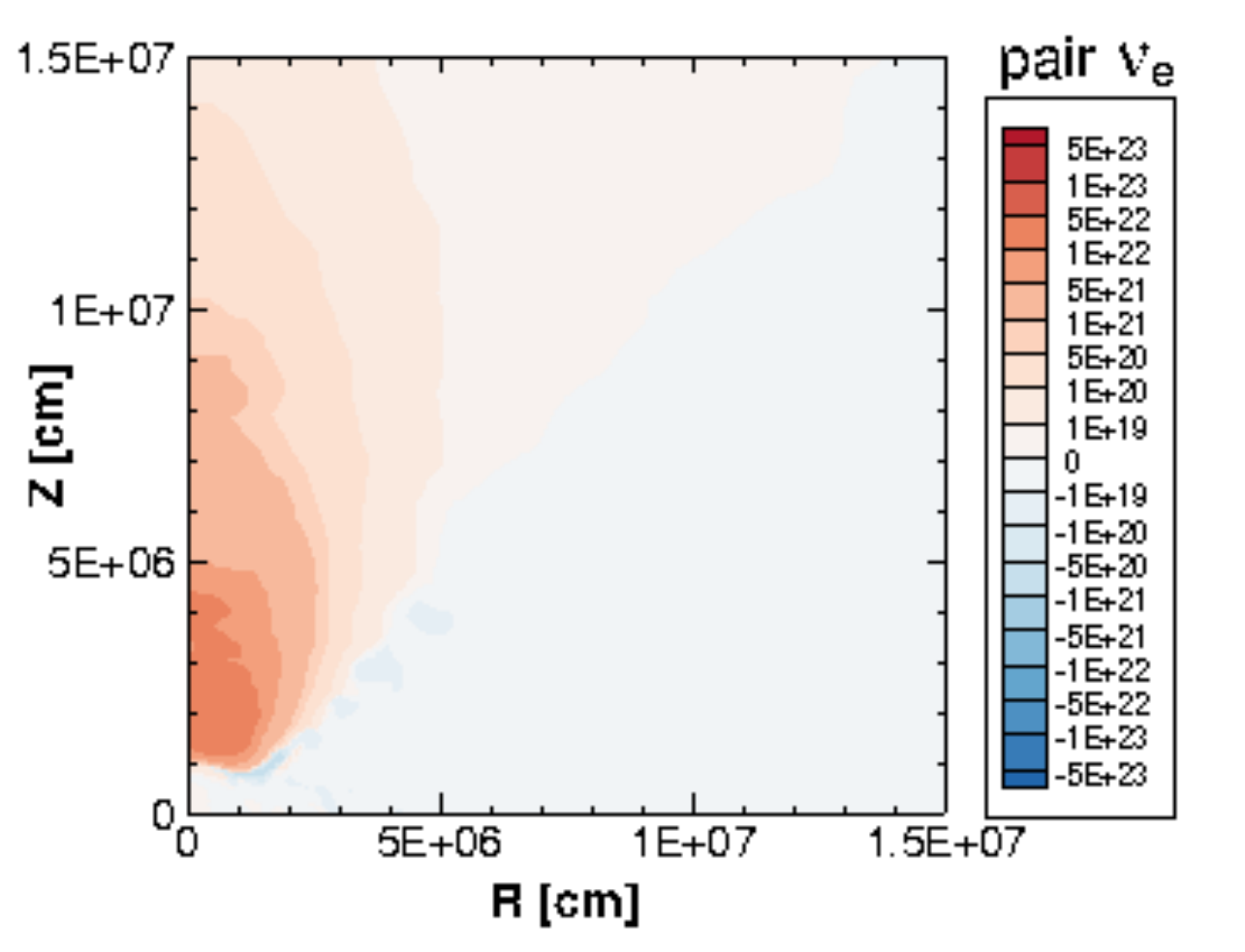}
\plotone{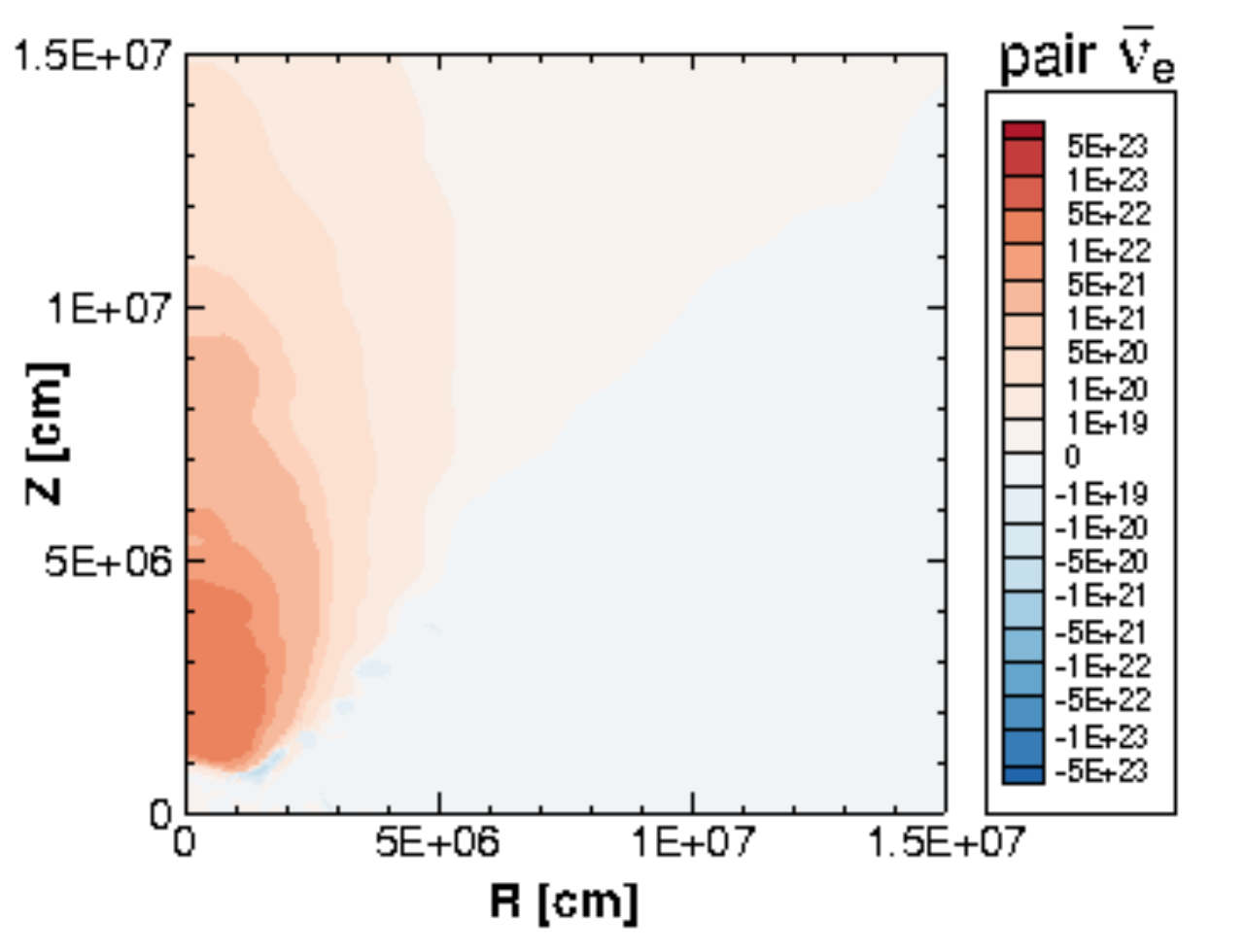}
\plotone{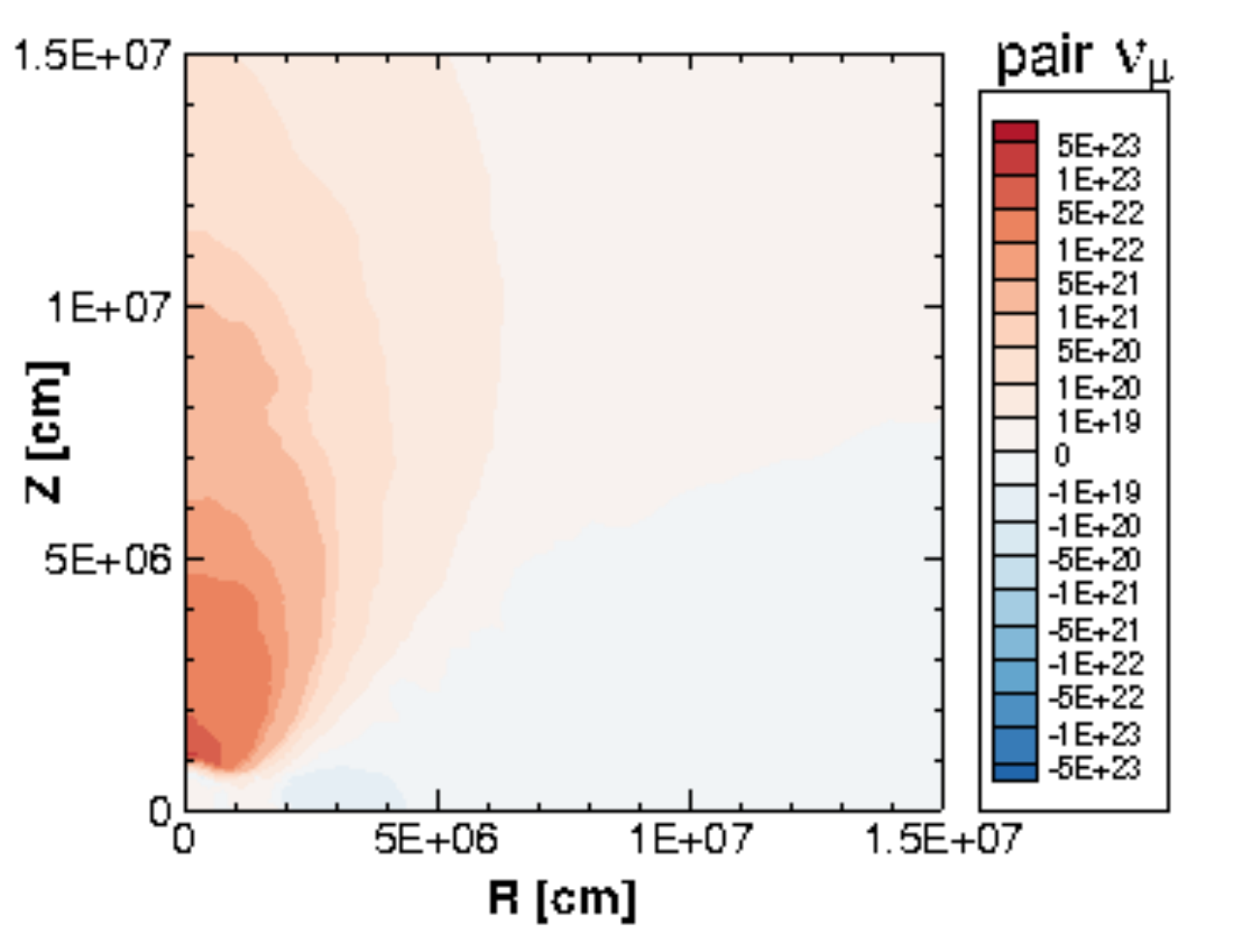}
\caption{Specific heating and cooling rates [erg\,g$^{-1}$s$^{-1}$] for each reaction are shown by contour plots in reddish (heating) and bluish (cooling) colors.  Contributions of heating and cooling are plotted for the charged current reactions for $\nu_e$ (upper left) and $\bar{\nu}_e$ (upper right) with nucleons, the pair creation and annihilation for $\nu_e$ (lower left), $\bar{\nu}_e$ (lower middle) and $\nu_\mu$ (lower right). 
\label{fig:nu_heating_t000_xxxx}}
\end{figure}

We show in Fig.\ref{fig:nu_heating_t000_xxxx} the specific heating rate for each neutrino reaction.  The heating by the charged current reactions with neutrons and protons extends in a region above the neutron star as well as the torus for electron-type (anti-)neutrinos.  The heating by the pair annihilation reaction is effective in a narrow region above the neutron star for which the matter density is low and neutrino fluxes are high for three species.  The contribution of heating by nucleon-nucleon bremsstrahlung is minor (not shown here) though it is important for emissions (see below).  

\subsection{Emission} \label{sec:nu-transfer_emis}

\begin{figure}[ht!]
\epsscale{0.4}
\plotone{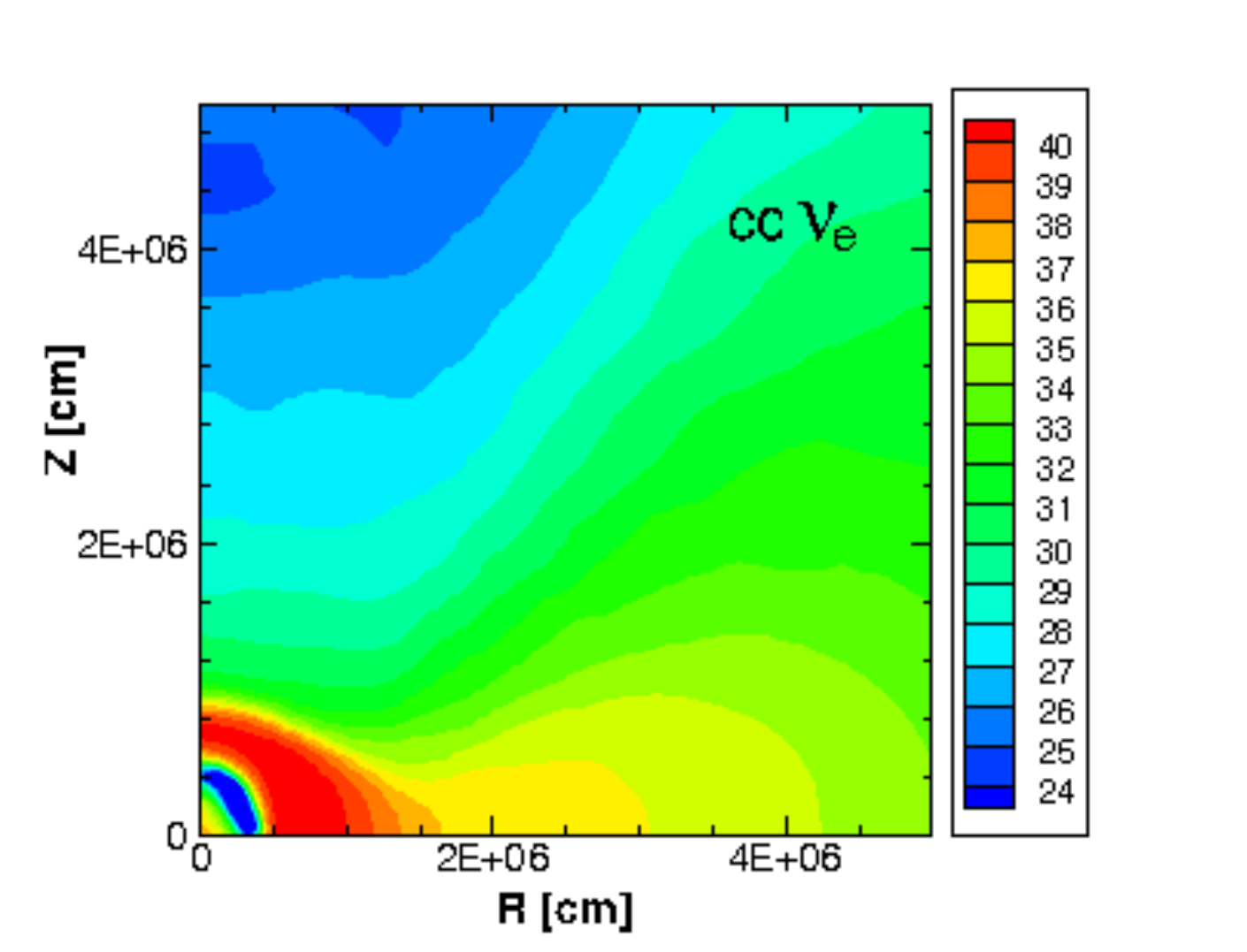}
\plotone{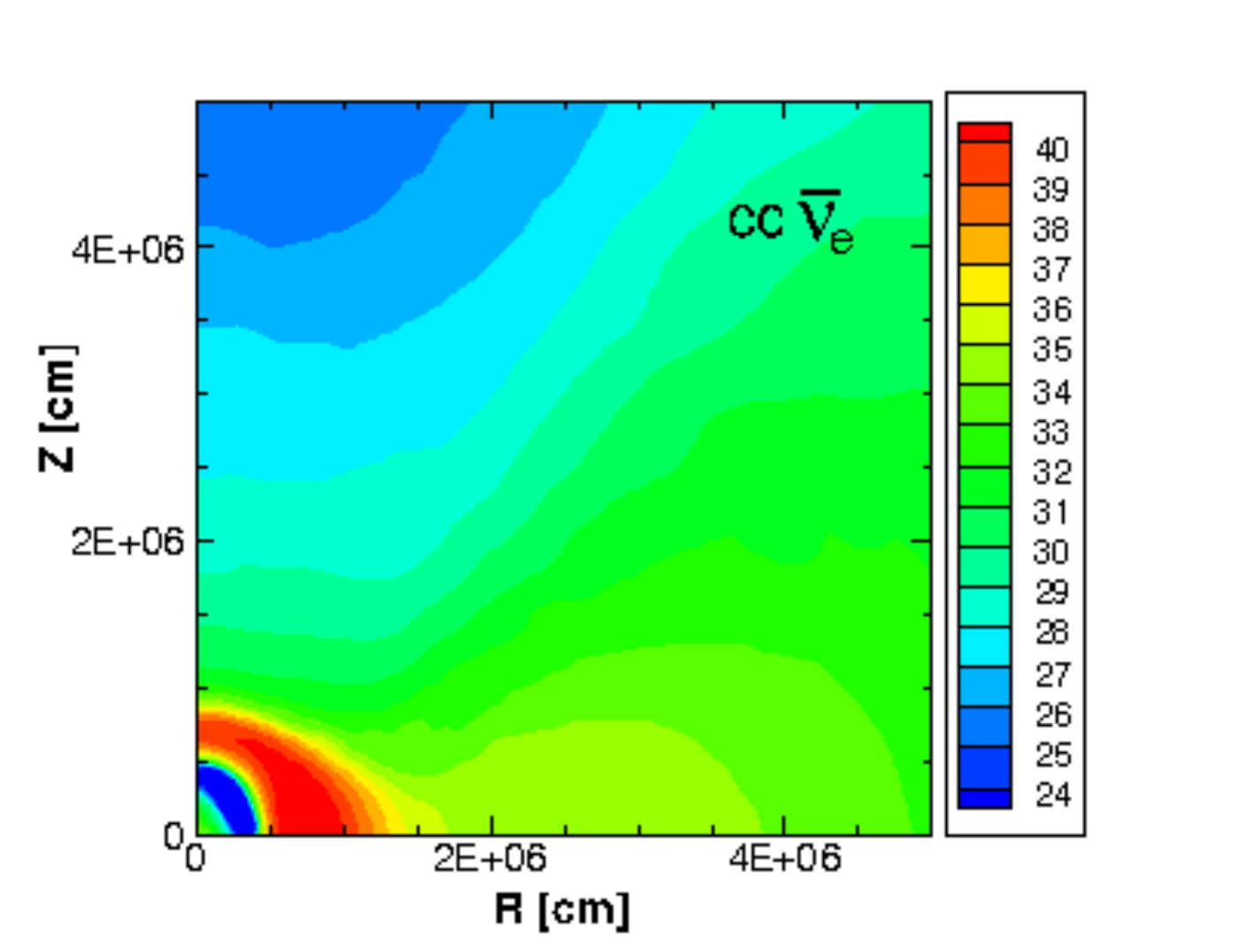}\\
\plotone{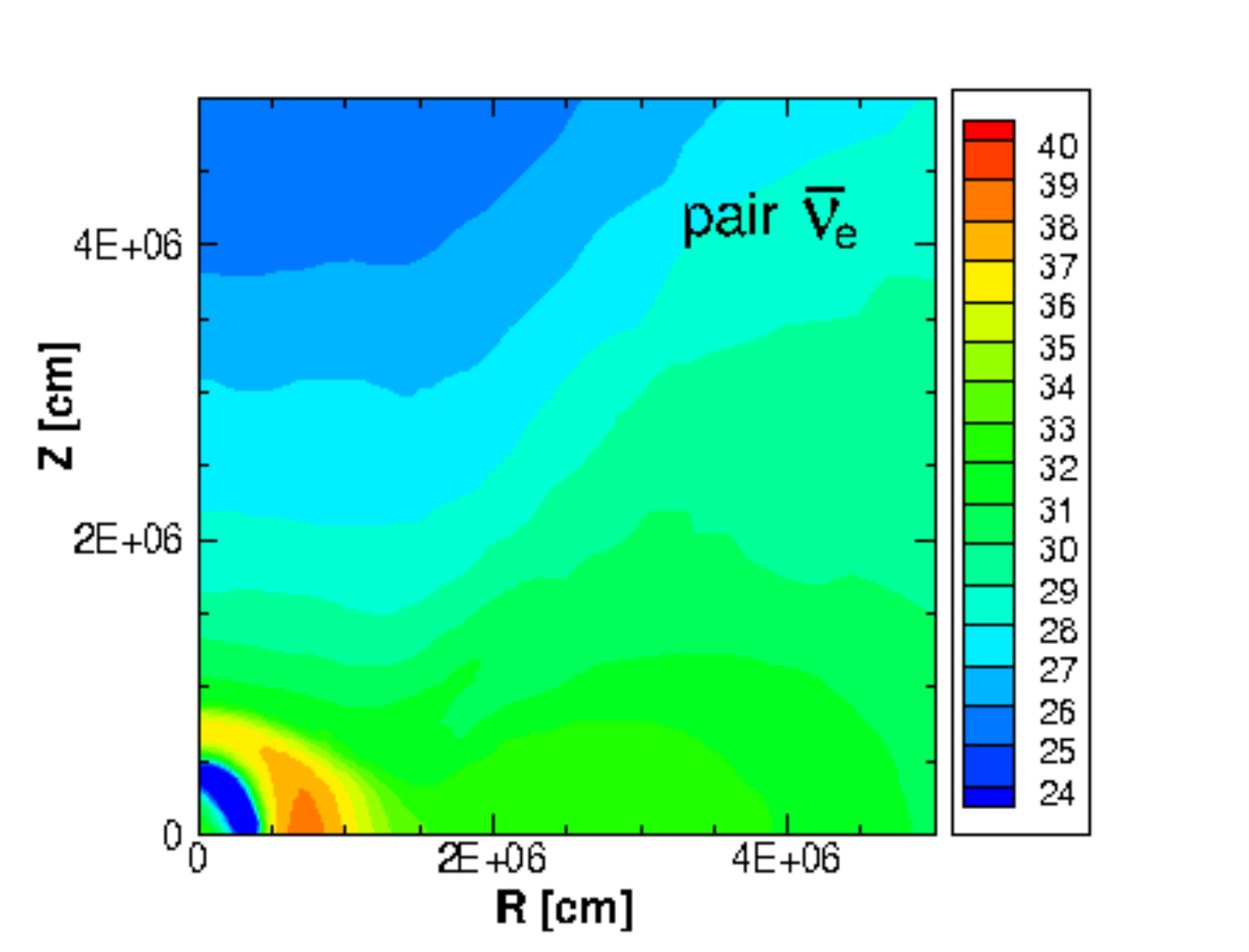}
\plotone{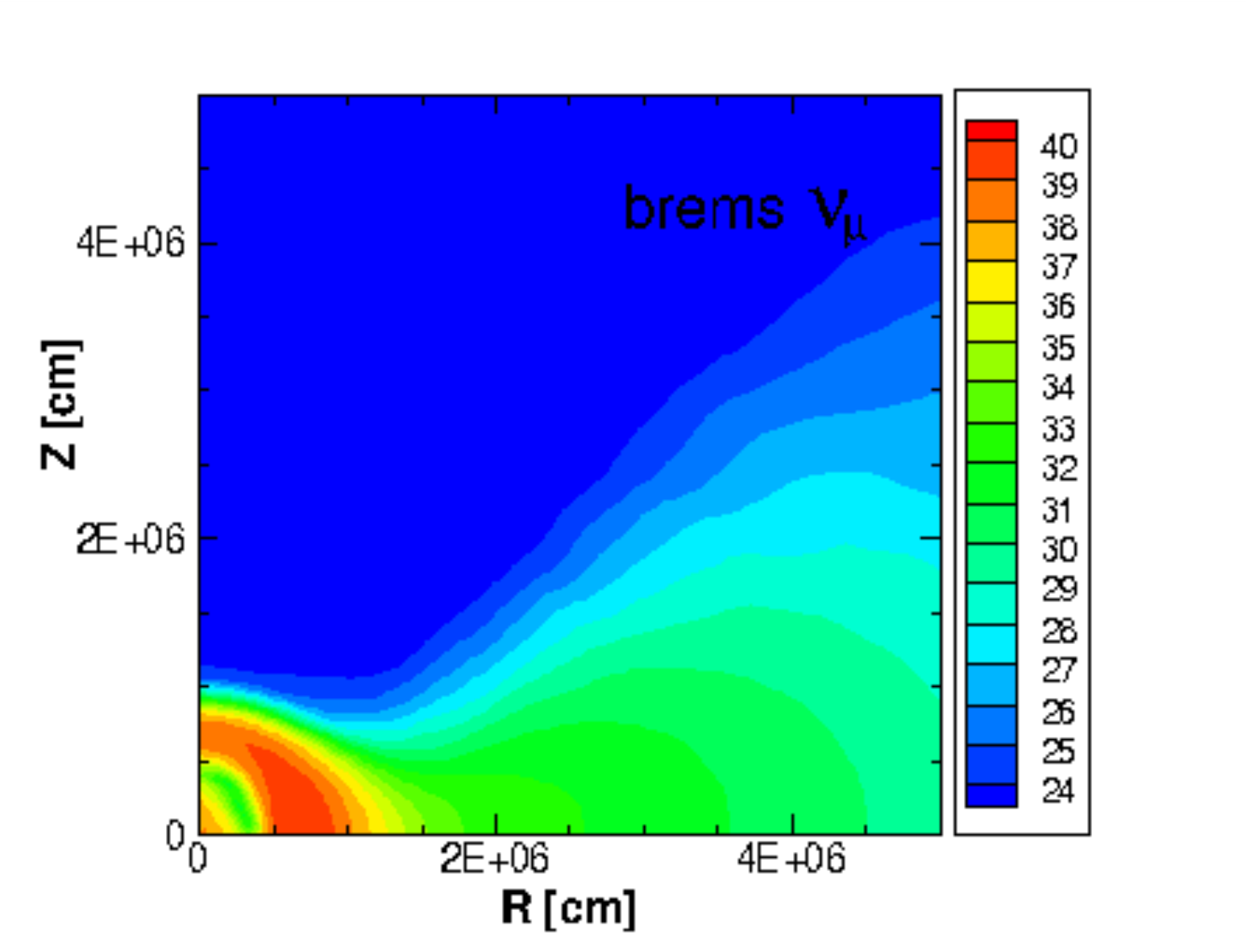}
\caption{Emission rates [erg\,cm$^{-3}$s$^{-1}$] in log scale for several reactions are shown by contour plots.  Contributions from the charged current reactions with protons (upper left) and neutrons (upper right), the pair creation (lower left) for $\bar{\nu}_e$ and the nucleon-nucleon bremsstrahlung (lower right) for $\nu_\mu$ are shown.  
\label{fig:nu_emission_t000_xxxx}}
\end{figure}

The intense heating in the region above the neutron star is caused by the influence of neutrino emission from the merger remnant.  We examine the neutrino emission rates for several reactions in Fig. \ref{fig:nu_emission_t000_xxxx}.  For  $\nu_e$ and $\bar{\nu}_e$, the charged current reactions with nucleons are dominant.  
The region of strong emissivity extends in the torus especially for $\nu_e$ due to the contribution of the electron capture on protons.  
The neutrino emission by pair creations proceeds mainly through the nucleon-nucleon bremsstrahlung for all species, which is more efficient than the electron-positron pair annihilation.  For $\mu$- and $\tau$-type neutrinos, the emission of pairs ($\nu_\mu$,  $\bar{\nu}_{\mu}$, $\nu_\tau$ and $\bar{\nu}_{\tau}$) also proceeds dominantly through the nucleon-nucleon bremsstrahlung.  These pairs of neutrinos contribute to the heating through the neutrino pair annihilation in the low matter density region while the electron-type (anti-)neutrinos additionally contribute to the heating through the absorption on nucleons off the region where the matter density is relatively high as shown above.  

\section{Evaluation of energy and angle moments} \label{sec:moment_evaluation}
\subsection{Angle moments} \label{sec:nu-transfer_moments}

\begin{figure}[ht!]
\epsscale{1.1}
\plotone{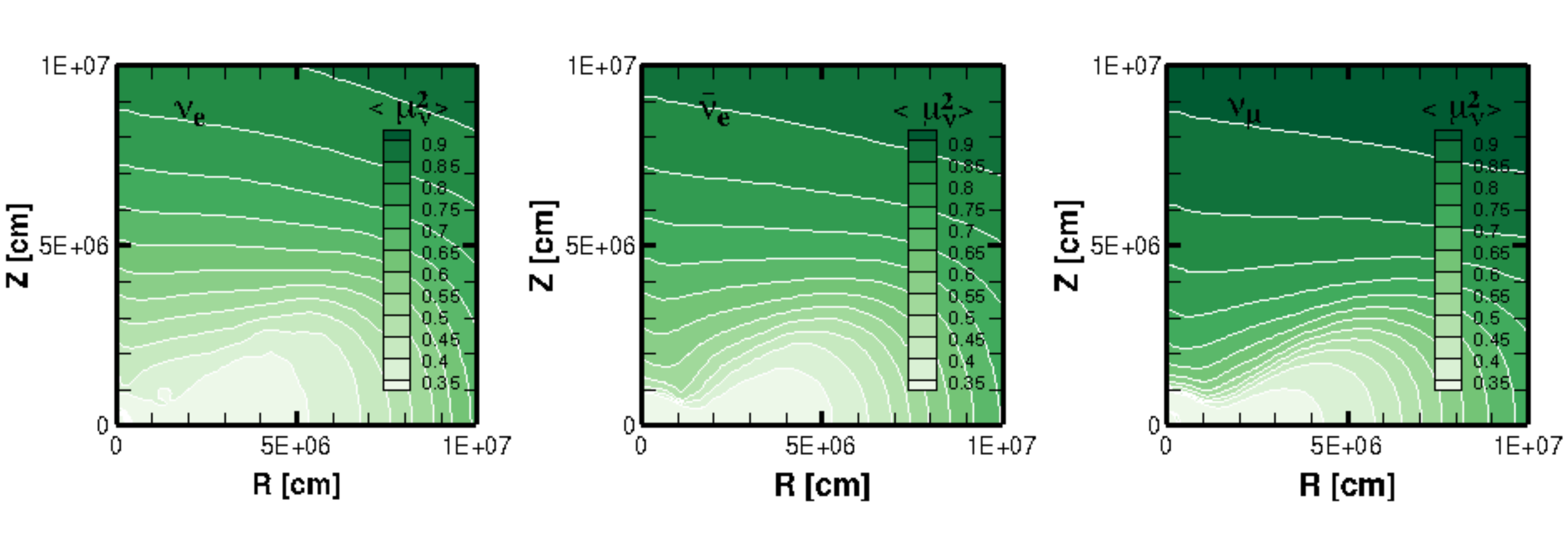}
\caption{The average value of squared angle moment, $\langle \mu_{\nu}^2 \rangle$, where the angle factor is defined as $\mu_{\nu}=\cos \theta_{\nu}$, are shown for $\nu_e$ (left), $\bar{\nu}_e$ (middle) and $\nu_\mu$ (right) by contour plots.  
\label{fig:nu_moment_t000_inx}}
\end{figure}

It is advantageous that the direct solution of the Boltzmann equation can provide the angle distributions for multi-energy at the whole spatial grids.  We examine the angle moments from the neutrino distribution function in five dimensions.  As an example, Fig. \ref{fig:nu_moment_t000_inx} displays the average value of the squared angle moment, $\langle \mu_{\nu}^2 \rangle$, where the angle factor is defined as $\mu_{\nu}=\cos \theta_{\nu}$, for three species.  The region with $\langle \mu_{\nu}^2 \rangle=\frac{1}{3}$ (isotropic) is not spherical but largely deformed including the torus along the equatorial plane.  The shape of contour lines follows the torus near the equator and becomes an oblate spheroid at the outer region.  The squared angle moment increases toward 1 (forward peak) at large distances.  

\begin{figure}[ht!]
\epsscale{0.35}
\plotone{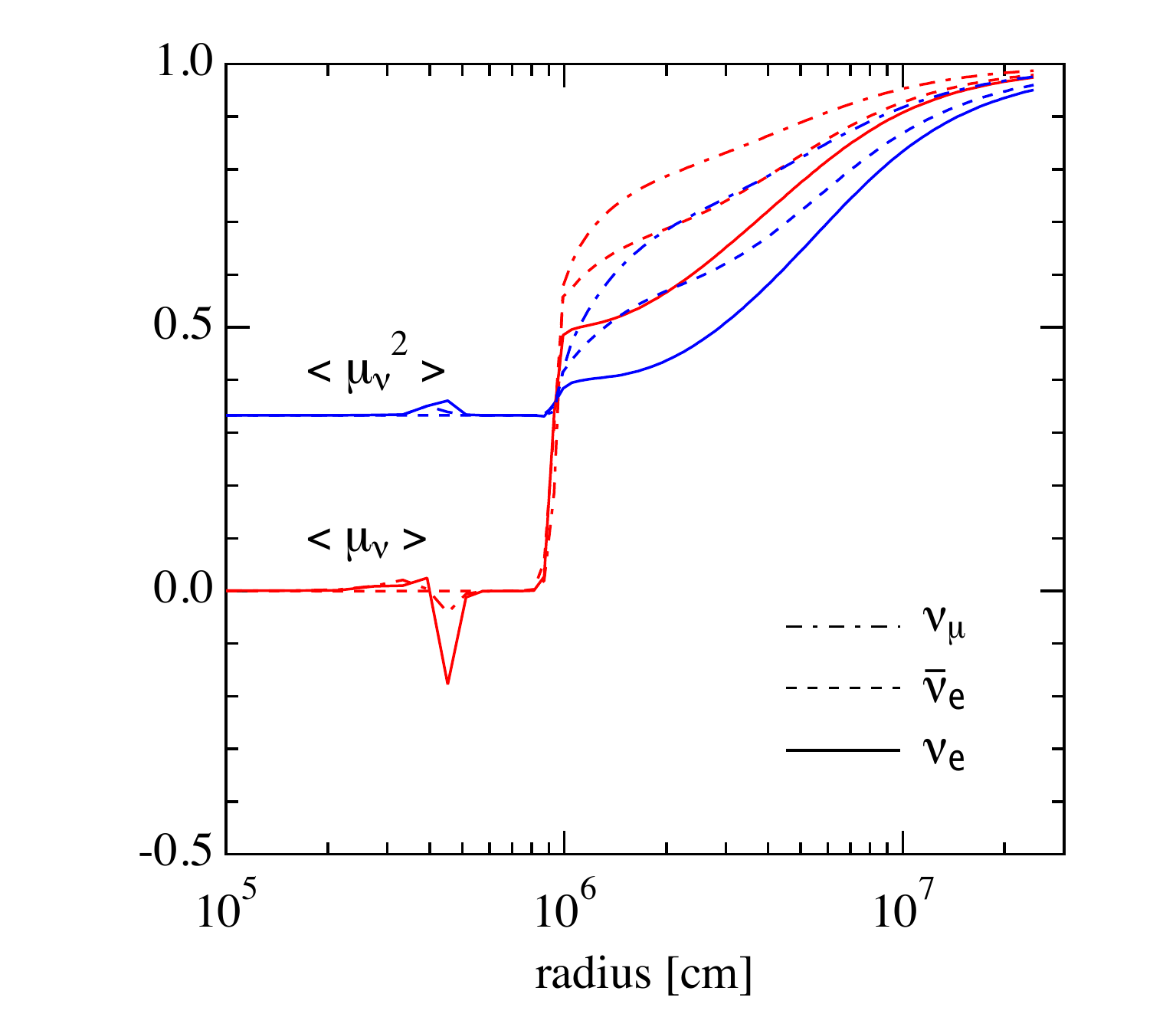}
\plotone{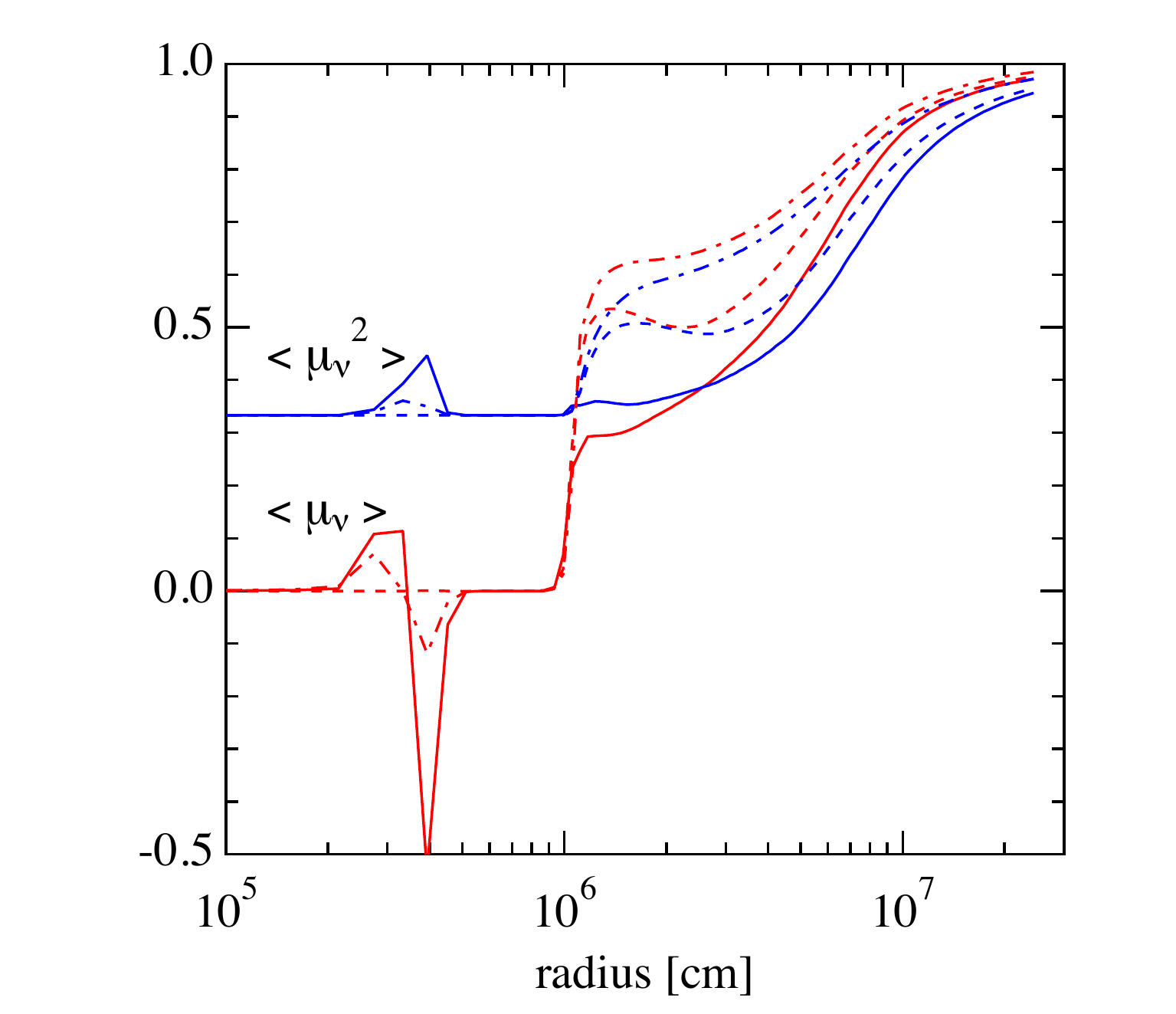}
\plotone{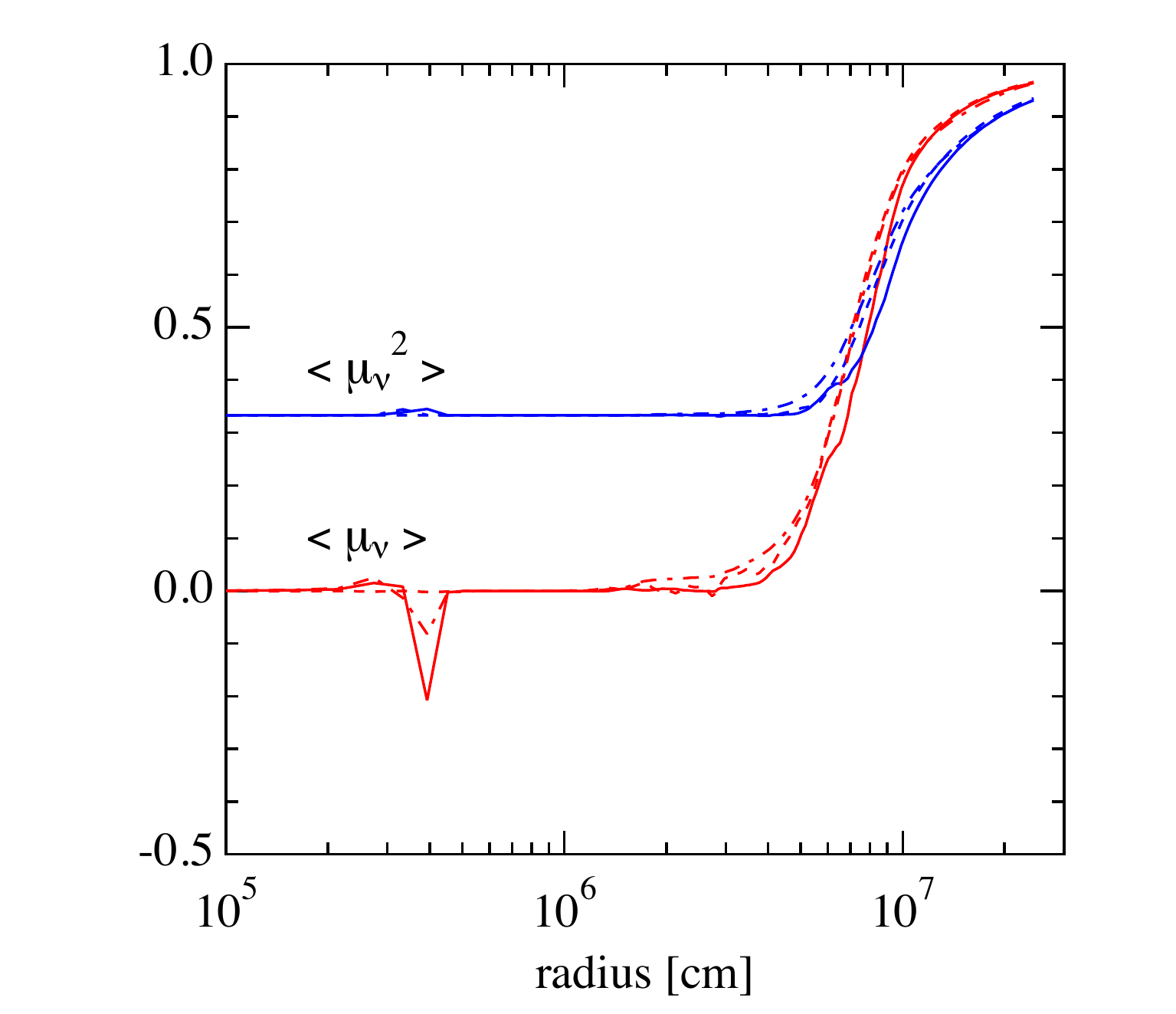}
\caption{The average values of angle moments $\langle \mu_{\nu} \rangle$ (red lines) and $\langle \mu_{\nu}^2 \rangle$ (blue lines) are shown 
for $\nu_e$ (solid line), $\bar{\nu}_e$ (dashed line) and $\nu_\mu$ (dash-dotted line) as a function of radius.  The profiles along three polar angles at 9.8 (above the neutron star), 48, (above the torus) and 90 (along the equator) degree from the $z$-axis are shown in left, middle and right panel, respectively.  
\label{fig:nu_moment_t000_radial_ithxx}}
\end{figure}

Figure \ref{fig:nu_moment_t000_radial_ithxx} displays the behavior of the radial distributions of angle moments, $\langle \mu_{\nu} \rangle$ and $\langle \mu_{\nu}^2 \rangle$ along three polar angles.  It is interesting to see that these quantities behave a non-monotonic manner depending on the direction.  The angle moments along the equator (right panel) remain at the isotropic values due to the neutrino trapping in the merger remnant and simply increase to the forward peak value.  There are some wiggles within 10 km due to inward and outward flows by the neutrino diffusion near the shell-like high temperature region inside the neutron star (See upper-right panel of Fig. \ref{fig:hydro_set_t000}).  Above the neutron star (left panel), the behavior is similar, but the transition occurs around 10 km.  
Along the direction above the extended torus (middle panel), the angle moments show complicated increase and decrease from isotropic to forward peak due to the shape of torus.  The angle moment $\langle \mu_{\nu} \rangle$ approaches nearly 1 at the outer edge, 250 km.  We note that the convergence of the forward peak value of the angle moments at large distances is achieved in the current simulation with high angular resolution.  The convergence with the various number of angle grids has been checked as described in Appendix \ref{sec:append_resolution}.  

\subsection{Eddington tensor} \label{sec:nu-transfer_tensor}

We analyze the Eddington tensor obtained from the simulation of the Boltzmann equation.  Here, the Eddington tensor can be directly evaluated by integrating the neutrino distribution functions (hereafter, the direct evaluation).  We compare it with the Eddington tensor by the closure relation in order to assess the validity of this approximation used in the moment formalism.  In the closure relation, the pressure tensor is evaluated by assuming relations with the lower moments; energy flux and energy density.  The definition of the Eddington tensor and the procedure to evaluate it using the closure relation can be found in Appendix \ref{sec:append_eddington}.  

Figure \ref{fig:nu_tensor_t000_inx_ie08_diag} displays the $rr$-component of the Eddington tensor for the neutrino energy of 34 MeV obtained by the direct evaluation, the closure relation, and their differences for three species.  
The profile of $rr$-component of the Eddington tensor by the direct evaluation (top) is deformed according to the shape of the merger remnant.  The region of the value with $\frac{1}{3}$, which corresponds to the isotropic distribution, is roughly located inside the neutrinosphere (see middle panel of Fig. \ref{fig:nu_sphere_dens_t000iexx}).  It increases to the limiting value of 1 (forward peak) away from the remnant.  The profile obtained by the closure relation (middle) in general resembles the profile by the direct evaluation.  The difference of $rr$-components between the closure relation and the direct evaluation is shown in the bottom panels.  Note that the plots show the absolute difference, not the relative one, from the direct evaluation.  Although the difference is small in most of region, there is sizable difference ($\sim$0.1) in the region close to both the neutron star and the torus where neutrinos from both two are important.  


\begin{figure}[ht!]
\epsscale{0.35}
\plotone{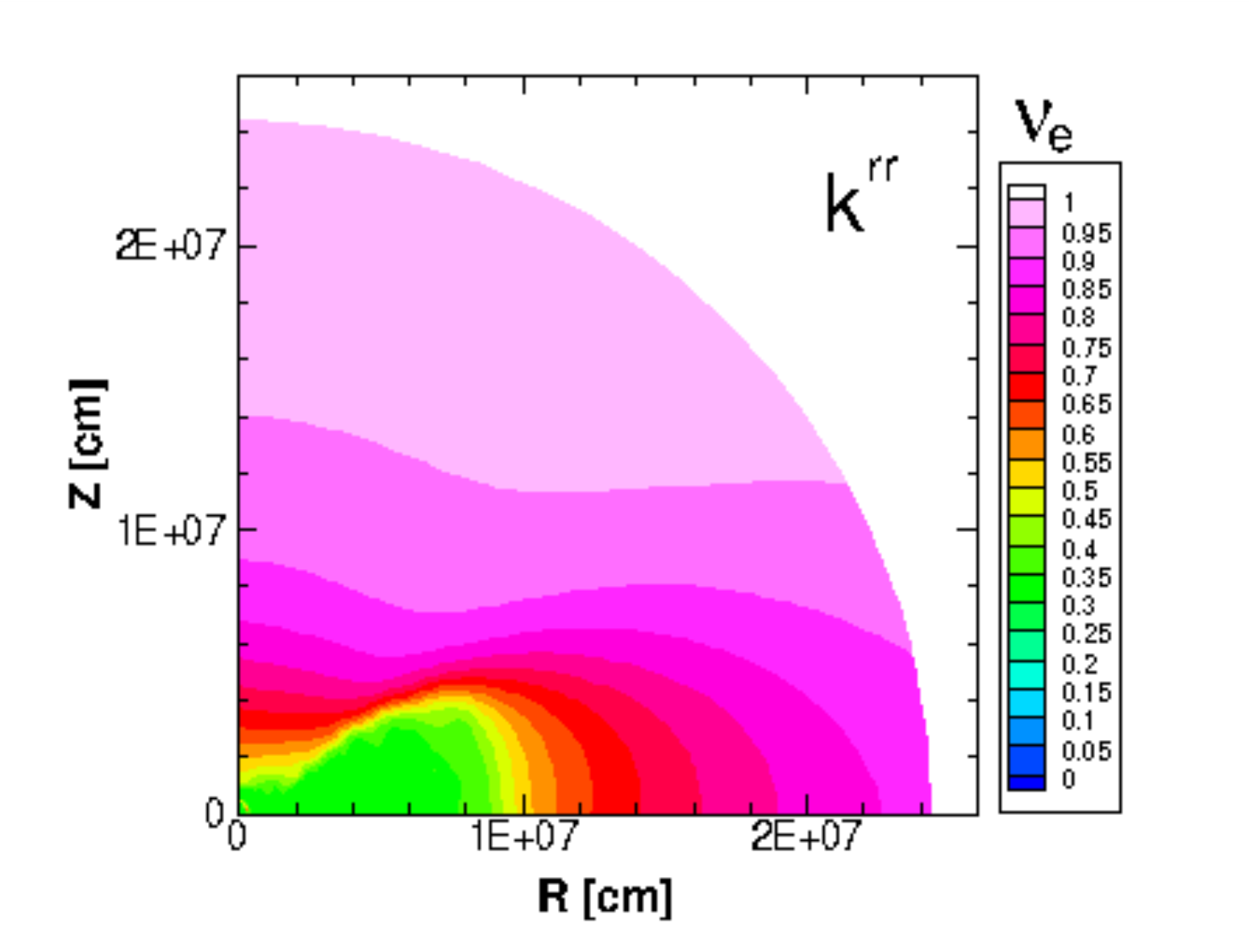}
\plotone{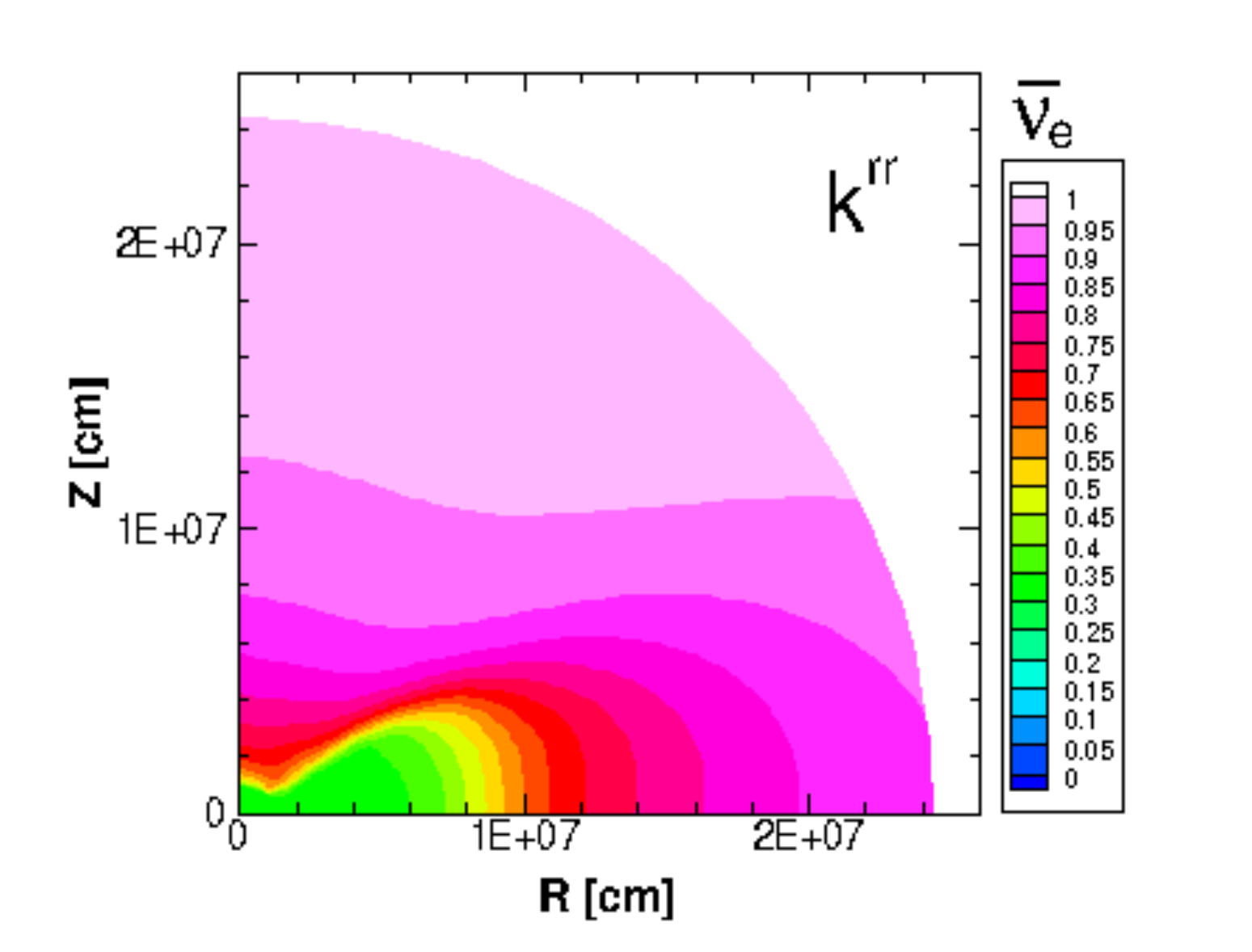}
\plotone{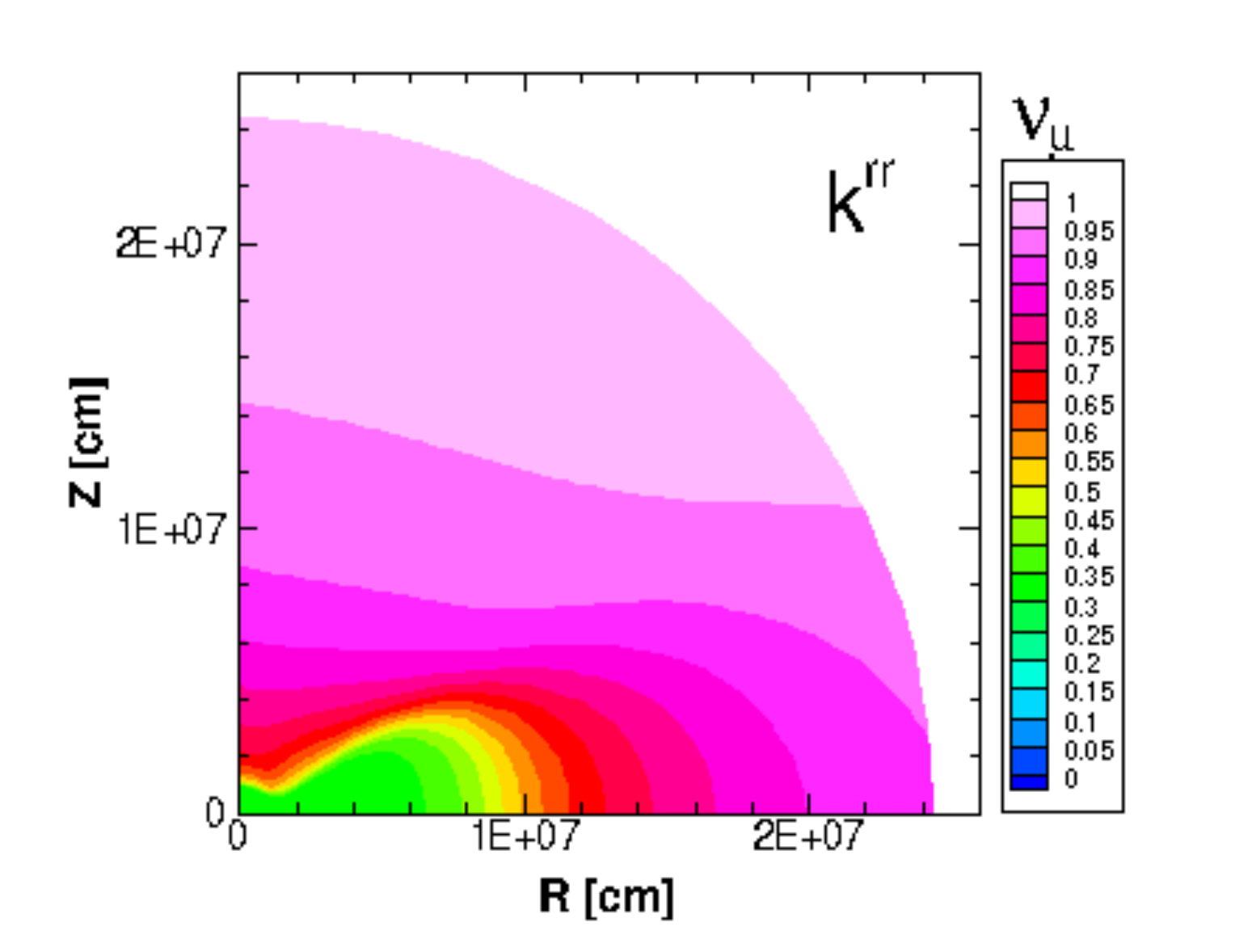}
\plotone{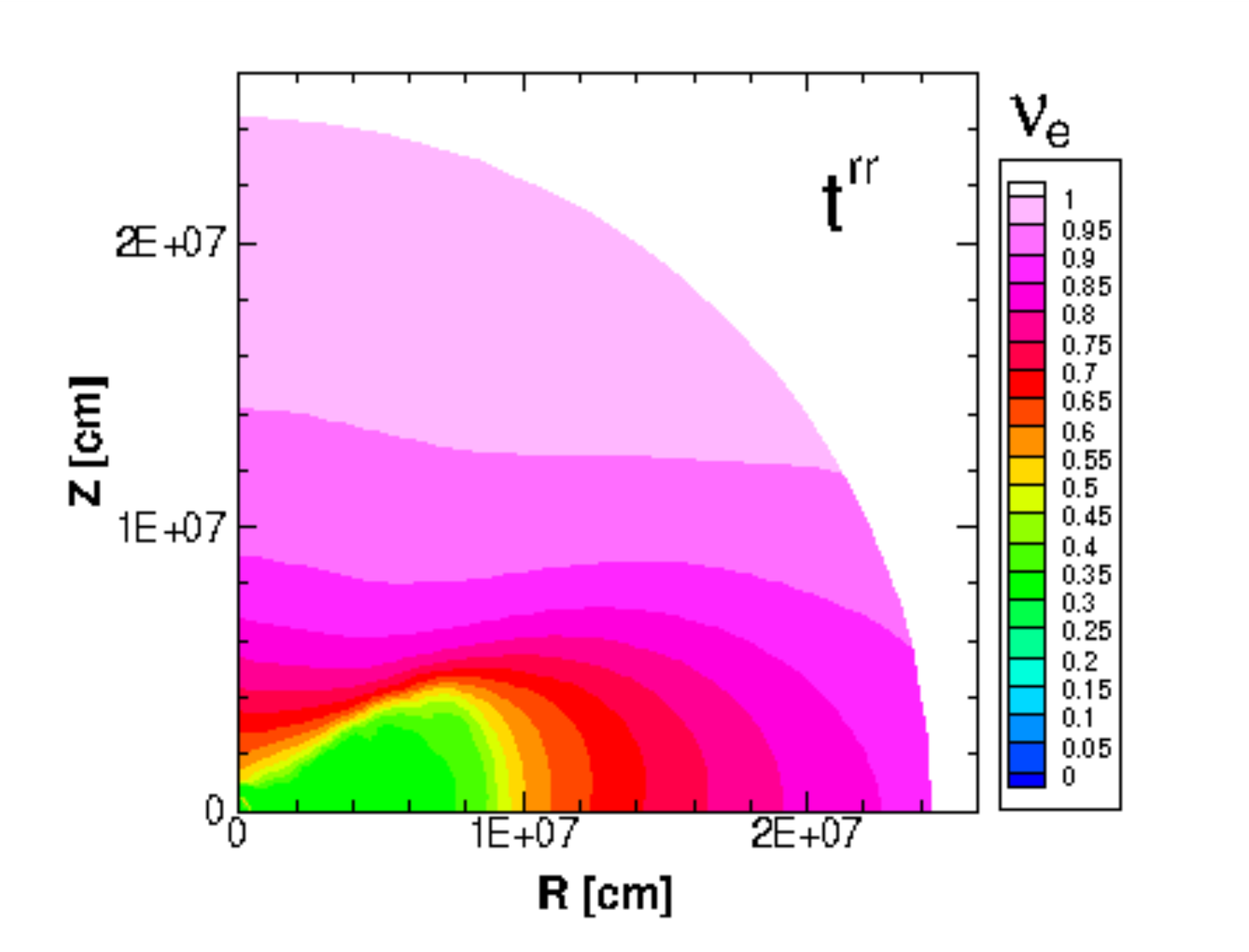}
\plotone{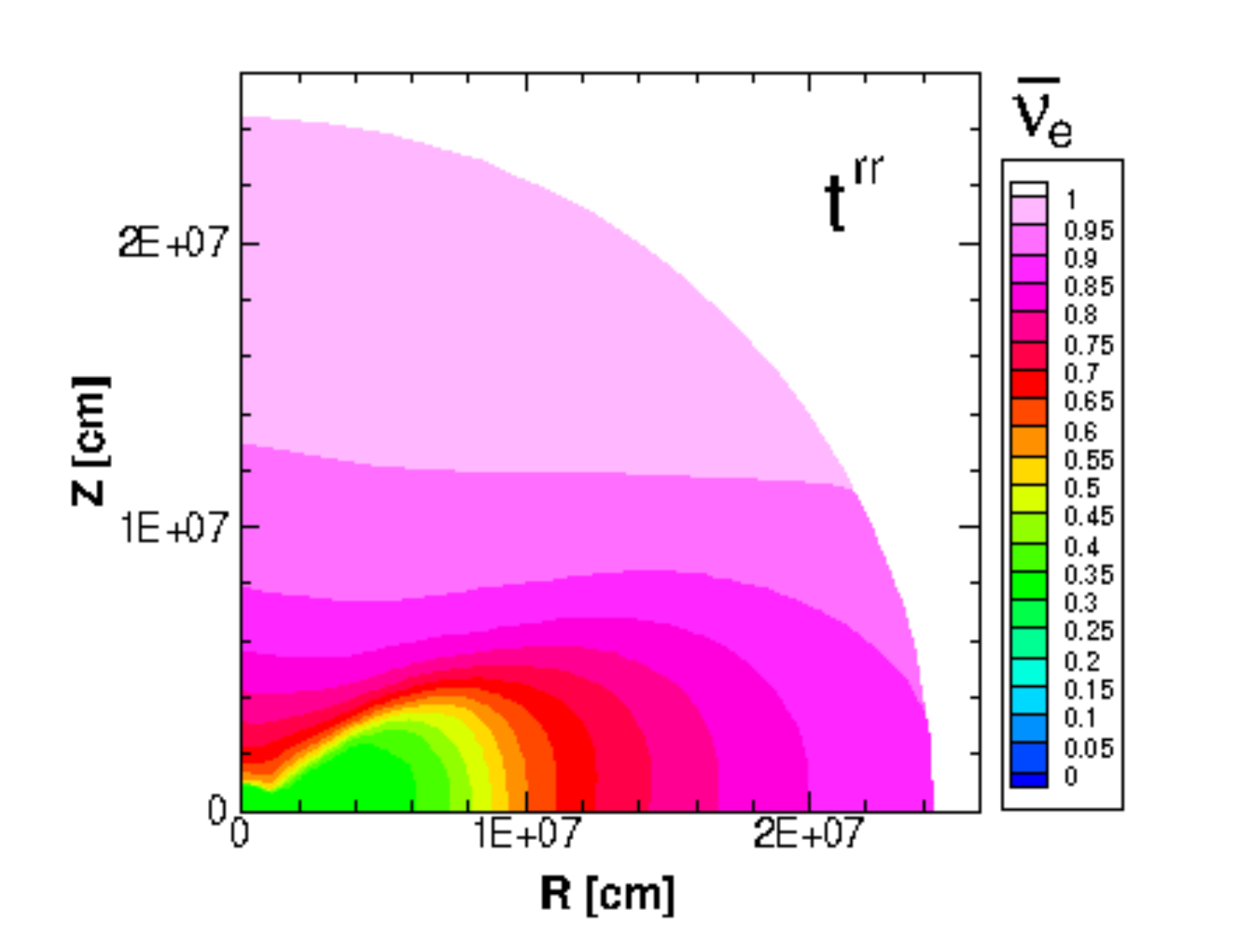}
\plotone{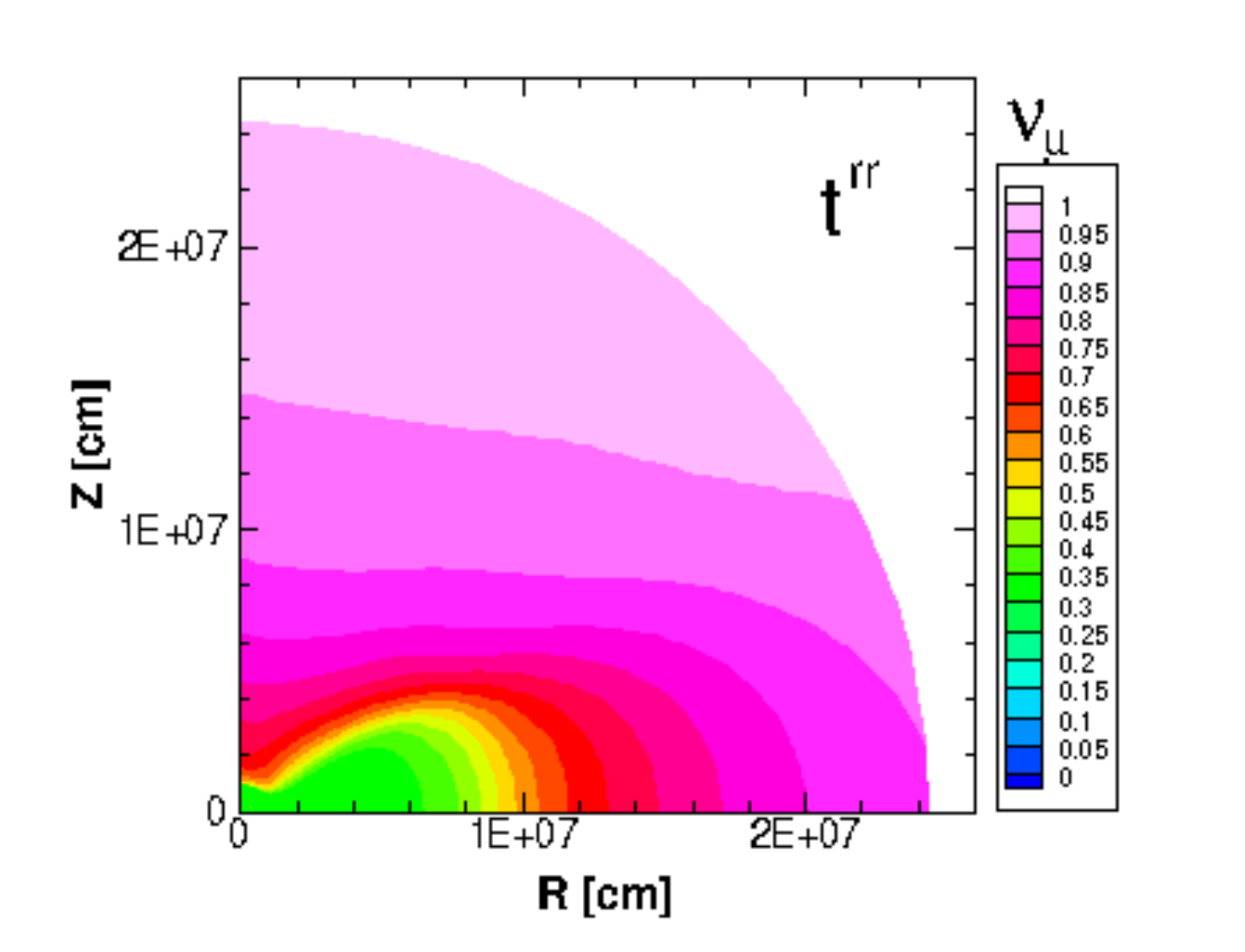}
\plotone{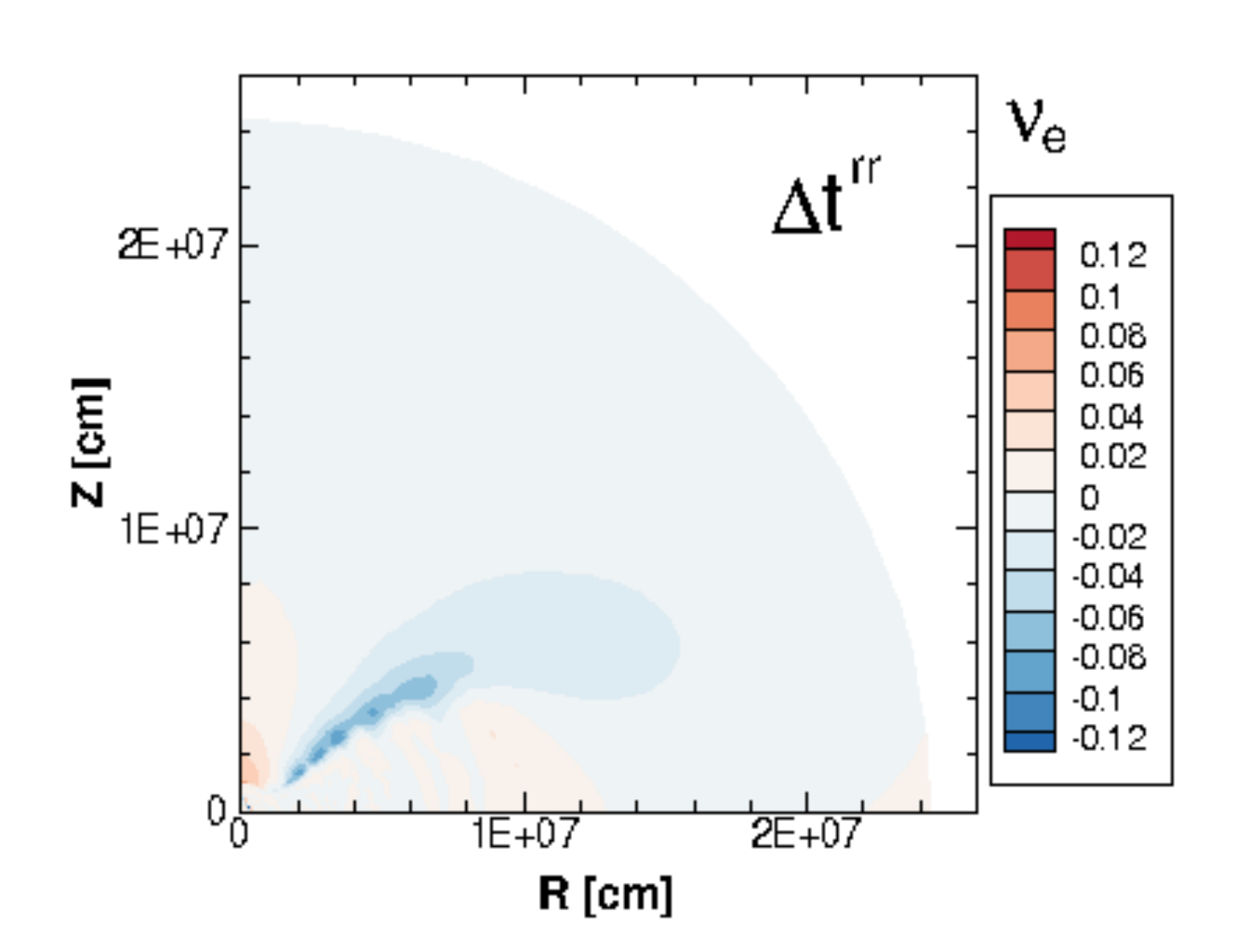}
\plotone{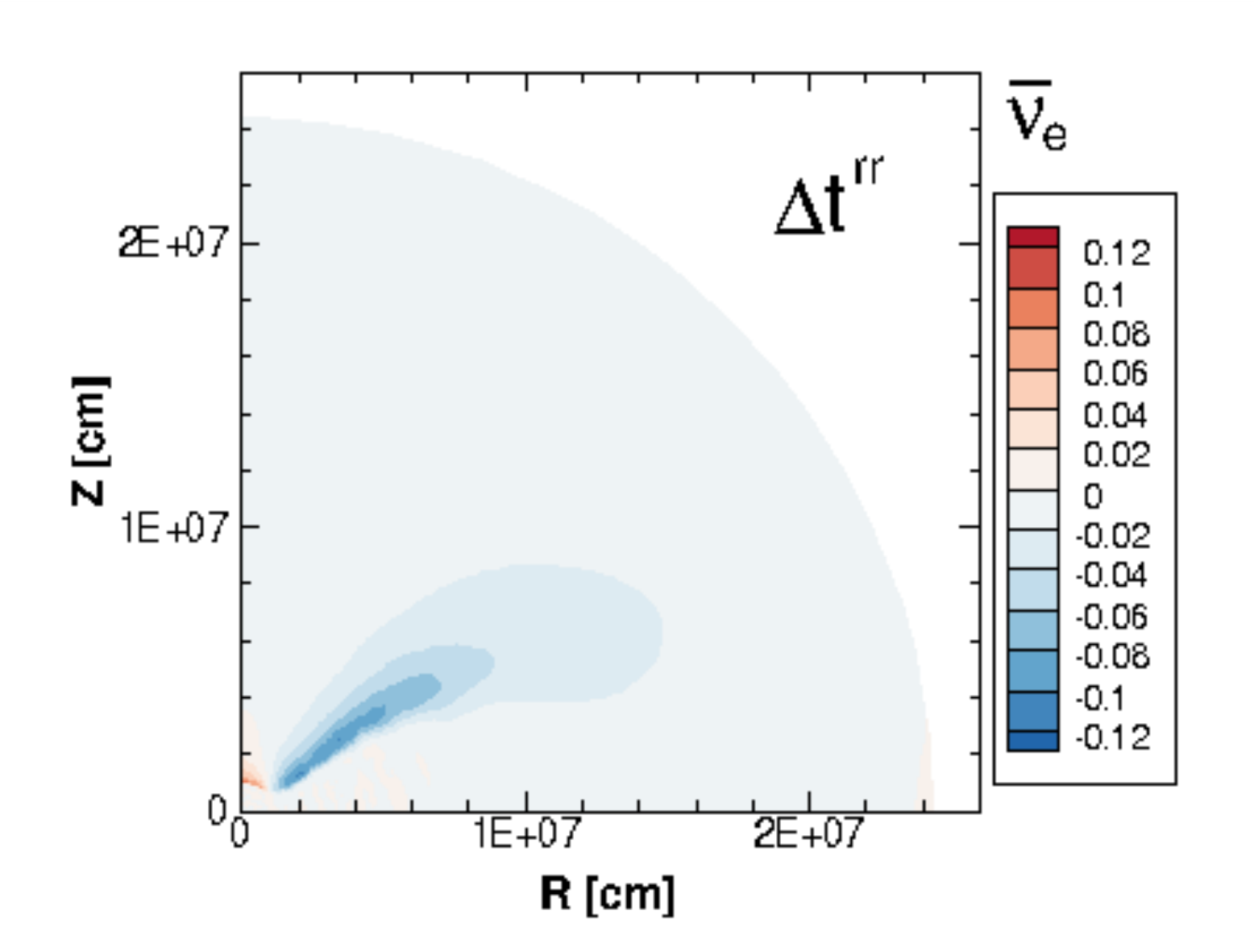}
\plotone{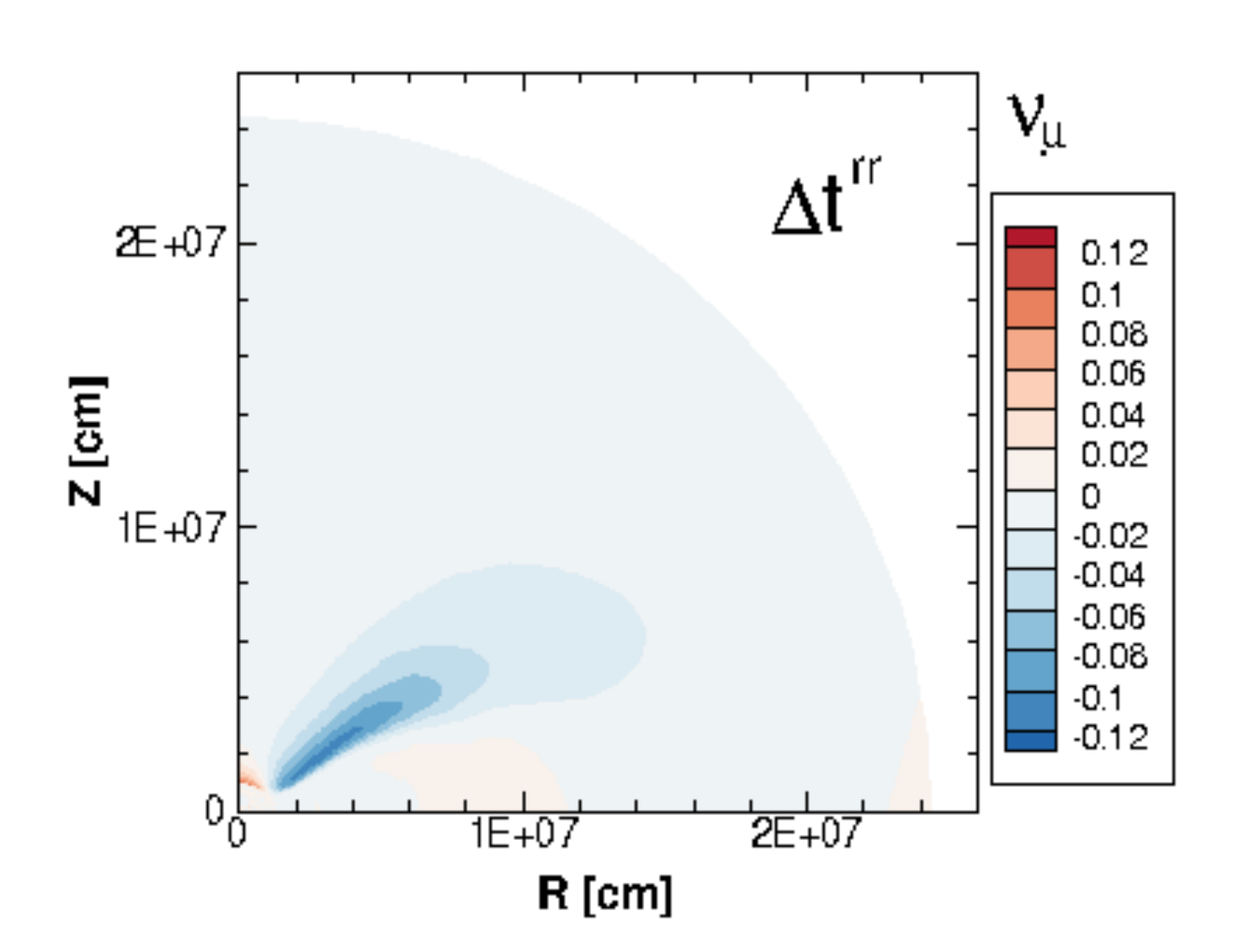}
\caption{The $rr$-component of the Eddington tensor evaluated by the neutrino distribution functions (top), the closure relation (middle) and the difference between them (bottom) for neutrino energy of 34 MeV are shown for $\nu_e$ (left), $\bar{\nu}_e$ (middle) and $\nu_\mu$ (right) by contour plots.  
\label{fig:nu_tensor_t000_inx_ie08_diag}}
\end{figure}

\begin{figure}[ht!]
\epsscale{0.35}
\plotone{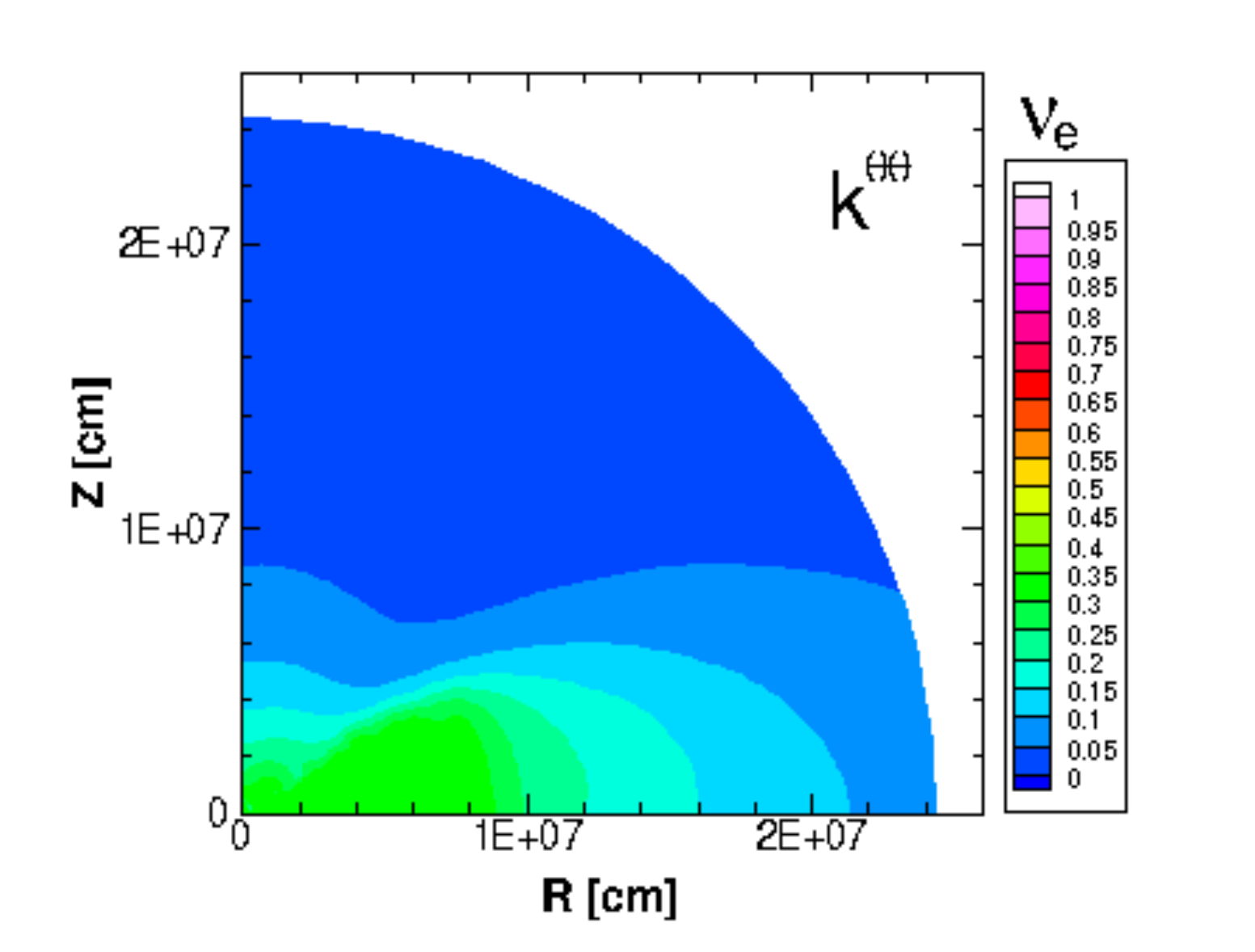}
\plotone{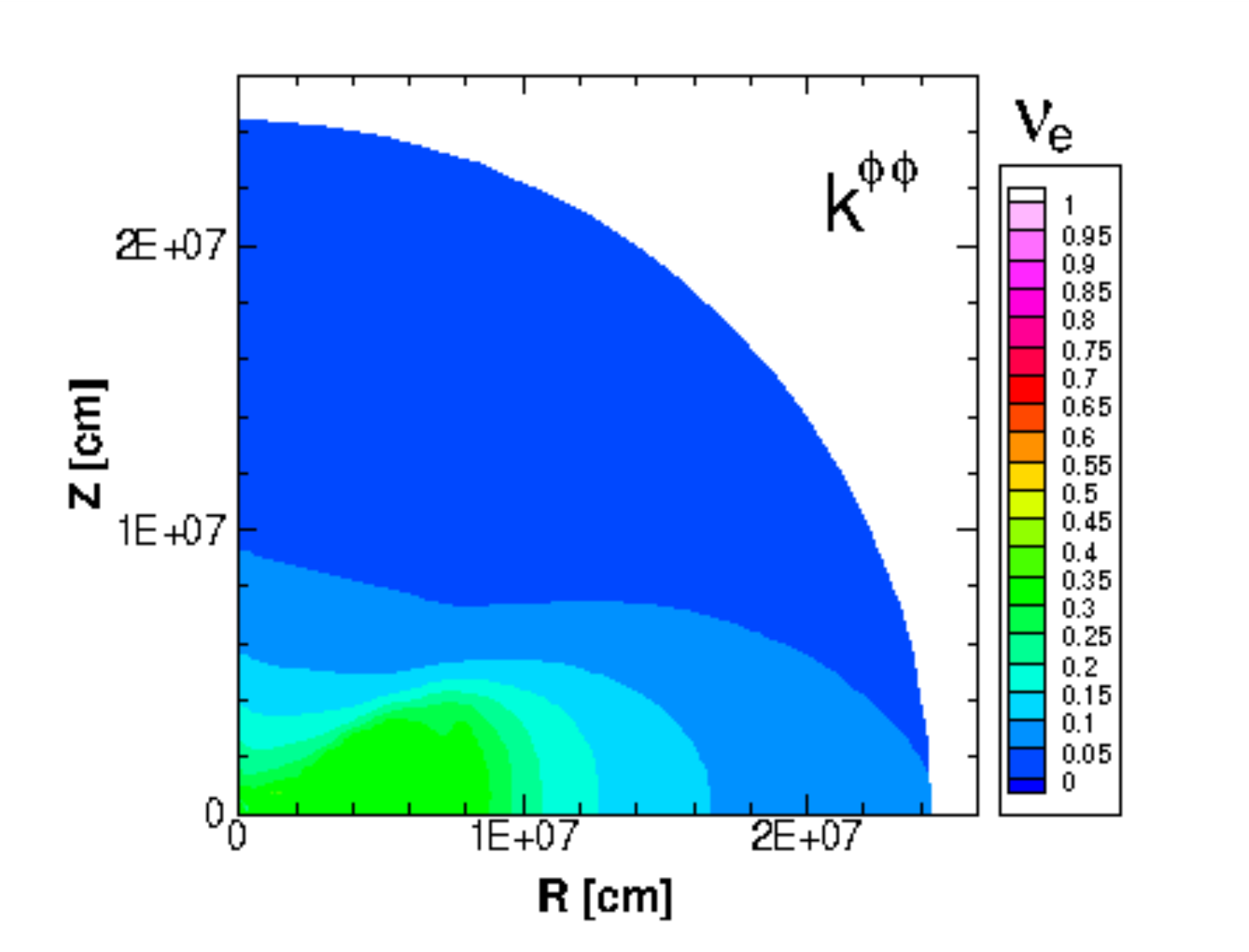}
\plotone{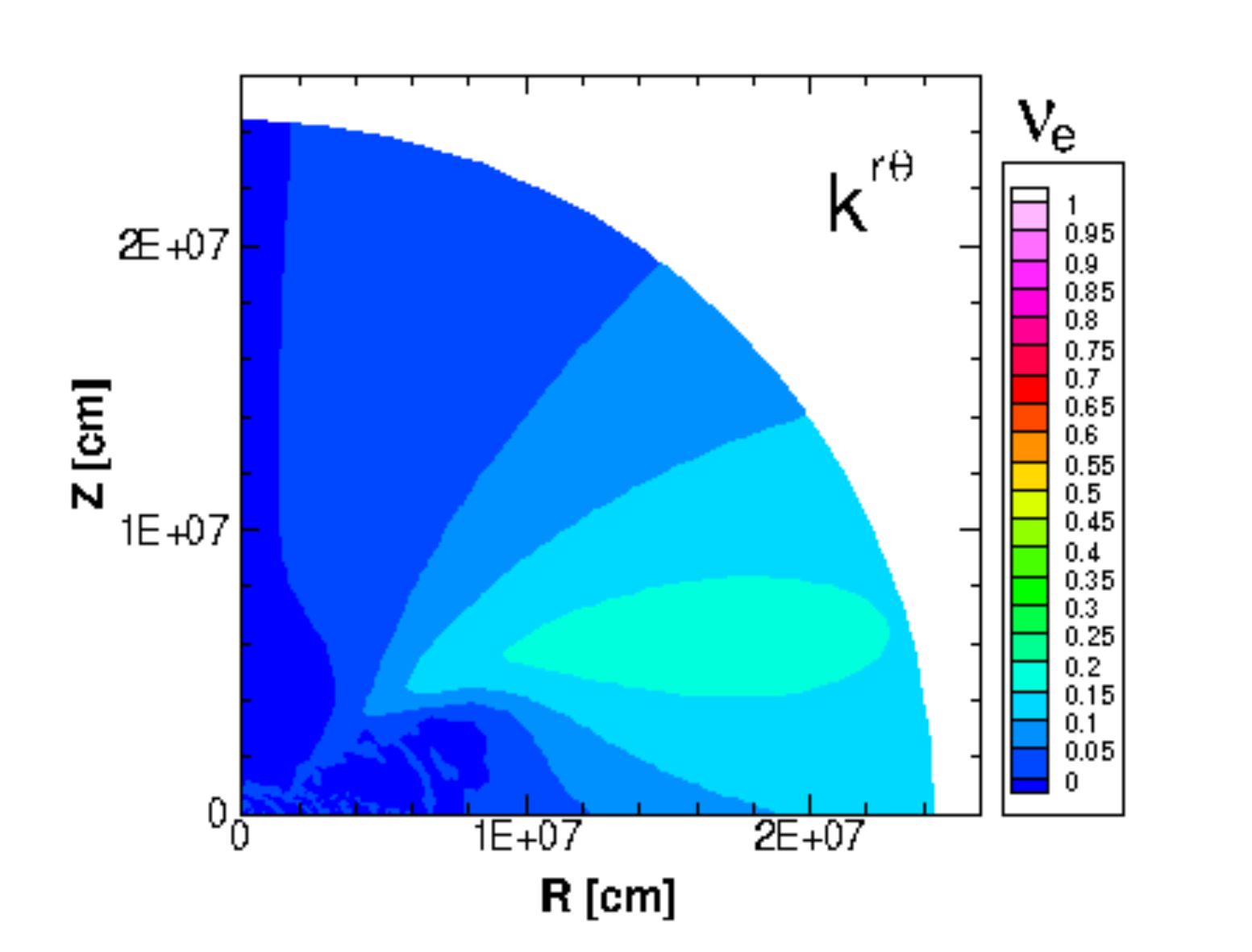}
\plotone{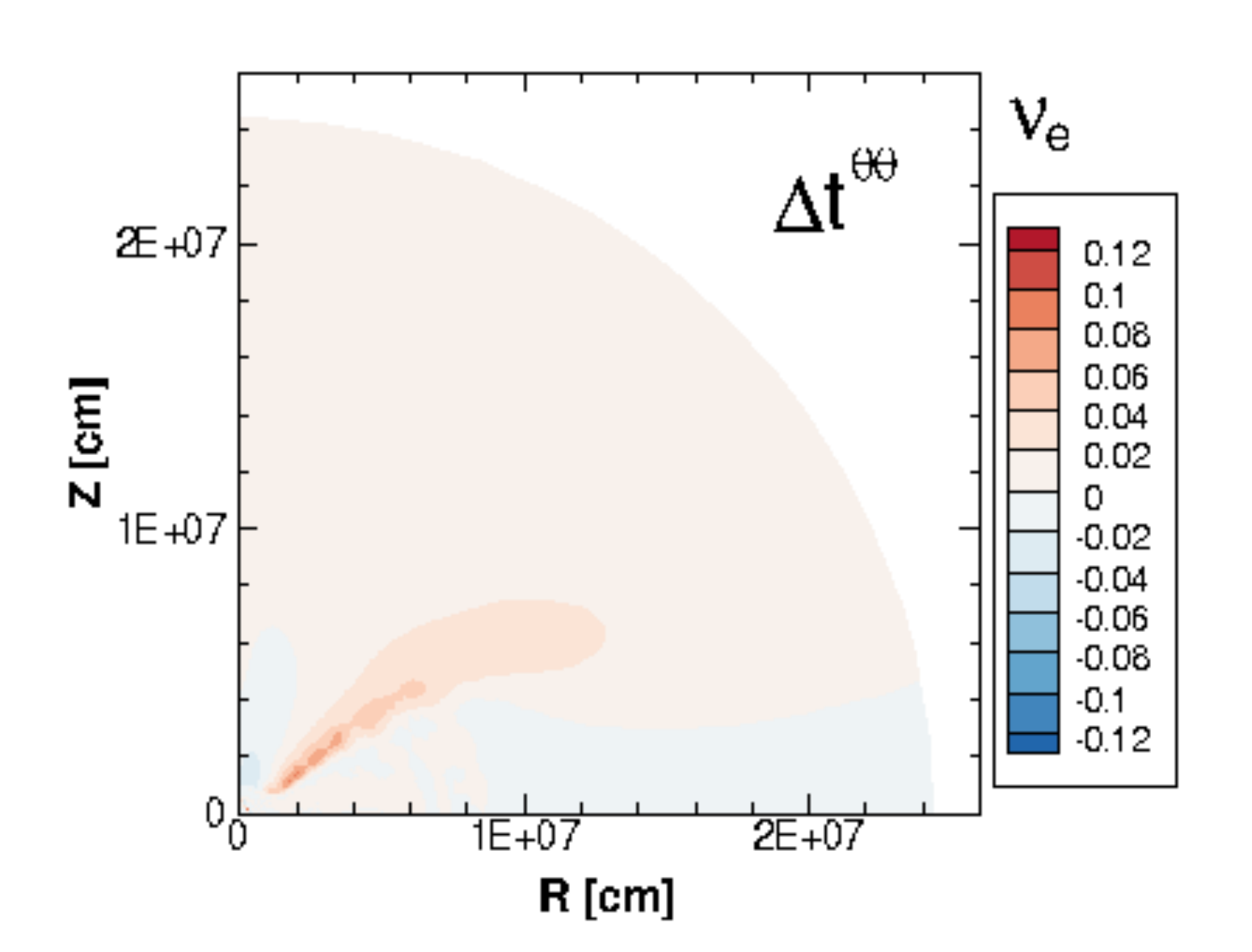}
\plotone{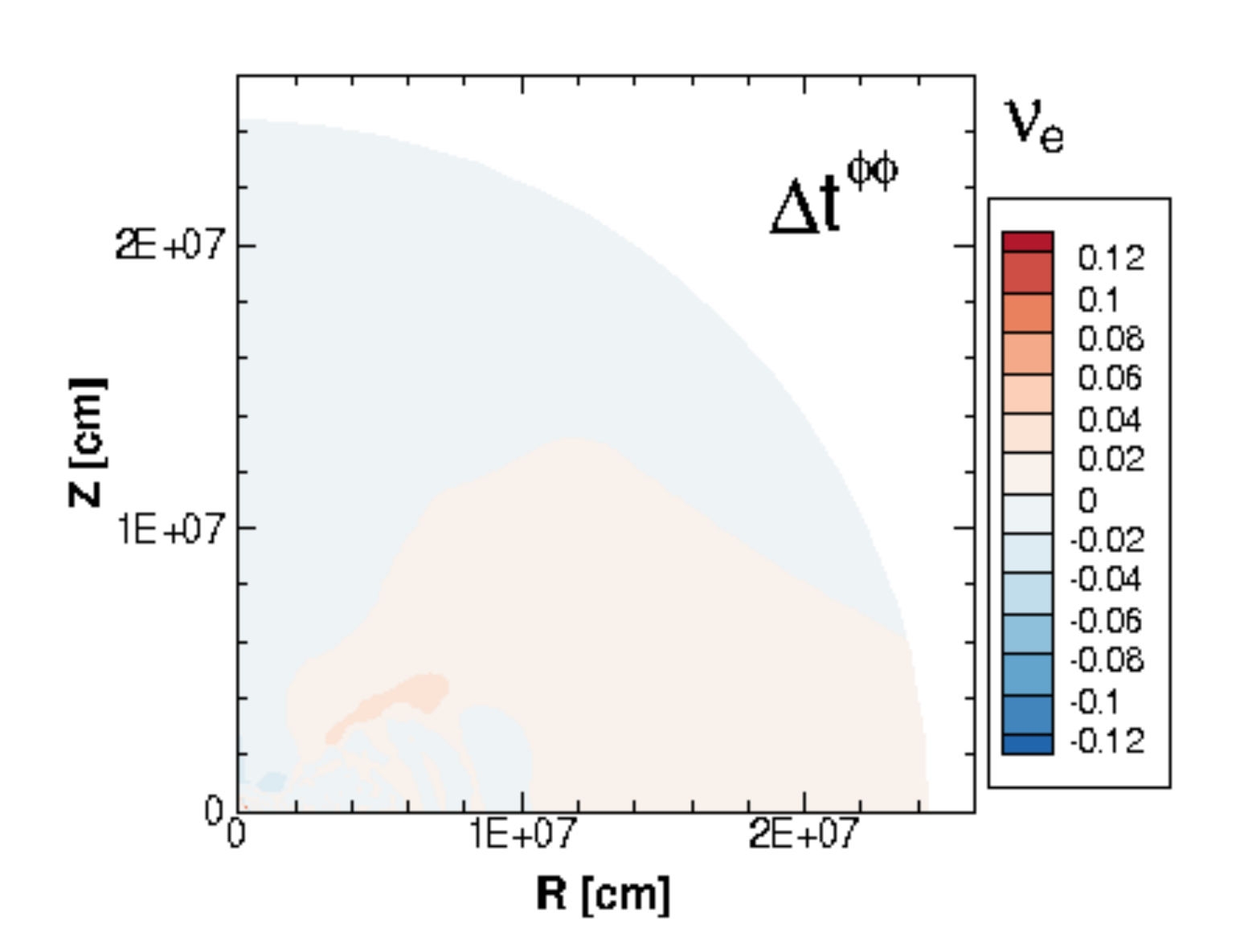}
\plotone{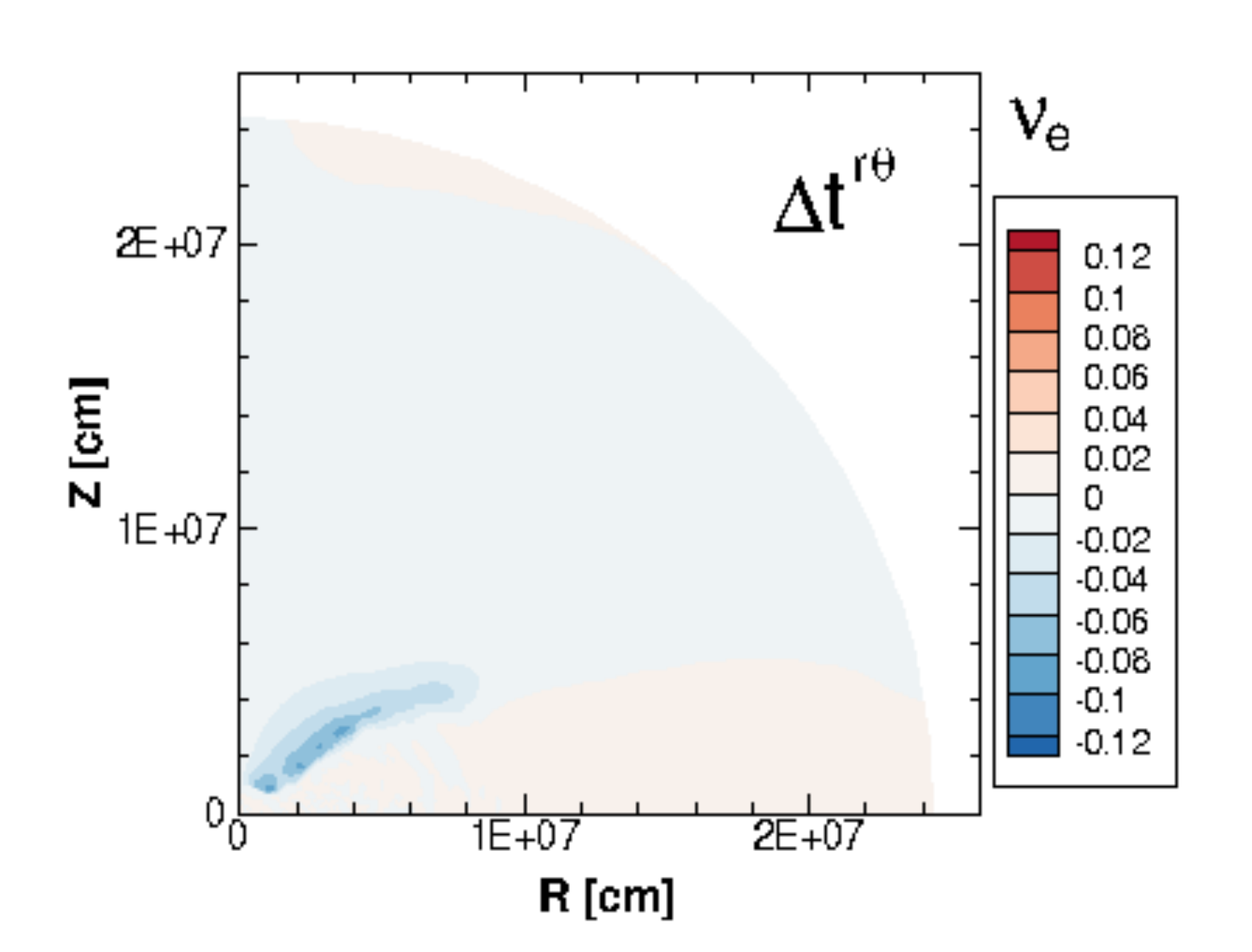}
\caption{The $\theta \theta$-, $\phi \phi$-, and $r \theta$-components of the Eddington tensor evaluated by the neutrino distribution functions (top) and the deviation of the values by the closure relation from those by the direct evaluation (bottom) for $\nu_e$ with neutrino energy of 34 MeV are shown by contour plots.  
\label{fig:nu_tensor_t000_in1_ie08_nondiag}}
\end{figure}

Figure \ref{fig:nu_tensor_t000_in1_ie08_nondiag} shows $\theta \theta$-, $\phi \phi$-, and $r \theta$-components of the Eddington tensor for $\nu_e$.  
The profiles of the $\theta \theta$- and $\phi \phi$-components are similar to each other and also to the $rr$-component.  The $\theta \theta$- and $\phi \phi$-components
have the isotropic value of $\frac{1}{3}$ inside the the merger remnant and decrease to 0 at large distances.  It is remarkable that the $r \theta$-component has non-zero finite values in rather outer regions.  The difference between the closure relation and the direct evaluation is generally small but noticeable in the region above the torus as in the case of the $rr$-component.  

\begin{figure}[ht!]
\epsscale{0.35}
\plotone{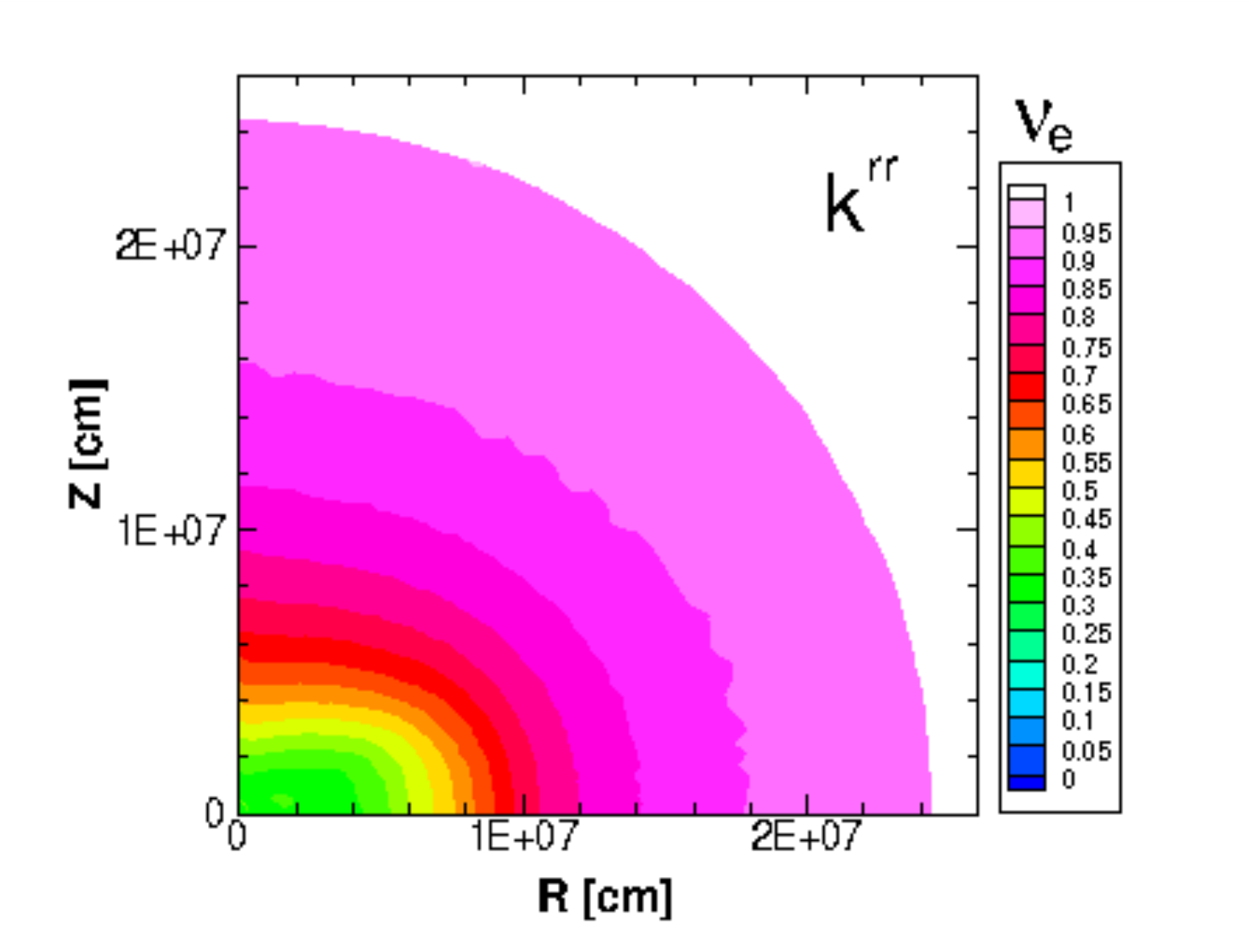}
\plotone{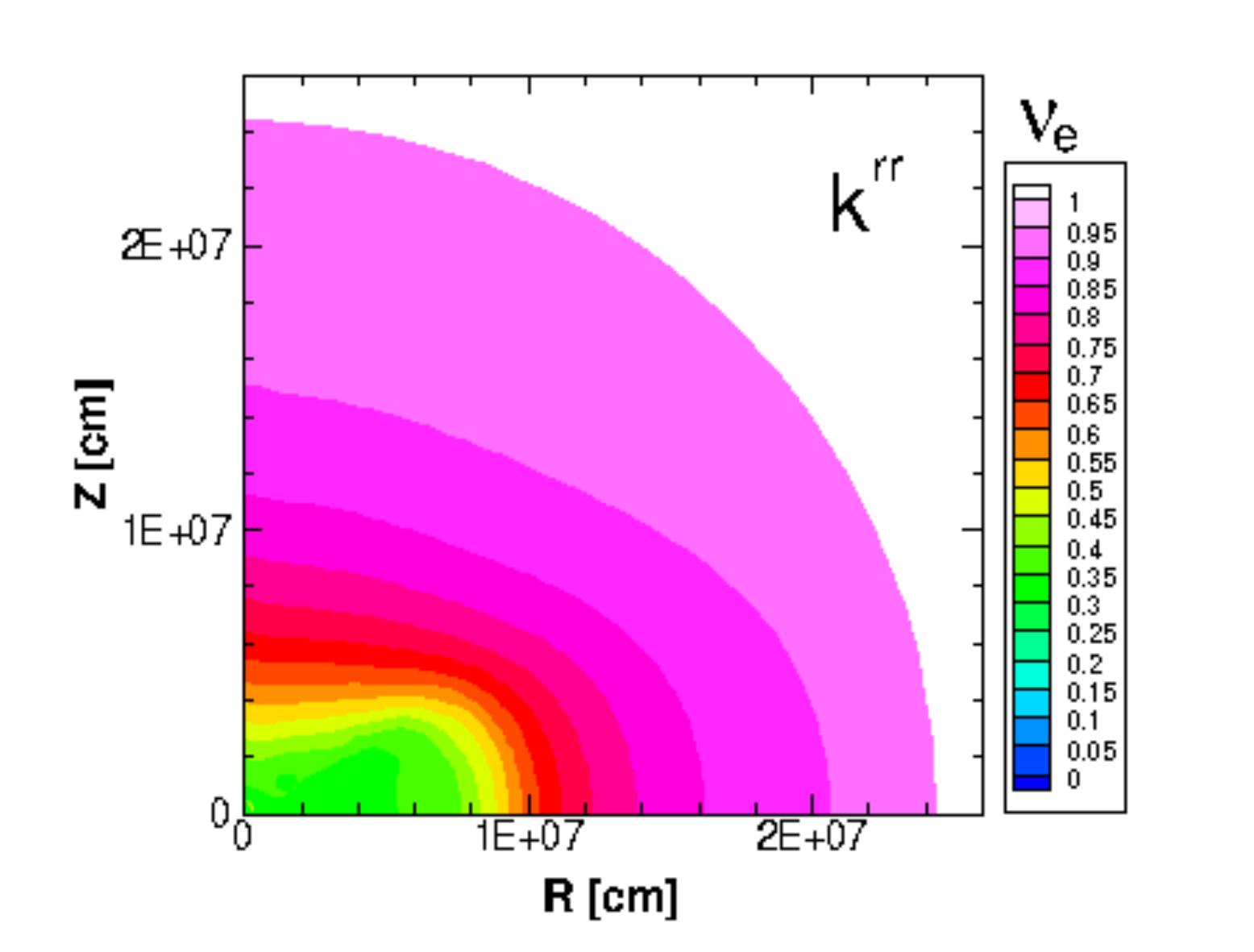}\\
\plotone{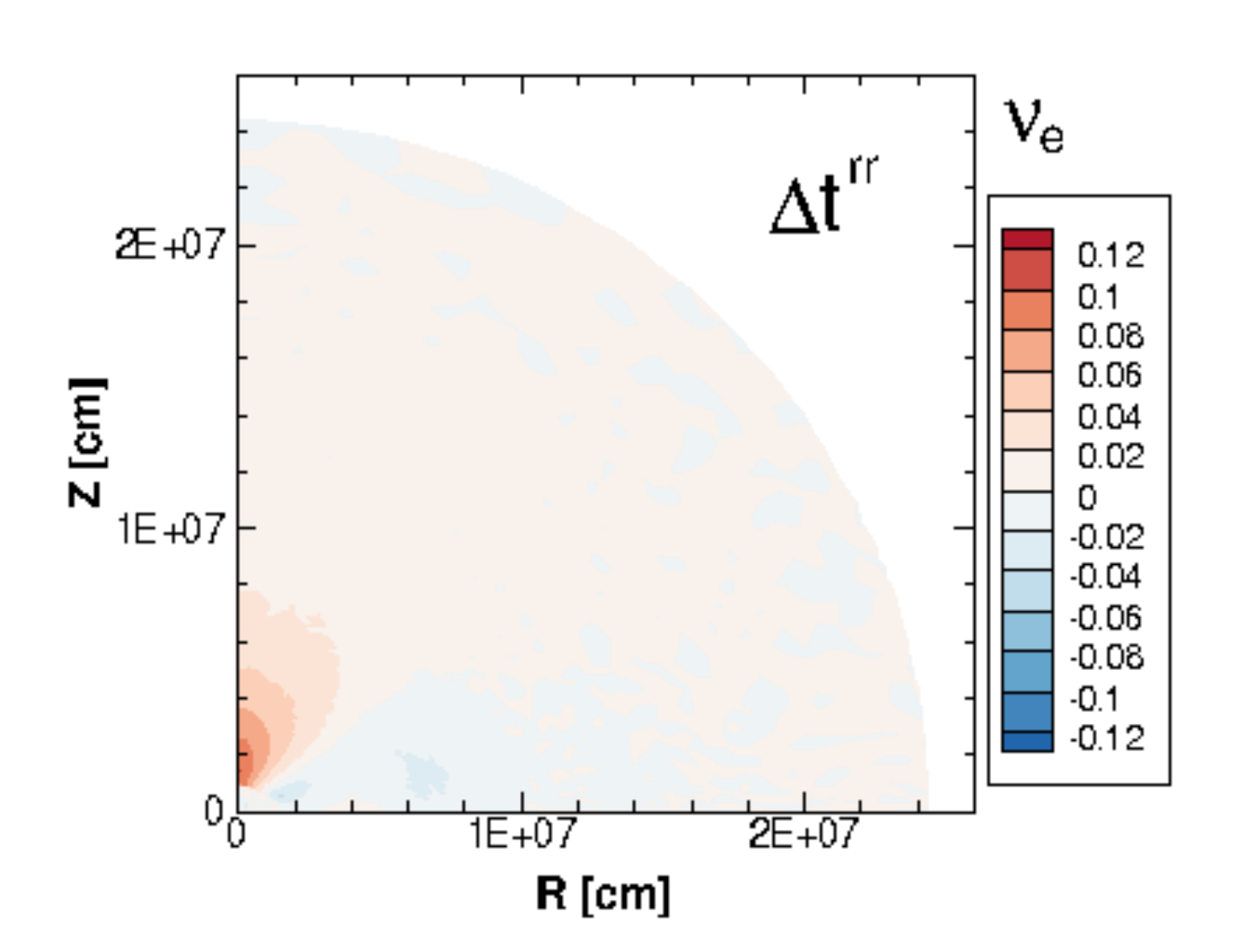}
\plotone{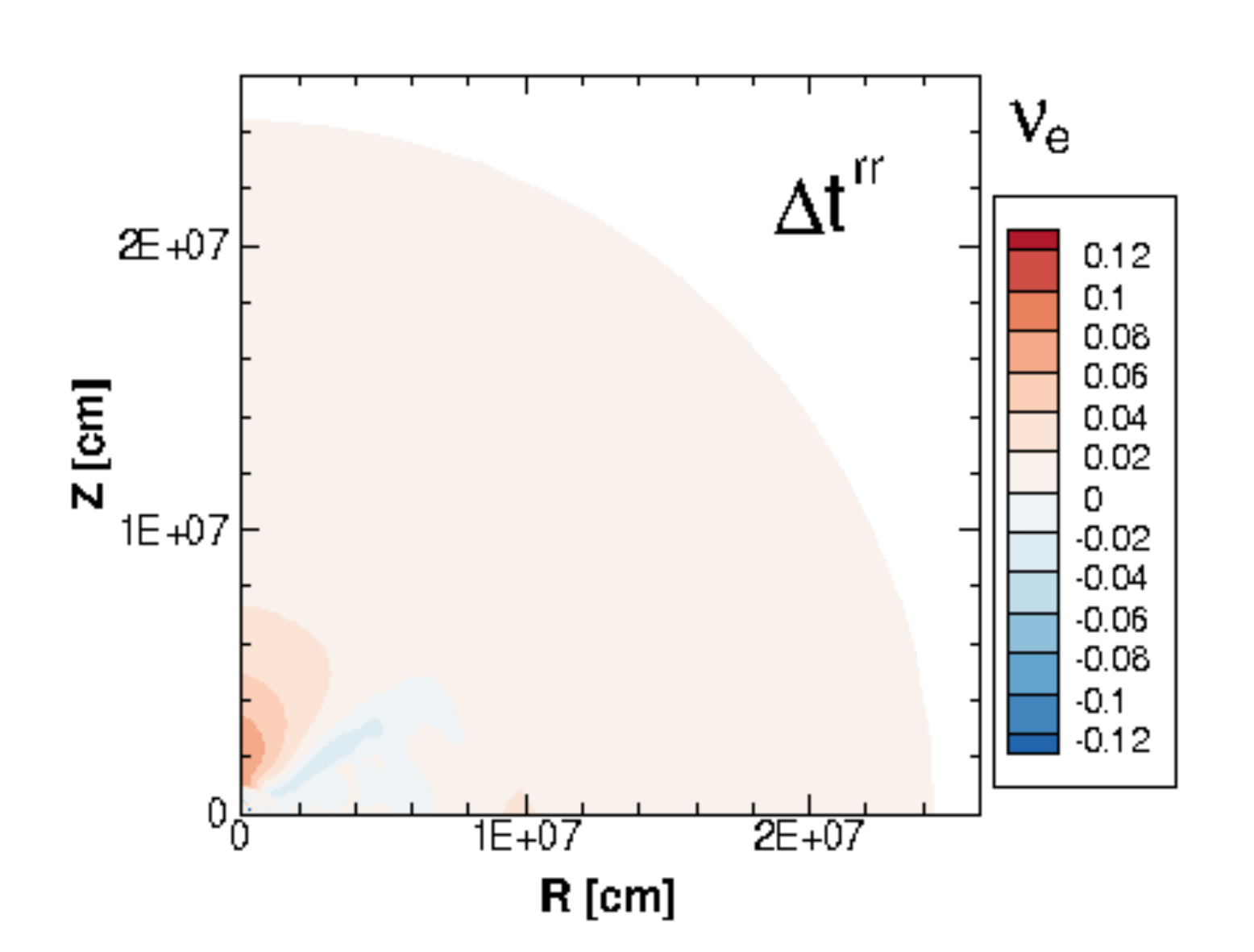}
\caption{The $rr$-component of the Eddington tensor obtained by the direct evaluation (top) and the deviation of the closure relation (bottom) for neutrino energy of 4.9 (left) and 13 (right) MeV are shown for $\nu_e$ by contour plots.  
\label{fig:nu_tensor_t000_inx_ie0x_diag}}
\end{figure}

The feature of the Eddington tensor with deformed shape substantially depends on the neutrino energy as we have seen the large deformation of neutrinosphere for high energies in Fig. \ref{fig:nu_sphere_dens_t000iexx}.  The validity of the closure relation accordingly depends on the energy range for each species.  
%
For low energies, the degree of deformation is rather small reflecting the small deformation of the neutrinosphere in the remnant.  Figure \ref{fig:nu_tensor_t000_inx_ie0x_diag} shows the $rr$-component of the Eddington tensor for $\nu_e$  with the neutrino energy of 4.9 and 13 MeV.  
For these energies, the deviation of the closure relation is small around the torus.  The overestimation above the remnant neutron star is noticeable instead.  

\begin{figure}[ht!]
\epsscale{0.35}
\plotone{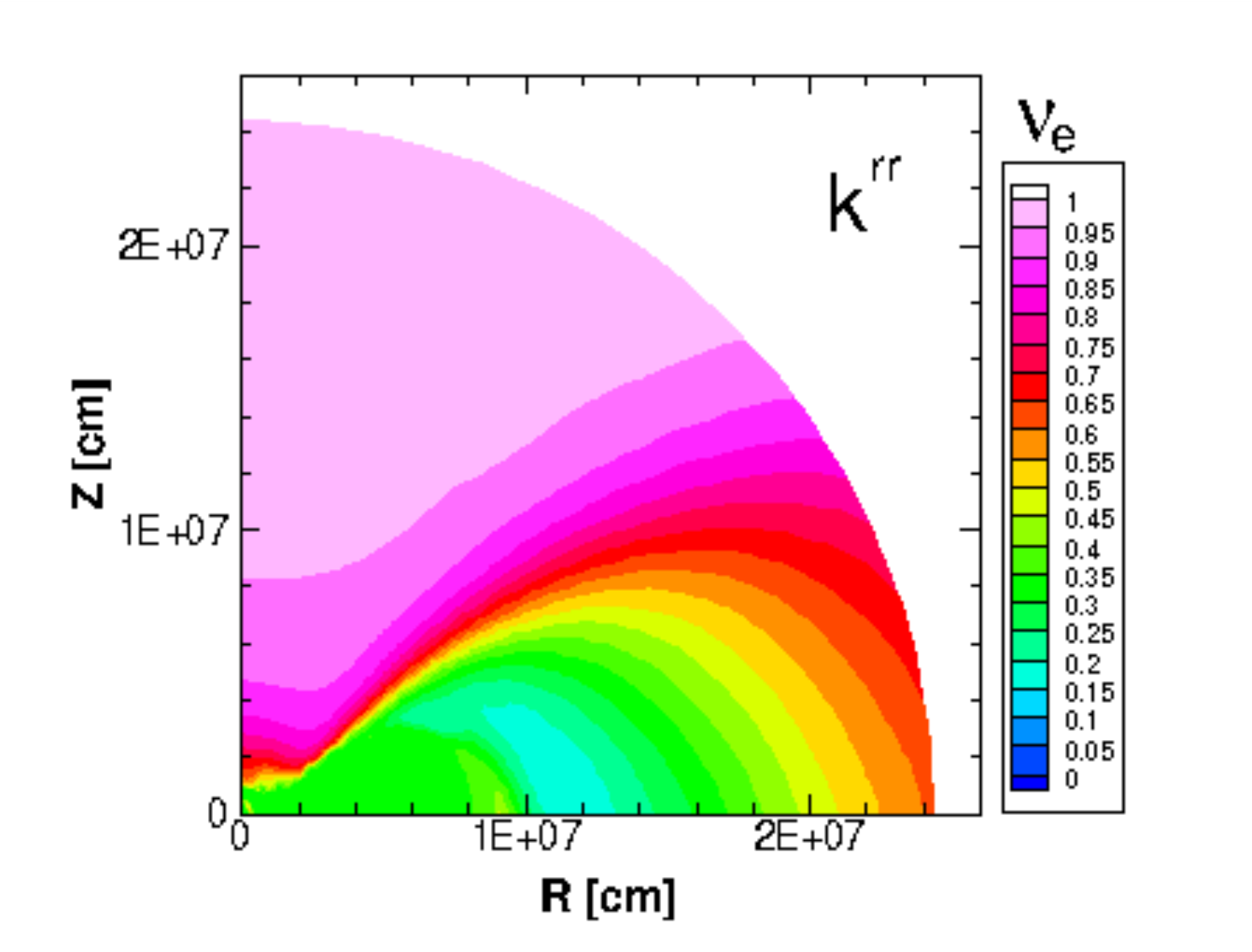}
\plotone{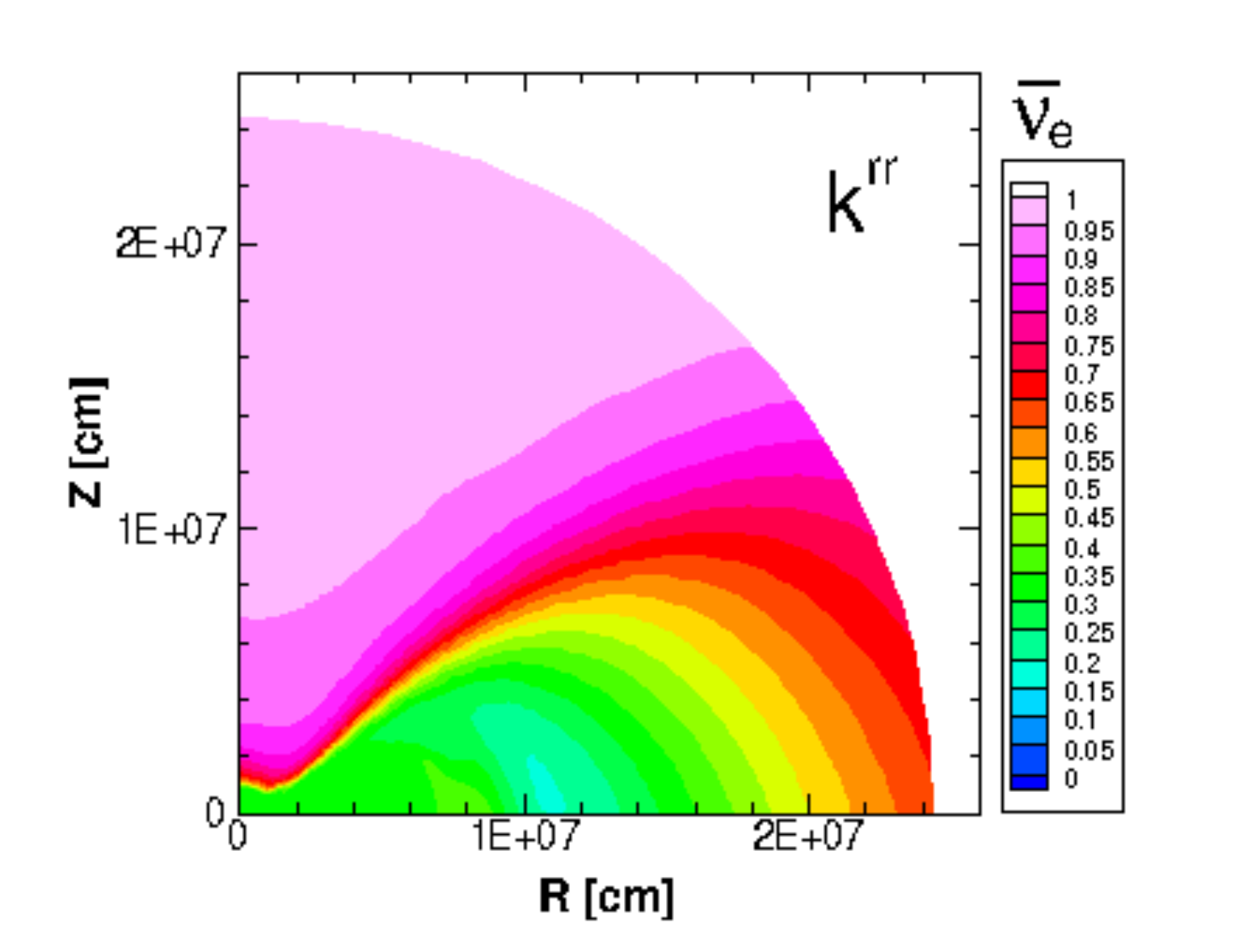}
\plotone{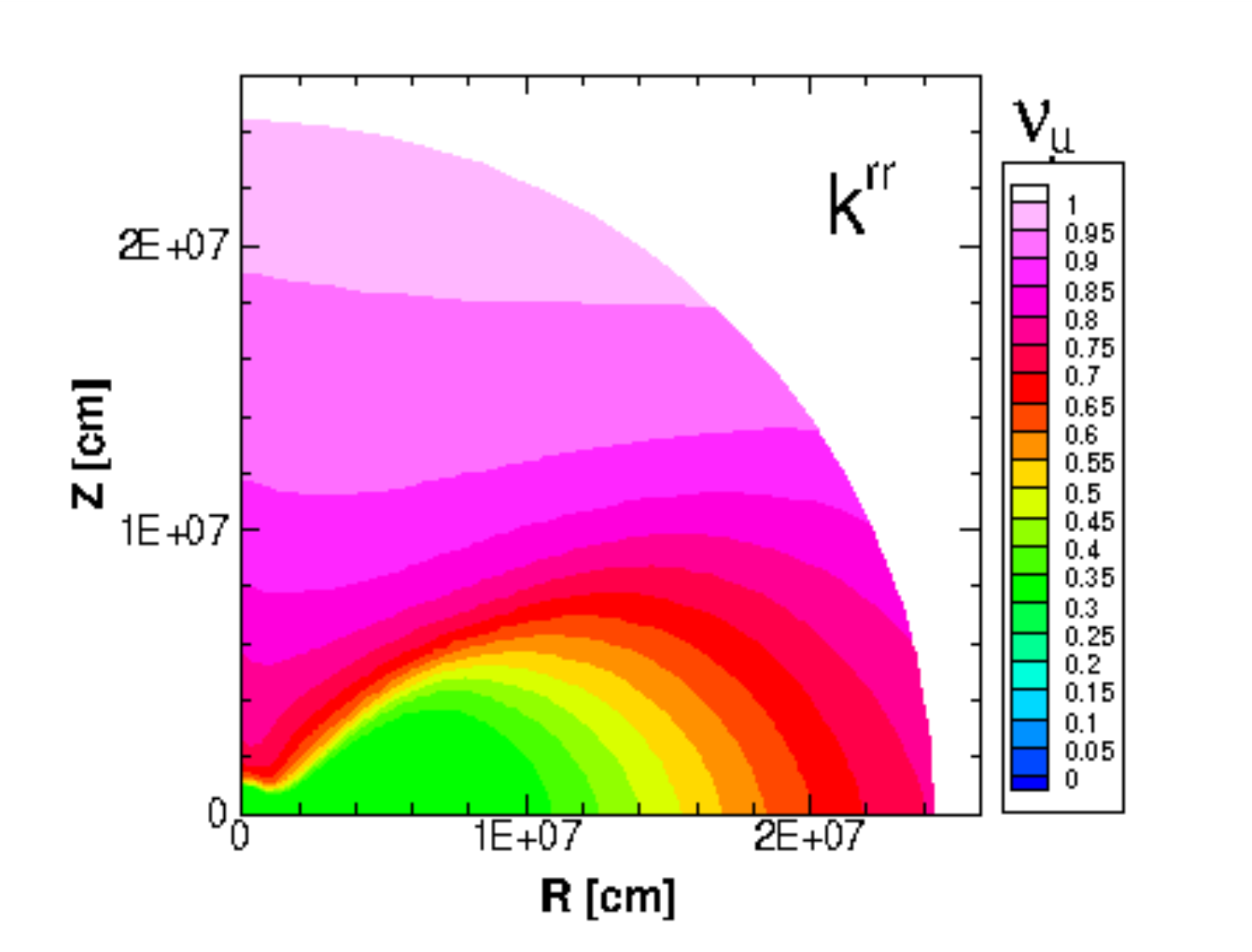}
\plotone{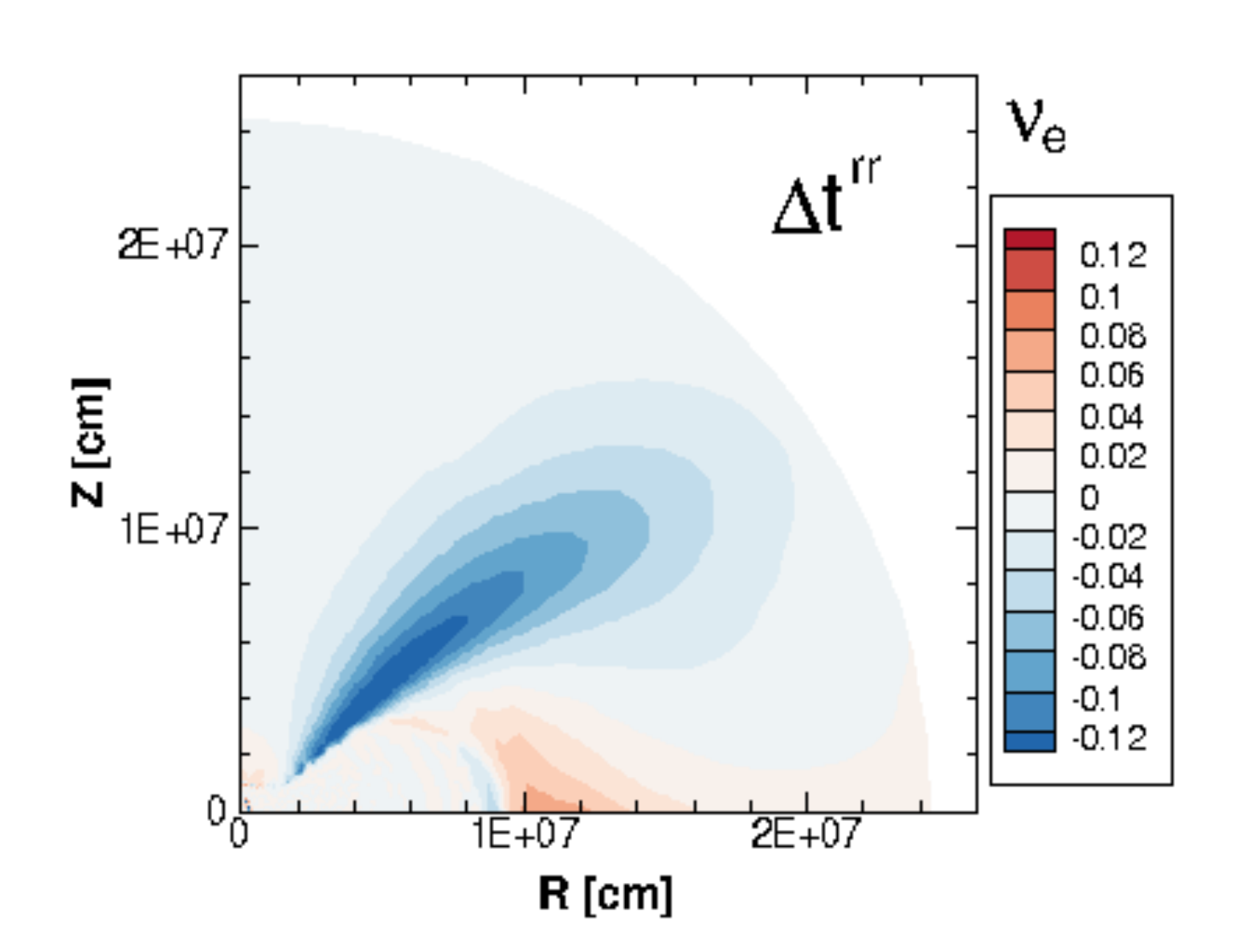}
\plotone{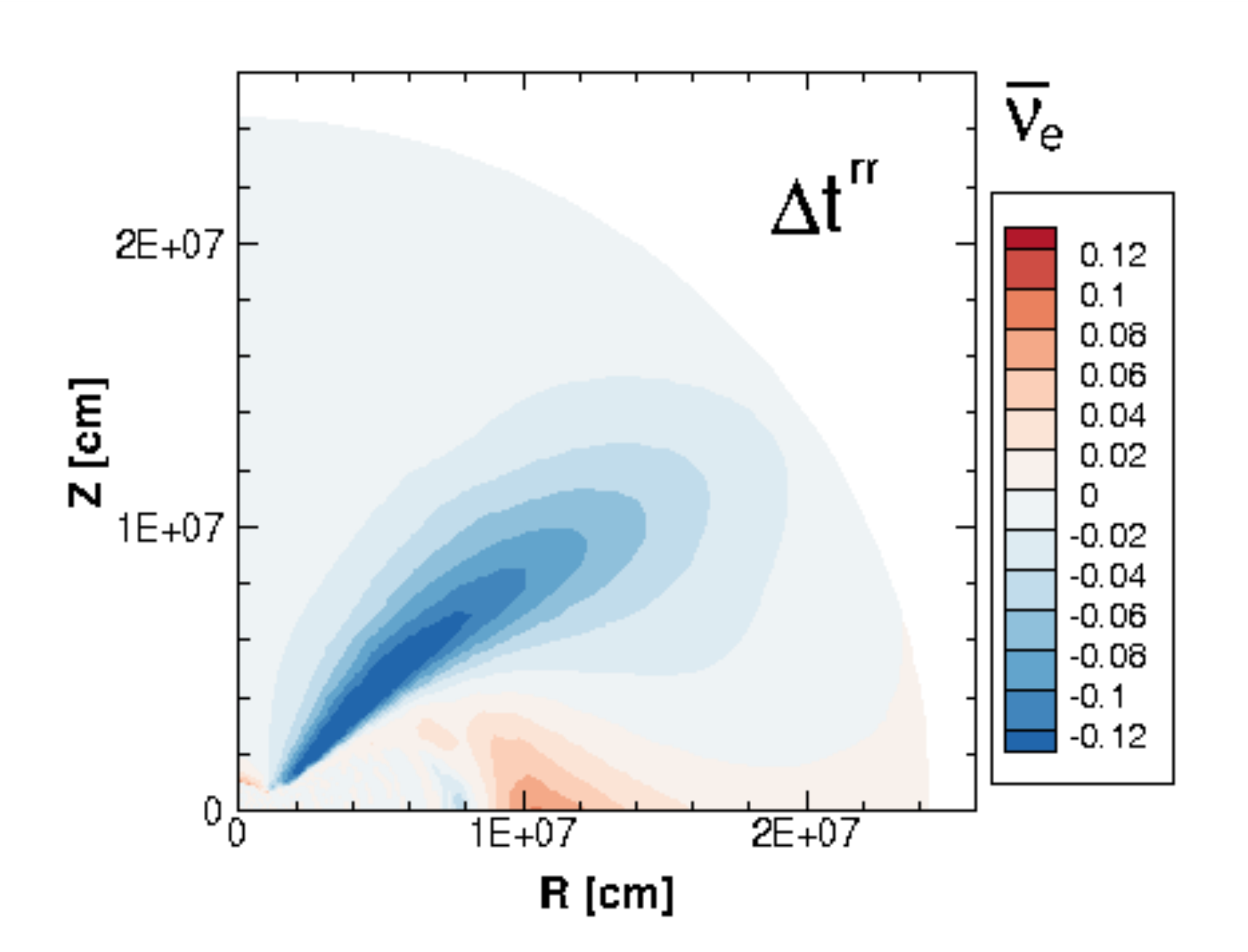}
\plotone{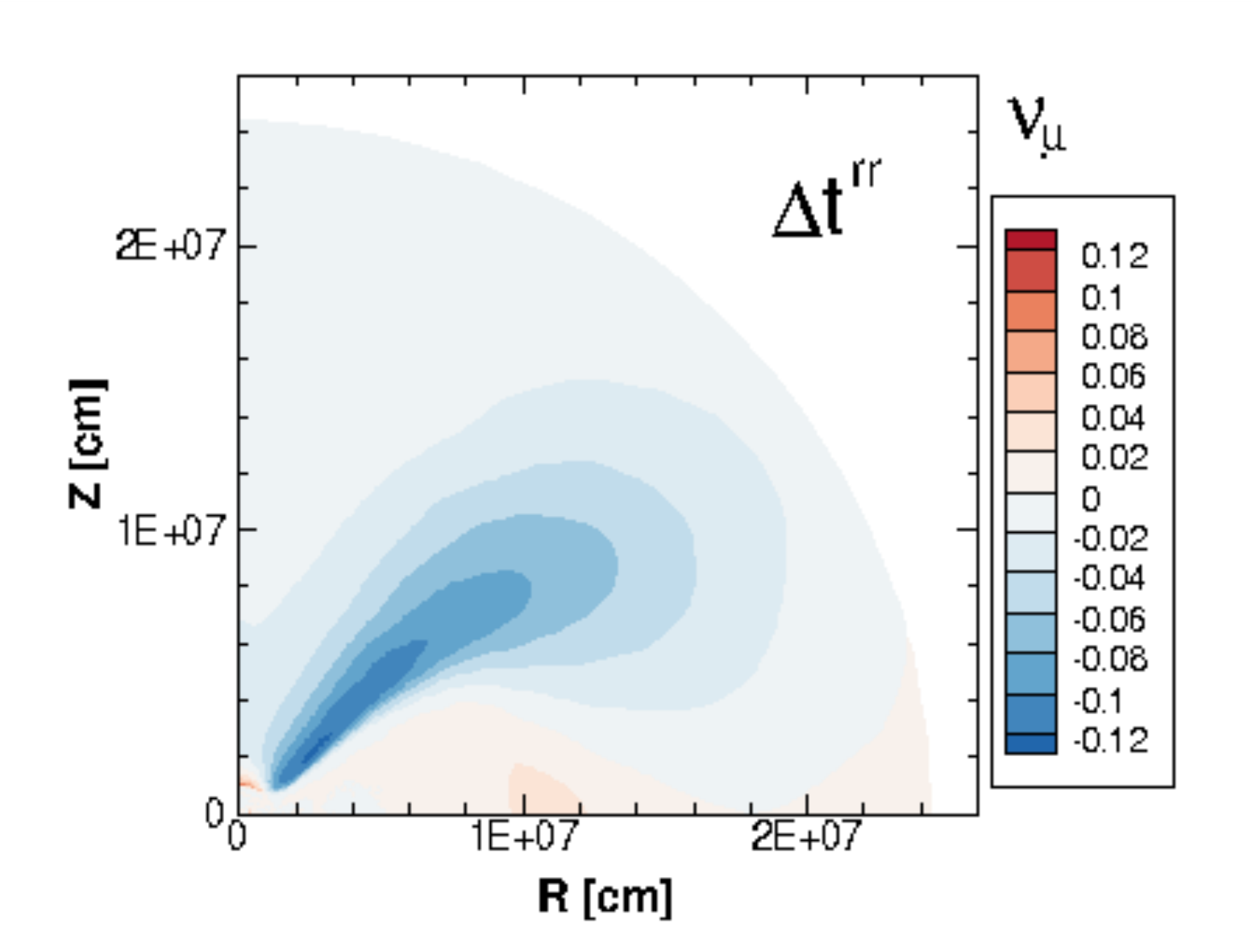}
\caption{The $rr$-component of the Eddington tensor obtained by the direct evaluation (top) and the deviation of the value by the closure relation (bottom) for neutrino energy of 89 MeV are shown for $\nu_e$ (left), $\bar{\nu}_e$ (middle) and $\nu_\mu$ (right) by contour plots.  
\label{fig:nu_tensor_t000_inx_ie10_diag}}
\end{figure}

Fig. \ref{fig:nu_tensor_t000_inx_ie10_diag} shows the $rr$-component of the Eddington tensor for three species with the neutrino energy of 89 MeV, which is higher than those examined in Figs. \ref{fig:nu_tensor_t000_inx_ie08_diag}, \ref{fig:nu_tensor_t000_in1_ie08_nondiag}, and \ref{fig:nu_tensor_t000_inx_ie0x_diag}.  The deformed profile of low values (isotropic i.e. $\sim\frac{1}{3}$) is extended further and has a large and thick shape, inside which there are some regions of non-isotropic values.  This is also a consequence of more deformed shape of the neutrinosphere for higher energies.  The shape also depends on the neutrino species,  showing a smaller torus for $\mu$-type neutrinos than the others.  While the $rr$-component for electron-type neutrinos rapidly increases to 1 along the $z$-axis, the increase is slow for $\mu$-type neutrinos.  Due to the large deformation, the deviation of the closure relation is apparently large for high energies as seen in the bottom figures.  The region of underestimation extends around the edge of torus.  
Further studies are necessary to clarify these differences around the torus.  

These trends of deformed shapes depending on the energy and species are well in accord with those of the shape of neutrinosphere as seen in Figs. \ref{fig:nu_sphere_xxxx_t000ie06} and \ref{fig:nu_sphere_dens_t000iexx}.  
In this respect, it requires more cautions in approximations of neutrino transfer for higher energies due to large anisotropies than lower energies.  It also suggests the importance of the numerical simulation in multi-energy group, which treats the neutrino transport in different energies.  

\begin{figure}[ht!]
\epsscale{0.35}
\plotone{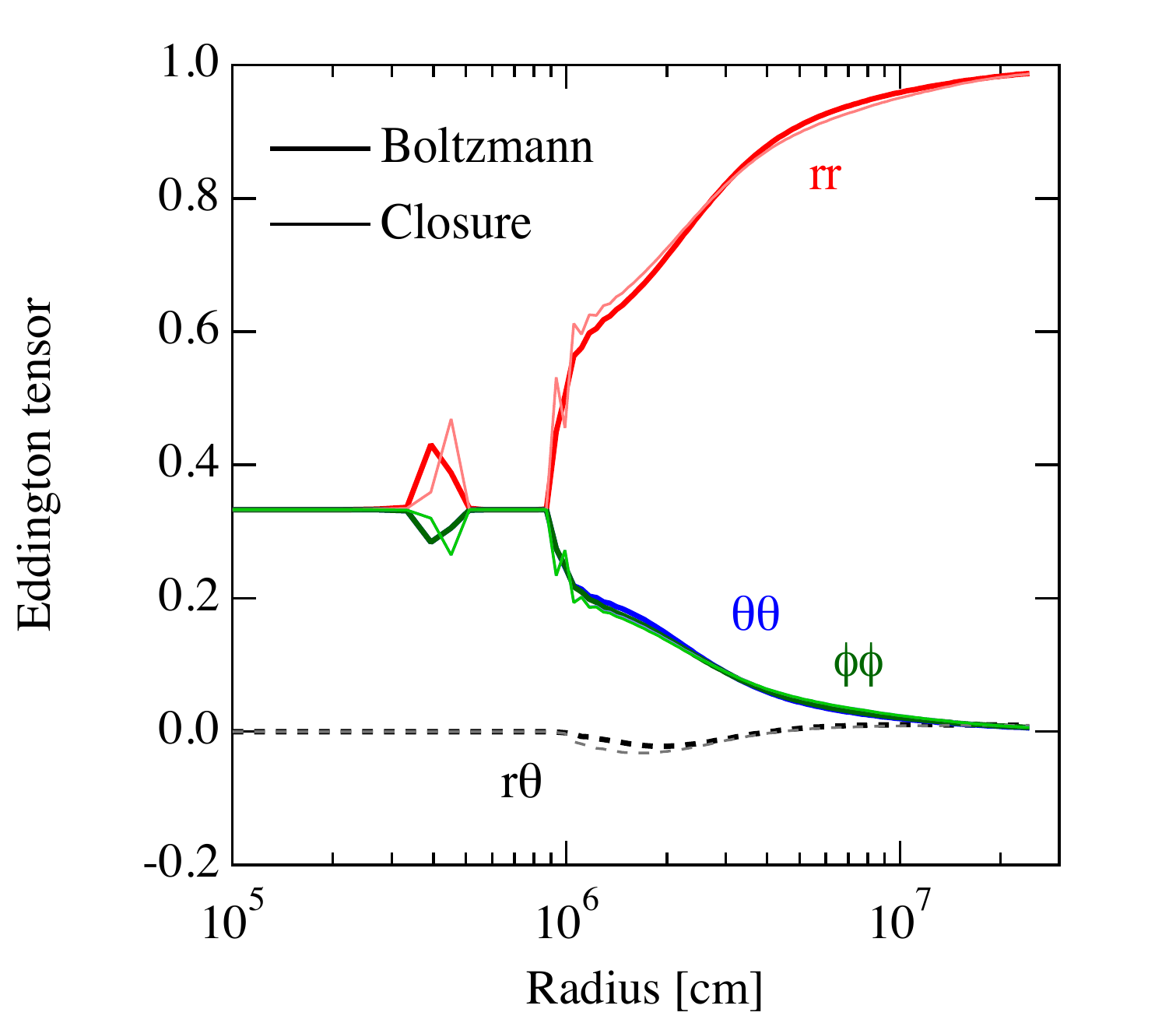}
\plotone{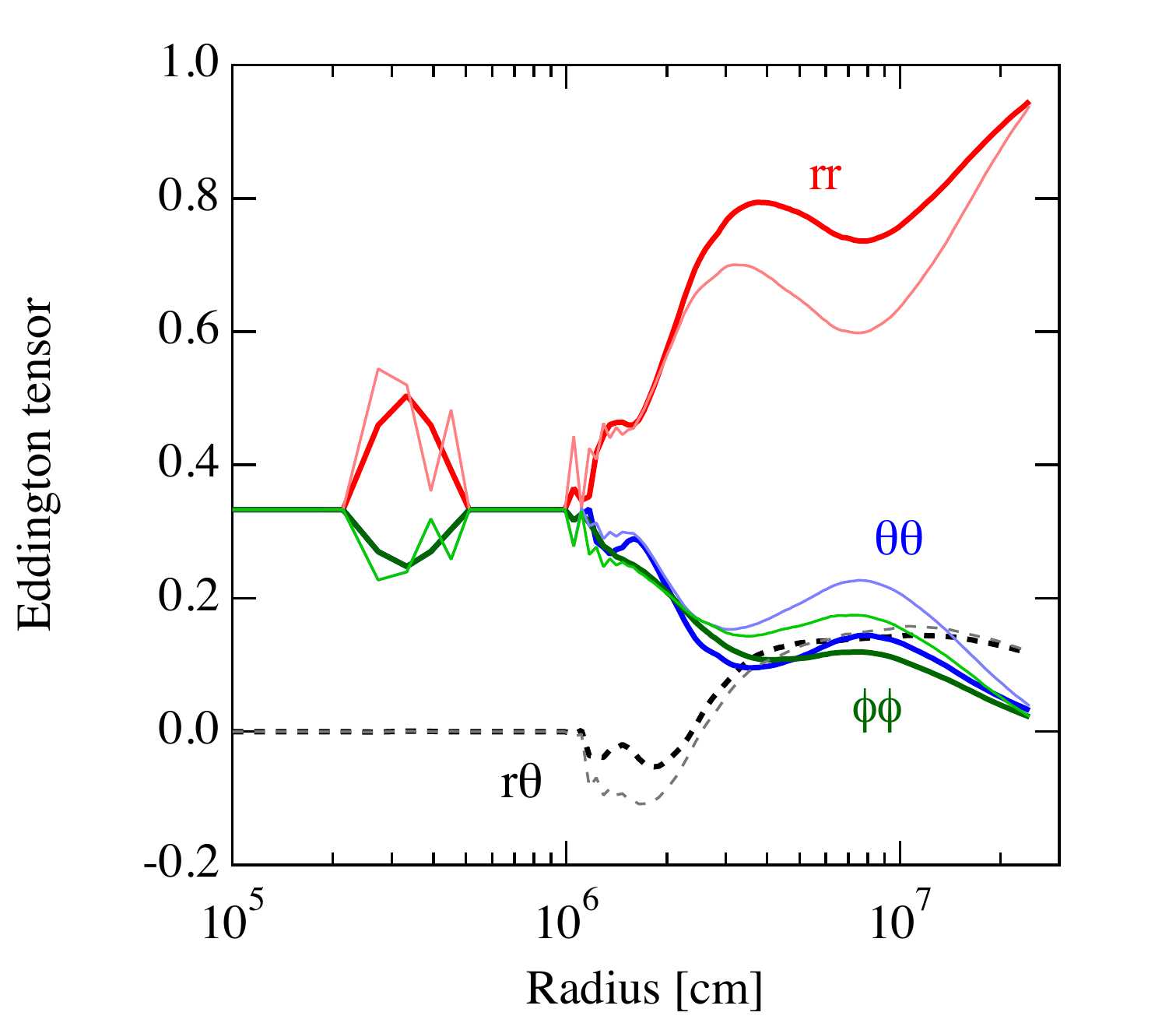}
\plotone{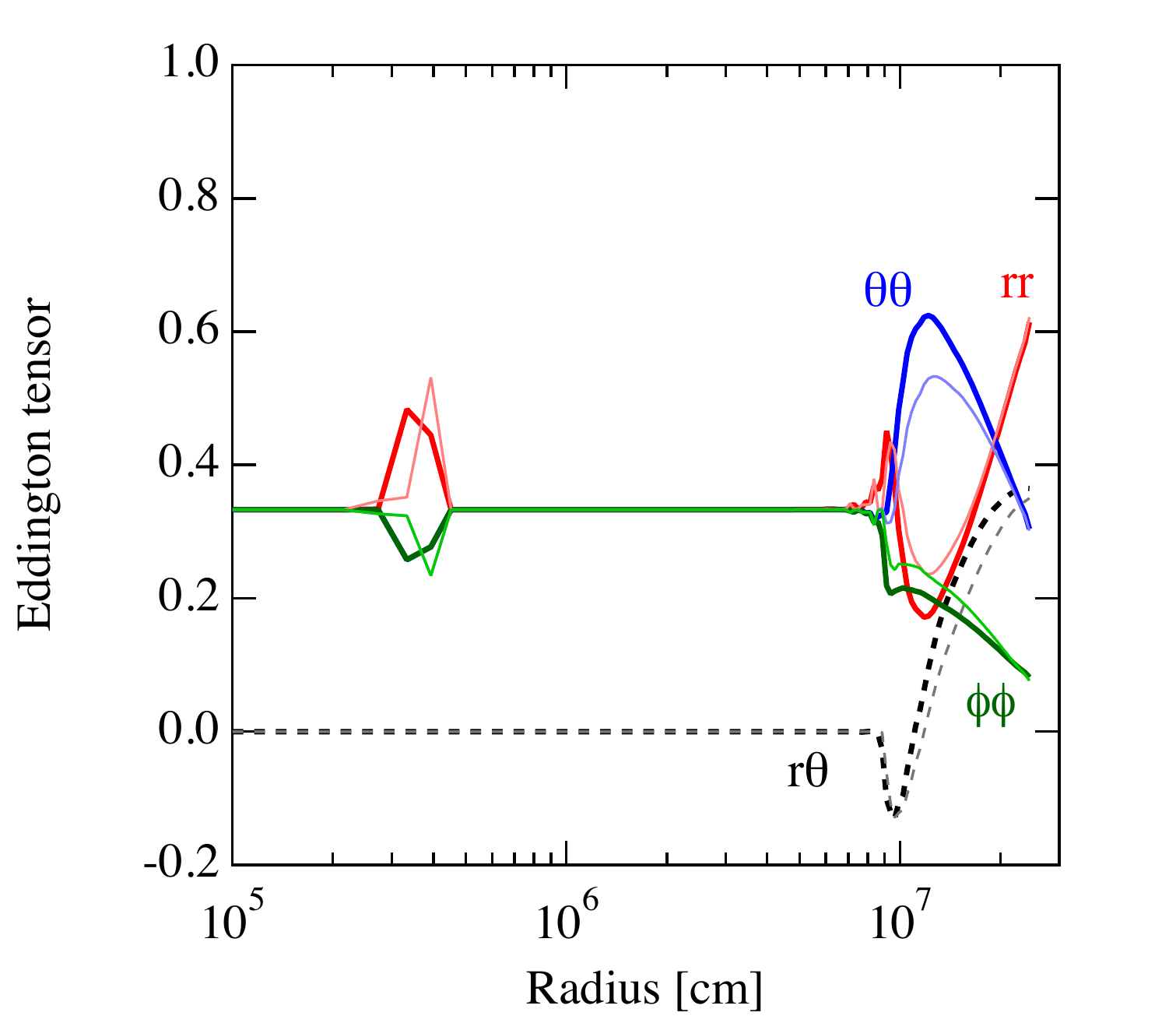}
\plotone{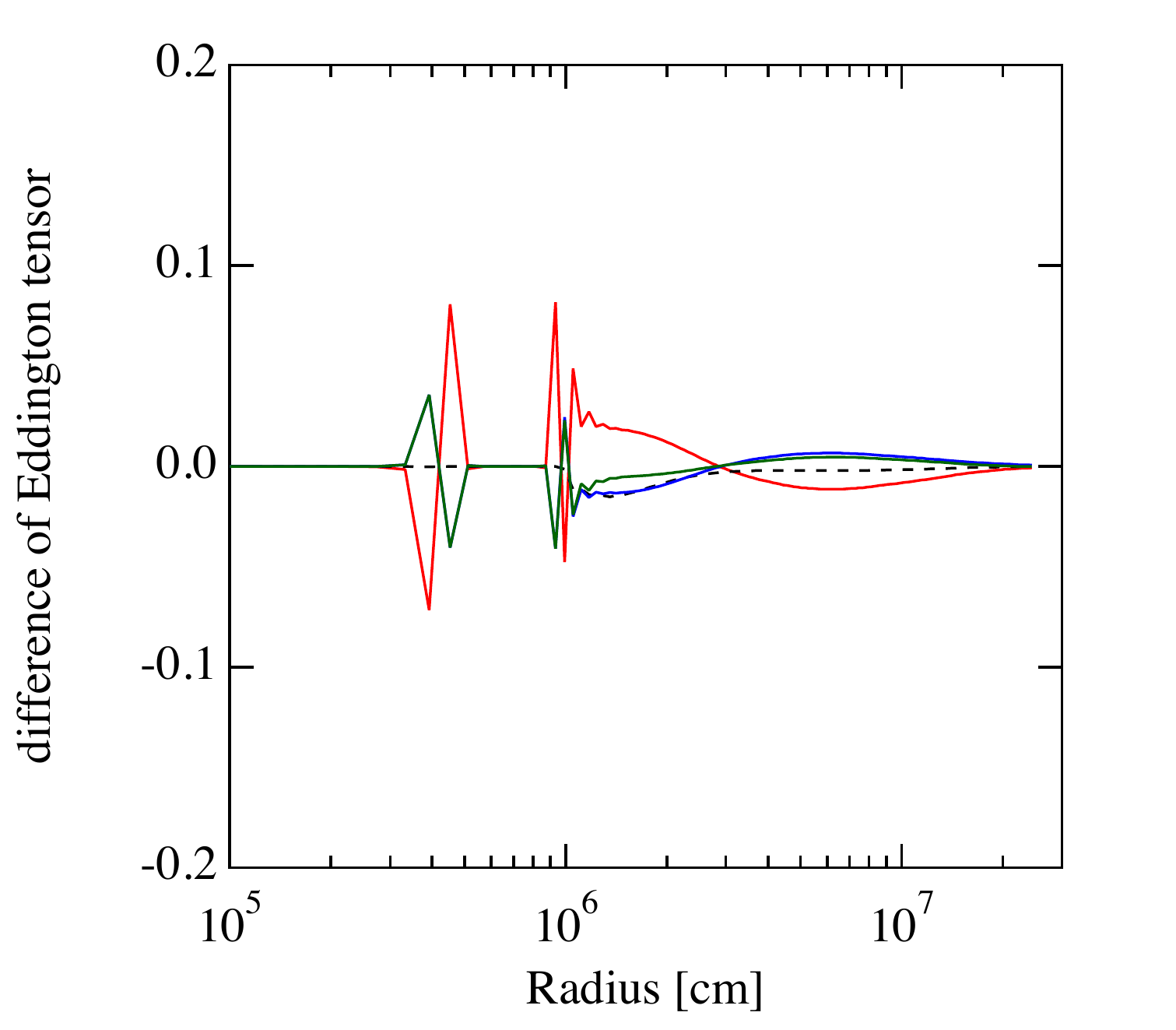}
\plotone{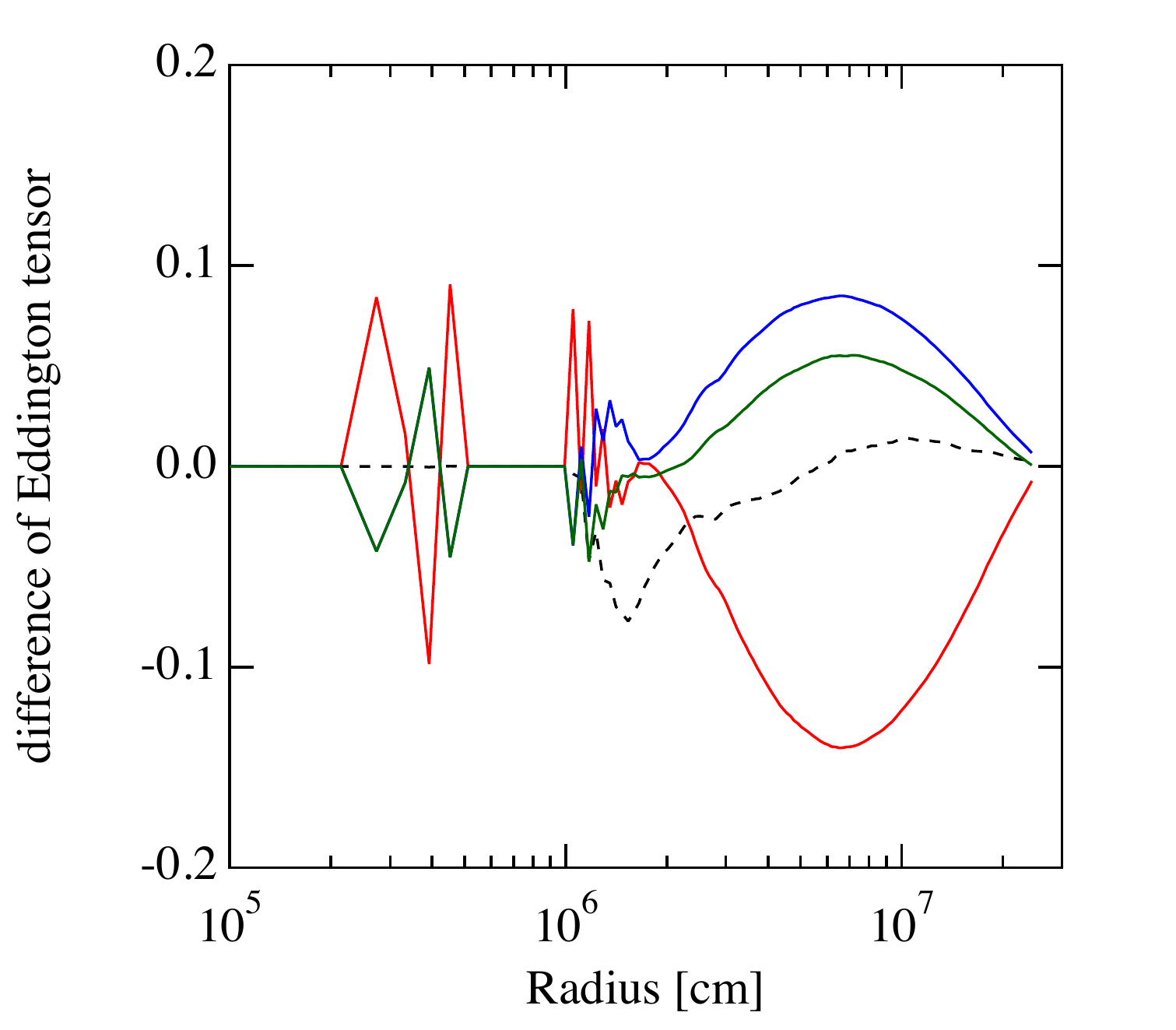}
\plotone{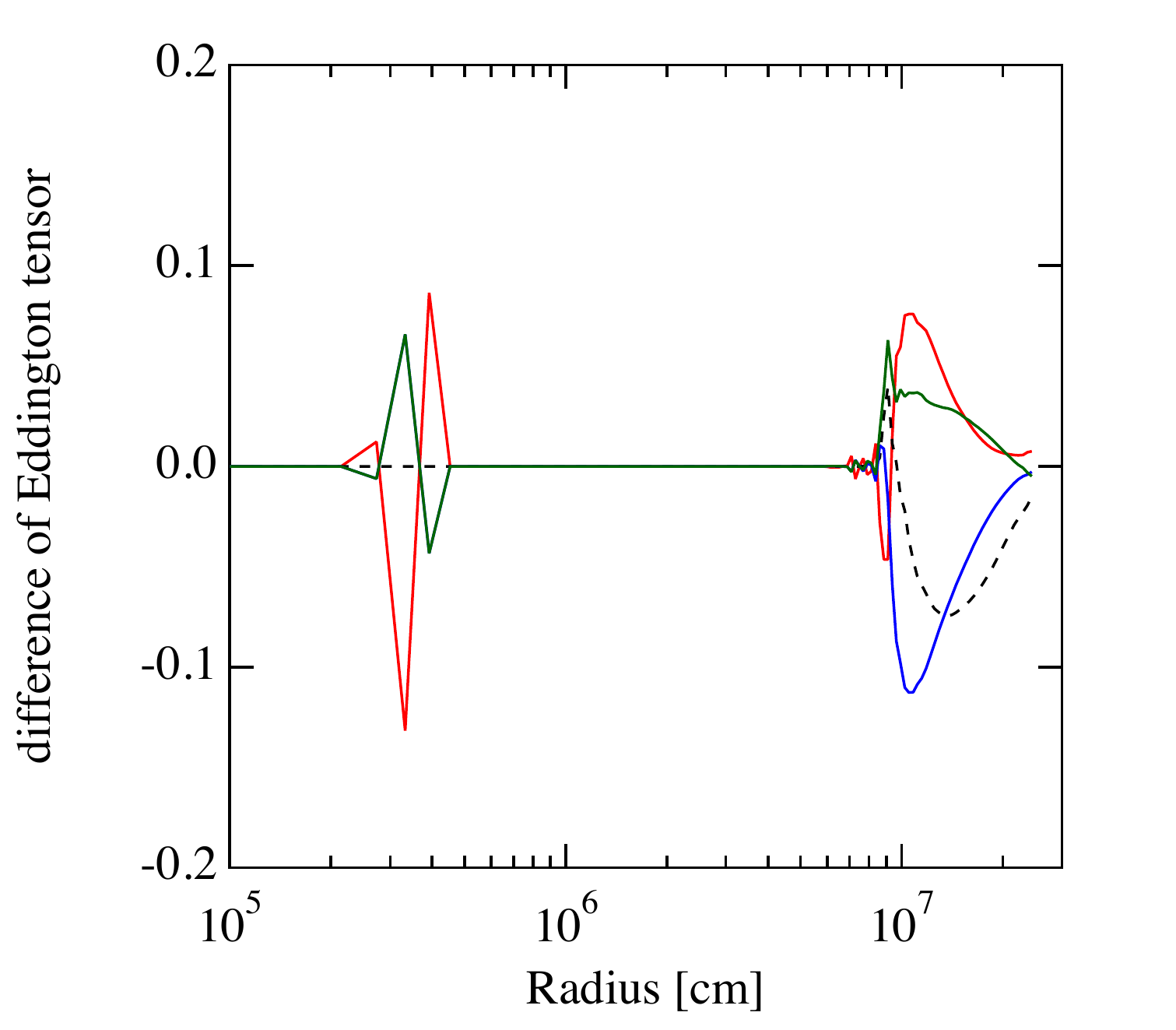}
\caption{Components of the Eddington tensor for $\nu_e$  obtained by the direct evaluation (thick line) and the closure relation (thin line) for neutrino energy of 89 MeV are shown as functions of radius in the top panels.  The diagonal $rr$- (red), $\theta \theta$- (blue), and $\phi \phi$- (green) components are shown together with the non-diagonal $r \theta$-components (black dashed-lines) along the polar angles at 9.8 (left), 48 (middle), and 90 (right) degree from the $z$-axis.  
The corresponding differences of the values by the closure relation from those by the direct evaluation are shown in the bottom panels.  
\label{fig:nu_tensor_t000_in1_ie10_radial}}
\end{figure}

The radial distributions of the Eddington tensor along different polar angles behave in various ways depending on the traversing profile of the remnant.  Figure \ref{fig:nu_tensor_t000_in1_ie10_radial} shows the components of the Eddington tensor as functions of radial distance along the three polar angles ($\theta=$ 9.8, 48, and 90 degree) for neutrino energy of 89 MeV.  While the radial variation near the $z$-axis (left panel) rapidly occurs from the isotropic value in the merger remnant\footnote{There are some wiggles around 3--5 km due to the same reason as we have seen in Fig. \ref{fig:nu_moment_t000_radial_ithxx}.  } to the forward-peaked value in the ambient dilute material, those in the equator (right) and around the edge of the torus (middle) behave in non-monotonic way because of the deformed profile.  The polar component of the flux of neutrinos contributes to sizable non-diagonal $r \theta$-components and affects the behavior of diagonal components.  The components of the Eddington tensor by the closure relation (thin lines) follows most of the trends in the components by the direct evaluation (thick lines), however, there are some regions of substantial difference as shown in the bottom panels.  The deviation can amount to over 0.1 in absolute value of the Eddington tensor, which is not negligible for the pressure evaluation of the moment scheme in the transitional regime between diffusive and transparent situations.  

%


\section{Snapshots from time evolution} \label{sec:evolution}

We examine the properties of neutrino transfer in the time sequence of the remnant of the neutron star merger at 30, 65 and 135 ms (see also Fig. \ref{fig:hydro_set_txxx}).  
As we will see, the feature of neutrino transfer remains similar in $\sim$100 ms, but 
the asymmetric features of neutrino quantities seen in the initial profile becomes gradually mild due to the shrinkage of the torus, in particular of the high temperature region of the torus.  

\begin{figure}[ht!]
\epsscale{0.35}
\plotone{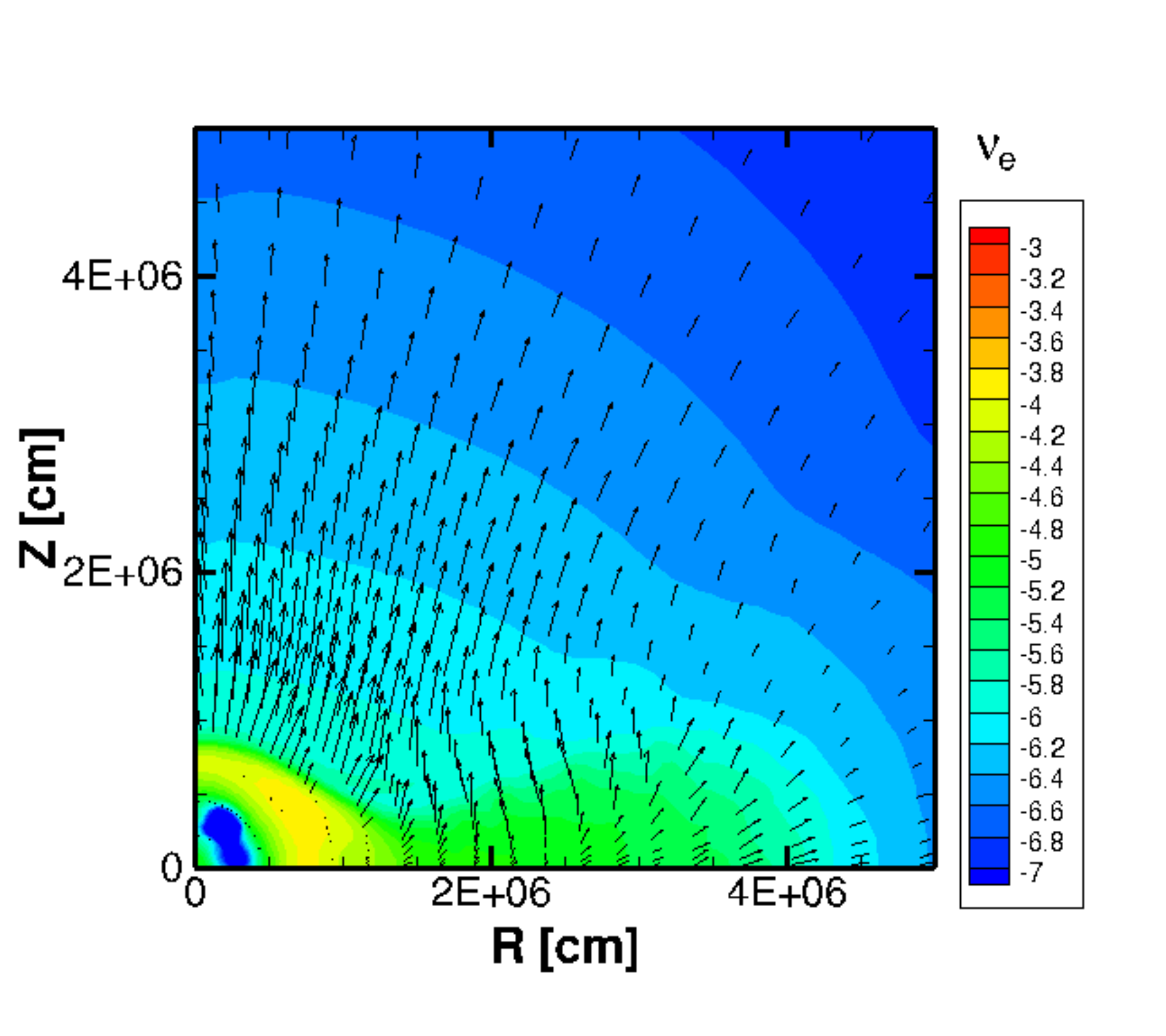}
\plotone{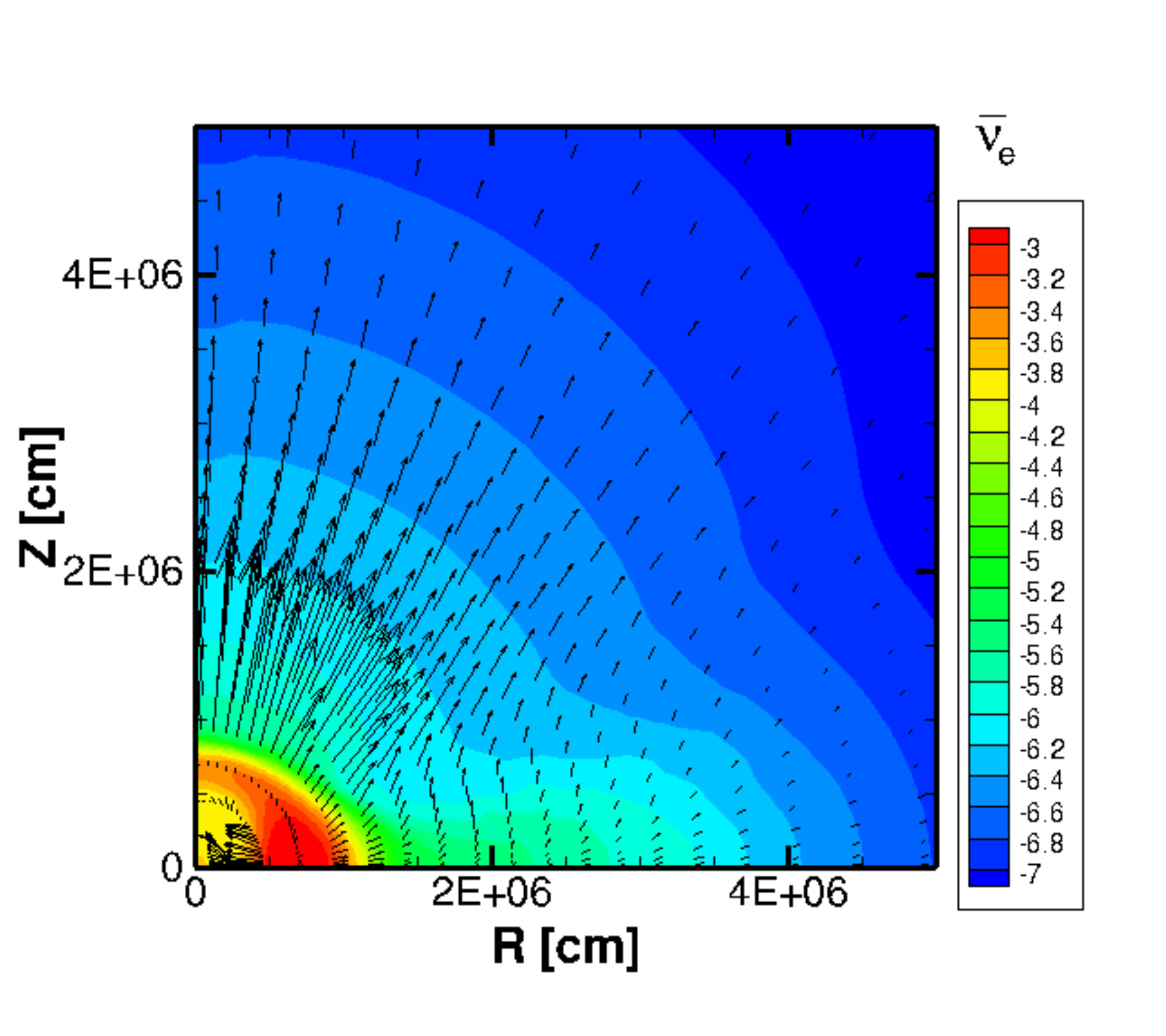}
\plotone{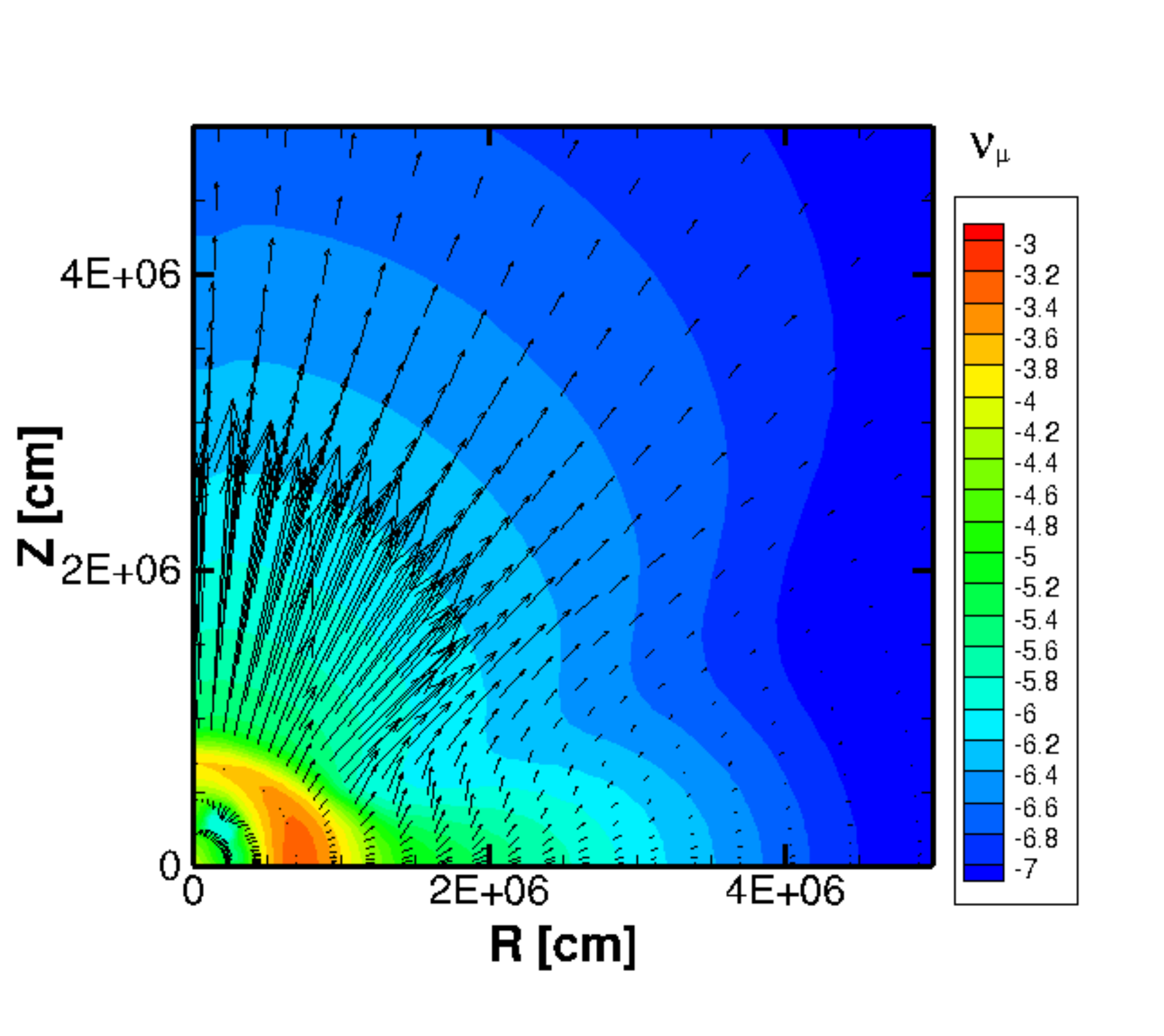}
\plotone{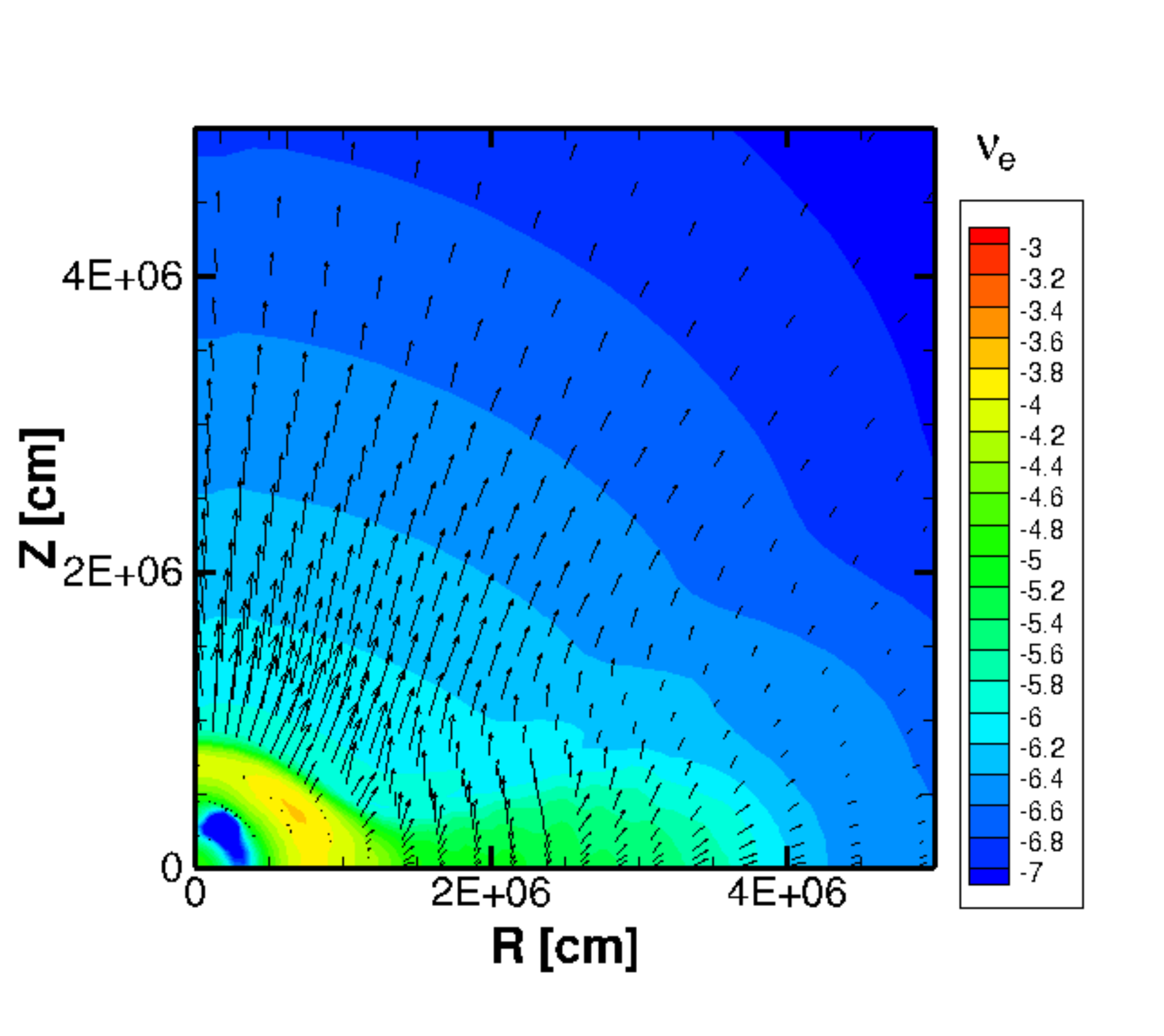}
\plotone{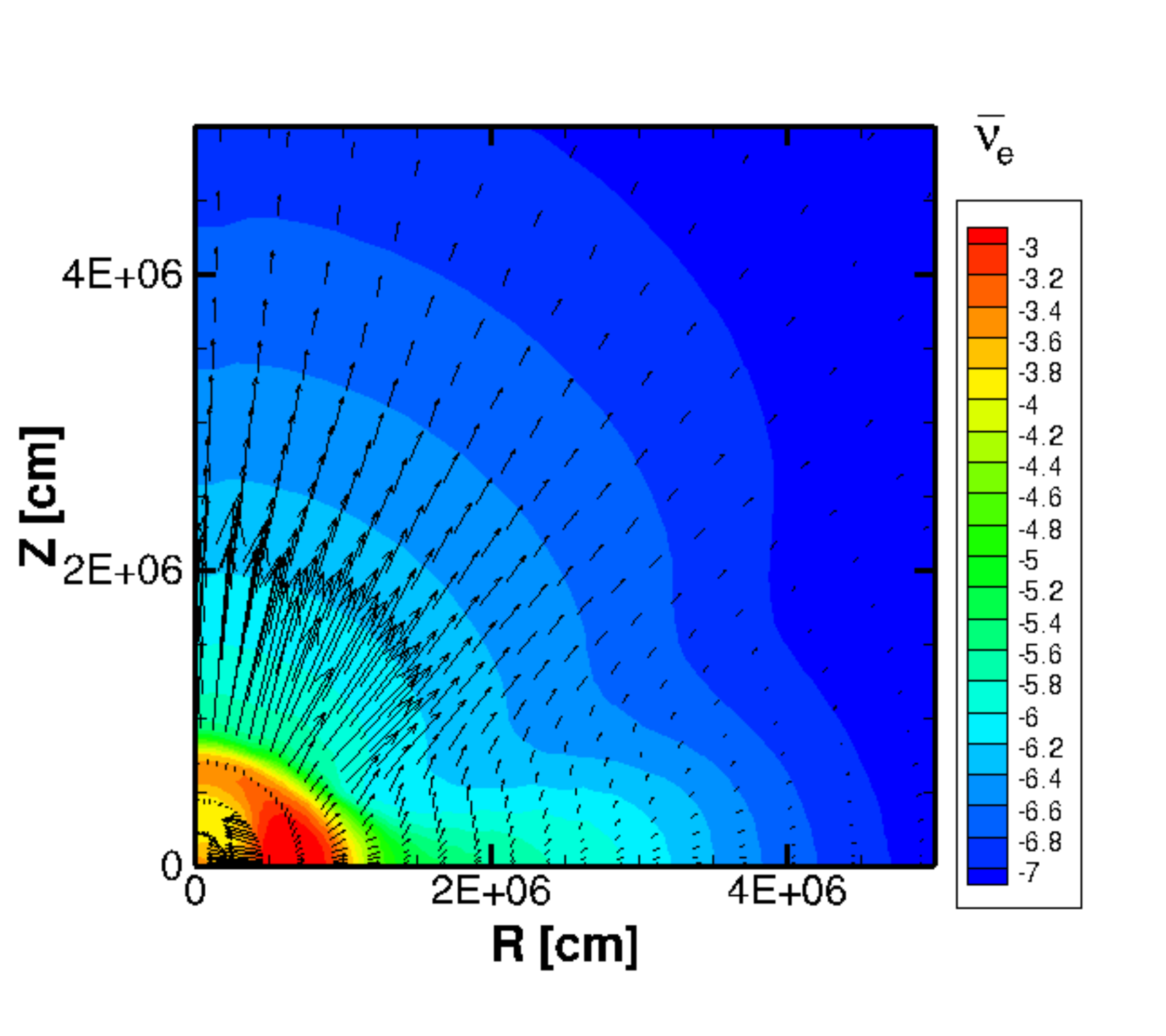}
\plotone{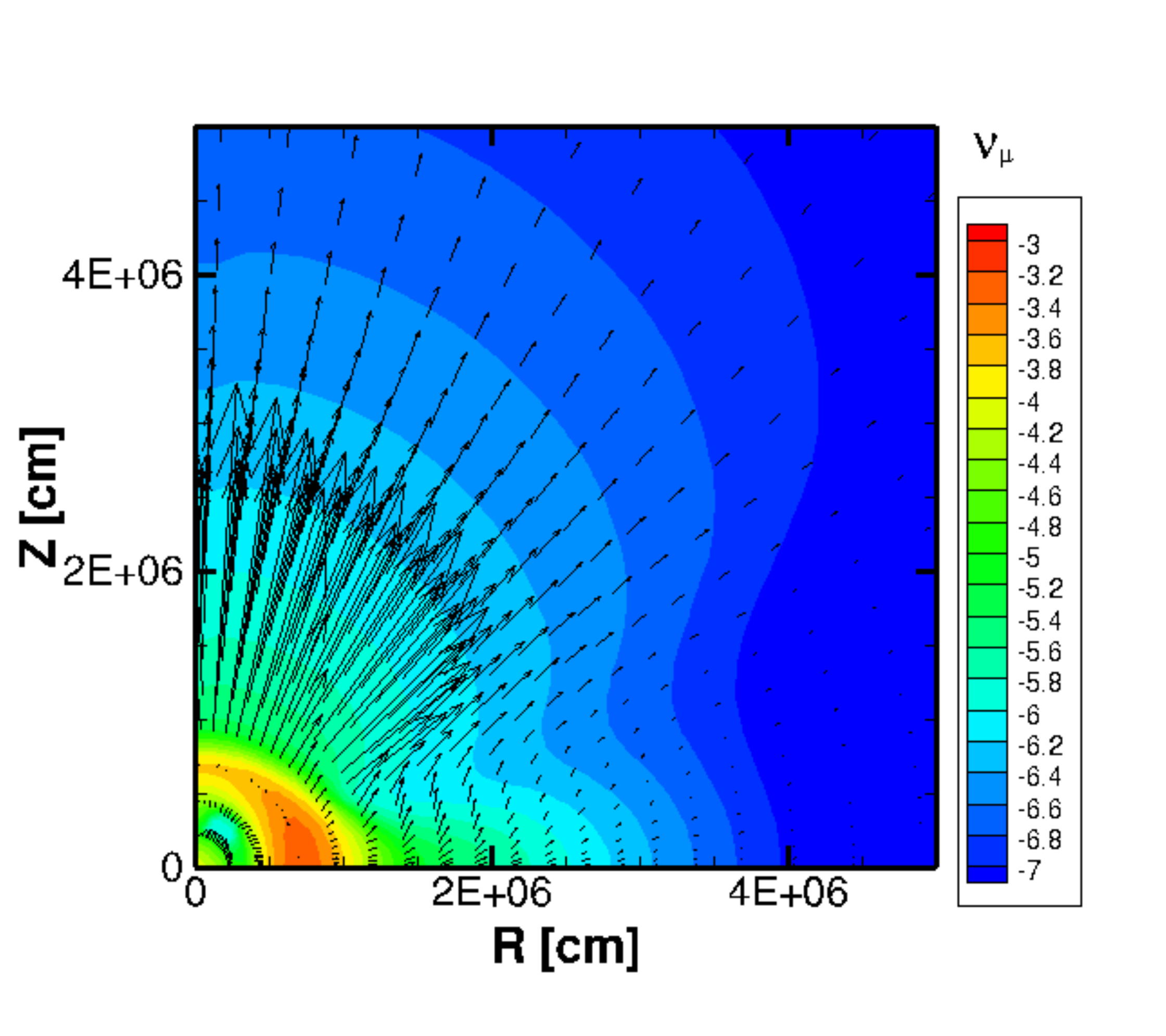}
\plotone{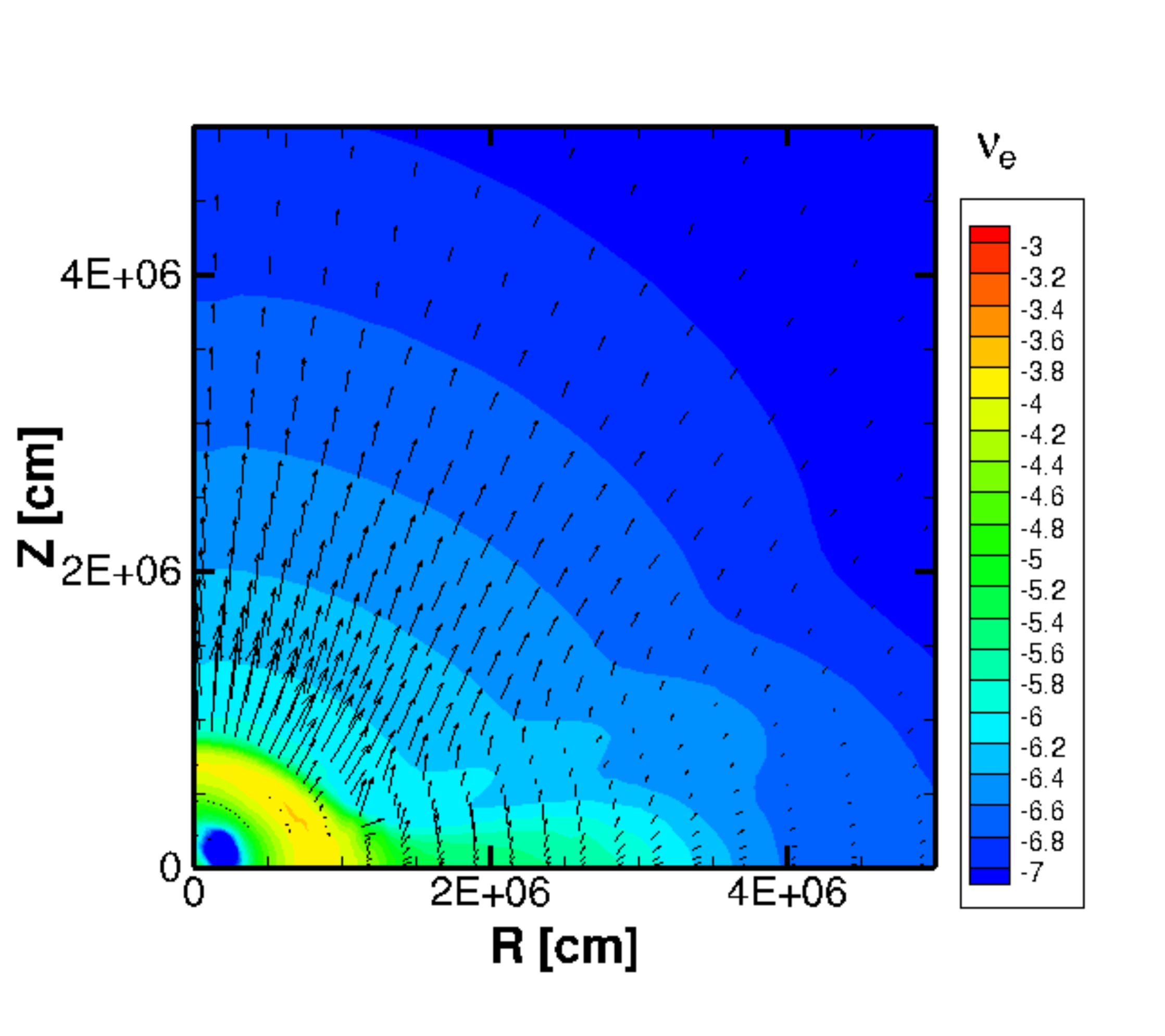}
\plotone{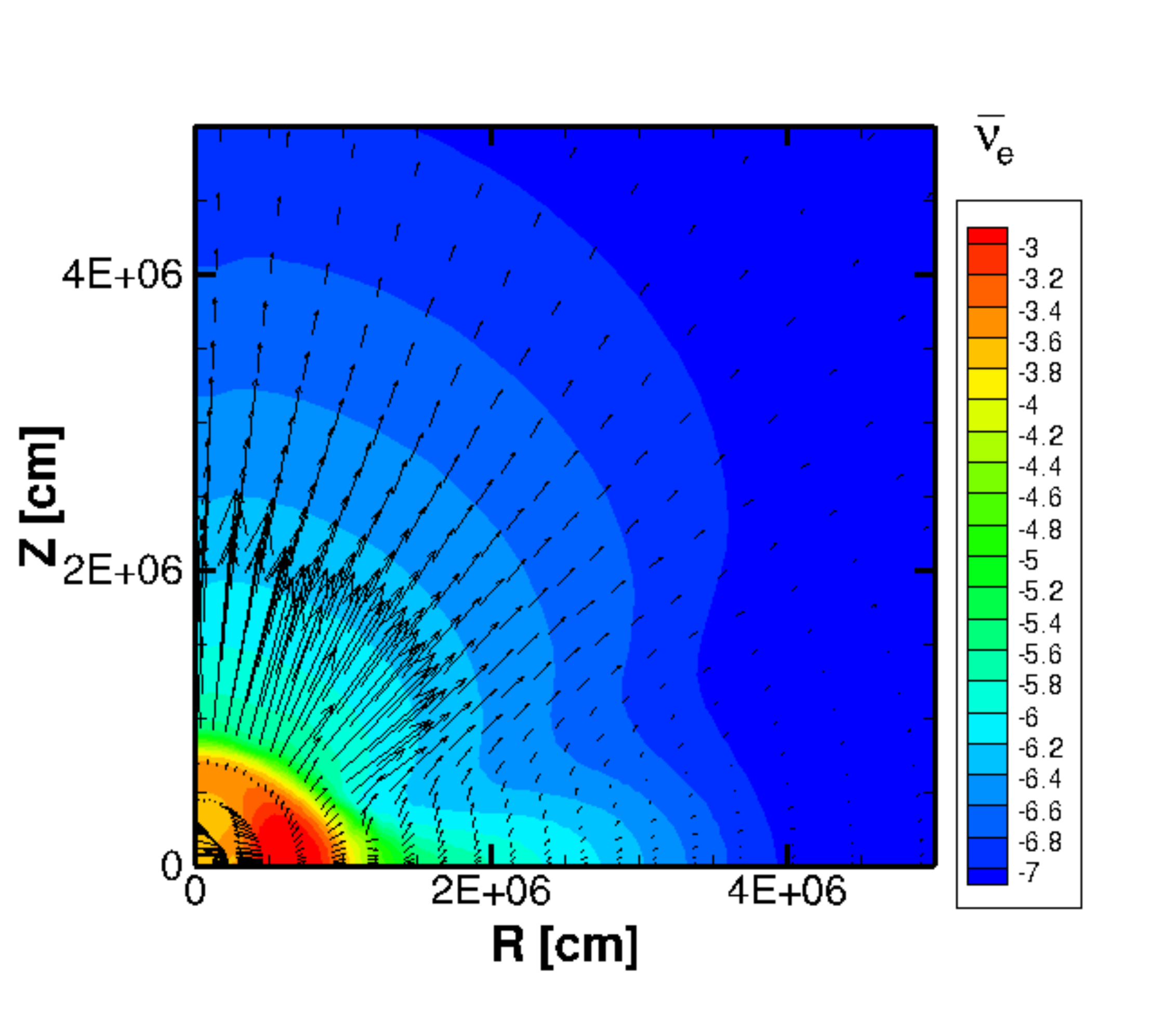}
\plotone{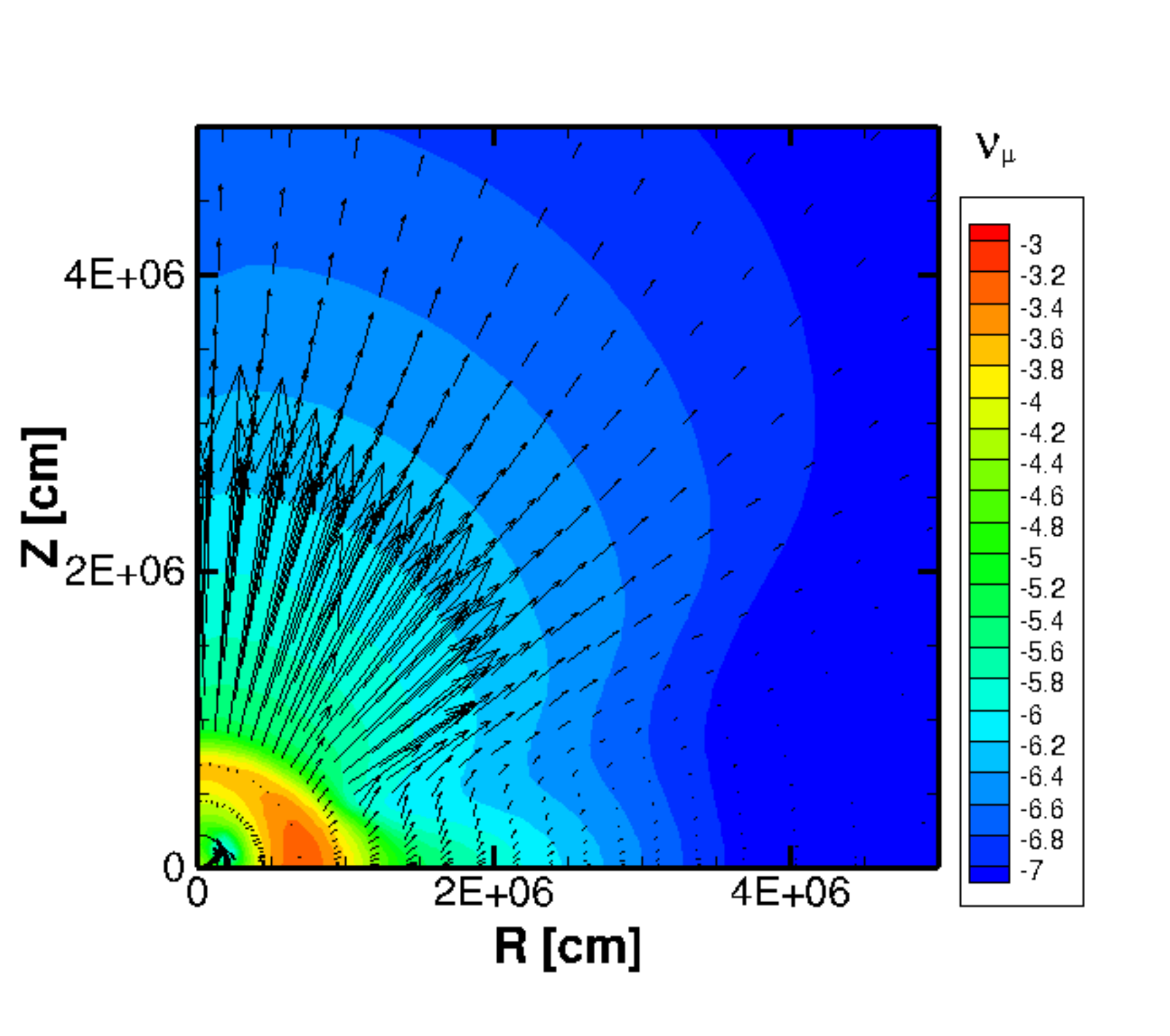}
%
\caption{Neutrino number density and flux are shown for $\nu_e$ (left), $\bar{\nu}_e$ (middle) and $\nu_\mu$ (right) for the profiles at 30 (top), 65 (middle) and 135 (bottom) ms.  The neutrino number densities in the log scale of fm$^{-3}$ are plotted by color maps.  The vectors of neutrino flux are plotted by arrows whose length are proportional to the magnitude.  
\label{fig:nu_densflux_txxxiny}}
\end{figure}

Figure \ref{fig:nu_densflux_txxxiny} displays the number density and flux of neutrinos for $\nu_e$ (left), $\bar{\nu}_e$ (middle) and $\nu_\mu$ (right) for the profiles at 30, 65 and 135 ms.  The asymmetric emission of neutrinos continues up to 135 ms.  
However, the region with high neutrino number densities
in the torus gradually shrinks according to the evolutionary change due to the cooling through neutrino emission (see Fig. \ref{fig:hydro_set_txxx}).  
The reduction of temperature induces this trend as we can see the region in and around the torus at the late profiles, although the structure of matter density profile slowly changes.  (See the bottom panels in Fig. \ref{fig:hydro_set_txxx}).  The neutrino emission from the remnant is gradually reduced accordingly.  

\begin{figure}[ht!]
\epsscale{0.35}
\plotone{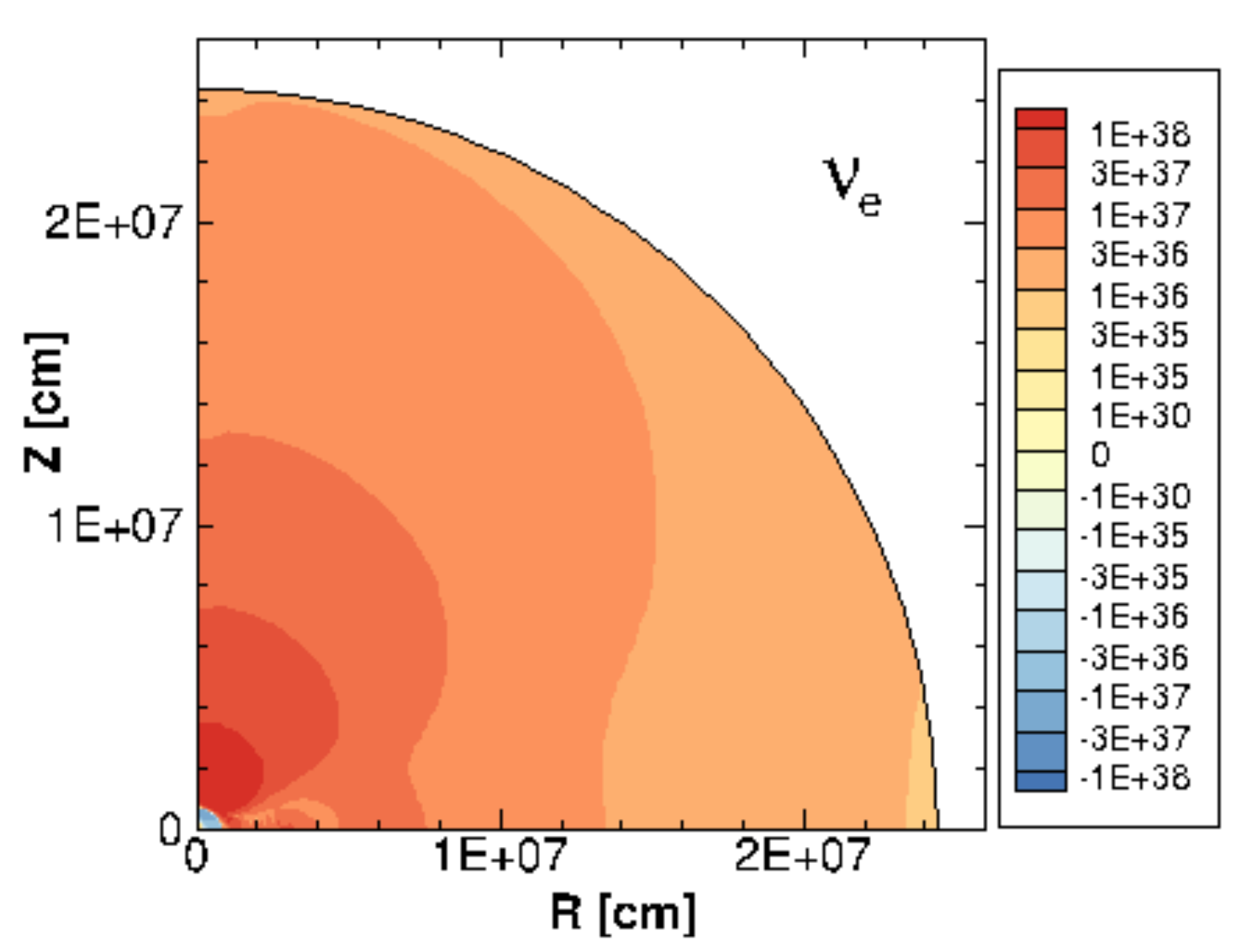}
\plotone{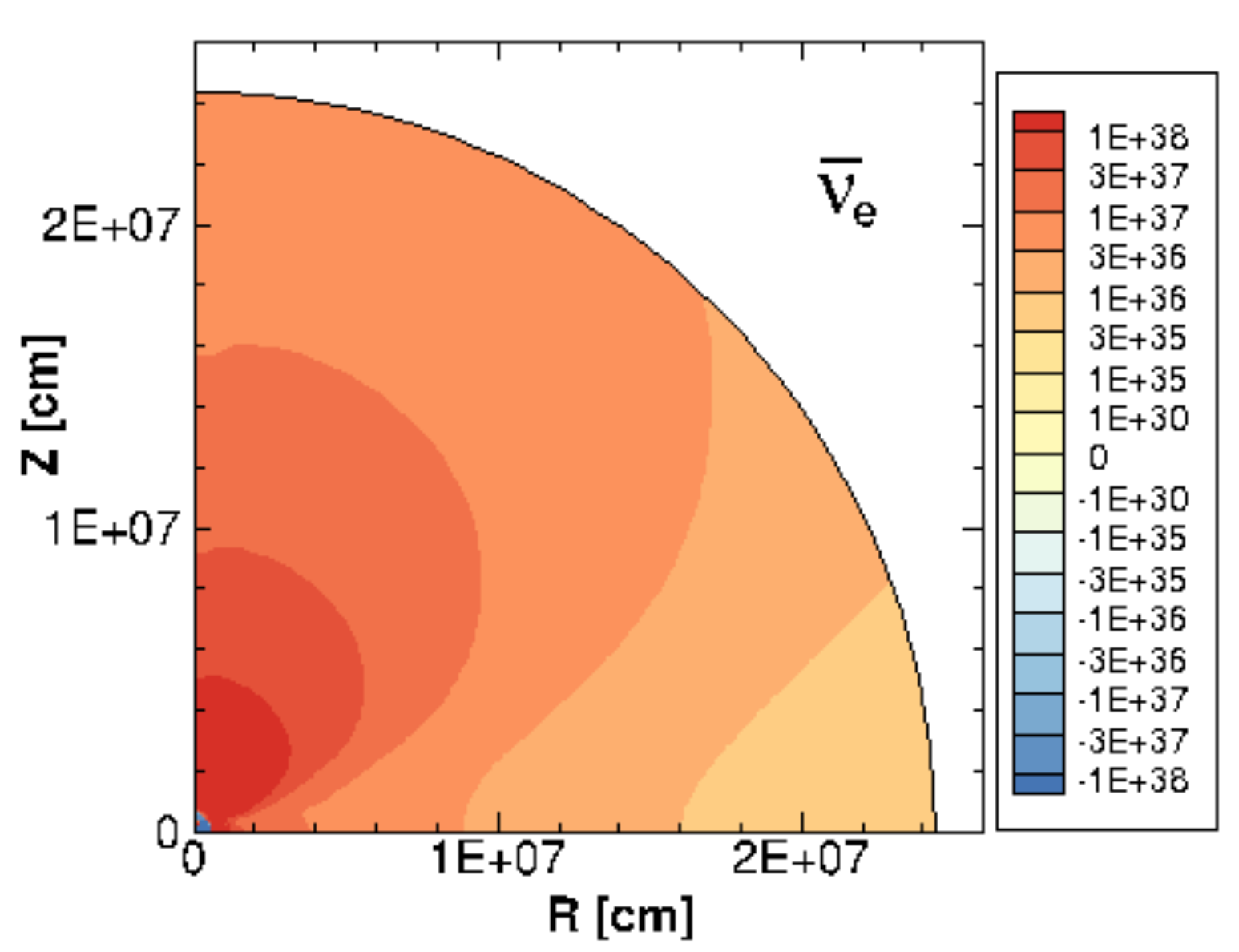}
\plotone{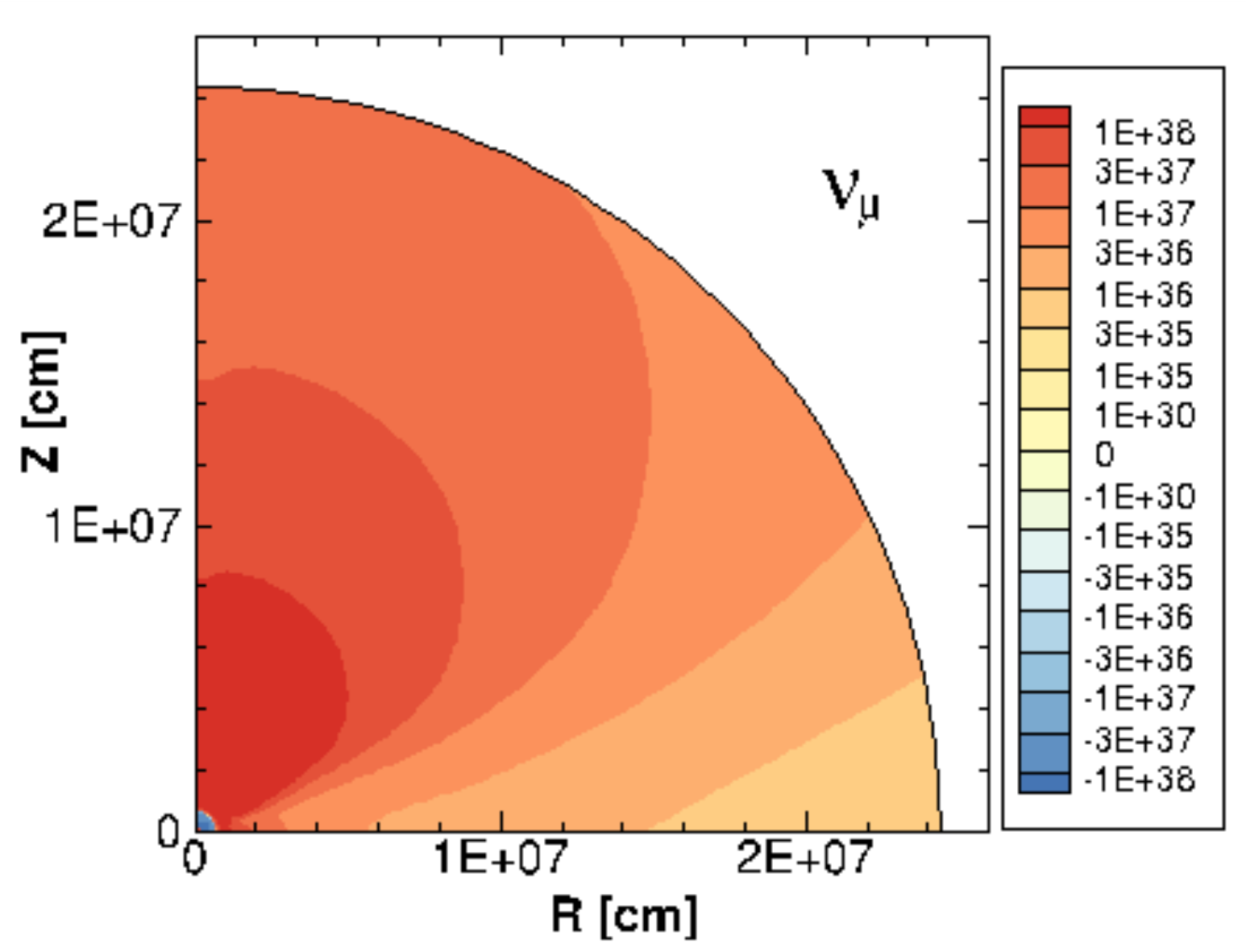}
\caption{The radial component of the neutrino energy fluxes [erg cm$^{-2}$ s$^{-1}$] at 135 ms are plotted for $\nu_e$ (left), $\bar{\nu}_e$ (middle) and $\nu_\mu$ (right).  
\label{fig:nu_eflux_t135iny}}
\end{figure}

The feature in time evolution of neutrino flux depends on the neutrino species.  The fluxes of $\bar{\nu}_e$ and $\nu_\mu$ is focused above the neutron star and this tendency is enhanced at the late profiles.  We show the radial component of the neutrino energy fluxes at 135 ms for three species by the contour plots in Fig. \ref{fig:nu_eflux_t135iny}.  Compared with the situation in Fig. \ref{fig:nu_eflux_t000iny}, the radial energy flux is focused in a narrower region away from the equator.  This tendency is stronger in $\bar{\nu}_e$ and $\nu_\mu$ than in $\nu_e$ because the emission of $\bar{\nu}_e$ and $\nu_\mu$ is confined in the high temperature region of the neutron star whereas the emission of $\nu_e$ continues also from the torus.  
However, the influence of the torus on the neutrino transfer in general becomes minor as the extension becomes small.  

\begin{figure}[ht!]
\epsscale{0.5}
\plotone{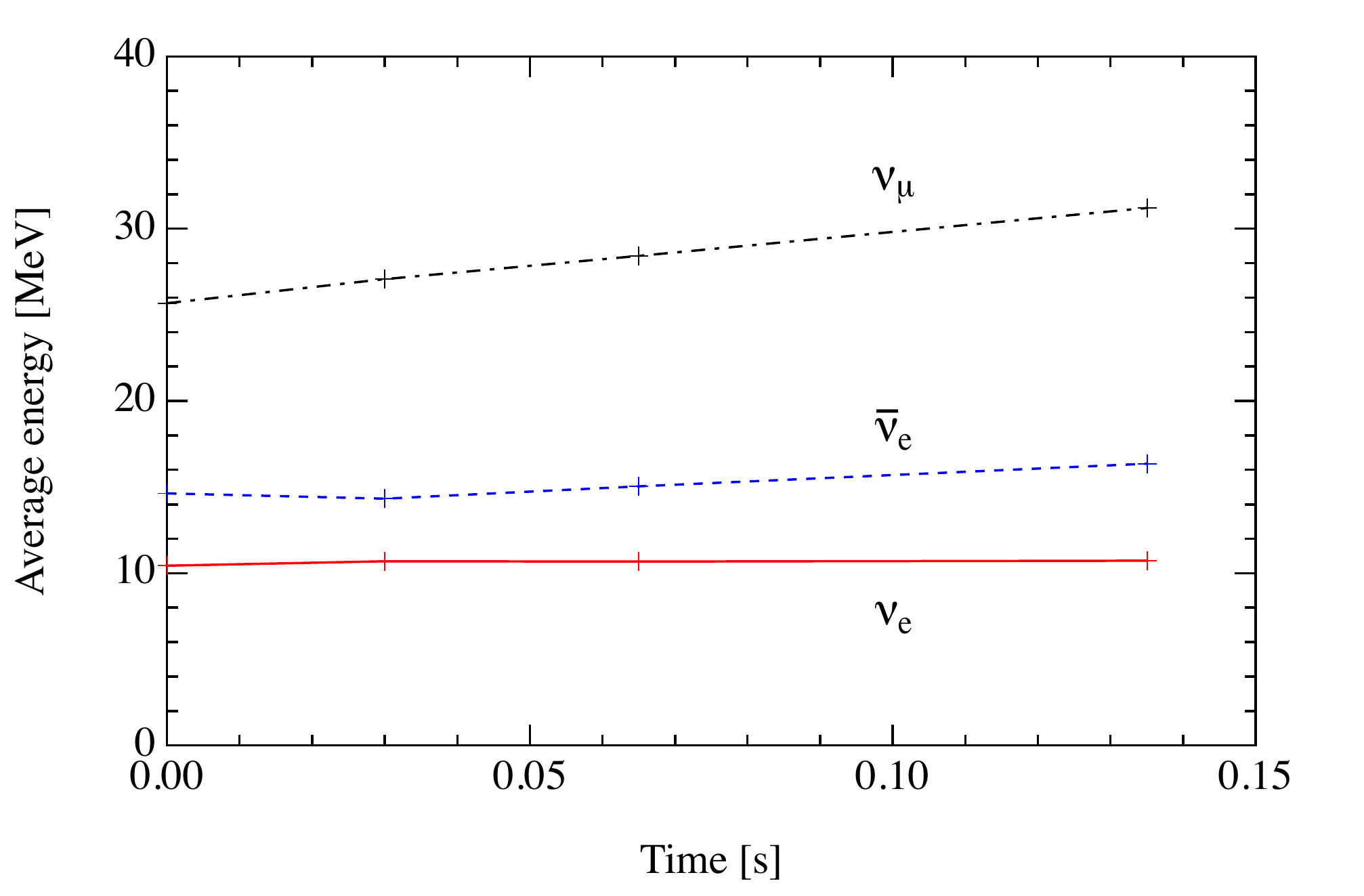}
\plotone{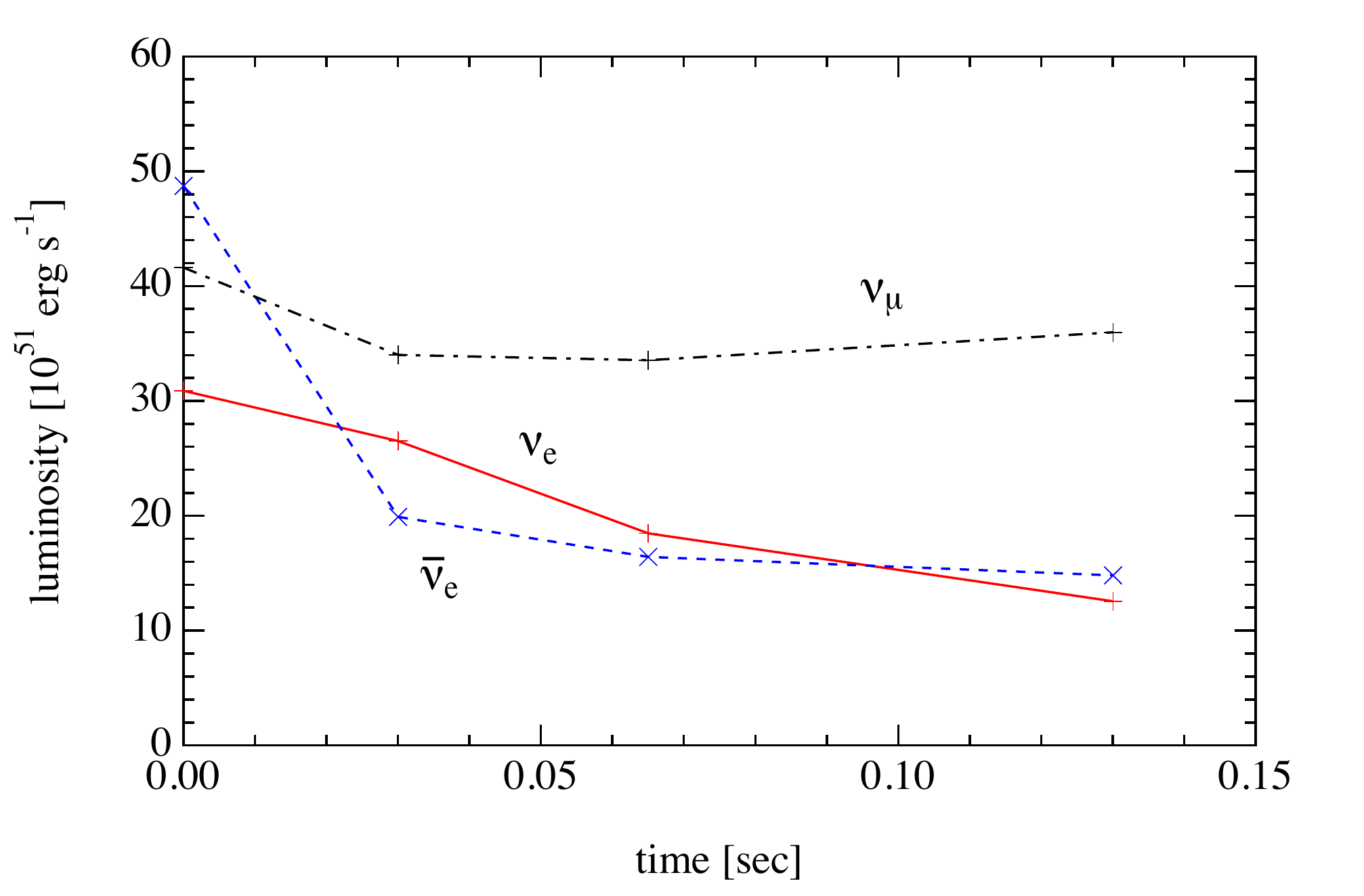}
\caption{Average energy and the total luminosity of neutrino for three species as a function of time.  The average energy is evaluated as the average over the solid angle in the calculated area.  The total luminosity is evaluated by the integral over the solid angle and scaled to cover the $4\pi$ solid angle.  
\label{fig:nu_lumieave_txxx_time}}
\end{figure}

Figure \ref{fig:nu_lumieave_txxx_time} shows the average energy and luminosity at the outer boundary for the four snapshots.  
The average energy is evaluated as the average over the solid angle.  
The luminosity is the integral over the solid angle and scaled for $4\pi$-coverage.  
The average energies for $\nu_\mu$ and $\bar{\nu}_e$ gradually increases whereas the one for $\nu_e$ remains almost the same.  On the other hand, the luminosity for $\nu_\mu$ slowly decreases and the ones for $\nu_e$ and $\bar{\nu}_e$ rapidly decrease.  

%

\begin{figure}[ht!]
\epsscale{0.35}
\plotone{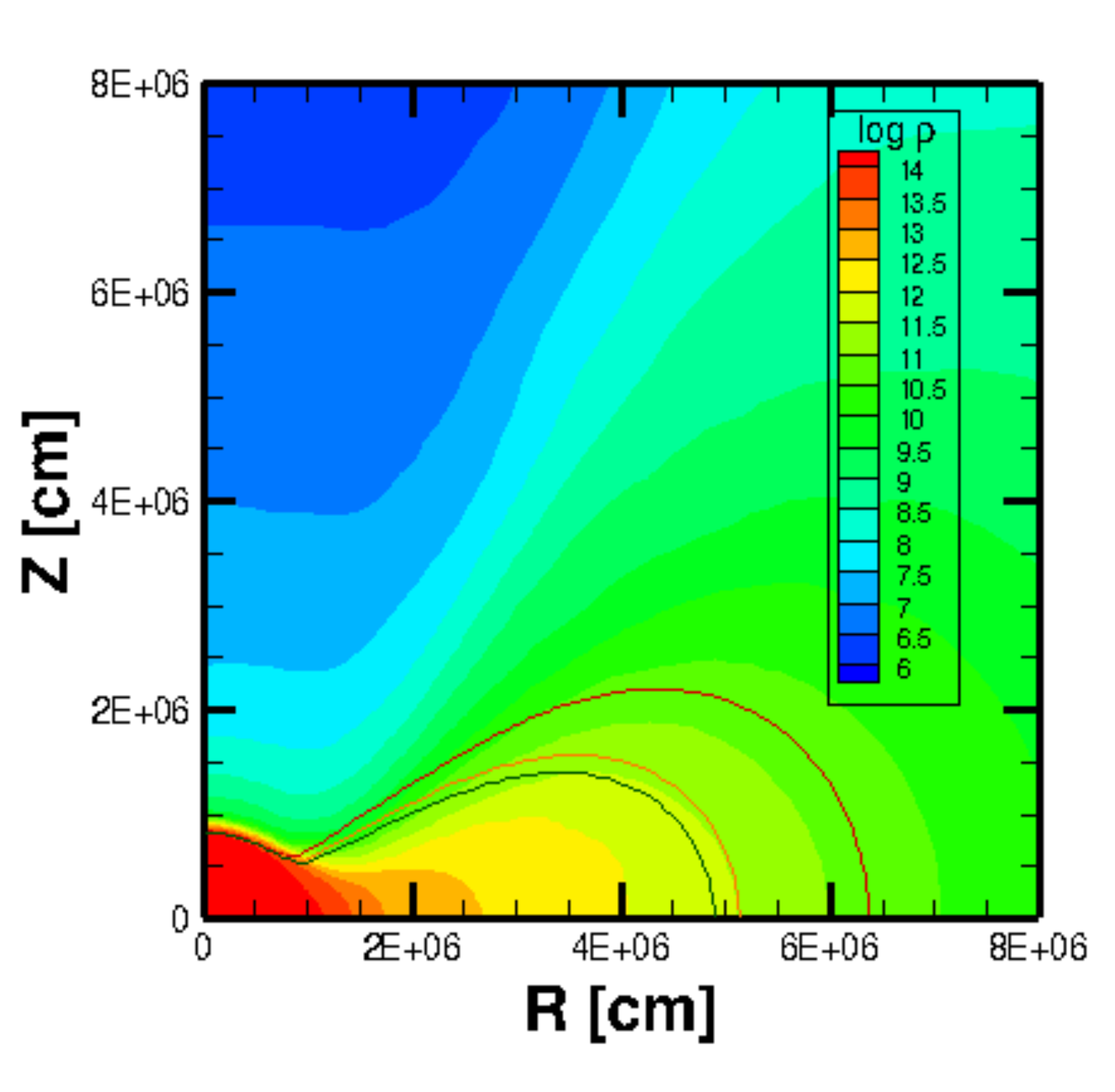}
\plotone{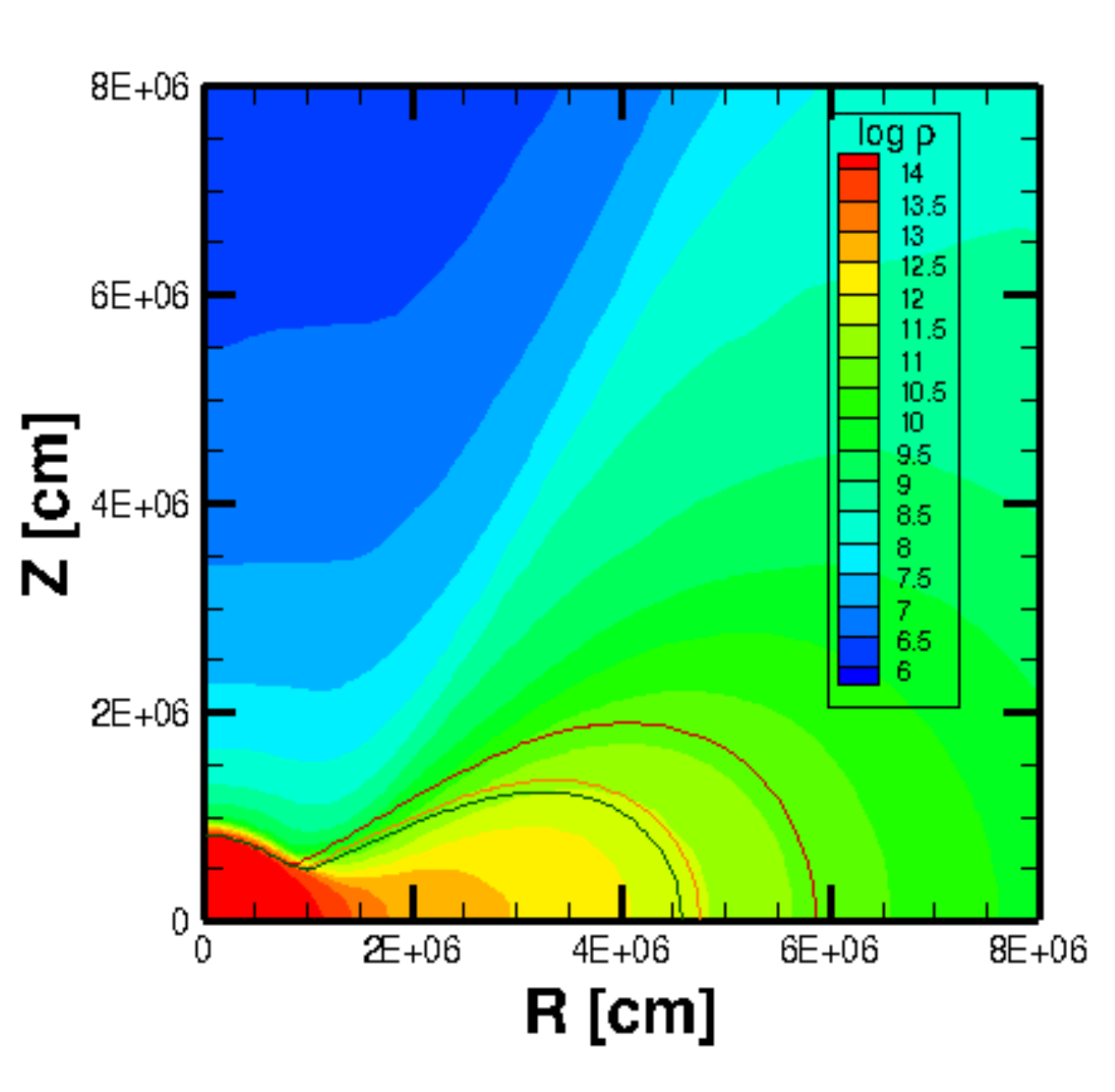}
\plotone{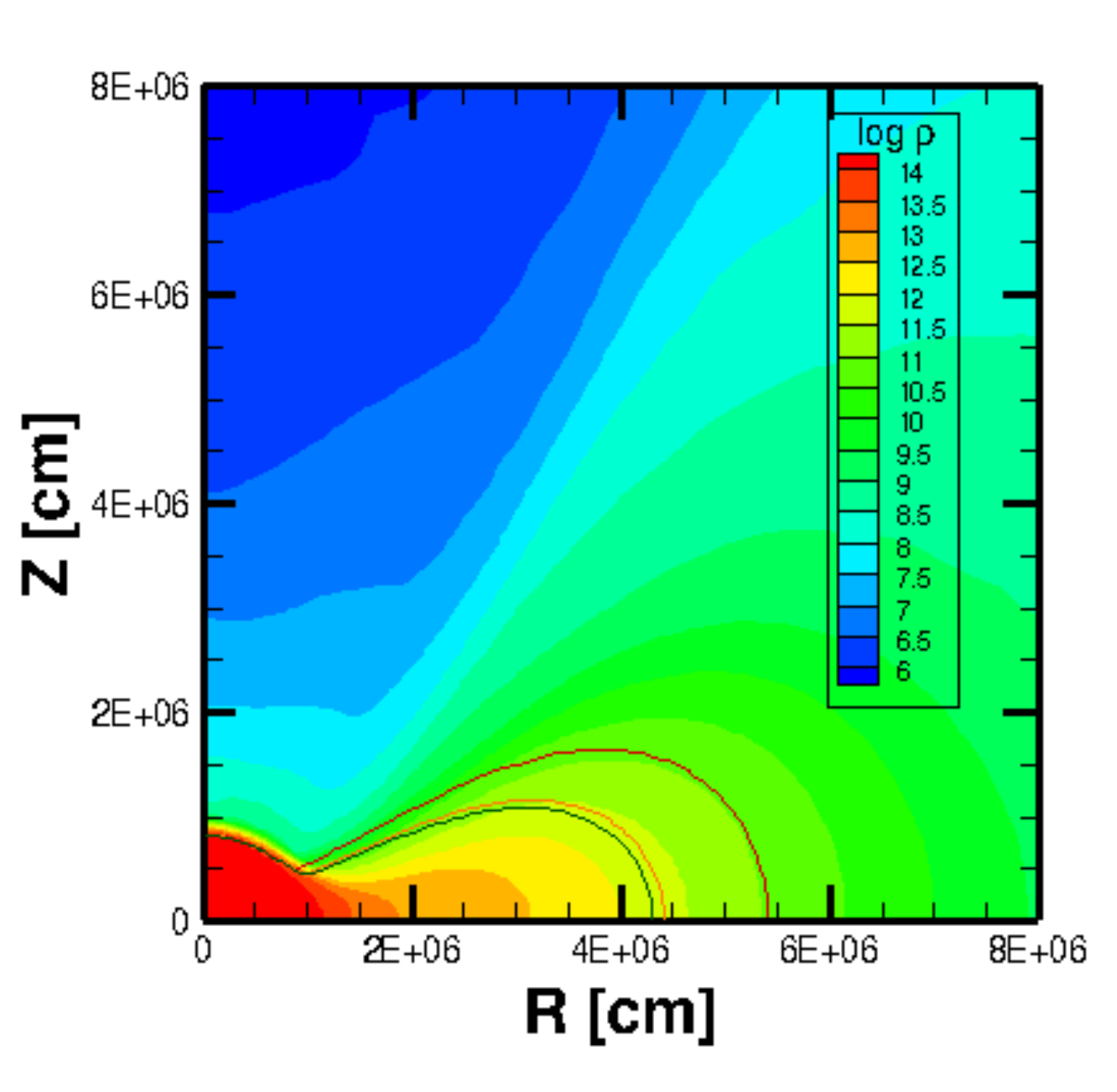}
\caption{Locations of the neutrinosphere at 30 (left), 65 (middle) and 135 (right) ms are drawn on the contour plots of rest-mass density.  The radial position of neutrinosphere is evaluated for neutrino energy of 13 MeV.  Three solid lines correspond to the neutrinosphere for $\nu_\mu$, $\bar{\nu}_e$ and $\nu_e$ in the order from inside to outside.  
\label{fig:nu_sphere_dens_txxxie06}}
\end{figure}

The evolutionary trend of neutrino fluxes is related with the shrinkage of region of neutrino emission and the extended structure of the neutrinosphere.  
Figure \ref{fig:nu_sphere_dens_txxxie06} shows the evolution of the neutrinosphere.  
The shape of neutrinosphere remains extended to the equator due to the presence of the torus although the degree of the extension gradually shrinks as the torus shrinks.  
As the matter density above the neutron star becomes low and the matter density around the equatorial region is still high even at the late stages, the enhancement of flux above the neutron star can happen due to the opaque condition in the equatorial directions.  Since the production of thermal neutrinos ($\bar{\nu}_e$ and $\nu_\mu$) continues in the high temperature region, the neutrino emission region becomes almost confined in the neutron star and the neutrino flux is focused due to the hindrance by the extended neutrinoshpere along the equatorial directions.  In contrast, the production of electron neutrinos continues through charged current reactions in both the neutron star and the torus, and thus, the neutrino flux is not so highly focused.  

\begin{figure}[ht!]
\epsscale{0.35}
\plotone{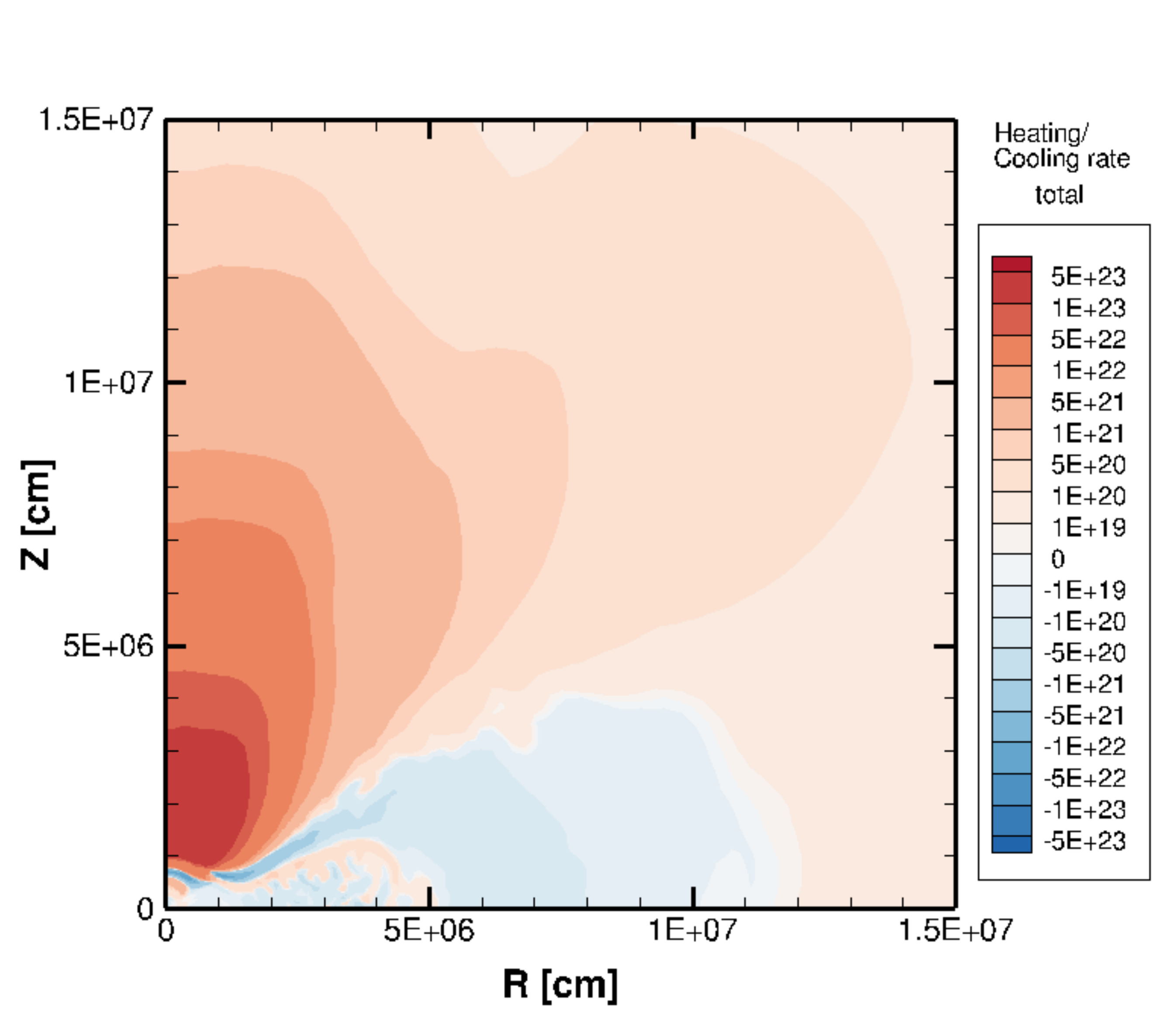}
\plotone{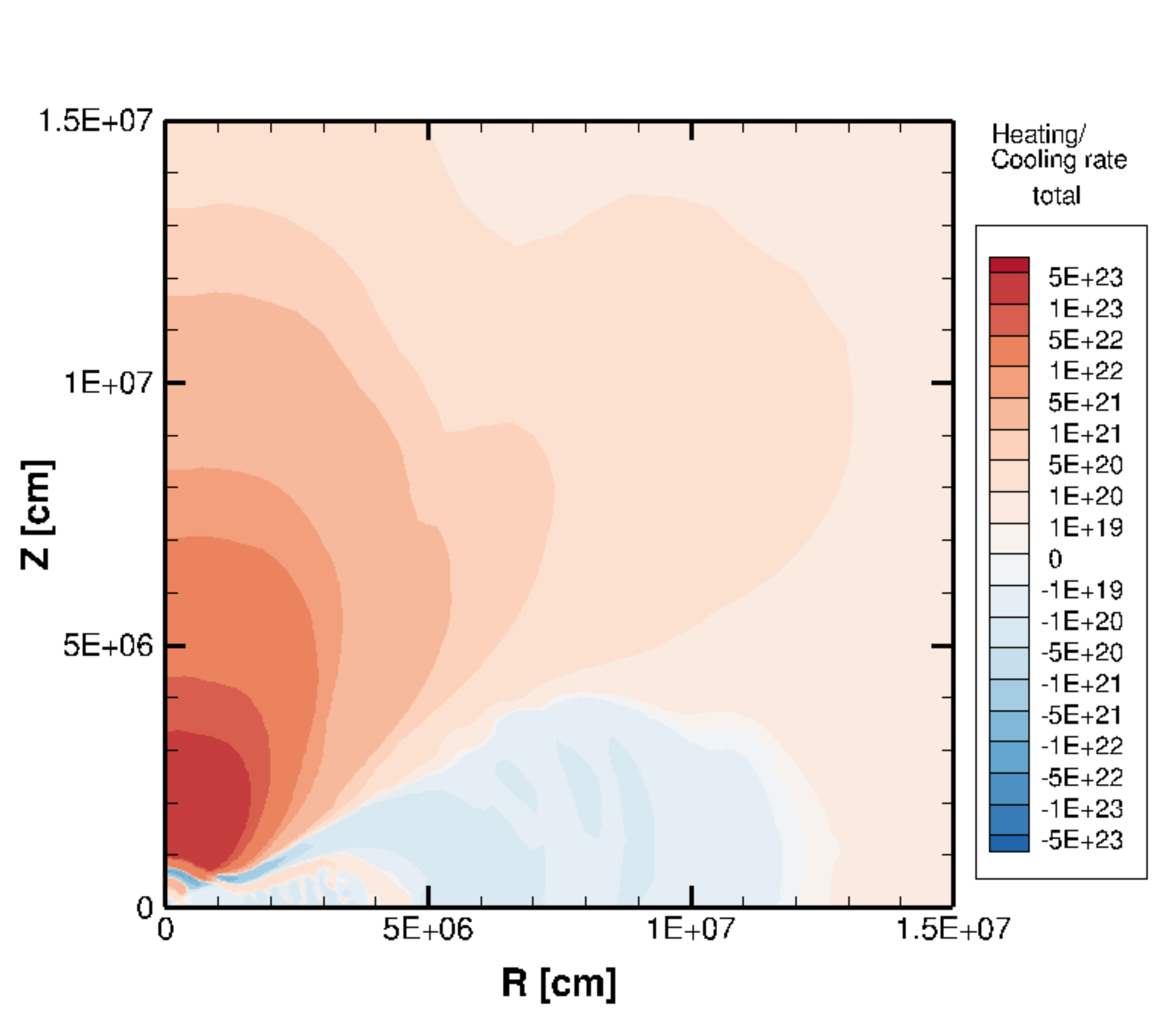}
\plotone{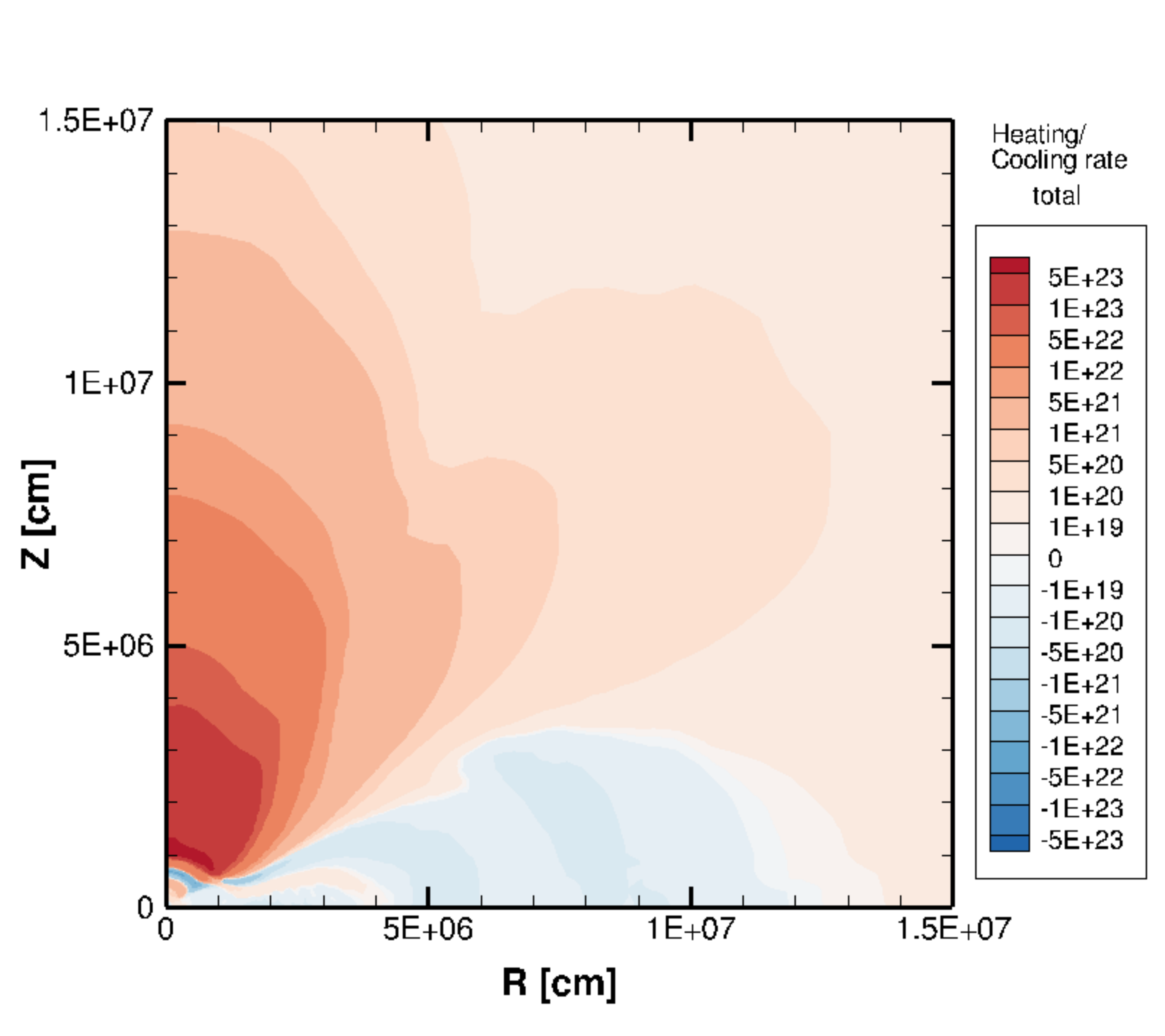}
\caption{Specific heating and cooling rates [erg g$^{-1}$ s$^{-1}$] at 30, 65 and 135 ms are shown by contour plots in reddish and bluish colors by left, middle and right panels, respectively.  
\label{fig:nu_heating_txxx}}
\end{figure}

\begin{figure}[ht!]
\epsscale{0.5}
\plotone{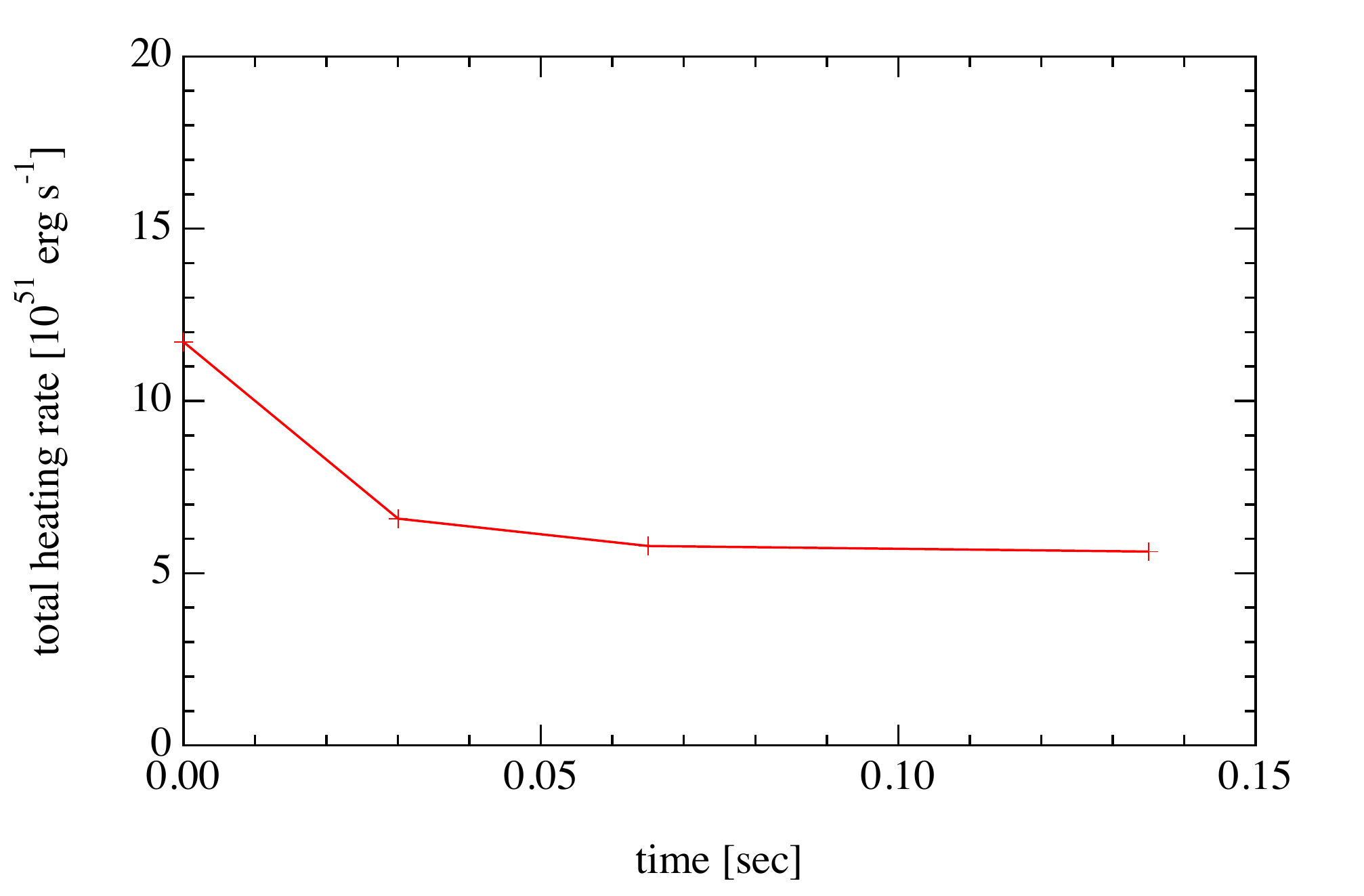}
\caption{Total heating rates [erg s$^{-1}$] is shown as a function of time.  The total heating rate is evaluated by the volume integral using the general relativistic factor and scaled for the  $4\pi$-coverage.  
\label{fig:nu_heating_total_time}}
\end{figure}

It is interesting to examine the time evolution of heating rate above the neutron star merger as a source of mass ejection (and associated radiation) in the region above the remnant.  
We show the evolution of heating and cooling rates by contour plots in Fig. \ref{fig:nu_heating_txxx}.  It is remarkable that the heating in the region just above the neutron star persists over 100 ms.  The total heating rate over the volume scaled for the $4\pi$-coverage is plotted as a function of time in Fig. \ref{fig:nu_heating_total_time}.  
The total heating rate amounts to over $5\times10^{51}$~erg s$^{-1}$ and stays constant even at the late phase.
This continuation of strong heating may influence the ejection of material along the $z$-axis.  
The corresponding heating rate in \cite{fuj17} is $\sim$(6--8)$\times10^{50}$\,erg\,s$^{-1}$ for $t\sim0.05$--0.15\,s, and thus, the heating rate in the late phase could be $\sim7$ times larger with a more sophisticated neutrino transfer method.
Detailed comparison of the two simulations has to be made with considerations of different numerical schemes as well as reaction rates and a separate study will be reported elsewhere.  


\section{Summary}\label{sec:summary}

We studied the properties of neutrino transfer in a remnant system of binary neutron star merger consisting of a massive neutron star and torus by solving the multi-dimensional Boltzmann equation.  Adopting snapshots of the merger remnant obtained in numerical relativity simulations \citep{sek15,fuj17} as the background profile of matter, we performed the numerical simulations of neutrino propagation with reactions to obtain the neutrino distributions in full dimensions with energy and angle dependence.  
The solution of the Boltzmann equation enabled us to examine the properties of neutrino transfer in such a deformed structure in detail.  

We revealed that the neutrino transfer is highly aspherical in the profiles of massive neutron star with an elongated torus.  We show that neutrinos are trapped inside the deformed neutron star and the torus extended along the equator.  The emission of neutrinos proceeds both from the neutron star and the extended torus.  The neutrino fluxes are focused above the neutron star since the torus is geometrically thick along the equator.  

We showed that the shape of the neutrinosphere is largely extended in the deformed neutron star with the geometrically thick torus.  The location of the neutrinosphere is determined by the matter distribution of the merger remnant with dependence on neutrino species and energy.  The large deformation of the neutrinosphere for high energy neutrinos enhances the focused flux above the neutron star near the $z$-axis.  

The aspherical features of neutrino transfer differ depending on the neutrino species due to different origins.
While the energy flux of $\nu_e$ is widely extended, the energy fluxes of $\bar{\nu}_e$ and $\nu_\mu$ are more focused above the neutron star.  
This is because the emission of $\nu_e$ proceeds in both the neutron star and the torus whereas the emission of $\bar{\nu}_e$ and $\nu_\mu$ occurs only in the high temperature region.  In addition, the energy flux 
is focused in the region above the neutron star 
having the optically thick condition of the torus in the equatorial region.  Therefore, the neutrino energy fluxes have different angular variations 
depending on the species.  

We found 
that neutrino heating proceeds dominantly just above the remnant neutron star by the neutrino energy fluxes focused along the $z$-axis.  
The heating occurs through the charged current reactions with nucleons and the pair annihilation of neutrinos.  The neutrinos for heating originate from the emission of the trapped neutrinos in the deformed remnant.  
While the charged current reactions is the main process for electron-type neutrinos, the nucleon-nucleon bremsstrahlung is additionally important 
for the emission of neutrino pairs, especially for $\mu$- and $\tau$-type neutrinos.  

Our simulations of the Boltzmann equation enables us to directly evaluate the angle moments and the Eddington tensors from the full information of space, angle, and energy distributions of neutrinos.  
%
We studied the angle moments and tensor components by the direct integration of the neutrino distribution functions to validate the approximate methods of neutrino transfer.  
We compared the Eddington tensor obtained by the closure relation with those by the direct evaluation.  We found 
that the difference between the two methods is generally small, 
but it becomes large for high energy neutrinos due to the contributions of fluxes both from the neutron star and the torus.  

We found that the general trend of the neutrino transfer is preserved for $\sim$100 ms after the merger by examining the neutrino transfer adopting the series of snapshots as the background.  As the torus along the equator shrinks, the aspherical features of neutrino emission become less drastic.  The neutrino heating above the neutron star persists for $\sim$100 ms in these snapshots.  

The direct evaluation of the neutrino distribution functions by the Boltzmann equation provides new information for the treatment of neutrino transfer in numerical simulations of the remnant of neutron star merger.  
The validation by the direct evaluation would be helpful to improve the approximate methods such as closure relations to provide the angle-averaged quantities in a truncated moment scheme and to gauge parameters of neutrino emission and absorption in simplified schemes.  

We examined the convergence due to the resolution of angle bins in the neutrino transfer.  We demonstrated that it is important to have the high angular resolution, as adopted in the current study, for the convergence of heating rates due to fluxes of neutrino pairs.  

It would be interesting to pursue the evolution of neutrino transfer in more longer evolution of remnant of the neutron star merger. 
Continuous neutrino heating may affect the mass ejection from the merger remnant.  
In its long-term evolution, the magnetic field is amplified in the neutron star and torus by the magnetohydrodynamical instabilities \citep[e.g.,][]{Price2006a,Balbus1991a}. The effective viscosity originating from the instabilities transports the angular momentum in the system. The viscosity also heats up the material in the remnant. In this way, it may modify the matter density and temperature structures of the remnant and affect the neutrino emission.

It will be also important to study differences of neutrino transfer further in various methods to painstakingly solve the Boltzmann equation.  The Monte Carlo method, for example, reveals the asymmetric neutrino emission and provides neutrino energy spectra and angle distributions \citep{sher15,fou20}, which enables examinations of the closure relation used in the M1 scheme \citep{fou18} as in the current study.  While these studies reveal the  luminosities and average energies of neutrino emission and the heating rates through pair annihilation, it is necessary to investigate quantitative differences arose from the methods with different setting and microphysics.  

We are preparing for the detailed comparison of the neutrino transfer with the hydrodynamical simulations so that we can validate and/or improve the approximate methods necessarily adopted for the long term evolution of numerical-relativity simulations.  These analyses will be separately reported elsewhere.  

\acknowledgments

This work is supported by
Grant-in-Aid for Scientific Research
(15K05093, 16H06341, 16K17706, 18K03642, 19K03837, 20H00158, 20H01905)
and 
Grant-in-Aid for Scientific Research on Innovative areas 
"Gravitational wave physics and astronomy:Genesis"
(17H06357, 17H06361, 17H06363, 17H06365)
from the Ministry of Education, Culture, Sports, Science and Technology (MEXT), Japan. 

For providing high performance computing resources, 
Computing Research Center, KEK, 
JLDG of NII, 
Research Center for Nuclear Physics, Osaka University, 
Yukawa Institute of Theoretical Physics, Kyoto University, 
Information Technology Center, University of Tokyo, 
and 
Max Planck Computing and Data Facility 
are acknowledged. 

This work was supported by 
MEXT as "Program for Promoting Researches on the Supercomputer Fugaku" 
(Toward a unified view of the universe: from large scale structures to planets, 
Simulation for basic science: from fundamental laws of particles to creation of nuclei), 
JICFuS, 
and
the Particle, Nuclear and Astro Physics Simulation Program (No. 2019-002, 2020-004) of Institute of Particle and Nuclear Studies, High Energy Accelerator Research Organization (KEK).

\appendix

\section{Eddington tensor}
\label{sec:append_eddington}

We briefly describe here the definition of angle and energy moments 
and the Eddington tensor with associated approximations.  
The energy density, flux and pressure tensor of neutrinos 
are evaluated by the neutrino distribution function, 
$f(\varepsilon, \Omega)$, 
for energy, $\varepsilon$, and angle, $\Omega$, as 
\begin{equation}
\label{eqn:energy}
E =
\int \frac{d \varepsilon~\varepsilon^2}{(2 \pi)^3} \int d \Omega ~
\varepsilon
f(\varepsilon, \Omega) ~, 
\end{equation}
\begin{equation}
\label{eqn:eflux}
F^{i} = 
\int \frac{d \varepsilon~\varepsilon^2}{(2 \pi)^3} \int d \Omega ~
\varepsilon n_{i}
f(\varepsilon, \Omega) ~, 
\end{equation}
\begin{equation}
\label{eqn:epressure}
P^{ij} = 
\int \frac{d \varepsilon~\varepsilon^2}{(2 \pi)^3} \int d \Omega ~
\varepsilon
n_{i}
n_{j}
f(\varepsilon, \Omega) ~, 
\end{equation}
respectively.  
Hereafter we use natural units with $\hbar=c=1$.  Here we suppress the spatial coordinate, $r$, $\theta$ and $\phi$, 
in the neutrino distribution function.  
$n_{i}$ and $n_{j}$ are the unit vectors 
for the three components of $r$, $\theta$ and $\phi$ directions, 
which are denoted by the subscripts, $i$ and $j$.  
The detailed definition of variables and the unit vectors 
in the spherical coordinate system can be found in \citet{sum12}.  

It is essential to handle the moment equations 
with the multi-energy group, for example, for core-collapse supernovae.  
We make analysis of the moments for energy zone 
in the discretized energy variable.  
The energy density for the energy zone, $\varepsilon_k$, 
with width, $\Delta \varepsilon_k$, 
is expressed by 
\begin{equation}
\label{eqn:energy_enu}
E(\varepsilon_k) = 
\frac{\Delta \varepsilon_k~\varepsilon_k^2}{(2 \pi)^3} \int d \Omega ~
\varepsilon_k
f(\varepsilon_k, \Omega) ~.  
\end{equation}
The pressure tensor for the energy zone 
is similarly expressed by 
\begin{equation}
\label{eqn:epressure_enu}
P^{ij}(\varepsilon_k) = 
\frac{\Delta \varepsilon_k~\varepsilon_k^2}{(2 \pi)^3} \int d \Omega ~
\varepsilon_k
n_{i}
n_{j}
f(\varepsilon_k, \Omega) ~.  
\end{equation}
The Eddington tensor for each energy zone 
can be evaluated by 
\begin{equation}
\label{eqn:Eddington-multienergy}
k^{ij}(\varepsilon_{k}) = 
\frac{\int d \Omega ~ n_{i} n_{j} f(\varepsilon_{k}, \Omega)}
     {\int d \Omega ~             f(\varepsilon_{k}, \Omega)} ~, 
\end{equation}
from the neutrino distribution function.  
Since we describe the neutrino distribution function 
by the Boltzmann equation in 6D, 
we can directly provide these angle moments.  

In the moment formalism with termination at a certain rank, 
the highest moment has to be determined by lower moments.  
For example, it is necessary to provide the second moment, 
i.e. the pressure tensor, 
through closure relation 
by a functional form of the energy density and flux.  
In a classic form of the closure relation by \citet{lev84}, 
the closure relation is given by flux vectors 
via a form of the tensor, $t^{ij}(\varepsilon_{k})$, 
\begin{equation}
\label{eqn:Pressure-closure-tensor}
P^{ij}(\varepsilon_k) = t^{ij}(\varepsilon_{k}) E(\varepsilon_k) ~, 
\end{equation}
to determine the pressure tensor.  
The tensor form of closure is given by 
\begin{equation}
\label{eqn:closure-tensor}
t^{ij}(\varepsilon_{k}) = 
  \delta^{ij}          \frac{1 - \chi}{2} 
+ u^i u^j  \frac{3\chi - 1}{2} ~.  
\end{equation}
$u^i$ is the unit vector defined by 
\begin{equation}
\label{eqn:unit-vector}
u^i = \frac{f^i}{f} ~, 
\end{equation}
using the normalized flux 
\begin{equation}
\label{eqn:normalized-flux}
f^i = \frac{F^i(\varepsilon_k)}{E(\varepsilon_k)} ~, 
\end{equation}
and its norm, $f$.  
$\chi$ is the Eddington factor 
which smoothly connects the two limiting cases 
from the diffusion regime ($\chi=1/3$) 
to the transparent regime ($\chi=1$).  
We adopt the expression as a function of $f$ 
by \citet{lev84}, 
\begin{equation}
\label{eqn:chi-function-f}
\chi =  \frac{1}{3} + \frac{2 f^2}{2+ \sqrt{4-3f^2}} ~, 
\end{equation}
in the current study.  
We examine 
the difference of the tensor form of closure 
from the Eddington tensor 
\begin{equation}
\label{eqn:tensor_deviation}
\Delta t^{ij} = t^{ij} - k^{ij}
\end{equation}
to check the quality of approximation 
by the closure relation.  

%

\section{Angular resolution}
\label{sec:append_resolution}

In order to assess the convergence of the grid resolution in angle mesh for neutrinos, we perform a series of simulations for additional selected models by changing the number of grids for angles.  We study the dependence of the behavior of moments, tensor, and heating rate on the angular resolution for the models of central core in the gravitational collapse of massive stars.  As we show below, the number of angle grid we adopted is sufficient for the current purpose to study the deformed neutron star with torus.  We take the procedure to obtain the stationary state of neutrino distribution function by solving the Boltzmann equation with the fixed background profile chosen from hydrodynamics.  We utilize the same setting of neutrino reactions as described in \S \ref{sec:boltzmann} using the Shen EOS \citep{she98a,she98b} matched with hydrodynamics.  We adopt the same number of grids for space as well as neutrino energy and azimuthal angle as in the case of profiles from neutron star merger, but change the number of neutrino polar angle grids, $N_{\theta_{\nu}}$, from 6, 12, 18, 24, 30, 36, 40, 44, 48, 52 to 56.  


A set of profiles is taken from the 2D numerical simulation \citep{sek12,kot12} of rotating collapse of a 100M$_\odot$ star by \citet{ume08}.  The core-collapse simulation revealed the dynamics of a collapsar; the formation of neutron star with a geometrically thick torus from the rotating massive star as a model of long gamma-ray bursts.  We adopt a profile of the central part including a neutron star with the torus-like extension formed after the core bounce.  Due to the rotation given in the initial condition, the massive proto-neutron star has an elongated shape with extended matter distribution.  This situation resembles the remnant in neutron star merger.  Therefore, the examination of angular resolution is applicable to the situation in the current study.  

\begin{figure}[ht!]
\epsscale{0.5}
\plotone{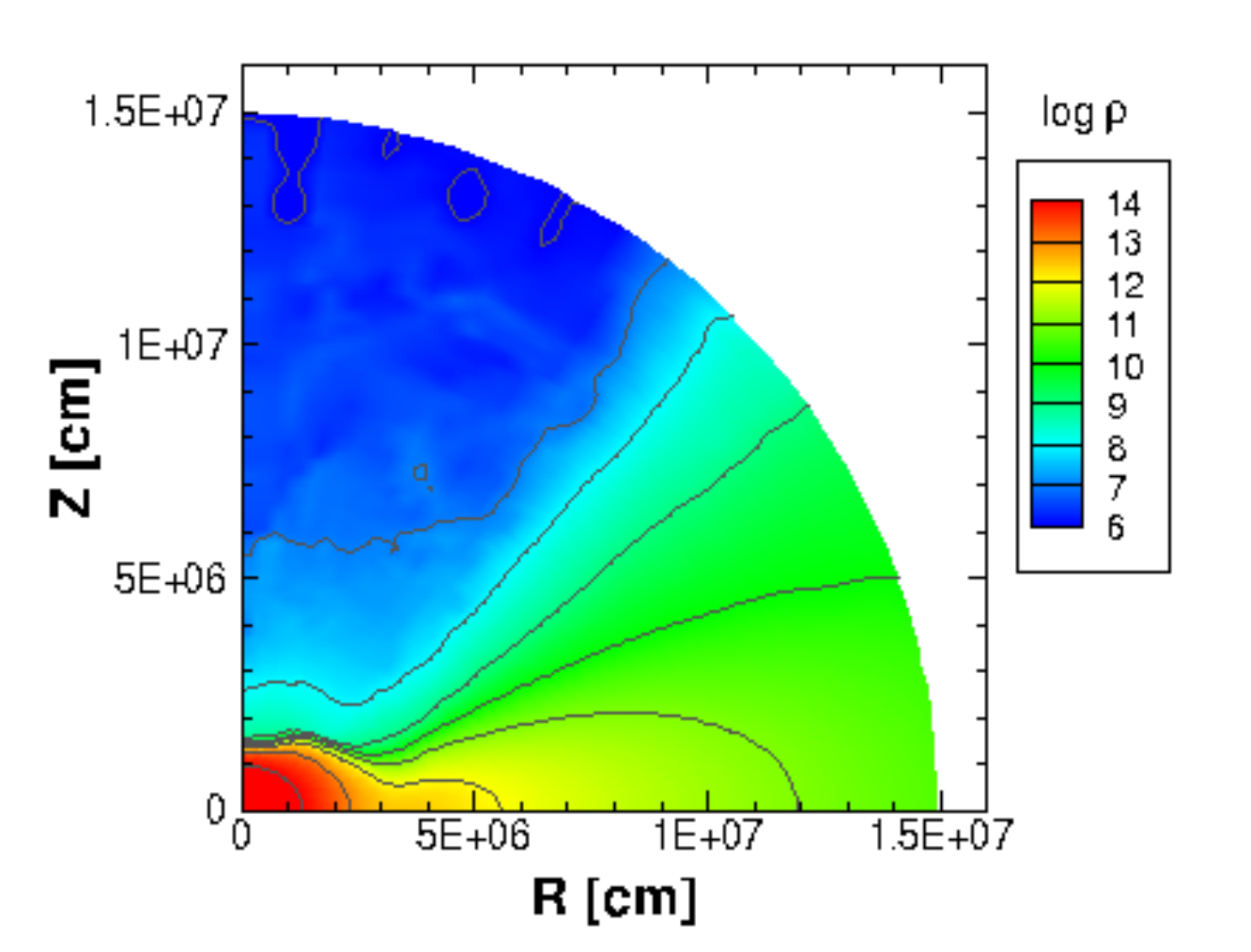}
\plotone{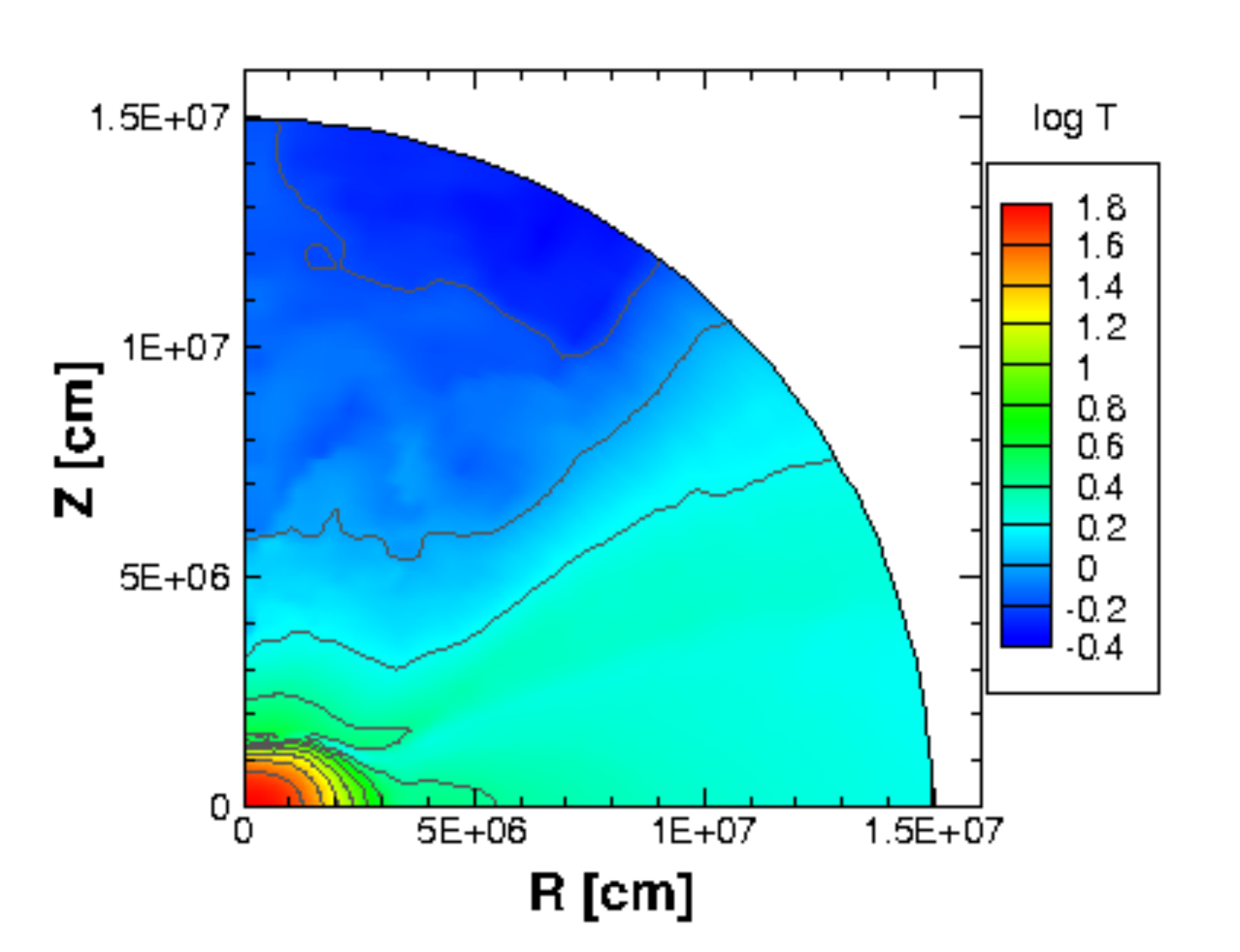}
\caption{Profiles of hydrodynamics quantities in the collapsar.  The rest-mass density [g cm$^{-3}$] (left) and temperature [MeV] (right) are shown in the plane of R and Z axes.  Note that the rest-mass density and temperature are plotted in log scale.  
\label{fig:sek_resolution_profile}}
\end{figure}

\begin{figure}[ht!]
\epsscale{0.45}
\plotone{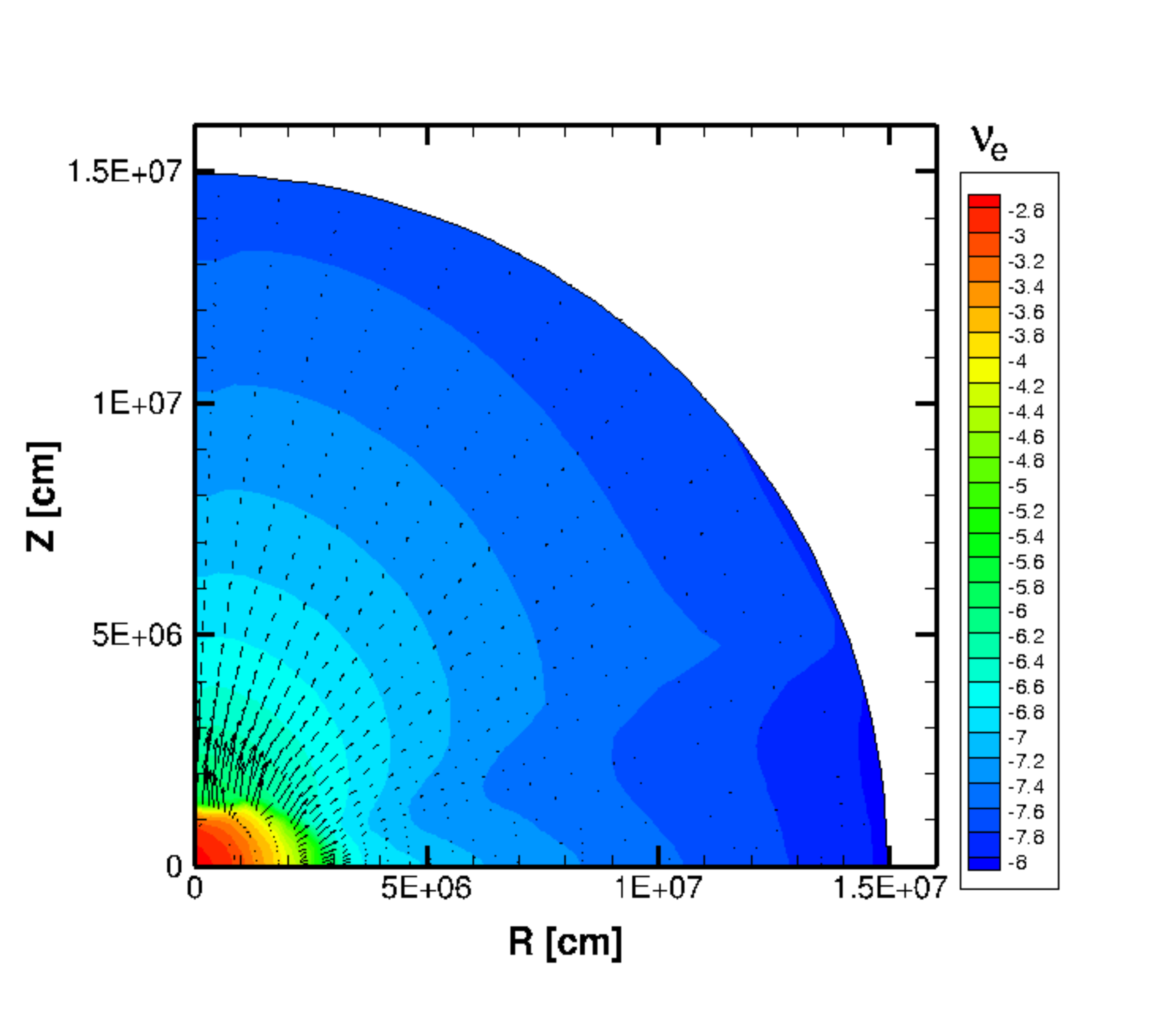}
\plotone{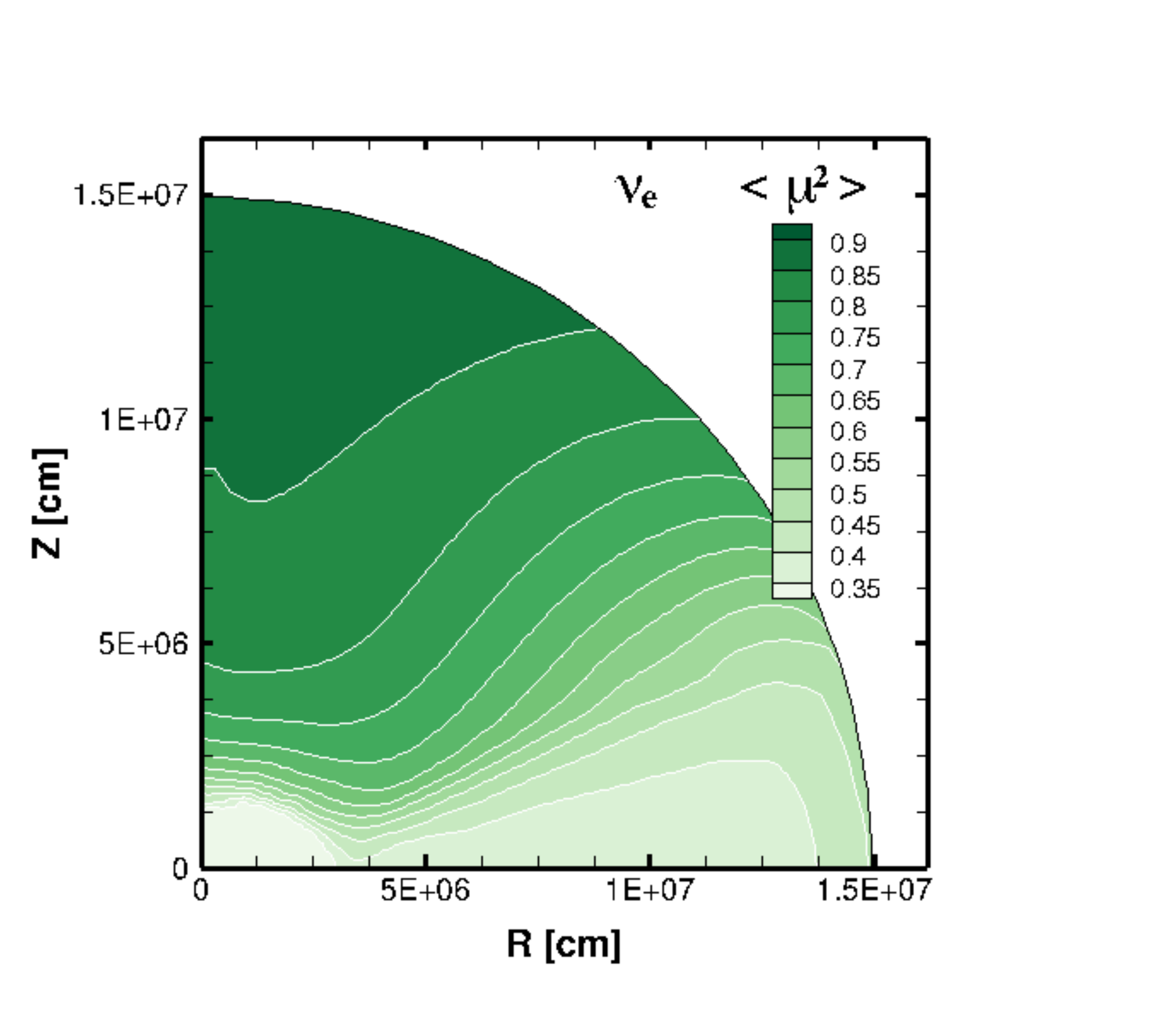}
\caption{Profiles of neutrino distributions in the collapsar.  The number density and flux (left) and the averaged squared angle moment, $\langle\mu_{\nu}^2\rangle$, (right) are shown for $\nu_e$ in the plane of R and Z axes.  Note that the number density of neutrinos [fm$^{-3}$] are plotted in log scale.  
\label{fig:sek_resolution_densflux}}
\end{figure}

We show, in Fig. \ref{fig:sek_resolution_profile}, the profiles of hydrodynamics quantities in the collapsar.  At the center, the massive proto-neutron star is born with the extended torus at high matter density and temperature.  The torus has a geometrically thick structure along the equator and the material is dilute along the $z$-axis.  The detailed information about the dynamics and snapshots can be found in \citet{sek12} and in \S 3.3.3 of \citet{kot12}.  We demonstrate the profiles of neutrino quantities obtained by the simulations of the Boltzmann equation with the highest angular resolution in Fig. \ref{fig:sek_resolution_densflux}.  The neutrino distributions and fluxes are non-spherical having abundant neutrinos in the extended neutron star with enhanced fluxes along the $z$-axis.  The neutrino distributions is deformed with nearly isotropic region ($\langle\mu_{\nu}^2\rangle\sim\frac{1}{3}$) extended along the equator.  
 
\begin{figure}[ht!]
\epsscale{0.5}
\plotone{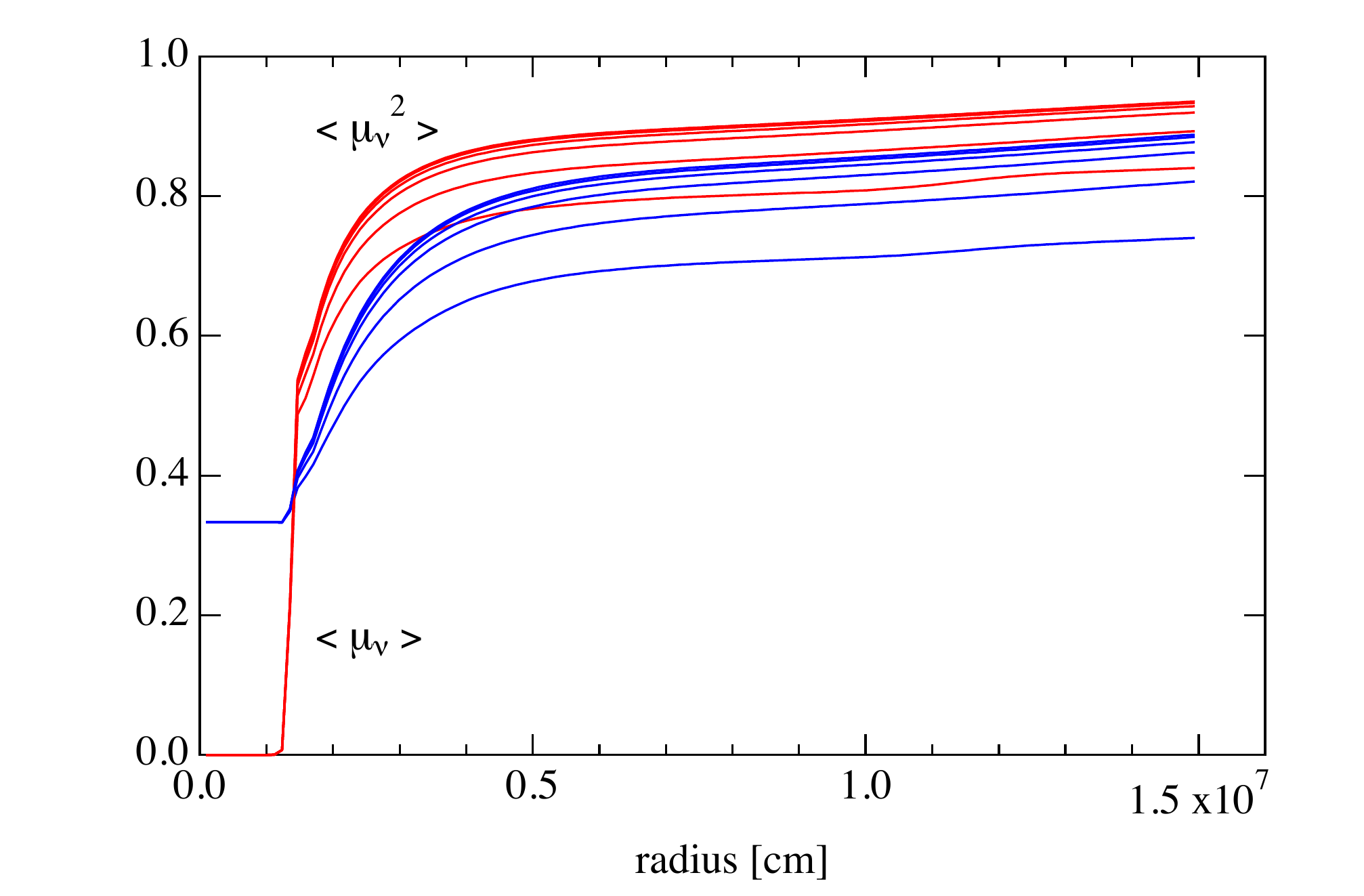}
\plotone{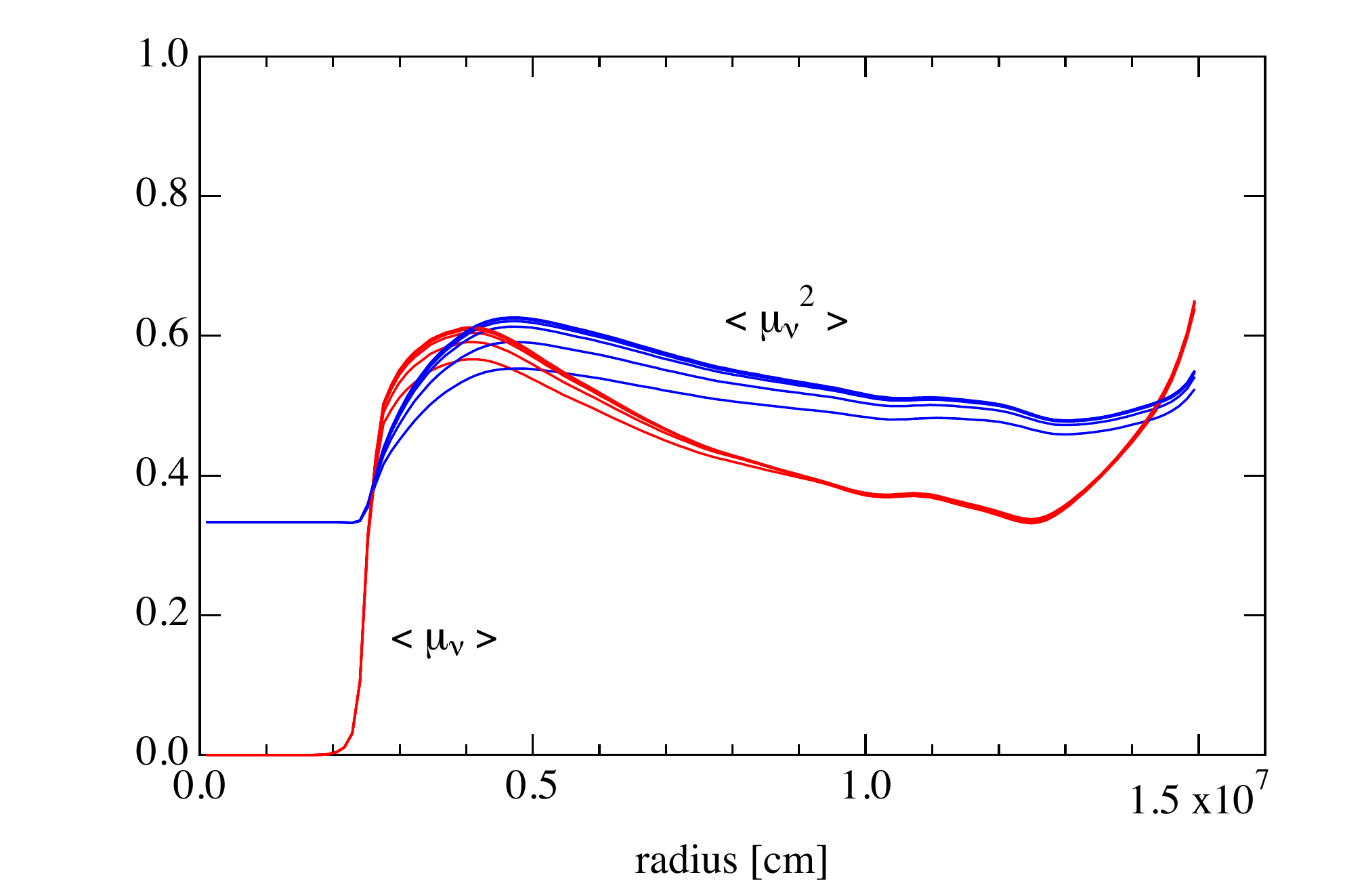}
\caption{Convergence of angle moments for the cases with the number of angle grids of 6, 12, 24, 36, 48 and 56 from bottom to up.  The average values of angle moments $\langle\mu_{\nu}\rangle$ (red) and $\langle\mu_{\nu}^2\rangle$ (blue) for $\nu_e$ are shown as functions of radius.  
The profiles along the polar angles at 2.0 and 69 degree from the $z$-axis are shown in left and right panels, respectively.  
\label{fig:sek_resolution_mom_radial}}
\end{figure}

\begin{figure}[ht!]
\epsscale{0.5}
\plotone{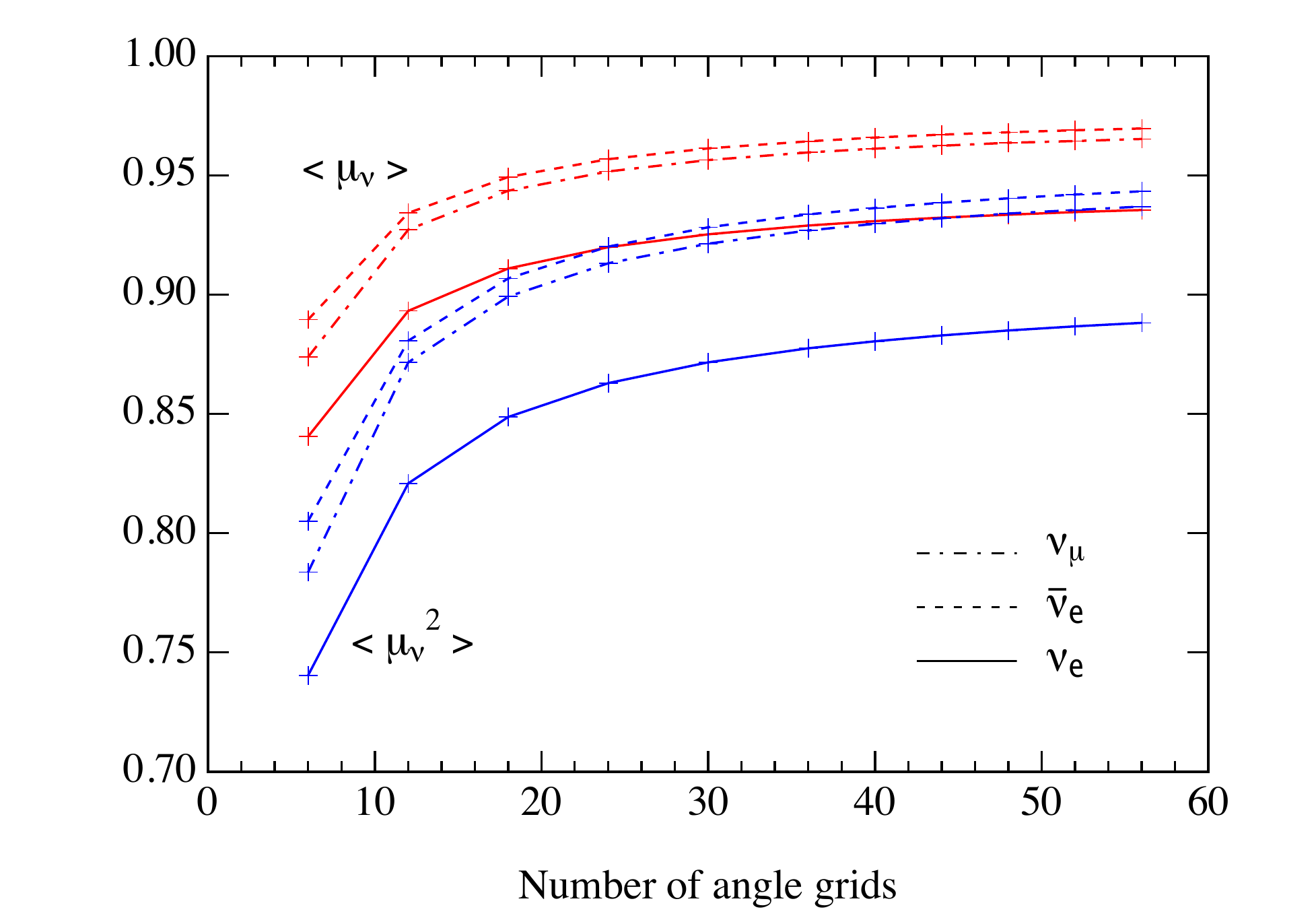}
\plotone{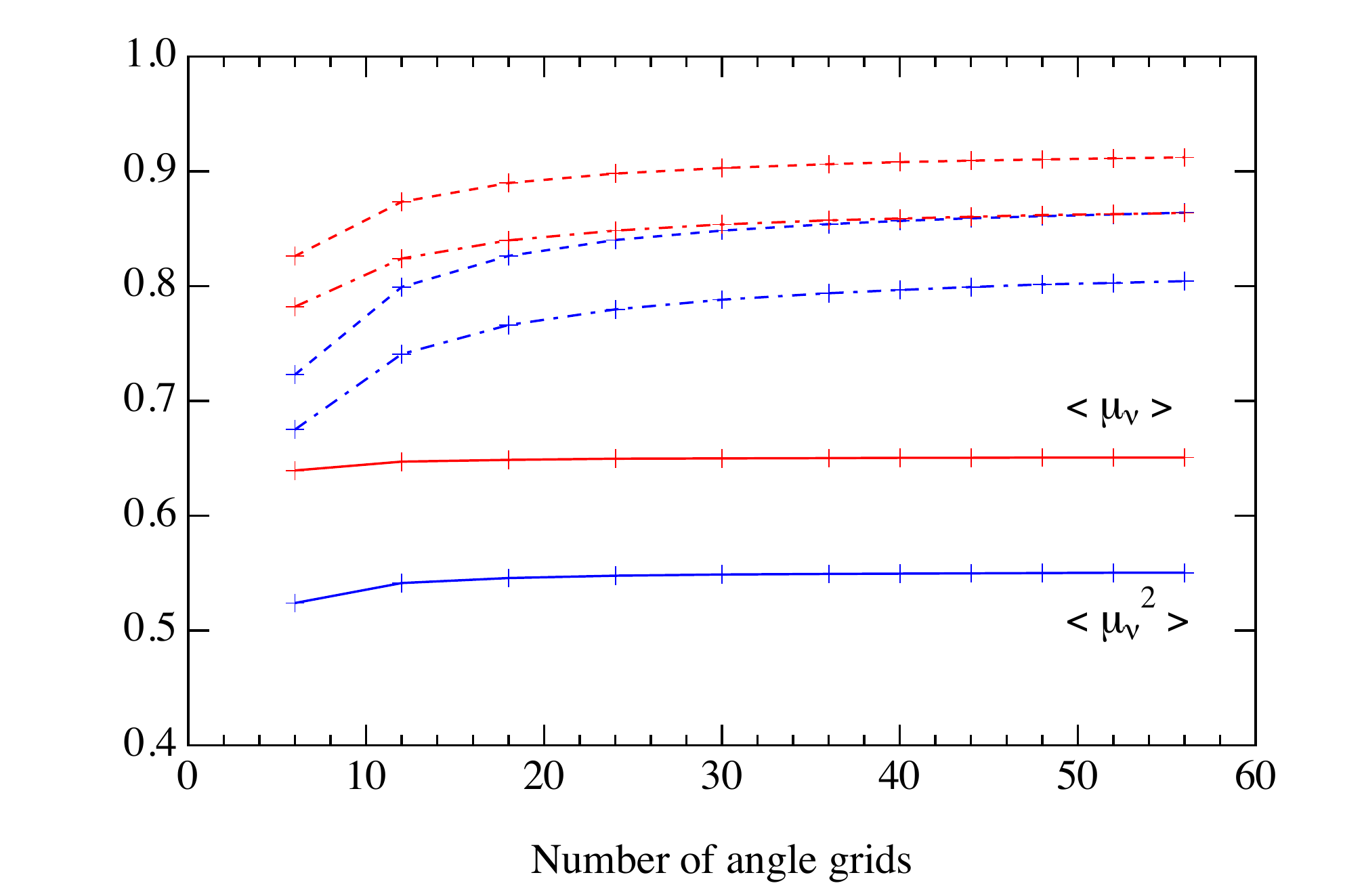}
\caption{The average values of angle moments $\langle\mu_{\nu}\rangle$ (red) and $\langle\mu_{\nu}^2\rangle$ (blue) at the outermost radial position are shown 
for $\nu_e$ (solid line), $\bar{\nu}_e$ (dashed line) and $\nu_\mu$ (dash-dotted line) as functions of the number of angle grid.  The values in the polar angles at 2.0 and 69 degree from the $z$-axis are shown in left and right panels, respectively.  
\label{fig:sek_resolution_mom}}
\end{figure}

We show in Fig. \ref{fig:sek_resolution_mom_radial} radial distributions of angle moments , $\langle\mu_{\nu}\rangle$ and $\langle\mu_{\nu}^2\rangle$ at two polar angles.  The angle moments increases smoothly along the $z$-axis (left) in the situation from the neutron star to the dilute material, while they behave in a non-monotonic way along the edge of geometrically thick torus.  It is apparent to see the convergence of the curve of angle moments by increasing the number of angle grids.  In order to show the variations of the angle moments in detail, we plot the values at the outermost radial position as a function of the number of grid in Fig. \ref{fig:sek_resolution_mom}.  The angle moments rapidly change for the small number of grids, but adequately converge for the cases over 40 grids.  We have checked the energy moments converges much faster than the angle moments.  

%

\begin{figure}[ht!]
\epsscale{0.5}
\plotone{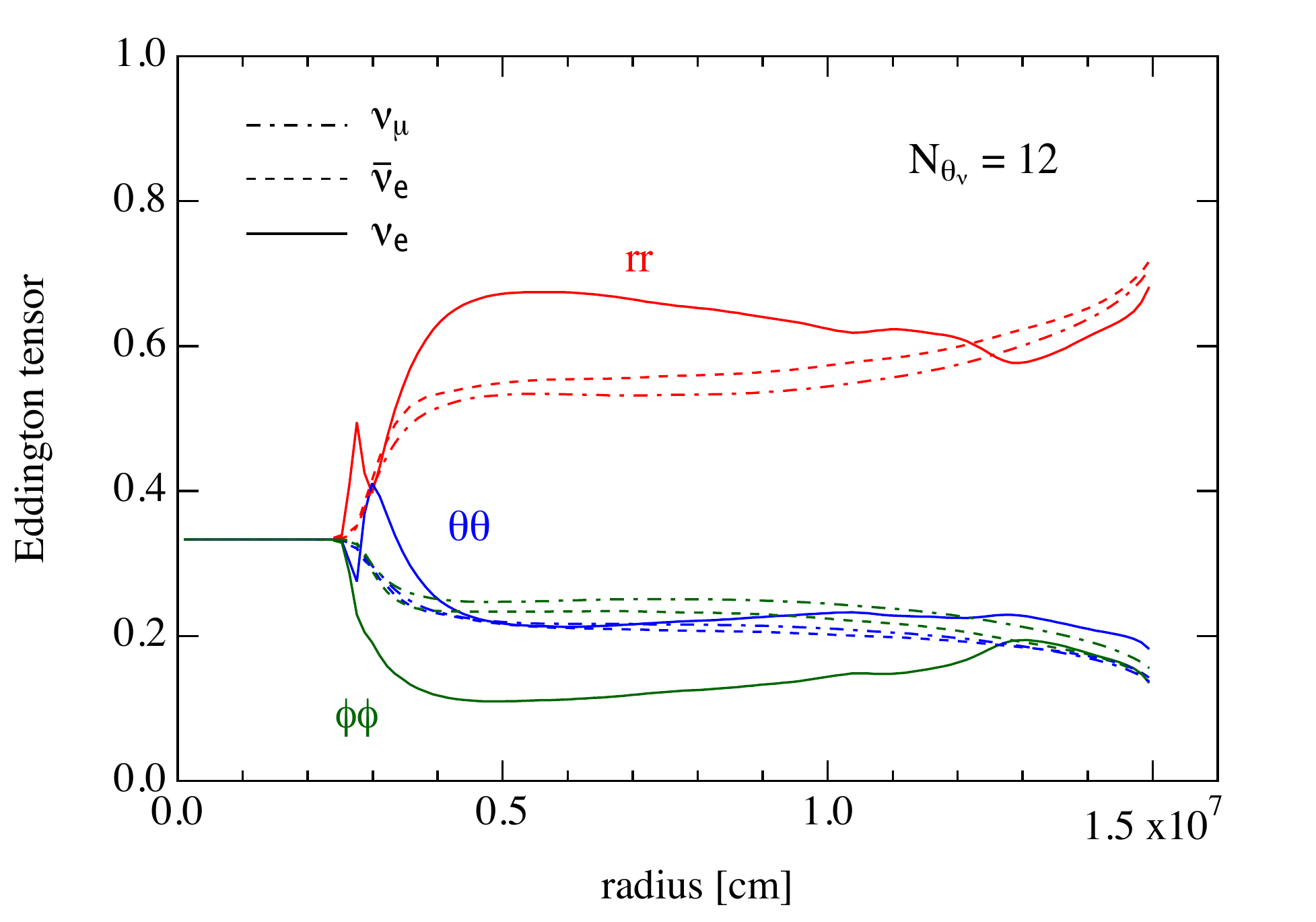}
\plotone{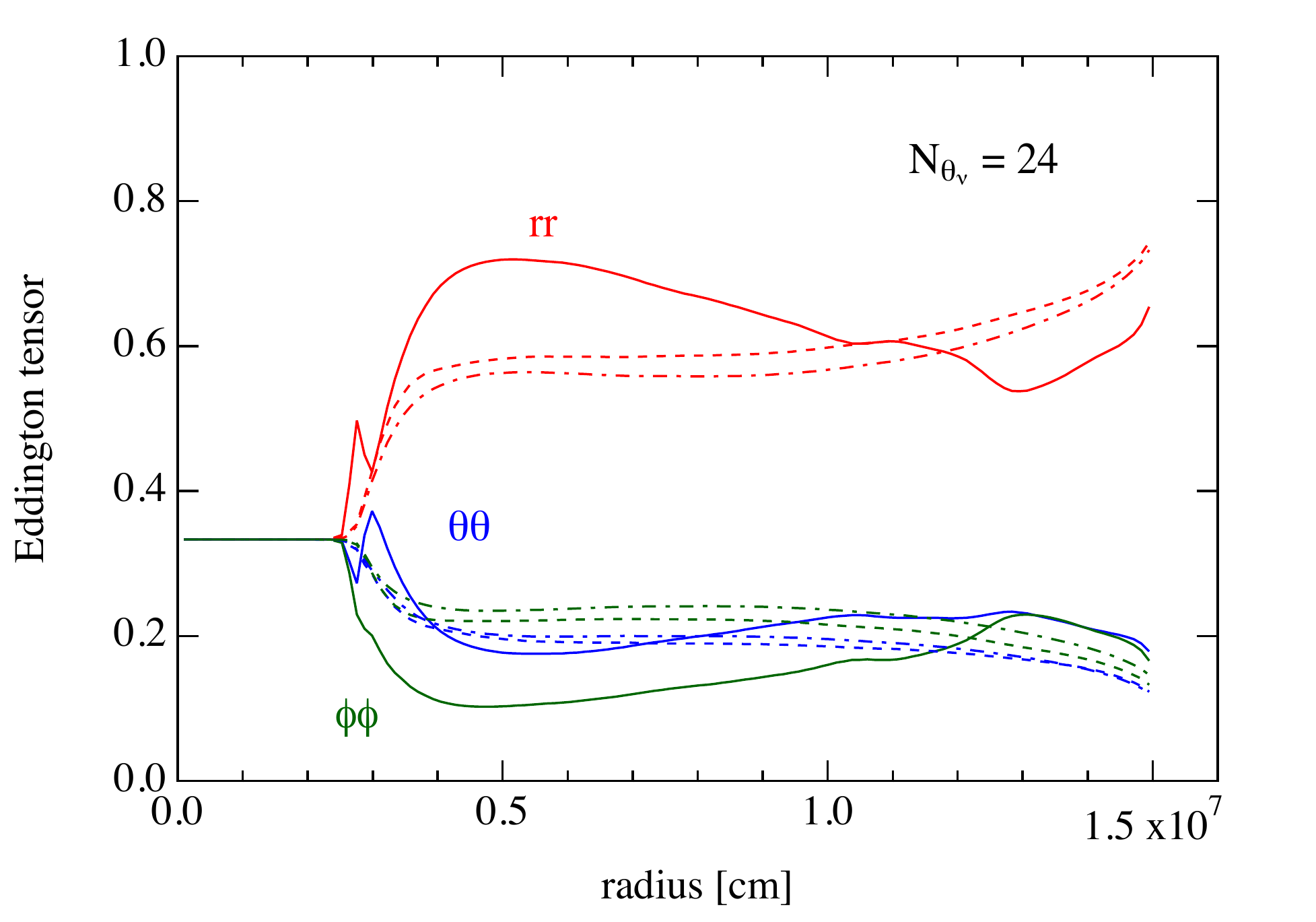}
\plotone{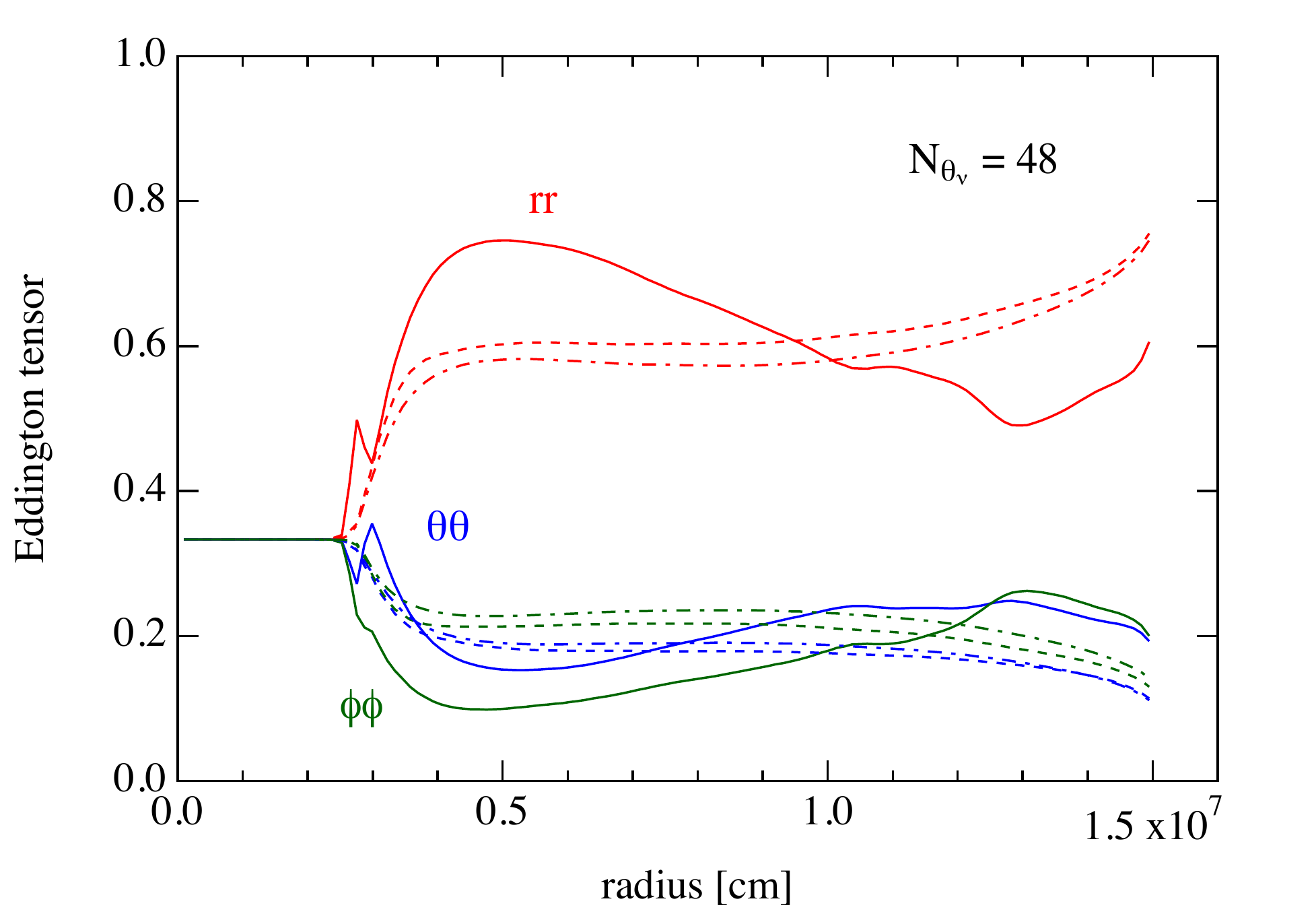}
\plotone{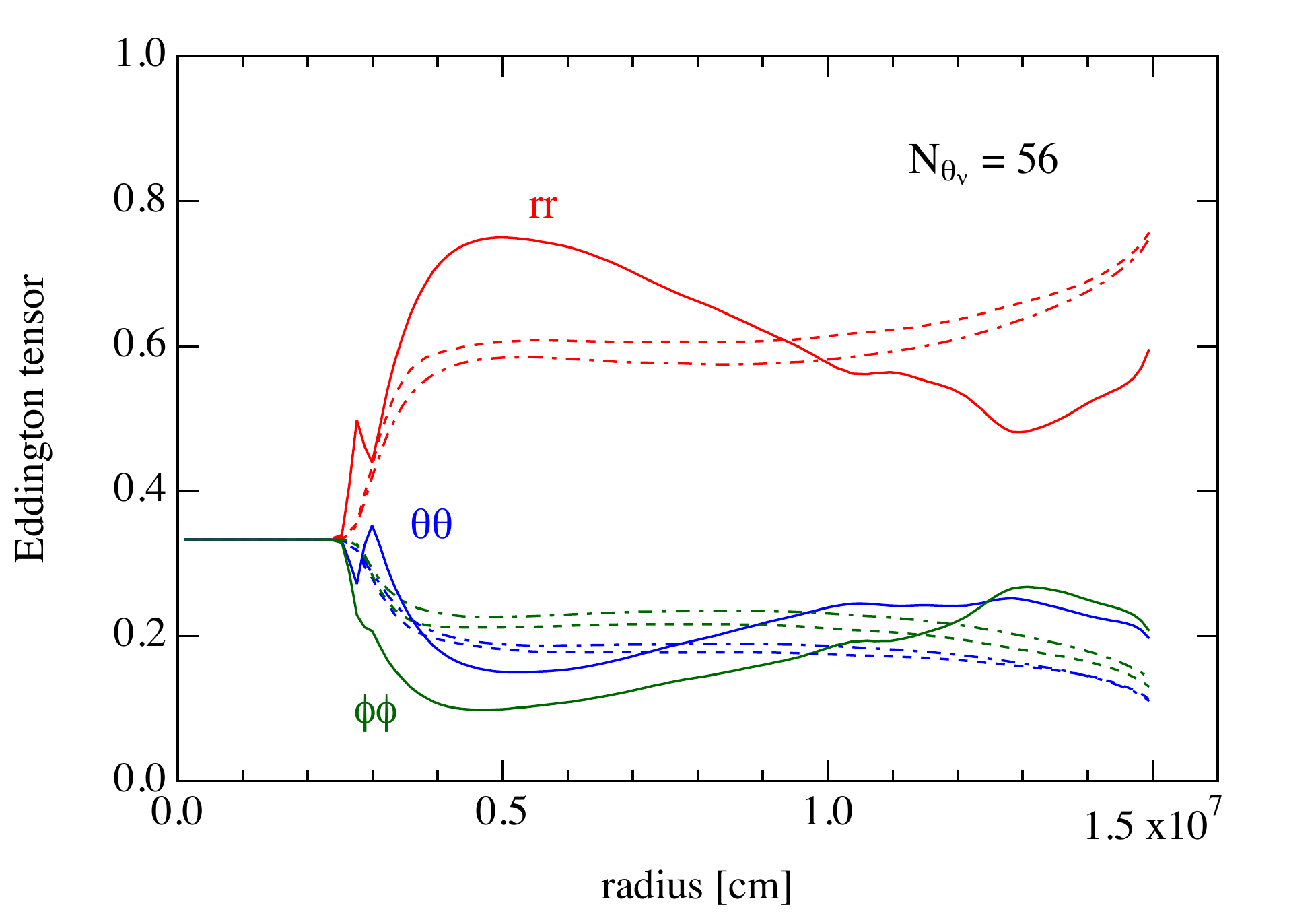}
\caption{Components of the Eddington tensor for $\nu_e$ (solid line), $\bar{\nu}_e$ (dashed line) and $\nu_\mu$ (dash-dotted line) obtained by the direct evaluation for neutrino energy of 34 MeV are shown as functions of radius for different setting of the number of angle grids.  The diagonal $rr$-, $\theta \theta$-, $\phi \phi$-components are shown in red, blue and green lines, respectively.  The distributions along the radial directions at 69 degree from the $z$-axis are shown for the case with the number of angle grid of 12 (top left), 24 (top right), 48 (bottom left) and 56 (bottom right).  
\label{fig:sek_resolution_mom_diag}}
\end{figure}

\begin{figure}[ht!]
\epsscale{0.5}
\plotone{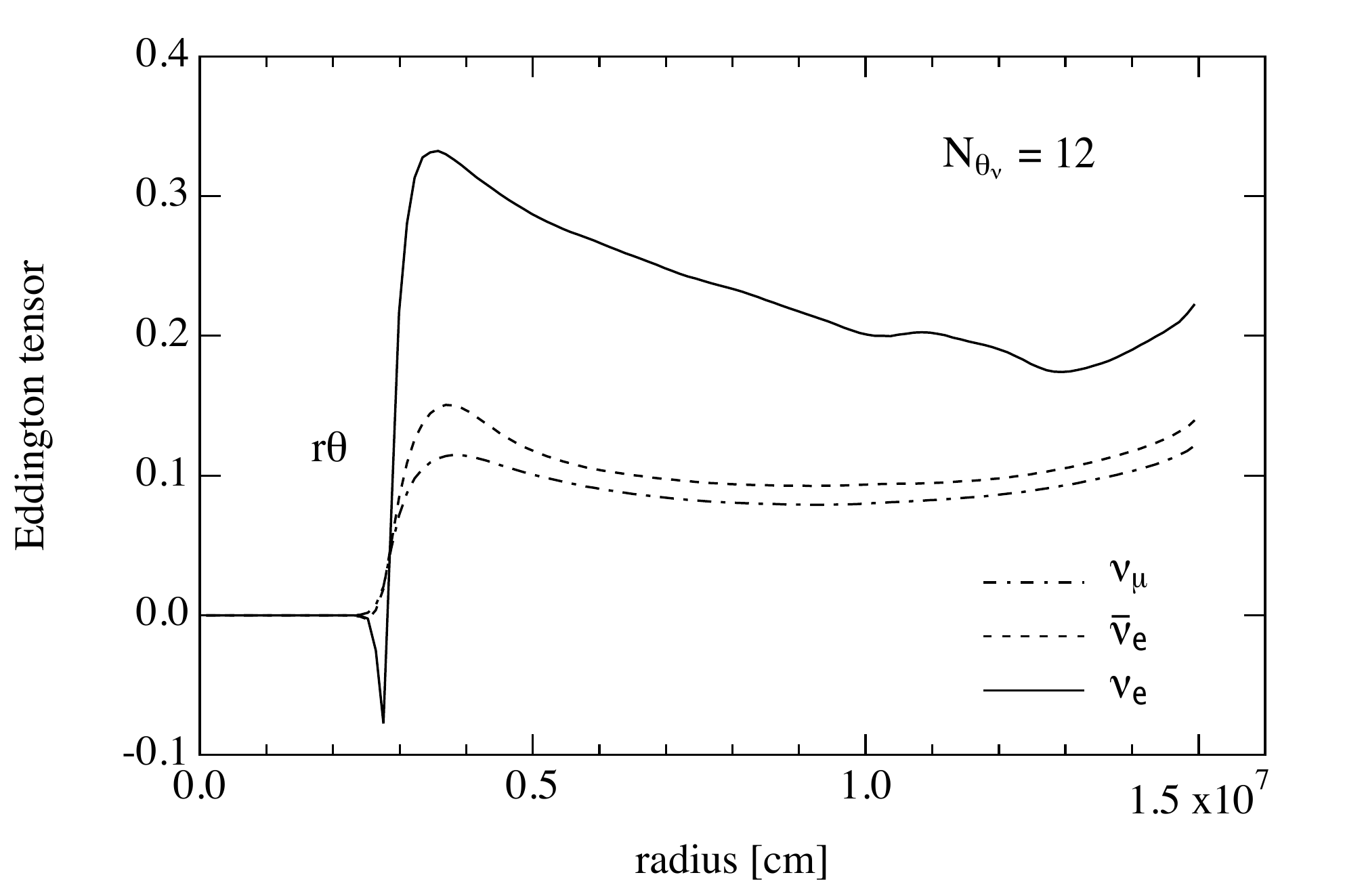}
\plotone{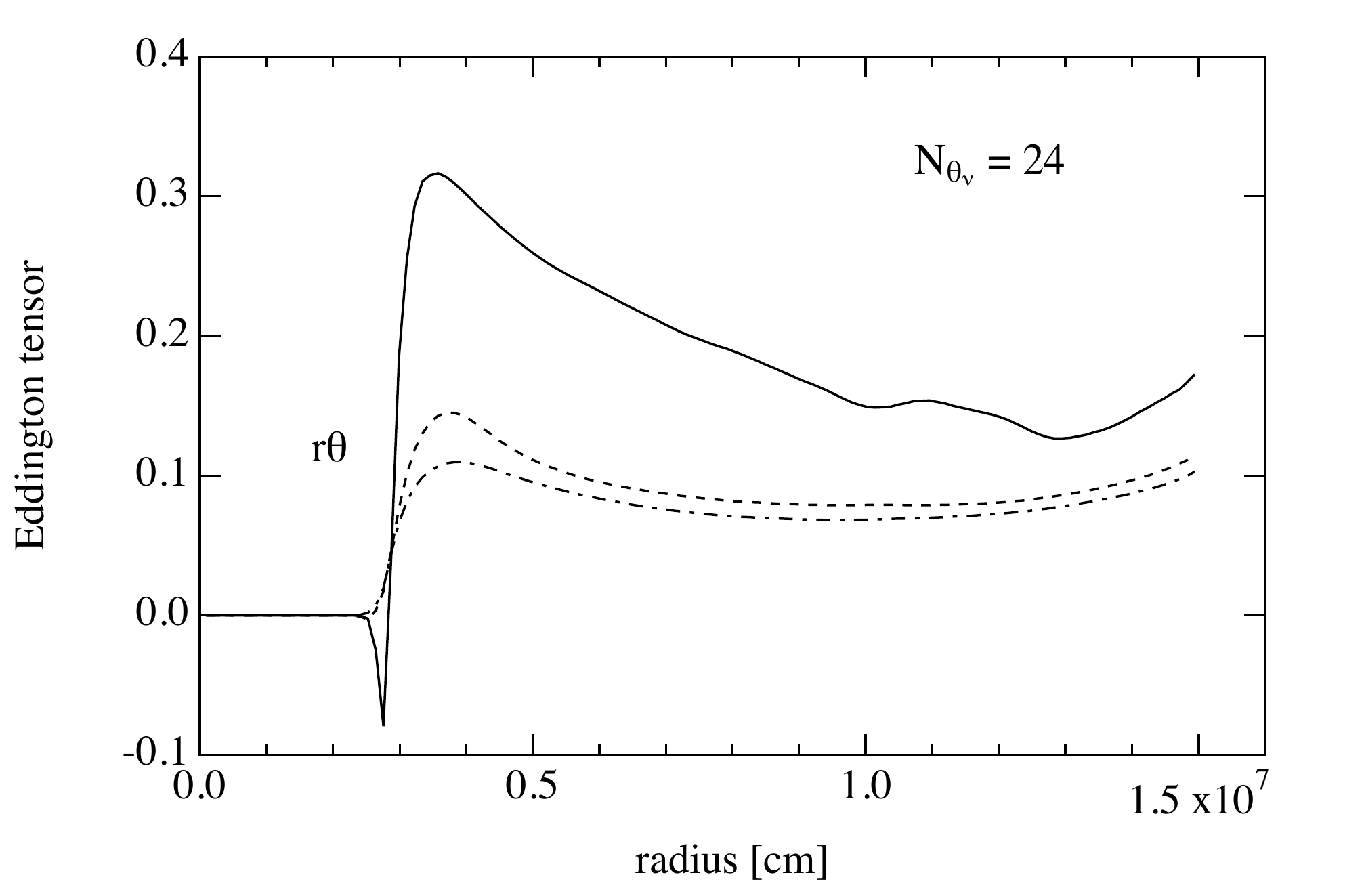}
\plotone{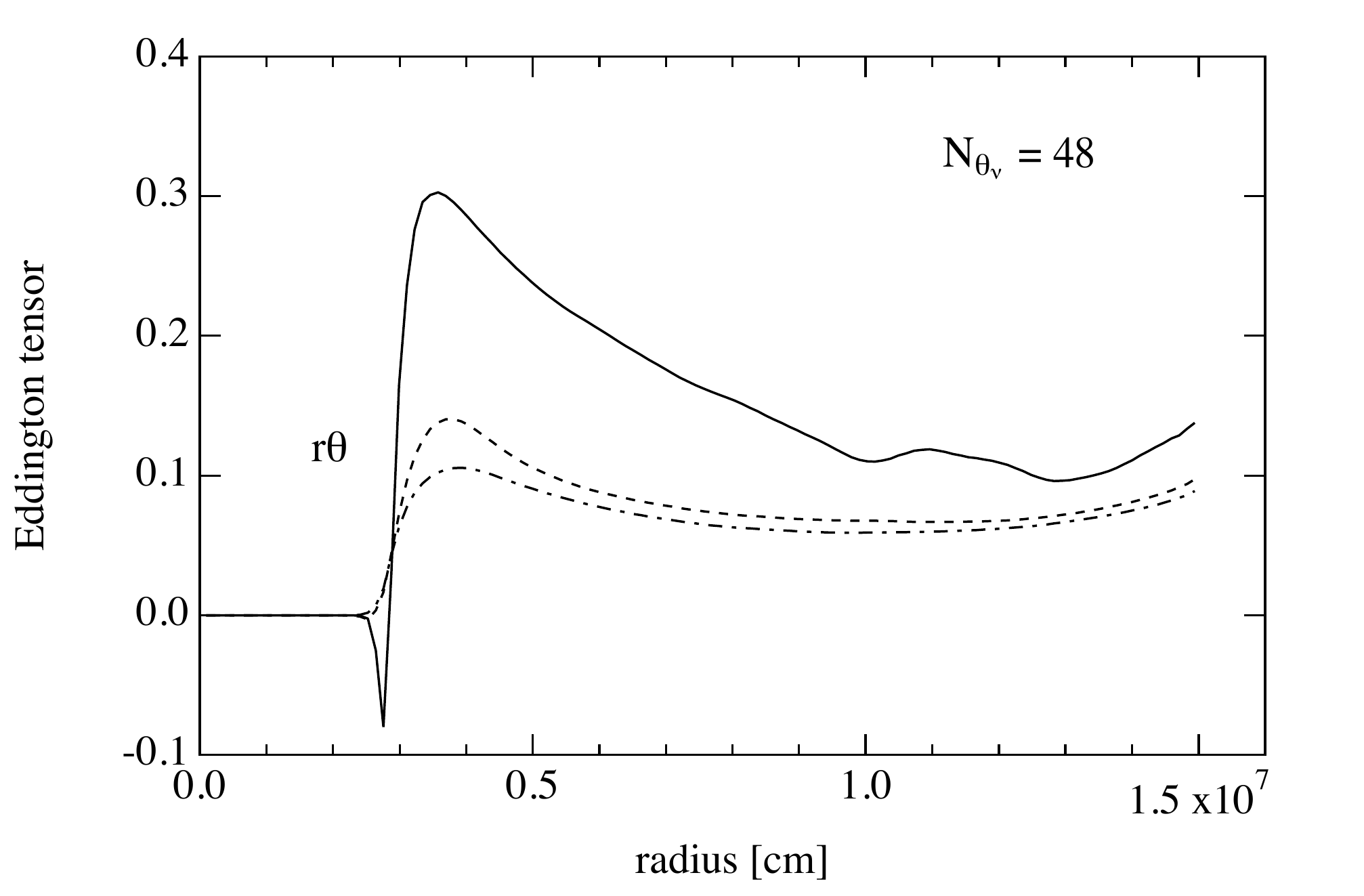}
\plotone{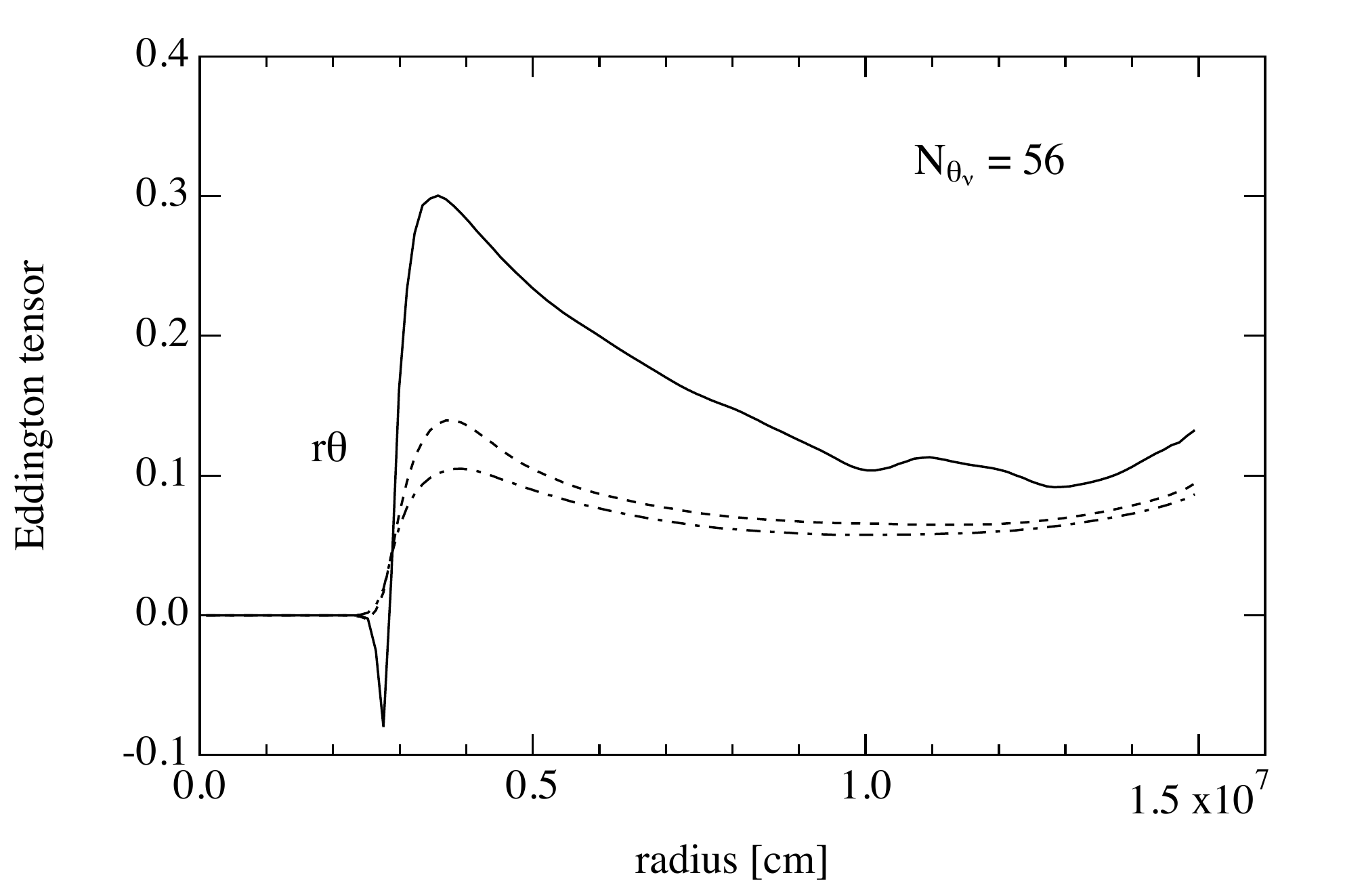}
\caption{Components of the Eddington tensor for $\nu_e$ (solid line), $\bar{\nu}_e$ (dashed line) and $\nu_\mu$ (dash-dotted line) obtained by the direct evaluation for neutrino energy of 34 MeV are shown as functions of radius for different setting of the number of angle grids.  The non-diagonal $r \theta$-component, is shown by black lines.  The non-diagonal $r \phi$-, $\theta \phi$-components are zero (not shown).   The distributions along the radial directions at 69 degree from the $z$-axis are shown for the case with the number of angle grids of 12 (top left), 24 (top right), 48 (bottom left) and 56 (bottom right).  
\label{fig:sek_resolution_mom_offd}}
\end{figure}

We show in Figs. \ref{fig:sek_resolution_mom_diag} and \ref{fig:sek_resolution_mom_offd} the radial distributions of the components of the Eddington tensor for three species along the radial directions at 69 degree from the $z$-axis (the edge of geometrically thick torus) for different settings of the number of angle grids.  The components of the Eddington tensor behave in a non-monotonic manner for $\nu_e$ and in a monotonic manner for $\bar{\nu}_e$ and $\nu_\mu$.  The overall feature remains similar for $\bar{\nu}_e$ and $\nu_\mu$ and is sensitive to the number of grids for $\nu_e$.  It converges well for the cases with 48 and 56 angle grids even in that case.  Relative differences of the values with 48 angle grids with respect to those with 56 angle grids are within 3\% and 7\% for diagonal and non-diagonal components, respectively.  Hence, we believe that the number of angle grids in the current study for neutron star merger is sufficient to examine the quantities of the Eddington tensor in detail.  

\begin{figure}[ht!]
\epsscale{0.35}
\plotone{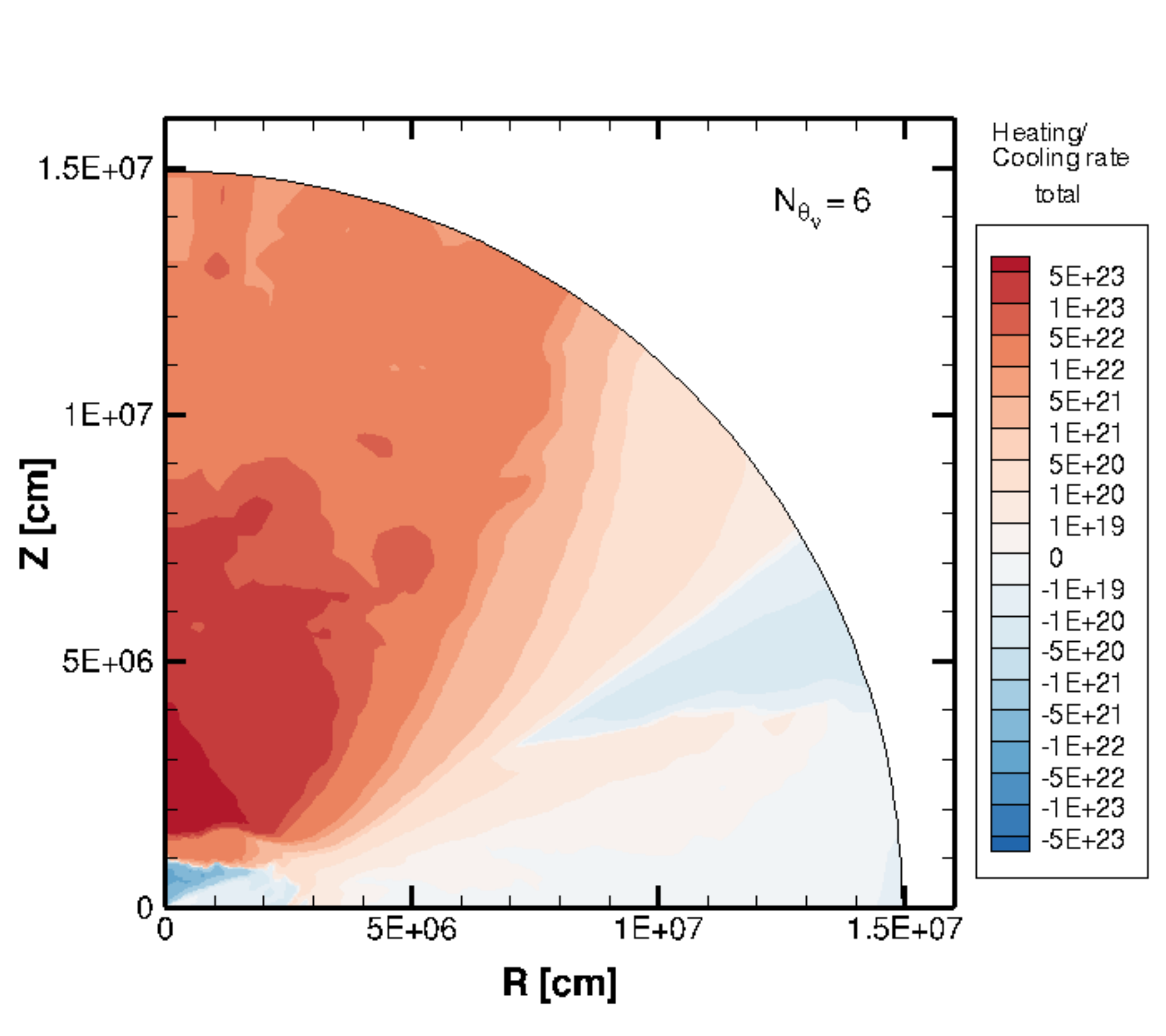}
\plotone{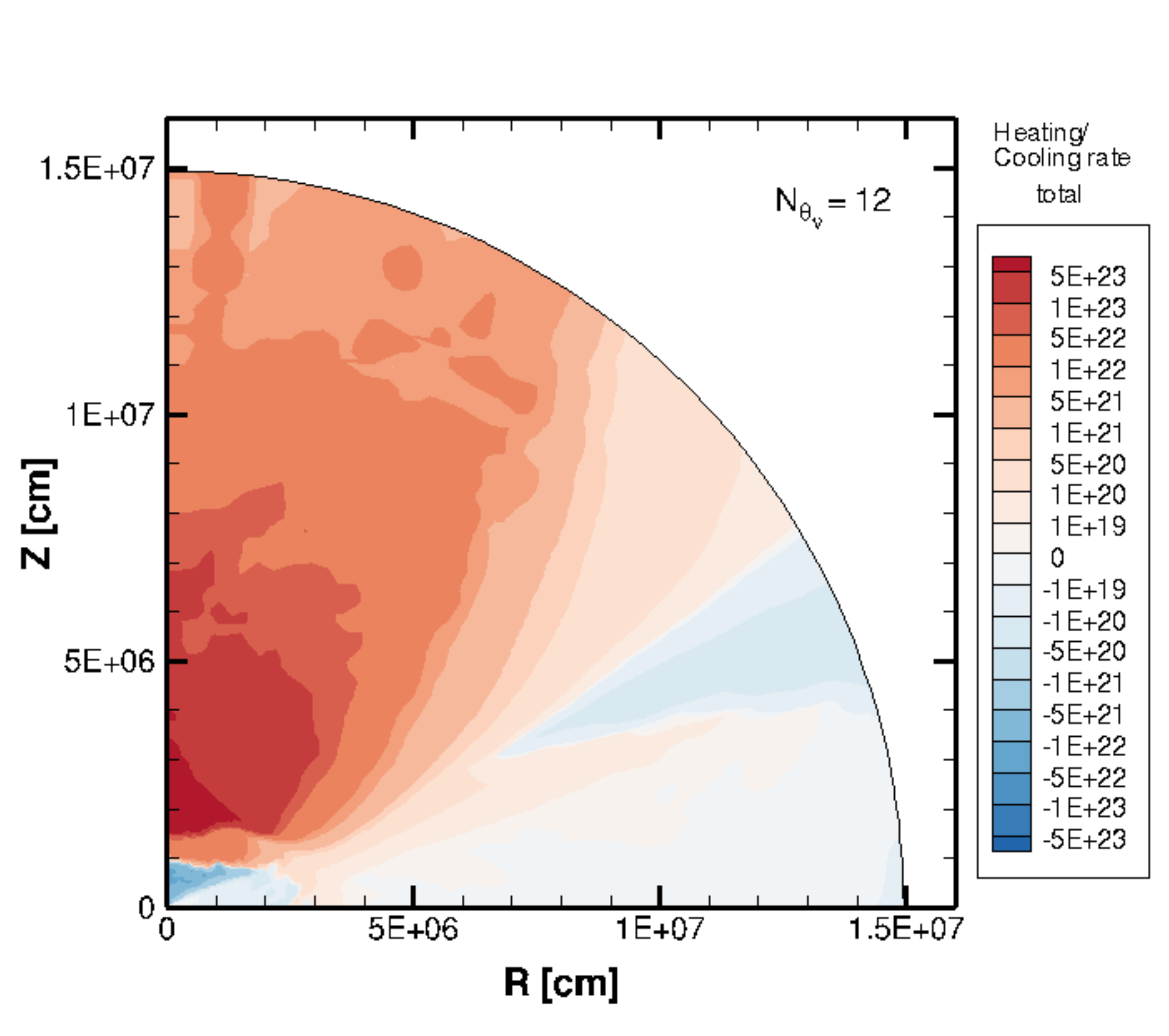}
\plotone{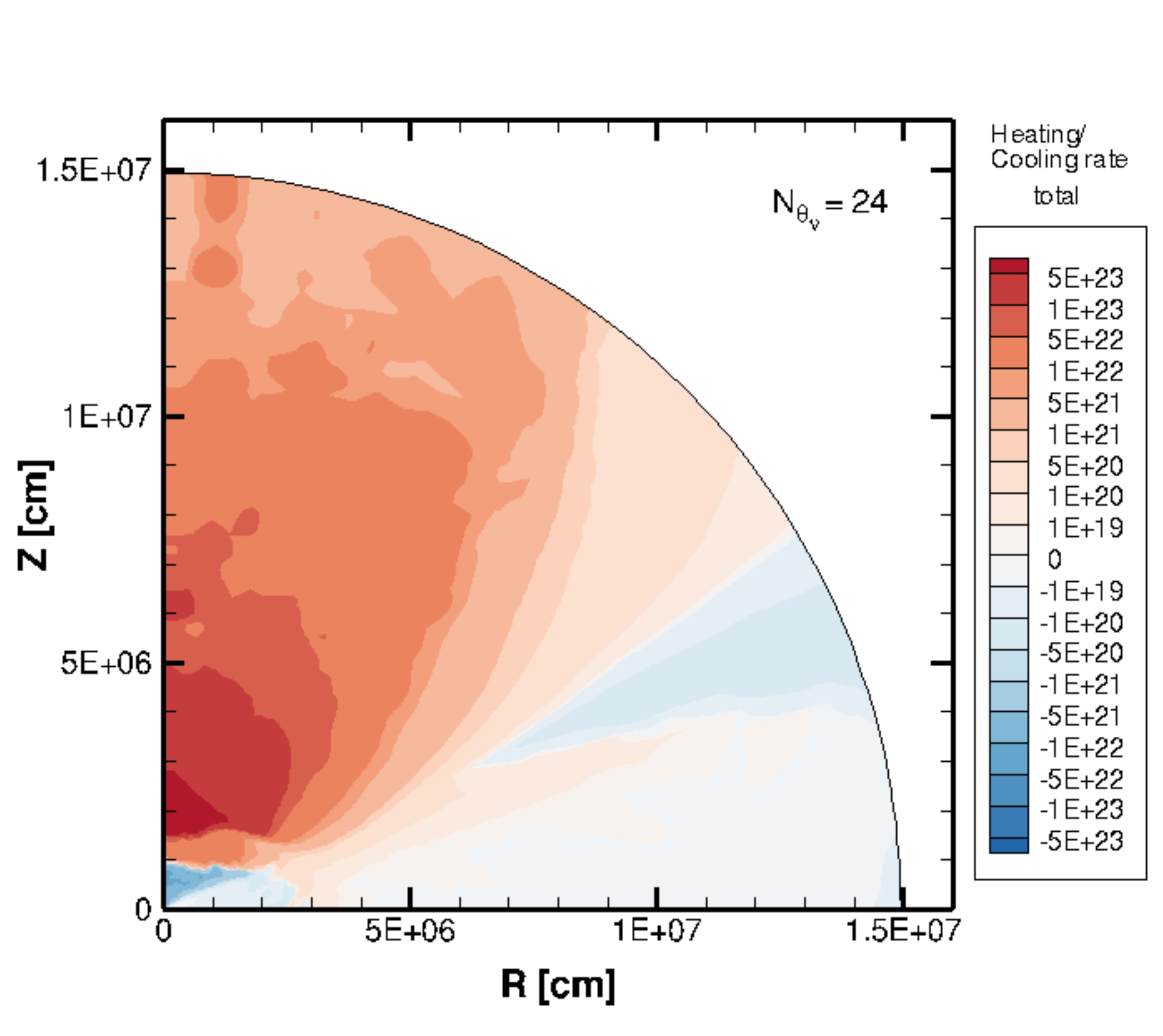}
\plotone{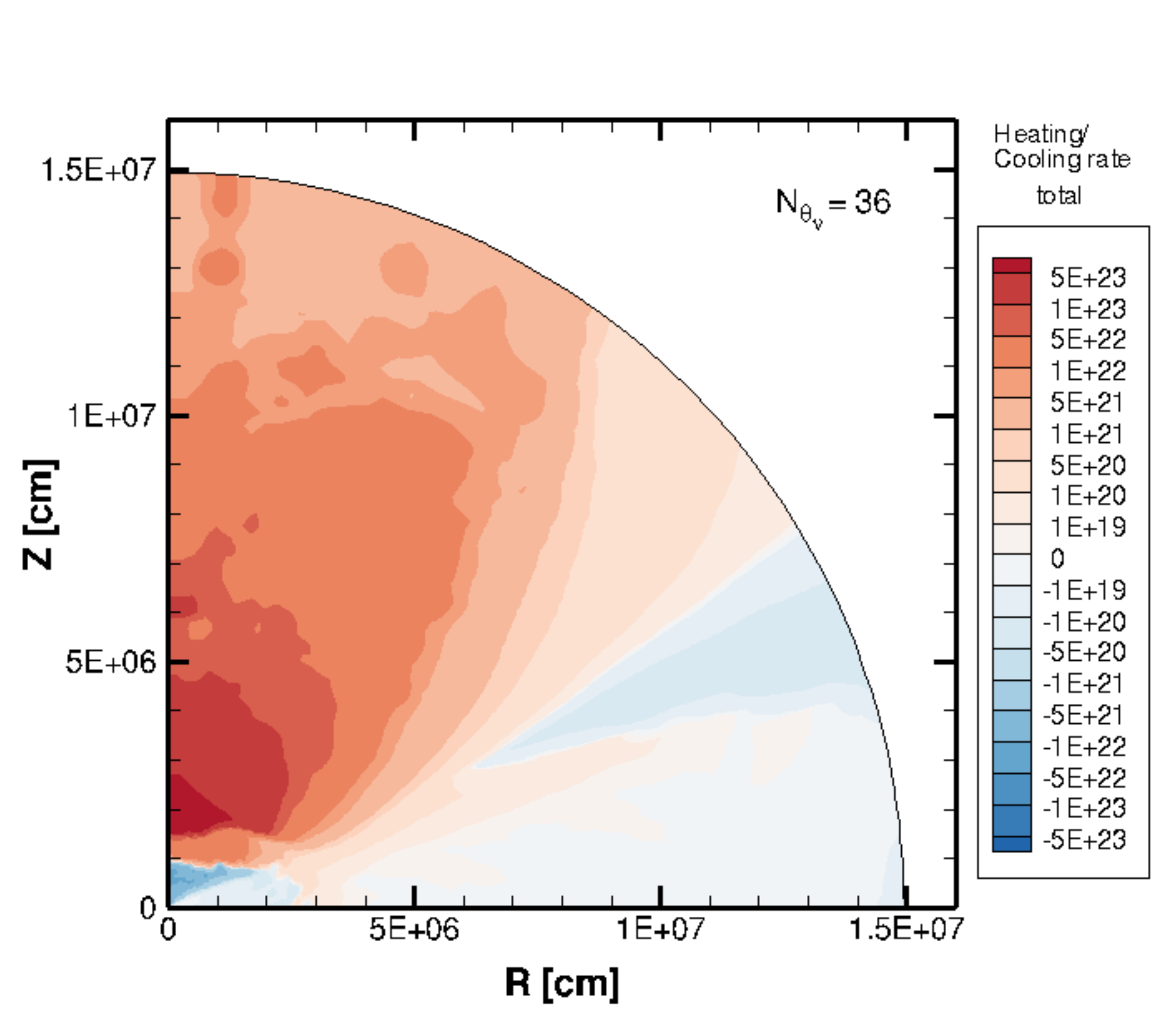}
\plotone{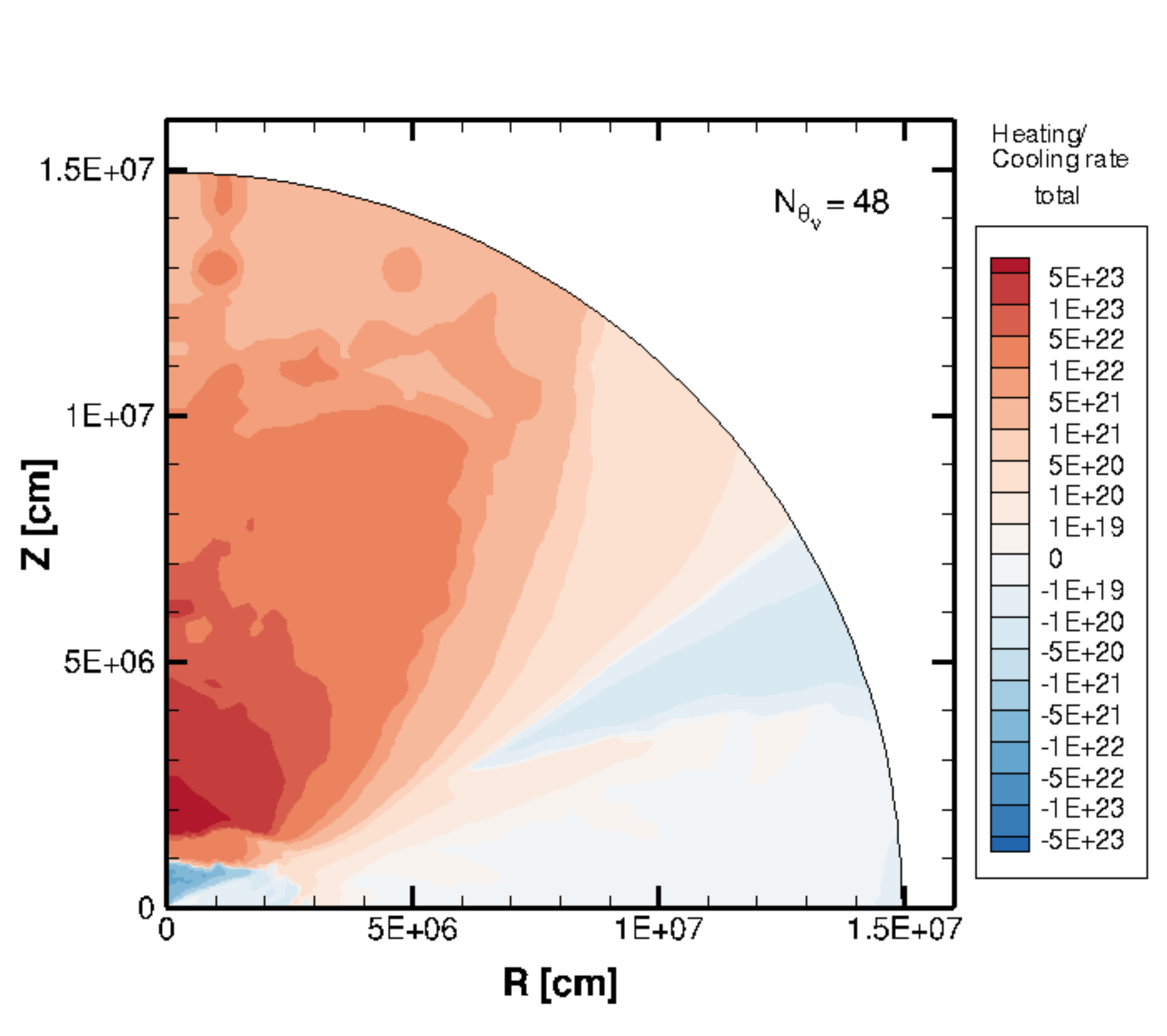}
\plotone{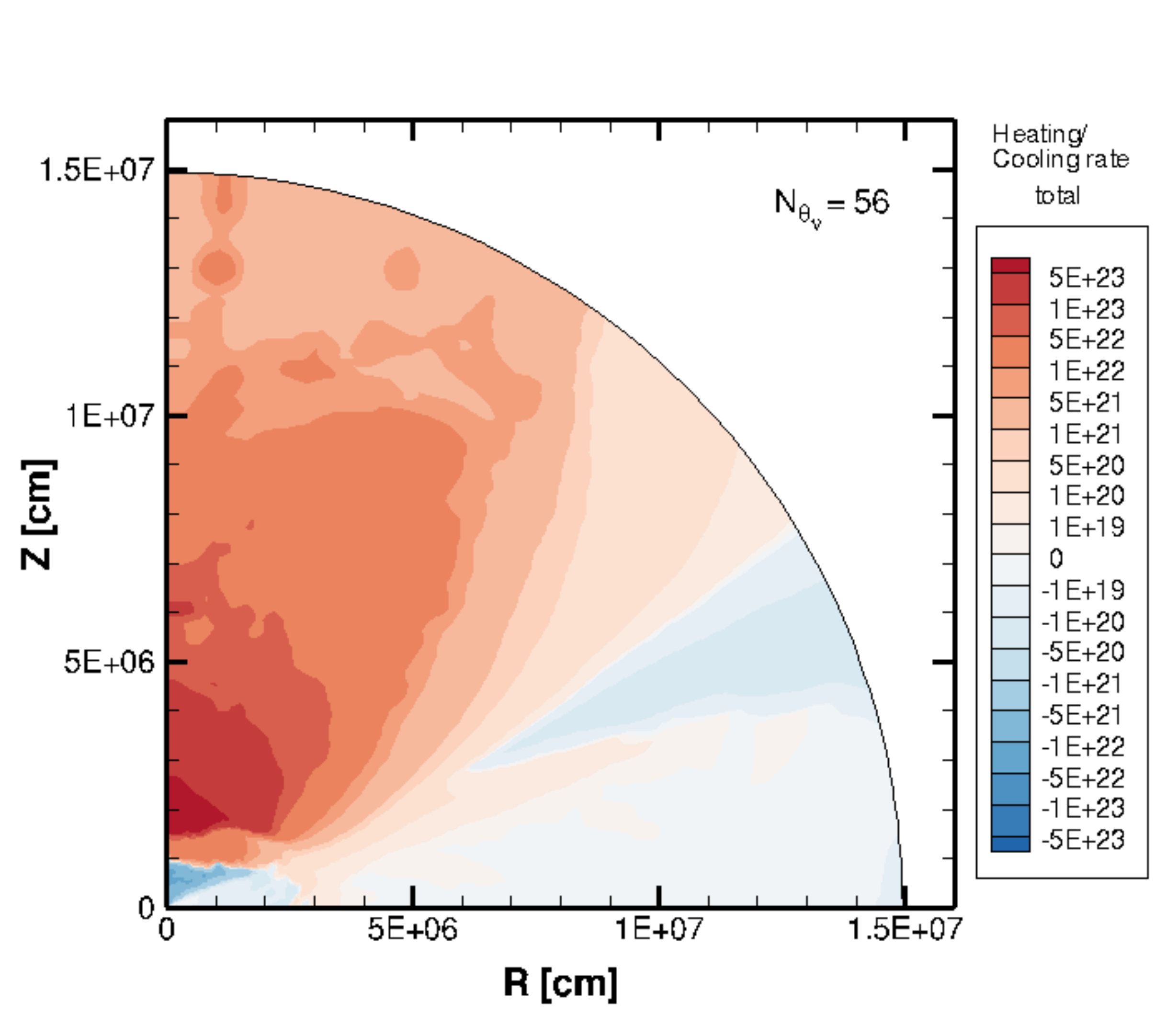}
\caption{Specific heating and cooling rates [erg g$^{-1}$ s$^{-1}$] are shown by contour plots in reddish (heating) and bluish (cooling) colors for different settings of the number of angle grids.  The panels show the results in 
the cases with the number of angle grids of 6 (top left), 12 (top middle), 24 (top right), 36 (bottom left), 48 (bottom middle) and 56 (bottom right).  
\label{fig:sek_resolution_heating}}
\end{figure}

The heating rate is important for the hydrodynamics such as mass ejection, which may contribute to the nucleosynthesis as well as jet formation.  The amount of heating through the pair annhilation of neutrinos and anti-neutrinos, which is a major contribution in this situation, is sensitive to the neutrino fluxes with detailed angular distributions.  Therefore, the angular resolution is crucial to the reliable evaluation.  We show in Fig. \ref{fig:sek_resolution_heating} the contour plot of heating rates for different settings of the number of angle grids.  The heating is dominant in the region just above the neutron star along the $z$-axis in a similar situation to the case of neutron star merger.  It is noticeable to see that stronger heating proceeds in the region above the neutron star for the case of the small number of angle grids.  The heating distribution converges for the large number of angle grids in the lower panels.  To see the convergence in a quantitative manner, we show the total heating rate for different settings of the number of angle grids.  The total heating rate is larger for lower angular resolution and decreases for higher angular resolution.  It nearly converges for the finest angular resolution in the current study at 56.  Relative difference of the value with 48 angle grids with respect to that with 56 angle grids is 1.3\%.  This is related with the sensitivity of pair-annihilation rate to the angle distribution.  It is, therefore, advisable to provide enough angular resolution to determine the heating rate in a reliable manner for the situation of a neutron star with extended torus such as the collapsar and neutron star merger.  


\begin{figure}[ht!]
\epsscale{0.5}
\plotone{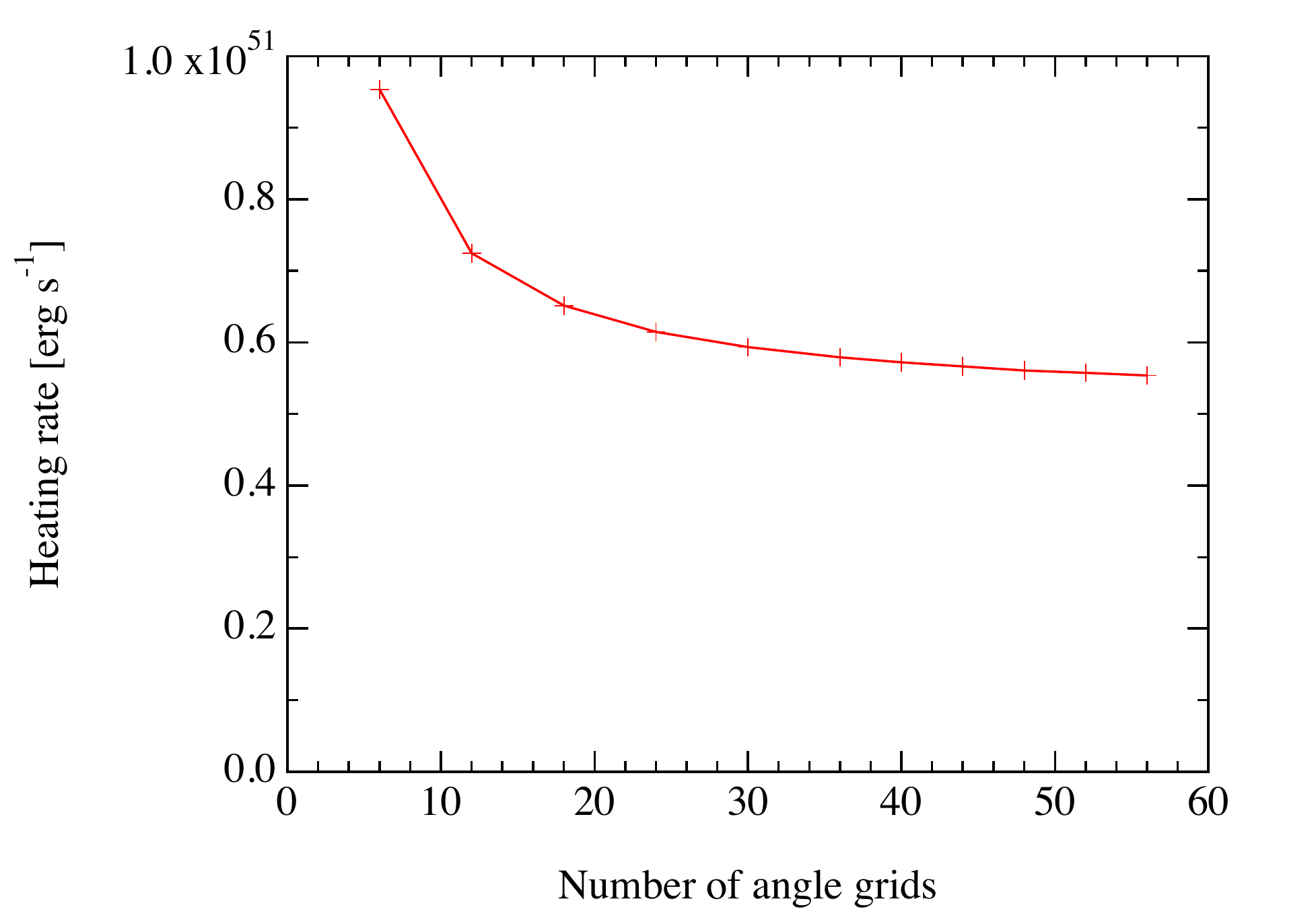}
\caption{Total heating rates [erg s$^{-1}$] for the remnant in the collapsar is shown as a function of the number of angle grids.  The total heating rate is evaluated by the volume integral for the $4\pi$-coverage.  
\label{fig:sek_resolution_heating_total}}
\end{figure}


Slow convergence of heating rate is caused by the extended geometry of material 
as seen in this study.  This is different from ordinary situations in core-collapse supernovae where the moderate resolution of angular resolution is tolerated to determine the heating rate to some extent.  Here we additionally demonstrate the dependence of the total heating rate for supernova core on the angular resolution for comparison.  Note that the high angular resolution is necessary to obtain the forward peak at large distances for all applications in principle.  The detailed study of angular resolution of the neutrino transfer by the Boltzmann equation is reported in \citet{sher17,iwa20}.  

We adopt a profile of supernova core at 150 ms after the core bounce taken from the core-collapse simulations in 2D \citep{tak12,tak14,hor14} of a 11.2M$_\odot$ star \citep{woo02}.  The case of 11.2M$_\odot$ star, which leads to explosions in the 2D and 3D simulations by \citet{tak12}, is an example of the configuration of propagating shock wave with a largely elongated shape.  We obtained the stationary neutrino distributions by solving the Boltzmann equation in the same procedure as in \citet{sum15} and the one above.  The EOS by \citet{lat91} with the incompressiblility of 180 MeV is used to match with the original simulations.  We adopt the number of polar angle grids of neutrinos from 6, 12, 18, 24, 30 to 36 with the same settings of 256 for radial and 64 for polar grids in space and of 14 for energy and 12 for azimuthal angle of neutrinos.  

The basic properties of neutrino transfer in the 3D profiles of supernova cores for the case of 11.2M$_\odot$ and 27M$_\odot$ are reported in \citet{sum15}.  The properties of neutrino distributions in the 2D supernova core is investigated in \citet{abbar19,abbar20} for the applications to collective neutrino oscillations.  



\begin{figure}[ht!]
\epsscale{0.5}
\plotone{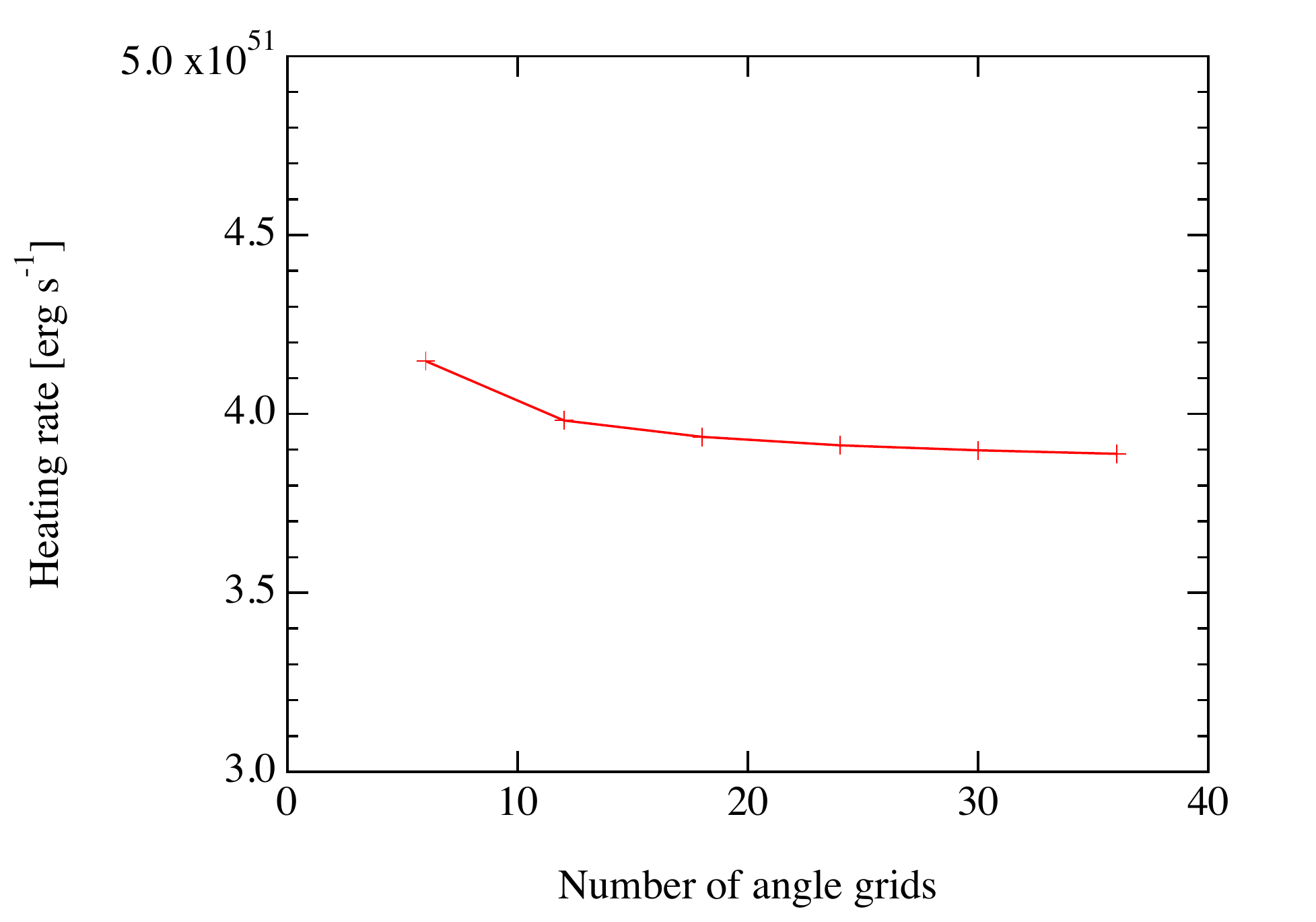}
\caption{Total heating rates [erg s$^{-1}$] for the supernova core is shown as a function of the number of angle grids.  The total heating rate is evaluated by the volume integral for the $4\pi$-coverage.  
\label{fig:tak_resolution_heating}}
\end{figure}

We show the total heating rates in Fig. \ref{fig:tak_resolution_heating} for different settings of the number of angle grids.  It is evident to see that the total heating rate does not change very much with increasing the number of angle grids.  Although there is a slight overestimation with the case of 6 angle grids, it remains flat for larger values.  This is clearly different from the situation in collapsar seen in Fig. \ref{fig:sek_resolution_heating_total}.  Hence, it is not crucial to increase the number of angle grids for the total heating in core-collapse supernovae.  

\end{document}